%% file: main.tex
\documentclass[aps,prx,longbibliography,twocolumn,floatfix,superscriptaddress,nofootinbib]{revtex4-2}
\usepackage{graphicx}
\usepackage{booktabs}
\usepackage{dcolumn}
\usepackage{bm}
\usepackage{amsmath}
\usepackage{amssymb}
\usepackage{mathtools}
\usepackage{braket}
\usepackage{xcolor}
\usepackage[normalem]{ulem}
\usepackage[version=4]{mhchem}
\usepackage[hidelinks]{hyperref}
\hypersetup{colorlinks,citecolor=blue,linkcolor=blue,urlcolor=blue}

\usepackage{parskip}

\newcommand{\Nood}{N_{\mathrm{OOD}}}          %
\newcommand{\cpd}[1]{\textbf{#1}}              %
\newcommand{\VClsix}{\ce{[VCl6]^3-}}
\newcommand{\VFsix}{\ce{[VF6]^3-}}
\newcommand{\CoMe}{\ce{[Co(CH3)4]-}}
\newcommand{\NiAllyl}{\ce{[Ni($\eta^3${-}C4H7)(NCCH3)2]+}}
\newcommand{\CrMorph}{\ce{[Cr(CO)5(HNC4H8O)]}}
\newcommand{\CrGly}{\ce{[Cr(CO)4(H2NCH2CO2)]-}}
\newcommand{\TiArene}{\ce{[Ti($\eta^6${-}C6H6)(AlCl4)2]}}
\newcommand{\CoCOtwo}{\ce{[Co(H2O)5(CO2)]^{2+/3+}}}
\newcommand{\CoCOthree}{\ce{[Co(H2O)5(CO2)]^{3+}}}
\newcommand{\boxedgraphics}[2]{{\setlength{\fboxsep}{4pt}\setlength{\fboxrule}{0.5pt}%
  \fbox{\includegraphics[#1]{#2}}}} %
\newcommand{\kcal}{\,kcal\,mol$^{-1}$}

\newcommand{\tdkdel}[1]{{\color{red}\sout{#1}}\hspace{0.15em}}
\definecolor{wernergreen}{rgb}{0,0.5,0}
\definecolor{sabreorange}{rgb}{0.85,0.4,0}

\newcommand{\wernerdel}[1]{{\color{wernergreen}\sout{#1}}\hspace{0.15em}}

\newcommand{\sabredel}[1]{{\color{sabreorange}\sout{#1}}\hspace{0.15em}}
\begin{document}

\preprint{Draft --- \today}

\title{Machine-learned active spaces enable quantum  calculations of spin-state energetics of transition-metal complexes on quantum hardware}

\author{Muhammad Arsalan Ali Akbar}
\affiliation{Department of Electrical and Computer Engineering, North Carolina State University, Raleigh, NC 27606, USA}
\affiliation{Department of Chemistry, North Carolina State University, Raleigh, NC 27695, USA}

\author{Thomas D. K\"uhne}
\affiliation{Center for Advanced Systems Understanding (CASUS), 02826 Görlitz, Germany}
\affiliation{Helmholtz-Zentrum Dresden-Rossendorf (HZDR), 01328 Dresden, Germany}
\affiliation{Institute of Artificial Intelligence, Technische Universit\"at Dresden, Helmholtzstra{\ss}e 10, 01069 Dresden, Germany\looseness=-3}

\author{Sabre Kais}
\thanks{These authors contributed equally and jointly supervised this work.}
\email{skais@ncsu.edu}
\affiliation{Department of Electrical and Computer Engineering, North Carolina State University, Raleigh, NC 27606, USA}
\affiliation{Department of Chemistry, North Carolina State University, Raleigh, NC 27695, USA}
\affiliation{Department of Physics, North Carolina State University, Raleigh, NC 27695, USA}

\author{Werner Dobrautz}
\thanks{These authors contributed equally and jointly supervised this work.}
\email{w.dobrautz@hzdr.de}
\affiliation{Center for Advanced Systems Understanding (CASUS), 02826 Görlitz, Germany}
\affiliation{Helmholtz-Zentrum Dresden-Rossendorf (HZDR), 01328 Dresden, Germany}
\affiliation{Center for Scalable Data Analytics and Artificial Intelligence (ScaDS.AI) Dresden/Leipzig, 01069 Dresden, Germany}
\affiliation{Technical University Dresden, 01069 Dresden, Germany}

\begin{abstract}

Multireference calculations on transition-metal complexes, on classical or
quantum hardware, depend on the choice of active space. The orbitals that
belong in it are those that are strongly entangled in the correlated
wavefunction, but measuring this entanglement requires the very correlated
calculation that the active space is meant to make affordable. Here we
present \texttt{SpinAware}, a machine-learning model that predicts
density-matrix renormalization group (DMRG) single-orbital entropies and
orbital-pair mutual information from the unrestricted Hartree--Fock (UHF)
solution of each spin state, without any correlated calculation. Each of the
20 highest occupied and 20 lowest unoccupied orbitals is described by 60
descriptors computed from the UHF solution, among them the differences between
its $\alpha$ and $\beta$ occupations and energies, its metal-$d$ and ligand
character, and its exchange coupling to the other orbitals, and a transformer
predicts the entropy of each orbital in the context of all others. Trained on
3\,454 spin-pure DMRG calculations on mononuclear Cr--Ni complexes, the model
transfers to experimentally characterized complexes with ligands, binding
topologies, oxidation states and metals absent from training. Over 11 spin states
it reproduces the DMRG entropy profiles with a mean absolute error of 0.027 and
combining the predicted entropies with the predicted mutual information gives
a candidate pool that contains 99\% of the DMRG-active orbitals of the
benchmark spin states while keeping, on average, only 15 of the 40 orbitals.
Natural-orbital occupation scans then select compact active spaces from this
pool for a nickel and a cobalt complex. Sample-based quantum diagonalization
with samples from IBM quantum processors reproduces the exact CASCI energies
of these spaces to within 0.3\kcal\ and their spin-state gaps to within
0.10\kcal\ for up to 32 qubits.

\end{abstract}

\maketitle

\section{Introduction}
\label{sec:intro}

Transition-metal complexes are at the heart of homogeneous catalysis,
metalloenzymes and molecular magnetism, and much of their chemistry is
decided by the near-degeneracy of their $d$ orbitals. Partially filled $d$
shells, metal--ligand covalency and several low-lying spin states give rise to
strong static correlation, which single-reference methods and approximate
density functionals describe unreliably, most visibly in spin-state
energetics~\cite{Harvey2004,Swart2013,Cirera2018,Radon2019}. Multiconfigurational methods such as complete-active-space
configuration interaction (CASCI) and self-consistent field
(CASSCF)~\cite{Roos1980,Szalay2012,LiManni2020chapter}, combined with a perturbative treatment of dynamic
correlation such as second-order $N$-electron valence state perturbation theory (NEVPT2)~\cite{Angeli2001}, address this problem, but their
accuracy and cost are both fixed by the choice of active space. Exact
diagonalization is routinely possible for about 20 orbitals, and the largest
nonrelativistic calculations, with about $10^{12}$ determinants, reach 23 to 24
orbitals~\cite{Vogiatzis2017,Gao2024}. High-accuracy approximate solvers such
as the density-matrix
renormalization group (DMRG)~\cite{White1992,Chan2011}, full configuration
interaction quantum Monte Carlo (FCIQMC)~\cite{Booth2009,Guther2020NECI,Dobrautz2019GUGA}
and selected configuration interaction~\cite{Holmes2016,Smith2017} extend the reach to
several tens of orbitals at a cost that still grows steeply with the size and
the entanglement of the space.

The choice of active space also determines the resources of a quantum
computation. Quantum
computers have been proposed as solvers for active-space
Hamiltonians~\cite{Reiher2017,McArdle2020,Bauer2020}, and in the
Jordan--Wigner mapping every spatial orbital requires two
qubits~\cite{JordanWigner1928}. Near-term algorithms such as sample-based
quantum diagonalization (SQD)~\cite{Kanno2023,RobledoMoreno2025} now treat
active spaces of tens of qubits on current processors, so the active space is
the quantity that decides whether a problem fits the hardware at all.

SQD is now being applied to chemically relevant problems.
It has been used to follow carbene formation from
diazirines in active spaces of up to 30 orbitals~\cite{Shivpuje2025}, extended
to variable molecular geometries and reaction paths with circuits of up to
124 qubits~\cite{FortinDeschenes2026}, and applied to open-shell 3$d$
complexes: for \CoCOtwo, one of the application systems of this work, SQD on an
IBM Heron processor reproduced coupled-cluster and heat-bath
configuration-interaction energies in active spaces of up to 50
qubits~\cite{David2026}. The active space also governs the preparation of the
initial state. Classifying electronic structures by their orbital
entanglement, M\"orchen \emph{et al.}\ showed that the hardest cases for state
preparation are those in which no small set of determinants dominates the
wavefunction, whatever orbital basis is used~\cite{Morchen2024}, and generative models, refined by
reinforcement learning, now construct ground-state preparation circuits for
molecular electronic-structure problems~\cite{KoziellPipe2026}. In all of these studies
the active space is an input, fixed beforehand by classical means (for the
cobalt complex by projection onto preselected atomic orbitals), and the quantum
result can be no better than that choice.

Active spaces are still most often chosen by chemical
intuition~\cite{Veryazov2011}. Automated schemes select orbitals from
projections onto predefined atomic orbitals in the atomic valence active
space (AVAS) method~\cite{Sayfutyarova2017},
from natural-orbital occupations of an approximate correlated
calculation~\cite{Pulay1988}, by machine-learning classification of candidate
active spaces~\cite{Jeong2020}, or from the orbital entanglement of
an approximate DMRG wavefunction, quantified by single-orbital entropies and
mutual information~\cite{legeza2003,rissler2006,boguslawski2012,stein2016}.
Automated selection has since become an active field in its own
right~\cite{Wardzala2026}: multi-step protocols such as the Active Space Finder
construct orbitals and select active spaces from approximate correlated
calculations and have been benchmarked for electronic excitation
energies~\cite{Shirazi2025}, and tensor-network approaches locate strongly
correlated regions from a matrix-product-state representation of the purities
of all orbital partitions, without high-order density
matrices~\cite{Mingare2026}.
Entanglement-based selection is the most direct criterion, because it asks the
correlated wavefunction which orbitals are correlated, but it requires a
preliminary DMRG calculation over a large orbital space. Projection-based
schemes are inexpensive but rely on a predefined list of atomic orbitals and
do not distinguish spin states of the same complex.

Machine learning, now widely used to explore chemical
space~\cite{Butler2018,vonLilienfeld2020,MedranoSandonas2026}, offers a way to obtain entanglement information at
mean-field cost. Golub \emph{et al.}~\cite{golub2021} showed that DMRG
single-orbital entropies of transition-metal systems can be learned from
integral-based orbital descriptors. Subsequent work has combined such
a predictor with reinforcement learning of the selection threshold, rewarded
by the agreement of NEVPT2 with DMRG energies~\cite{Osaro2026}, and has built
graph networks on orbital-correlation features that predict correlation energies and
spin-state gaps of iron(II) complexes~\cite{Quebedeaux2026}. Agents driven by
large language models now choose and refine the active spaces of CASSCF
calculations~\cite{PerezSanchez2026} and select electronic-structure methods
from correlation diagnostics~\cite{Li2026}. Most
of these methods, however, have been trained or tested on main-group
molecules, small transition-metal species or closed-shell states, and
open-shell systems remain underrepresented in the data sets used to develop
machine-learned electronic-structure methods~\cite{Karton2026}.

A model that predicts orbital entanglement for open-shell transition-metal
complexes has to capture three features. First, the active space depends on
the spin state: the high- and low-spin states of the same complex share their
nuclear framework but need different active spaces, so the representation
must resolve the spin state at the level of the orbitals. Second, orbital
entanglement is collective: whether an orbital is important depends on the
other orbitals with which it is coupled, so the model should see the frontier
orbitals together rather than each one in isolation. Third,
single-orbital entropies alone do not say which orbitals are correlated
\emph{with each other}, information that is needed to keep correlated pairs
together in the active space. Beyond these features, the usefulness of any such model rests
on its transfer to chemistry beyond its training data, which requires a
benchmark whose difficulty is defined independently of the model.

Here we introduce \texttt{SpinAware}, a model that predicts DMRG single-orbital
entropies and orbital-pair mutual information of transition-metal complexes
from the unrestricted Hartree--Fock (UHF) solution of each spin state, and we
use it to build active spaces for classical and quantum multireference
calculations. The work makes four contributions. (i)~A spin-resolved
representation of every orbital by 60 descriptors computed from the UHF
solution, most of them matrix elements of the electronic Hamiltonian
(Sec.~\ref{sec:features}). (ii)~A transformer that predicts the entropy of
each orbital in the context of all other frontier orbitals, together with a
pairwise network for the mutual information (Sec.~\ref{sec:model}).
(iii)~An out-of-distribution benchmark of
experimentally characterized complexes whose difficulty is assigned from
composition and connectivity alone (Sec.~\ref{sec:benchmark}). (iv)~A complete
workflow from the predictions to an active space and on to a quantum
processor: a candidate pool from predicted entropies and mutual information,
a natural-orbital occupation number (NOON) scan that selects the final active space, a test
of the resulting spin-state gaps against coupled-cluster theory, and SQD on IBM
quantum hardware (Sec.~\ref{sec:results}). \texttt{SpinAware} reproduces the
DMRG entropy profiles of 11 spin states with a mean absolute error (MAE) of 0.027,
and the candidate pool contains 99\% of the DMRG-active orbitals while keeping
on average 15 of the 40 frontier orbitals. For a nickel and a cobalt complex,
SQD with samples from IBM processors reproduces the exact spin-state gaps of
the selected spaces to within 0.1\kcal\ for up to 32 qubits.

\section{Methods}
\label{sec:methods}

\texttt{SpinAware} takes a single UHF calculation for each spin state of a
complex as input and returns, for a window of 20 occupied and 20 virtual
orbitals around the frontier, the single-orbital entropy of every orbital and
the mutual information of every orbital pair. Section~\ref{sec:features}
describes how each orbital is represented by 60 descriptors computed from the
UHF solution. Section~\ref{sec:model} describes the two networks: a
transformer that predicts the entropy of each orbital in the context of all
other window orbitals (Sec.~\ref{ssec:model1}), and a pairwise network that
predicts the mutual information (Sec.~\ref{ssec:model2}).
Section~\ref{sec:data} summarizes the training data, the DMRG reference
calculations and the training, and Section~\ref{sec:benchmark} introduces the
out-of-distribution benchmark on which the models are evaluated in
Section~\ref{sec:results}.

\subsection{Physics-informed orbital descriptors}
\label{sec:features}

\begin{figure}[tbp]
  \centering
  \includegraphics[width=\columnwidth]{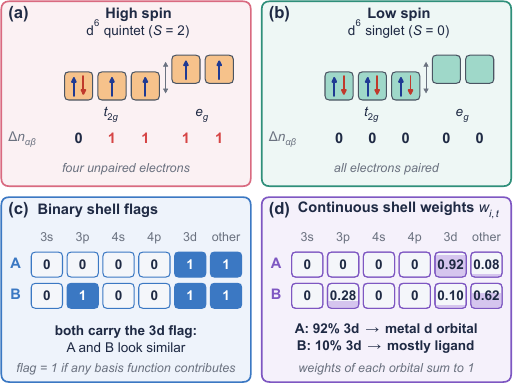}
\caption{Spin and composition descriptors.
    \textbf{a},\textbf{b},~Spin-occupation difference $\Delta n_{\alpha\beta}$
    for the $d$ orbitals of a $d^6$ ion in the high-spin (quintet, \textbf{a})
    and low-spin (singlet, \textbf{b}) states; the ones mark the four unpaired
    electrons of the quintet, whereas every occupied $d$ orbital of the singlet
    is paired.
    \textbf{c},\textbf{d},~Binary shell flags (\textbf{c}) assign the same code
    to a metal-centered and a ligand-dominated orbital; continuous shell
    weights (\textbf{d}) separate them.}
  \label{fig:features}
\end{figure}

The single-orbital entropies $s^{(1)}_i$ and the mutual information
$I_{ij}$ that decide whether an orbital belongs in the active space
require a correlated calculation over a large orbital space,
which is precisely the cost that automated active-space selection seeks
to avoid. We therefore describe every orbital only through quantities
available from a single UHF
calculation, and we choose them so that each one is
either a matrix element of the electronic Hamiltonian or a direct
consequence of one: the
one-electron energy of an orbital, the cost of doubly occupying it, its
composition in terms of metal and ligand functions, its spin
polarization, and its two-electron coupling to the other orbitals.
Our representation builds on the integral-based orbital encoding
introduced for machine-learned active-space selection by Golub
\emph{et al.}~\cite{golub2021} and extends it by spin
polarization, a continuous measure of metal--ligand mixing, the full
pattern of exchange coupling, and the environment of the orbital window
together with the molecular charge and spin state. This yields 60
descriptors per orbital, collected in Table~\ref{tab:features}
and illustrated in Figs.~\ref{fig:features} and \ref{fig:features2}.
After defining the reference wavefunction and the exchange matrix
(Sec.~\ref{sec:window}), we introduce the descriptors in six groups, one
per subsection (Secs.~\ref{sec:scalars} to \ref{sec:boundary}).

\subsubsection{Reference wavefunction and the exchange matrix}
\label{sec:window}

For a complex of total charge $Q$ and spin multiplicity $M=2S+1$, the
UHF equations are solved in the def2-SVP basis~\cite{weigend2005} at the
target multiplicity. This gives spin orbitals
$\phi^\sigma_i(\mathbf r)=\sum_\mu c^\sigma_{\mu i}\chi_\mu(\mathbf r)$,
$\sigma\in\{\alpha,\beta\}$, with energies $\varepsilon^\sigma_i$ and
occupations $n^\sigma_i\in\{0,1\}$.
In transition-metal complexes,
the static correlation that the active space
has to describe originates almost entirely in the frontier region, that is, among the orbitals whose energies lie close
to the gap between the highest occupied molecular orbital (HOMO) and the
lowest unoccupied molecular orbital (LUMO). We therefore retain a window
$\mathcal W$ of $N=40$ orbitals: the twenty occupied $\alpha$ orbitals
highest in energy, HOMO$-19$ to HOMO, and the twenty virtual $\alpha$
orbitals lowest in energy, LUMO to LUMO$+19$. Dynamic correlation is spread
over all orbitals and is added afterwards by perturbation theory
(Sec.~\ref{sec:data}). {All orbitals below the window are kept
doubly occupied as a frozen core, and those above it are discarded from
the DMRG calculations that provide the reference entropies.} Every descriptor refers to an orbital
$i\in\mathcal W$.

Beyond the mean-field level, orbitals are coupled to one another through
the two-electron integrals of the Hamiltonian. Of these, we use the
exchange integral, which appears in several descriptors below,
\begin{equation}
  \begin{aligned}
  K_{ij}=\iint{}& \phi_i(\mathbf r_1)\,\phi_j(\mathbf r_1)\,
  \frac{1}{|\mathbf r_1-\mathbf r_2|}\\
  &\times\phi_j(\mathbf r_2)\,\phi_i(\mathbf r_2)\;
  d\mathbf r_1\, d\mathbf r_2 \;\ge 0 ,
  \end{aligned}
  \label{eq:K}
\end{equation}
which we evaluate for the $\alpha$ and $\beta$ coefficient sets and
average, setting $K_{ii}\equiv0$.
$K_{ij}$ is
the self-repulsion of the overlap density $\phi_i\phi_j$. It is
therefore large only for orbitals that occupy the same region of space,
and it vanishes for orbitals localized on distant fragments, however close their
energies may be.
$K_{ij}$ is
also the Hamiltonian matrix element that connects a reference
determinant to the determinant in which an electron pair is transferred
from $i$ to $j$. In a two-orbital model, the weight of the doubly
excited configuration, and hence the entanglement between $i$ and $j$,
grows with the ratio of $K_{ij}$ to the energy separation of the two
orbitals. $K_{ij}$ thus indicates which orbitals can become correlated,
and the energy gaps decide which of them are.
The full $N\times N$ matrix $\mathbf K$ is
used both to build descriptors and, in the model, as a structural prior
on orbital--orbital interactions (Section~\ref{sec:model}).

\subsubsection{One-electron and on-site descriptors}
\label{sec:scalars}

The first group of descriptors describes each orbital in isolation through the
diagonal elements of the Hamiltonian. The one-electron integral
\begin{equation}
  h_{ii}=\int \phi_i(\mathbf r)\Big[-\tfrac12\nabla^2
  -\sum_A \frac{Z_A}{|\mathbf r-\mathbf R_A|}\Big]\phi_i(\mathbf r)\,
  d\mathbf r
  \label{eq:h}
\end{equation}
is strongly negative for compact orbitals
held close to a highly charged nucleus, such as metal $3d$ functions,
and less negative for diffuse ligand orbitals. The self-Coulomb integral
\begin{equation}
  J_{ii}=\iint
  \frac{|\phi_i(\mathbf r_1)|^2\,|\phi_i(\mathbf r_2)|^2}
       {|\mathbf r_1-\mathbf r_2|}\,
  d\mathbf r_1\, d\mathbf r_2
  \label{eq:J}
\end{equation}
is the repulsion between two electrons placed in the same orbital
and plays the role of an on-site Hubbard repulsion: when it is large
compared with the energy needed to move an electron to a nearby orbital,
double occupancy is penalized, as in the partially filled $d$ shells of
transition-metal ions.
The spatial extent
$\langle r^2\rangle_i=\sum_\mu |c_{\mu i}|^2\,\langle\chi_\mu|r^2|\chi_\mu\rangle$
gives a complementary one-electron measure of the same compactness.

The canonical orbital energy $\varepsilon_i$ (of the $\alpha$ set)
places the orbital on the energy axis.
Three further
labels, taken from Golub \emph{et al.}~\cite{golub2021}, describe the frontier
position independently of molecule size.
The first is the signed distance from the HOMO,
$k_i=i-i_\mathrm{HOMO}$, rescaled as
$p_i=[1+\exp(-k_i/100)]^{-1}$. The second is the occupation code $o_i$,
which takes the values $0.25$, $0.50$ and $0.75$ for virtual, singly
and doubly occupied orbitals.
The third is a bonding label $b_i=\pm1$,
which records whether $\varepsilon_i$ lies below or above the mean
energy of the window.

\subsubsection{Spin polarization}
\label{sec:spin}

The active space of a transition-metal complex depends on its spin
state. In a high-spin octahedral $d^6$ ion, four electrons are unpaired
and all five $d$ orbitals are partially occupied. In the low-spin state
the $t_{2g}$ set is filled, and correlation is dominated instead by the
$t_{2g}\rightarrow e_g$ near-degeneracy and by metal--ligand covalency.
{A $d$ orbital has nearly the same shape, composition and
energy whether it holds a spin pair or a single electron, so the
one-electron and on-site descriptors of Section~\ref{sec:scalars} change
only modestly between the two states. Yet the distinction matters for
correlation: a doubly occupied orbital has both spin channels filled and is
comparatively inert, whereas a singly occupied orbital has an open spin
channel into which electrons can fluctuate.}
A spin-blind
representation would describe both states with the same input for the
same nuclear framework, and no model built on it could predict that
their active spaces differ. We therefore encode the spin state at the
level of each orbital, through two complementary quantities
(Fig.~\ref{fig:features}a,b).

{The first is the spin-occupation difference
\begin{equation}
  \Delta n_{\alpha\beta}(i)=n^\alpha_i-n^\beta_i ,
  \label{eq:dn}
\end{equation}
which takes only two values. It is $1$ for an orbital that holds one
unpaired $\alpha$ electron, and $0$ for an orbital that is either doubly
occupied ($n^\alpha_i=n^\beta_i=1$) or empty. For a high-spin $d^6$
ion it marks the four unpaired electrons, whereas in the low-spin state
it is 0 throughout (Fig.~\ref{fig:features}a,b).
For the same nuclear
framework, where the shape-based descriptors barely differ, this
descriptor separates the two states.}

The second quantity is the spin splitting
$\Delta\varepsilon_{\alpha\beta}(i)\equiv\varepsilon^\alpha_i-\varepsilon^\beta_i$
between the $\alpha$ and $\beta$ orbital energies. In
Hartree--Fock theory, and to the extent that the $\alpha$ and $\beta$
orbitals share the same spatial form, the Coulomb terms cancel in this
difference and
\begin{equation}
  \Delta\varepsilon_{\alpha\beta}(i)
  \simeq -\sum_{u\in\mathrm{SOMO}} K_{iu}.
  \label{eq:deps}
\end{equation}
The splitting of an orbital is thus the total exchange it feels with the
unpaired electrons: it vanishes in a closed-shell state, is largest for
orbitals that share space with the singly occupied metal orbitals, and
is small for ligand orbitals remote from the spin density.

The two descriptors are complementary: $\Delta n_{\alpha\beta}$ is a
binary flag, whereas $\Delta\varepsilon_{\alpha\beta}$ is a continuous
measure of how strongly exchange acts on each orbital. A molecular spin
label $(M-1)/4$, shared by all orbitals, completes the group. Because
$\Delta\varepsilon_{\alpha\beta}$ and $\Delta n_{\alpha\beta}$ are given
for every orbital of the window, the model learns both that a complex is
high-spin and how the $\alpha$ and $\beta$ orbital energies differ along
the window.

\subsubsection{Metal--ligand composition}
\label{sec:ao}

In transition-metal complexes, static correlation is concentrated in
orbitals with substantial metal $d$ character, and the degree of
metal--ligand mixing (covalency) controls how strongly a $d$-derived
orbital is correlated. A simple way to encode atomic-orbital character
is a set of binary flags, one for each of the twelve shell types
$t\in\{1s,\ldots,5s,2p,\ldots,5p,3d,4d,5d\}$. Each flag is set to one
if any basis function of that type contributes to the orbital, and we
retain these twelve flags~\cite{golub2021}. Binary flags, however,
cannot express \emph{how much} of an orbital belongs to a given shell:
a nearly pure metal $3d$ orbital and a ligand $p$ orbital with a small
$3d$ admixture receive the same flags (Fig.~\ref{fig:features}c,d).
Since canonical orbitals are delocalized,
almost every orbital carries some small coefficient on almost every
shell type, and the flags tend to be set for all orbitals alike.

We resolve this with continuous shell weights derived from
normalized squared AO coefficients. The weight of shell type
$t$ in orbital $i$ is
\begin{equation}
  w_{i,t}=
  \frac{\sum_{\mu\in t}\tfrac12\big(|c^\alpha_{\mu i}|^2+|c^\beta_{\mu i}|^2\big)}
       {\sum_{\mu}\tfrac12\big(|c^\alpha_{\mu i}|^2+|c^\beta_{\mu i}|^2\big)},
  \qquad \sum_t w_{i,t}=1 .
  \label{eq:aoweights}
\end{equation}
The $3d$ weight $w_{i,3d}$ measured against the
ligand $p$ weights is a direct, continuous measure of covalency.

\begin{figure}[tbp]
  \centering
  \includegraphics[width=\columnwidth]{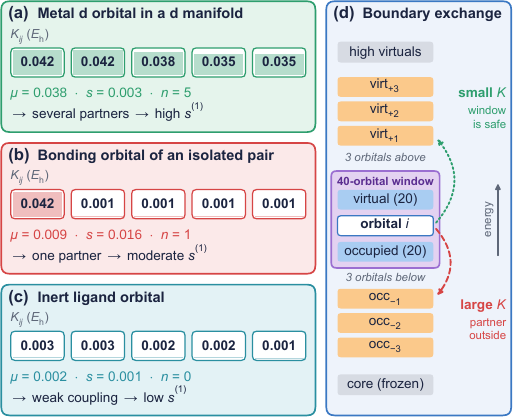}
  \caption{Exchange descriptors.
    \textbf{a}--\textbf{c},~Rows of the exchange matrix and their statistics for
    three archetypal orbitals: a metal $d$ orbital in a correlated manifold
    (\textbf{a}), coupled with comparable strength to several partners; the
    bonding orbital of an isolated pair (\textbf{b}), coupled strongly to a
    single partner; and an inert ligand orbital (\textbf{c}), weakly coupled to
    all orbitals. The leading couplings of \textbf{a} and \textbf{b} are
    equal; the mean $\mu_i$, spread $s_i$ and partner count $n_i$
    (Eq.~\ref{eq:kstats}) separate them. Values are schematic, and the partner counts use $\kappa=0.01\,E_\mathrm{h}$ for illustration.
    \textbf{d},~Boundary exchange: exchange integrals between a window orbital
    and the three orbitals immediately below and above the 40-orbital window
    reveal correlated partners outside the window.}
  \label{fig:features2}
\end{figure}

\subsubsection{Exchange connectivity of the orbital network}
\label{sec:exchange}

Static correlation is collective: an orbital becomes strongly entangled
because it is nearly degenerate with, and strongly coupled to, other
orbitals, not because of its own properties alone. The row
$\{K_{ij}\}_{j\neq i}$ of the exchange matrix defines a weighted network
over the window, and we describe the position of each orbital in this
network at three levels of detail. The ten largest couplings, sorted in
descending order~\cite{golub2021}, identify the strongest partners.

These values alone, however, cannot tell apart physically different
situations, {as three archetypes illustrate
(Fig.~\ref{fig:features2}a--c). A metal $d$ orbital in a partially filled
shell (A) is coupled with comparable strength to the other $d$ orbitals
of the same center and to little else. A bonding orbital of an isolated
two-center bond (B) is coupled strongly to a single partner, its
antibonding counterpart, and only weakly to everything else. An orbital
of a peripheral ligand (C) is weakly coupled to all orbitals.
Yet A and B can have the same leading coupling, so the
top of a sorted list of couplings cannot separate them.}

{We therefore summarize the whole exchange row by four
statistics,
\begin{equation}
  \begin{aligned}
  \mu_i&=\frac{1}{N}\sum_{j}K_{ij}, &
  s_i&=\Big[\frac{1}{N}\sum_{j}\big(K_{ij}-\mu_i\big)^2\Big]^{1/2},\\
  S_i&=\sum_{j}K_{ij}, &
  n_i&=\sum_{j}\Theta\big(K_{ij}-\kappa\big),
  \end{aligned}
  \label{eq:kstats}
\end{equation}
the mean, standard deviation and sum of the couplings, and the number of
partners whose coupling exceeds a threshold $\kappa$ ($\Theta$ is the
Heaviside step function). The sums run over all $N=40$ window orbitals
with $K_{ii}\equiv0$, and $\kappa=0.05\,E_\mathrm{h}$.} $S_i$ measures the
total exchange connectivity of the orbital, and $n_i$ the number of strong
partners.
We also characterize the shape of the distribution. With the normalized
couplings $\tilde K_{ij}=K_{ij}/S_i$ we define
\begin{equation}
  C_i=\max_j\tilde K_{ij},
  \qquad
  \mathrm{PR}_i=\Big(\sum_{j}\tilde K_{ij}^{\,2}\Big)^{-1}.
  \label{eq:pr}
\end{equation}
The participation ratio $\mathrm{PR}_i$, familiar as a measure of the
delocalization of vibrational and electronic states~\cite{bell1970},
is the effective number of exchange partners of orbital $i$. It is
close to one for a pair-like orbital and grows towards $N-1$ when the
coupling is spread evenly. The concentration $C_i$ is the fraction of
the coupling carried by the dominant partner.

In an idealized model in which orbital A couples with strength $K_0$ to
$m$ partners, orbital B with the same strength to a single partner, and
orbital C with a small strength $\delta\ll K_0$ to all $N-1$ orbitals
(Supplementary Section~S3), the leading coupling is $K_0$ for both A and
B, whereas $S_i$, $n_i$ and $\mathrm{PR}_i$ are $m$ times larger for A
($m=4$ for a $d$ orbital coupled to the other four $d$ orbitals of its
center). The spectator C is separated from both by its small
$\max_j K_{ij}$ and $S_i$ and by $n_i=0$, although its participation
ratio, $N-1$, is the largest of the three. The shape descriptors $C_i$
and $\mathrm{PR}_i$ are therefore meaningful only together with the
magnitudes $S_i$ and $\max_j K_{ij}$, which is why both kinds of
statistic are supplied to the model.

\begin{table}[tbp]
\centering
\caption{Orbital descriptors. Indices refer to the input vector of each
window orbital; descriptors marked $^\dagger$ are molecular quantities
shared by all orbitals of a complex. The order follows the input vector of
the trained model.}
\label{tab:features}
\footnotesize
\setlength{\tabcolsep}{3pt}
\def\cellf#1{\begin{tabular}[t]{@{}l@{}}#1\end{tabular}}
\begin{tabular}{@{}lll@{}}
\toprule
Index & Descriptor & Physical content \\
\midrule
0      & $h_{ii}$ (Eq.~\ref{eq:h})        & \cellf{kinetic + nuclear-\\attraction energy} \\
1      & $J_{ii}$ (Eq.~\ref{eq:J})        & \cellf{on-site repulsion,\\compactness} \\
2      & $\varepsilon_i$                  & orbital energy \\
3      & $\langle r^2\rangle_i$           & spatial extent \\
4      & $p_i$                            & \cellf{position relative\\to the HOMO} \\
5      & $o_i$                            & \cellf{occupation (virtual /\\singly / doubly)} \\
6      & $b_i$                            & \cellf{lower / upper half\\of the window} \\
7--18  & shell flags                      & presence of shell type $t$ \\
19--30 & $w_{i,t}$ (Eq.~\ref{eq:aoweights}) & \cellf{weight of shell type $t$,\\covalency} \\
31     & $(M-1)/4^\dagger$                & spin label \\
32     & $\Delta\varepsilon_{\alpha\beta}$ (Eq.~\ref{eq:deps}) & \cellf{$\alpha$--$\beta$ energy\\splitting} \\
33     & $\Delta n_{\alpha\beta}$ (Eq.~\ref{eq:dn}) & \cellf{$\alpha$--$\beta$ occupation\\difference} \\
34--37 & \cellf{$\mu_i, s_i, S_i, n_i$\\(Eq.~\ref{eq:kstats})} & exchange connectivity \\
38     & $\Delta_\mathrm{HL}^\dagger$     & HOMO--LUMO gap \\
39     & $g_i$ (Eq.~\ref{eq:gap})         & \cellf{gap to dominant\\exchange partner} \\
40--49 & ten largest $K_{ij}$             & \cellf{strongest exchange\\couplings} \\
50--53 & \cellf{max, sum of $K_{ik}$,\\$k\in\mathcal B_-$, then $\mathcal B_+$} & \cellf{coupling beyond\\the window} \\
54--55 & \cellf{count of $K_{ik}>0.05\,E_\mathrm{h}$,\\$k\in\mathcal B_-$, $\mathcal B_+$} & \cellf{partners beyond\\the window} \\
56     & $C_i$ (Eq.~\ref{eq:pr})          & \cellf{concentration of\\exchange coupling} \\
57     & $\mathrm{PR}_i$ (Eq.~\ref{eq:pr}) & \cellf{effective number of\\exchange partners} \\
58     & $M^\dagger$                      & spin multiplicity \\
59     & $Q^\dagger$                      & total charge \\
\bottomrule
\end{tabular}
\end{table}

\subsubsection{Energetic relation to the frontier and to exchange partners}
\label{sec:gaps}

Exchange coupling indicates which orbitals can interact, but whether the
interaction produces strong correlation depends on their energy
separation, which we describe by two energy differences. The molecular
HOMO--LUMO gap $\Delta_\mathrm{HL}=\varepsilon_\mathrm{LUMO}-
\varepsilon_\mathrm{HOMO}$ is a global indicator of near-degeneracy, as
small gaps generally accompany larger correlated spaces. The second
descriptor links the energetic and two-electron pictures for each
orbital,
\begin{equation}
  g_i=\big|\varepsilon_i-\varepsilon_{j^\ast}\big|,
  \qquad j^\ast=\arg\max_{j} K_{ij},
  \label{eq:gap}
\end{equation}
which is the energy separation between an orbital and its strongest
exchange partner. A large $K_{ij^\ast}$ together with a large $g_i$
marks the bonding and antibonding partners of a chemical bond, such as
the $\sigma/\sigma^\ast$ pair of a metal--ligand bond or the $d/\pi^\ast$
pair of back-bonding, whereas a small $g_i$ indicates mixing within a
single quasi-degenerate manifold, such as the $d$ shell itself. The pair
$(K_{ij^\ast},g_i)$ therefore separates these two mechanisms of static
correlation.

\subsubsection{Coupling beyond the window and global descriptors}
\label{sec:boundary}

Any finite orbital window may cut through a correlated subspace, a
well-known difficulty of active-space
selection~\cite{Veryazov2011,stein2016}: an orbital just inside the
window can be strongly coupled to one just outside it, and the resulting
active space is then incomplete. A representation confined to the window
implicitly assumes that all coupled partners lie inside it. This holds
for compact complexes with simple monodentate ligands, but it can fail
for systems with delocalized correlation, such as
macrocyclic~\cite{LiManni2018}, salen-type~\cite{Wouters2014} or
multinuclear complexes~\cite{Kurashige2013,Sharma2014,LiManni2020compression,LiManni2021FeS,Dobrautz2021stochCASSCF},
whose active spaces extend over the ligand framework or over several
metal centers. For Fe(II) porphyrin, for example, including the full $\pi$ system in
the active space changes the order of the spin
states~\cite{LiManni2018}. Although the orbitals outside the window take
no part in the correlated calculation, their exchange integrals with the
window orbitals are fully defined (Fig.~\ref{fig:features2}d).
We make the edges of the window visible by
evaluating Eq.~\eqref{eq:K} between each window orbital and the three
orbitals immediately below ($\mathcal B_-$) and above ($\mathcal B_+$)
the window. For each side we record
$\max_{k\in\mathcal B_\pm}K_{ik}$ and $\sum_{k\in\mathcal B_\pm}K_{ik}$,
together with the number of couplings above $0.05\,E_\mathrm h$. When
these quantities are small, orbital $i$ has its partners inside
$\mathcal W$. When they are large, the window is missing a coupled
partner, and the active space suggested for that orbital should be
interpreted with caution or the window enlarged.

Finally, the multiplicity $M$ and the total charge $Q$ are appended to
every orbital, so that the orbital-level descriptors are read in the
context of the global electron count and spin state. All reference
calculations are carried out for the isolated ion, at the charge implied
by the metal oxidation state and the formal ligand charges.
Every descriptor is obtained from a single mean-field calculation, at a
small fraction of the cost of the correlated calculation it replaces, and
none depends on a predefined list of elements or ligands.

\subsection{Model architecture}\label{sec:model}

\subsubsection{Physical motivation}\label{ssec:motivation}

The single-orbital entropy $s_i^{(1)}$ quantifies how much orbital $i$
is entangled with the rest of the electronic system.  It is defined
through the von~Neumann entropy of the one-orbital reduced density
matrix~$\rho_i$, obtained by tracing out all degrees of freedom except
orbital~$i$ from the DMRG wavefunction:
\begin{equation}\label{eq:s1}
  s_i^{(1)} \;=\; -\operatorname{Tr}\!\bigl[\rho_i \ln \rho_i\bigr],
\end{equation}
with the bounds $s_i^{(1)}\in[0,\ln 4]$.
An orbital with $s_i^{(1)}\approx 0$ is effectively doubly occupied or
empty and can safely be excluded from the active space, while an orbital
with $s_i^{(1)}\gtrsim 0.1$ carries significant entanglement and should
be retained.

A fundamental property of $s_i^{(1)}$ is that it is \emph{not} an
intrinsic property of orbital~$i$ alone.  Rather, it measures how
orbital~$i$ is entangled with every other orbital in the system.
Consider a concrete example from the benchmark of this work. In the
singlet of \VClsix\ (\cpd{1}), one window orbital has a DMRG entropy of
$s^{(1)}=0.099$, just below the threshold $\tau=0.10$ used to define the
active space, and would be discarded on its own. Its mutual information
with a single partner orbital ($s^{(1)}=0.142$) is, however, 0.028,
fourteen times larger than its mutual information with any other orbital
(at most 0.002). The two orbitals form a correlated pair, and excluding
the first while retaining its partner breaks this pair. This example illustrates a
general principle: the importance of an orbital depends on the context
provided by every other orbital in the molecule.

A model that maps the descriptors of each orbital to an entropy
independently of the other orbitals cannot represent such pairs: the
features of the partner never enter the prediction, and two orbitals that
matter only together cannot be retained or discarded together. We
therefore process all $N$ orbitals of the window jointly, so that the
predicted entropy of each orbital depends on the descriptors of every
other orbital, and we guide this exchange of information by the exchange
integrals $K_{ij}$ (Sec.~\ref{sec:window}).

\subsubsection{Model~1: Single-orbital entropy prediction}
\label{ssec:model1}

\paragraph{Problem statement.}

Given the descriptor vectors $\mathbf{x}_i \in \mathbb{R}^{60}$
(Section~\ref{sec:features}) of the $N=40$ window orbitals of a UHF
calculation, Model~1 predicts the DMRG single-orbital entropies
$s_i^{(1)}$ (Eq.~\ref{eq:s1}) of all $N$ orbitals simultaneously.

\paragraph{Architecture overview.}

\begin{figure*}[p]
  \centering
  \includegraphics[width=\textwidth]{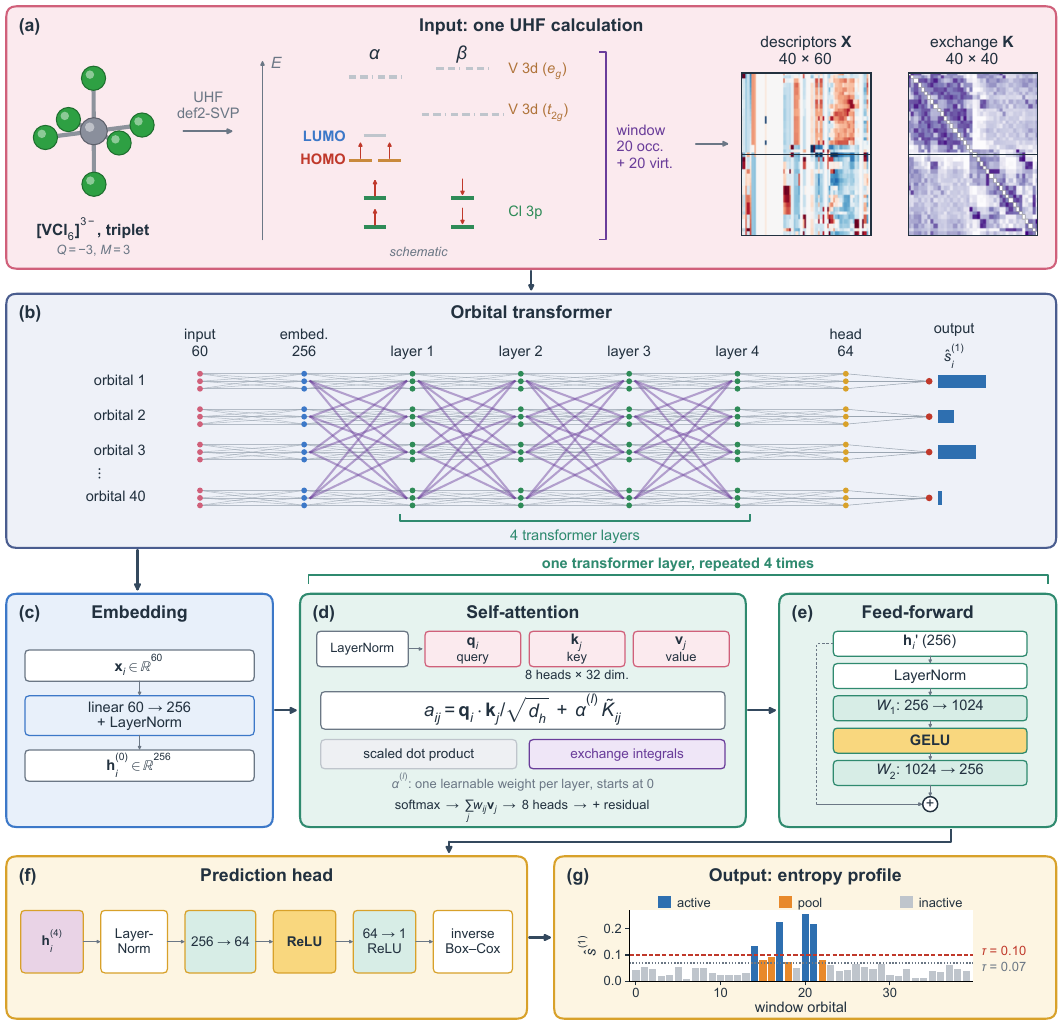}
  \caption{Architecture of the orbital transformer for single-orbital entropy
    prediction. \textbf{a},~Input from one UHF/def2-SVP calculation on the
    triplet of \VClsix\ (\cpd{1}): structure, schematic $\alpha$ and $\beta$
    orbital levels (V(III), $d^2$) with HOMO, LUMO and the window of 20
    occupied and 20 virtual orbitals, descriptor matrix $\mathbf X$ (40
    orbitals $\times$ 60 descriptors, including $M$ and $Q$) and exchange
    matrix $\mathbf K$.
    \textbf{b},~Network overview: the same weights act on every window
    orbital, through the embedding (60 $\to$ 256), four transformer layers in
    which the orbitals exchange information by self-attention
    (purple lines) and the prediction head, which returns $\hat s^{(1)}_i$. \textbf{c},~Orbital embedding with layer
    normalization. \textbf{d},~Self-attention: after layer
    normalization each orbital is projected to query, key and value vectors
    (8 heads $\times$ 32 dimensions); the scaled dot product is complemented by the normalized exchange integrals with a
    learnable weight $\alpha^{(l)}$ per layer, shared by the heads and
    initialized at zero. \textbf{e},~Feed-forward network (256 $\to$ 1024
    $\to$ 256, GELU); both sub-layers are pre-norm with residual connections.
    \textbf{f},~Prediction head shared by all 40 orbitals (256 $\to$ 64 $\to$
    1), applied after a final layer normalization and followed by the inverse
    Box--Cox transformation. \textbf{g},~Illustrative entropy profile with the
    selection thresholds $\tau = 0.10$ and 0.07.}
  \label{fig:arch}
\end{figure*}

The model comprises three stages (Figure~\ref{fig:arch}):
(i)~an embedding layer that projects each orbital's 60 descriptors
into a high-dimensional representation space;
(ii)~a stack of $L$~transformer layers with self-attention, through
which the orbitals exchange information; and
(iii)~a prediction head that maps each orbital's context-enriched
representation to a scalar entropy estimate.

\paragraph{Stage~1: Orbital embedding.}

Each orbital's descriptor vector $\mathbf{x}_i\in\mathbb{R}^{60}$ is
projected into a $d$-dimensional representation space:
\begin{equation}\label{eq:embed}
  \mathbf{h}_i^{(0)}
  = \operatorname{LayerNorm}\!\bigl(
      \mathbf{W}_{\mathrm{emb}}\,\mathbf{x}_i + \mathbf{b}_{\mathrm{emb}}
    \bigr),
  \qquad
  \mathbf{W}_{\mathrm{emb}} \in \mathbb{R}^{d\times 60},
\end{equation}
with $d=256$ and layer normalization~\cite{Ba2016LayerNorm}.
At this stage, the representations
$\{\mathbf{h}_i^{(0)}\}_{i=1}^N$ are still independent: no
information has flowed between orbitals.

\paragraph{Stage~2: Self-attention.}

The core of the architecture is a stack of $L=4$ transformer
layers.  Each layer~$l$ updates the orbital representations through
two sub-steps: multi-head self-attention with an exchange-integral
bias, followed by a position-wise feedforward network (FFN).

\paragraph{Self-attention with exchange-integral bias.}

We use multi-head scaled dot-product
attention~\cite{Vaswani2017Attention} with $n_{\mathrm{heads}}=8$ heads
of dimension $d_h=d/n_{\mathrm{heads}}=32$. In each head, the query, key
and value vectors $\mathbf{q}_i$, $\mathbf{k}_j$ and $\mathbf{v}_j$ are
linear projections of the layer-normalized orbital representations, and
the attention logit between orbitals~$i$ and~$j$ is
\begin{equation}\label{eq:attn}
  a_{ij}
  = \frac{\mathbf{q}_i \cdot \mathbf{k}_j}{\sqrt{d_h}}
    + \alpha^{(l)}\, \tilde K_{ij},
\end{equation}
where $\tilde K_{ij}=K_{ij}/K_\mathrm{s}$ is the exchange integral between
orbitals~$i$ and~$j$ divided by a fixed, dataset-wide scale
$K_\mathrm{s}$. The scalar $\alpha^{(l)}$ is a learnable parameter of
layer~$l$, shared by all heads and initialized to zero, so that training
starts from standard attention. The attention weights $w_{ij}$ are the
softmax of $a_{ij}$ over all orbitals~$j$, and the outputs
$\sum_j w_{ij}\mathbf{v}_j$ of the heads are concatenated and linearly
projected.

The second term supplies the
exchange-coupling matrix as an additional input to the attention
logits, and since $\alpha^{(l)}$ is unconstrained, it can increase or
decrease the relative logit of a pair. The exchange integral supplies
pair-resolved information beyond the query--key dot product, but it is
not itself an entanglement measure.

\paragraph{Attention and the exchange bias.}
Through attention, each orbital receives information from all other
window orbitals already in the first layer, and the four layers update
these representations nonlinearly. Each orbital thus also receives
information from its correlated partners (Sec.~\ref{ssec:motivation}),
which a model that scores each orbital independently~\cite{golub2021}
cannot do. The pairwise network of Model~2 adds an explicit prediction of
their mutual information. Retraining Model~1 without the exchange bias, or with the exchange
matrices of each molecule randomly permuted over its orbitals, gives the
same accuracy (Supplementary Table~S23), so the exchange information
reaches the model through the exchange descriptors of each orbital
(Table~\ref{tab:features}).

\paragraph{Feedforward network and residual connections.}

After the attention sub-layer, each orbital's representation passes
through the FFN:
\begin{equation}\label{eq:ffn}
  \mathrm{FFN}(\mathbf{z})
  = \mathbf{W}_2\,\operatorname{GELU}\!\bigl(
      \mathbf{W}_1\,\mathbf{z} + \mathbf{b}_1
    \bigr) + \mathbf{b}_2,
\end{equation}
with $\mathbf{W}_1\in\mathbb{R}^{d_{\mathrm{ff}}\times d}$ and
$\mathbf{W}_2\in\mathbb{R}^{d\times d_{\mathrm{ff}}}$, where
$d_{\mathrm{ff}}=4d=1024$, GELU is the Gaussian error linear unit, and dropout with rate 0.1 is applied to the
attention weights and inside the FFN. Each transformer layer applies
pre-normalization residual connections~\cite{Xiong2020PreNorm}, where
$\mathrm{MHA}_i$ denotes the multi-head attention output
for orbital~$i$:
\begin{align}
  \mathbf{h}_i' &= \mathbf{h}_i^{(l-1)}
    + \mathrm{MHA}_i\!\bigl(\operatorname{LayerNorm}(\mathbf{h}^{(l-1)})\bigr),
    \label{eq:res1}\\[4pt]
  \mathbf{h}_i^{(l)} &= \mathbf{h}_i'
    + \mathrm{FFN}\!\bigl(
        \operatorname{LayerNorm}(\mathbf{h}_i')
      \bigr).
    \label{eq:res2}
\end{align}

\paragraph{Stage~3: Prediction head.}

After $L=4$ attention layers, the final representations
$\{\mathbf{h}_i^{(L)}\}_{i=1}^N$ are passed through a
shared prediction head:
\begin{equation}\label{eq:head}
  \begin{aligned}
  \hat{y}_i
  = \operatorname{ReLU}\!\Bigl(
      &\mathbf{w}_{\mathrm{out}}^{\!\top}
      \operatorname{ReLU}\!\bigl(
        \mathbf{W}_{\mathrm{hid}}\,
        \operatorname{LayerNorm}(\mathbf{h}_i^{(L)})
        + \mathbf{b}_{\mathrm{hid}}
      \bigr)\\
      &+ b_{\mathrm{out}}
    \Bigr),
  \end{aligned}
\end{equation}
where $\mathbf{W}_{\mathrm{hid}}\in\mathbb{R}^{64\times 256}$ and
$\mathbf{w}_{\mathrm{out}}\in\mathbb{R}^{64}$, and ReLU is the rectified linear unit.  The final ReLU
activation enforces the physical constraint $\hat{y}_i \geq 0$,
since $s_i^{(1)}\geq 0$ by definition. During training the outer ReLU
is replaced by the identity, and the output is clamped at zero in evaluation
and inference, which gives the same function as Eq.~\eqref{eq:head} while
keeping a non-zero gradient for every orbital during training.

The prediction $\hat{y}_i$ is in a transformed target space.  The
true entropy is recovered via inverse Box--Cox transformation, as
described below.

\paragraph{Target preprocessing.}\label{sssec:boxcox}

The distribution of $s^{(1)}$ values across the training set is
heavily right-skewed: the majority of orbitals have near-zero
entropy, while a small fraction carry $s_i^{(1)}>0.5$.  Direct
regression on raw $s^{(1)}$ would be dominated by the trivial
near-zero predictions, with the model learning to predict
``approximately zero'' for all orbitals.

We apply a Box--Cox transformation~\cite{BoxCox1964} with parameter
$\lambda$ followed by min--max scaling to $[0,1]$:
\begin{equation}\label{eq:boxcox}
  y_i
  = \operatorname{MinMaxScale}\!\left(
      \frac{(s_i^{(1)} + \epsilon)^{\lambda} - 1}{\lambda}
    \right),
\end{equation}
with $\lambda = 0.1377934$, the same for all five models of the final
ensemble, and $\epsilon = 10^{-8}$. The transformation is fitted on the training complexes
and held fixed for validation, testing and benchmark predictions.
The Box--Cox transform amplifies distinctions among small $s^{(1)}$
values, where the boundary between important and unimportant orbitals
lies, while compressing differences among large values that are less
decision-relevant.

At inference, predicted values are mapped back to the physical
entropy scale:
\begin{equation}\label{eq:inv_boxcox}
  \hat{s}_i^{(1)}
  = \max\!\Bigl(0,\;
      \bigl(\lambda\cdot\hat{y}_i^{\,\prime} + 1\bigr)^{1/\lambda}
      - \epsilon
    \Bigr),
\end{equation}
where $\hat{y}_i^{\,\prime}$ is the inverse min--max-scaled
prediction.

\paragraph{Training objective.}

The model is trained by minimizing the MAE on the
transformed targets over batches of molecules:
\begin{equation}\label{eq:loss_s1}
  \mathcal{L}_{s^{(1)}}
  = \frac{1}{|\mathcal{B}|}
    \sum_{m\in\mathcal{B}}
    \frac{1}{N_m}
    \sum_{i=1}^{N_m}
    \bigl|\hat{y}_i^{(m)} - y_i^{(m)}\bigr|,
\end{equation}
where $\mathcal{B}$ is a training batch and $N_m$ is the number of
orbitals in the window for molecule~$m$. Compared with a squared
error, the absolute error is less dominated by the few orbitals with
very large entropy. The model contains approximately
$3.19\times 10^6$ trainable parameters (3\,189\,125). Training details are given in Sec.~\ref{sec:data}.

\subsubsection{Model~2: Mutual-information prediction}
\label{ssec:model2}

\paragraph{Motivation: beyond single-orbital entropies.}
Mutual information has previously been predicted by machine learning
for strongly correlated transition-metal systems by Golub \emph{et
al.}~\cite{golub2023mi}. Here we predict it from spin-resolved UHF
descriptors and use it to complete the entropy-based selection.

The single-orbital entropy identifies orbitals that are correlated with
\emph{something}. However, it does not reveal which orbitals are correlated with
\emph{each other}. This pairwise information is carried by the mutual
information
\begin{equation}\label{eq:mi}
  I_{ij} = \tfrac12\bigl(s_i^{(1)} + s_j^{(1)} - s_{ij}^{(2)}\bigr),
\end{equation}
where $s_{ij}^{(2)} = -\operatorname{Tr}[\rho_{ij}\ln\rho_{ij}]$ is the
two-orbital von~Neumann entropy of the two-orbital reduced density
matrix~$\rho_{ij}$. By subadditivity $I_{ij}\geq 0$, with $I_{ij}=0$ if and
only if orbitals~$i$ and~$j$ are uncorrelated, and $I_{ij}\le \ln 4$.
This convention, with the prefactor $\tfrac12$, is used throughout (for all labels,
predictions and thresholds).

The mutual information provides two capabilities that single-orbital
entropy alone cannot: it identifies low-entropy partners of strongly
coupled orbitals, which an entropy threshold would discard although they
belong to a correlated pair, and it identifies blocks of mutually
correlated orbitals whose individual entropies lie close to the
threshold. Both are used to build the candidate pool of
Sec.~\ref{ssec:pool}.

\paragraph{Pairwise architecture.}

The mutual-information model is a separate network that scores every pair
$(i,j)$, $i<j$, of the window. Its input combines the descriptors of the two
orbitals with the exchange structure around them. Let
$\tilde{\mathbf x}_i\in\mathbb R^{60}$ be the standardized descriptor vector
of orbital $i$ and $\tilde{\mathbf k}_i\in\mathbb R^{40}$ its standardized
row of the exchange matrix, which records how strongly the orbital couples to
every orbital of the window. To make the prediction independent of the order
of the two orbitals, the pair is ordered so that $a$ denotes the orbital with
the larger descriptor norm and $b$ the other. The pair representation is
\begin{equation}\label{eq:z_pair}
  \begin{aligned}
  \mathbf z_{ij} = \bigl[\,&
    \tilde{\mathbf x}_a \,\|\, \tilde{\mathbf x}_b \,\|\,
    |\tilde{\mathbf x}_a-\tilde{\mathbf x}_b| \,\|\,
    \tilde{\mathbf x}_a\odot\tilde{\mathbf x}_b \,\|\\
  & \tilde{\mathbf k}_a \,\|\, \tilde{\mathbf k}_b \,\|\,
    |\tilde{\mathbf k}_a-\tilde{\mathbf k}_b| \,\|\,
    \tilde{\mathbf k}_a\odot\tilde{\mathbf k}_b \,\|\,
    \ln|K_{ij}|
  \,\bigr]\in\mathbb R^{401},
  \end{aligned}
\end{equation}
where $\|$ denotes concatenation and $\odot$ the element-wise product. The
absolute difference and the product encode how similar the two orbitals are
and along which dimensions both are large, so that the network does not have
to rediscover these interactions. The exchange rows tell the network with
which \emph{other} orbitals each partner is coupled, and the direct exchange
integral $K_{ij}$, entered on a logarithmic scale, supplies the two-electron
coupling of the pair itself. On the training data, $\ln|K_{ij}|$ alone
correlates with $\ln I_{ij}$ with a coefficient of
0.56, which is why it is entered directly.

The vector $\mathbf z_{ij}$ is projected to a hidden dimension of 384
(linear layer, layer normalization, GELU) and processed by three gated
residual blocks,
\begin{equation}\label{eq:resblock}
  \mathbf u \leftarrow \mathbf u
  + \sigma(\mathbf W_g\mathbf u)\odot f(\mathbf u),
\end{equation}
in which $f$ consists of two linear layers with layer normalization, GELU
activation and dropout (rate 0.15), and the sigmoid gate $\sigma$ lets each
block decide how much it modifies the representation. A two-layer head maps
the result to a scalar $\hat y_{ij}$, which predicts $\ln(I_{ij}+\epsilon)$
with $\epsilon=10^{-7}$, and the mutual information is recovered as
$\hat I_{ij}=\max(0,\,e^{\hat y_{ij}}-\epsilon)$. The network has
$1.57\times10^{6}$ trainable parameters (1\,565\,185).

\paragraph{Training objective.}

The distribution of $I_{ij}$ is even more skewed than that of $s^{(1)}$: most
pairs have $I_{ij}\approx0$, and appreciable mutual information is
concentrated in a small number of physically important pairs. We therefore
regress the logarithm of $I_{ij}$ and weight each pair by its mutual
information,
\begin{equation}\label{eq:loss_mi}
  \mathcal L_I = \frac{1}{|\mathcal B|}\sum_{(ij)\in\mathcal B}
  w_{ij}\,\ell_\delta\!\bigl(\hat y_{ij},\,\ln(I_{ij}+\epsilon)\bigr),
  \quad
  w_{ij}\propto 1+100\,I_{ij},
\end{equation}
where $\ell_\delta$ is the Huber loss~\cite{Huber1964} with $\delta=2$ and the
weights are normalized to unit mean. The logarithmic target resolves the
small values that decide whether a weakly correlated orbital is rescued, and
the weights keep the rare strongly coupled pairs from being dominated by the
many uncorrelated ones. The labels $I_{ij}$ are obtained from the same DMRG
calculations as the entropy labels, from the one- and two-orbital reduced
density matrices.

Unlike Model~1, the pairwise network sees the rest of the molecule only
through the exchange rows $\tilde{\mathbf k}_a$ and $\tilde{\mathbf k}_b$.
Sharing the transformer backbone between the two tasks, so
that the pair representation is built from context-enriched orbital
representations $\mathbf h_i^{(L)}$, is a natural extension that we leave to
future work.

\subsubsection{Inference}\label{ssec:inference}

At inference time, the trained models are applied to molecules for which
no DMRG calculation has been performed. A UHF single-point calculation,
which takes minutes for the systems considered here, provides the orbital
energies, occupations and expansion coefficients and the integrals
$h_{ii}$, $J_{ii}$ and $K_{ij}$. From it, the 60 descriptors of each of
the $N=40$ window orbitals (Sec.~\ref{sec:features}) and the exchange
matrix $\mathbf K$ are computed in seconds. After the preprocessing
described below, Model~1 returns the $N$ entropies, mapped back to the
physical scale by Eq.~\eqref{eq:inv_boxcox}, and Model~2 the mutual
information of all 780 orbital pairs.

\paragraph{Preprocessing.}
For Model~1, the stored min--max transformation is applied to input
columns 0--3, 30, 31, 44--49, 58 and 59 of Table~\ref{tab:features}.
All other columns retain their supplied representation. The exchange matrix
is divided by the stored scale $K_\mathrm{s}=2.6162941\,E_\mathrm{h}$.
These same fitted transformations are used without modification at inference.

\paragraph{From predictions to an active space.}
The predictions are not used with a single entropy threshold. A fixed
threshold works when the entropy distribution is bimodal, but for many
transition-metal states the profile is flat near the threshold, and the
outcome of a hard cut is then decided by small, chemically insignificant
deviations (Sec.~\ref{ssec:transfer}). We therefore proceed in two steps. The
predicted entropies and mutual information are first combined into a compact
candidate pool designed for completeness (Sec.~\ref{ssec:pool}). The final
active space is then selected from the pool by a NOON
scan that relies only on the correlated wavefunction itself
(Sec.~\ref{ssec:noon}).

\subsection{Data and computational details}
\label{sec:data}

\paragraph{Training data.}
The models must learn how the orbital structure of a single UHF reference
relates to orbital entanglement across the chemistry of first-row
transition-metal complexes. The training set was therefore built to cover
the main factors that control static correlation: the metal and its
$d$-electron count, the spin state, the ligand field and the metal--ligand
bond length. It comprises 3\,454 spin-pure calculations on mononuclear
complexes of Cr, Mn, Fe, Co and Ni (Supplementary Table~S3): 2\,105
constructed complexes at equilibrium geometries, 807 with metal--ligand bonds
stretched by factors of 1.5 and 2.0 to sample increasingly strong static
correlation,  and 542
complexes taken from experimental crystal structures. Spin multiplicities
range from 1 to 6 and total charges from $-9$ to $+3$, all but nine of them between $-5$ and $+3$. None of the benchmark complexes
of Sec.~\ref{sec:benchmark} is part of the training data.

For every complex
and spin state, the UHF reference and the 60 descriptors of
Sec.~\ref{sec:features} were computed in the def2-SVP
basis~\cite{weigend2005}, and the reference labels (single-orbital
entropies and mutual information) were obtained from DMRG
calculations~\cite{White1992,Chan2011} with \textsc{block2}~\cite{Zhai2023}
in the same 40-orbital window, starting from the UHF occupation pattern and
increasing the bond dimension over 16 sweeps from 250 and 500 to a final
$M=1024$ with decreasing noise, with $S_z$ symmetry and the window
orbitals in their canonical (orbital-energy) order.
The single-orbital entropy of orbital $i$ is the entropy of DMRG site $i$
with its four local states, and $I_{ij}$ refers to pairs of such sites. All calculations use $M_s = S$
($M_s = 0$ for singlets), the spin projection of the UHF occupation. Because
the entropies depend on $M_s$, the labels refer to this component.

States whose UHF reference deviates from the exact
$\langle\hat S^2\rangle$ by more than 0.5, a looser filter than the
criterion applied to the benchmark (Sec.~\ref{sec:benchmark}), and DMRG calculations that had not converged were
removed, so that every label describes the intended spin state.
Because $S_z$-adapted DMRG fixes only the spin projection, the total
spin of every DMRG wavefunction was verified by evaluating the exact
$\langle\hat S^2\rangle$, including the overlaps between the $\alpha$ and
$\beta$ orbitals of the UHF basis and the contribution of the occupied
orbitals below the window. For states in which the $S_z$-adapted DMRG
approached a different total spin, an energy penalty on $\hat S^2-S(S+1)$ was
added to the Hamiltonian, which drives the DMRG to the intended spin state. In
all cases the descriptors and the labels refer to the same UHF reference.
The spin contamination $w=[\langle\hat S^2\rangle-S(S+1)]/(2S+2)$ is below 0.10
for every label (below 0.05 for the penalty calculations), and calculations
without a converged spin-pure state were not included. The resulting training
set comprises 3\,454 spin-pure states, 787 of them obtained with the spin
penalty.

Spin-adapted solvers, such as SU(2) DMRG~\cite{Sharma2012spin,Keller2016},
which is also available in \textsc{block2}, or FCIQMC in the graphical unitary
group approach~\cite{Paldus1974,Shavitt1977,Dobrautz2019GUGA,Dobrautz2021stochCASSCF},
require the same spatial orbitals for both spins. The
$S_z$-adapted formulation with a spin penalty was used instead, because the
descriptors and labels are defined in the unrestricted orbitals of the UHF
reference. Because the $\alpha$ and $\beta$ window orbitals span different
spaces, an exact spin eigenstate generally cannot be represented in the window,
so $\langle\hat S^2\rangle$ can only approach $S(S+1)$. Over all benchmark
labels, the largest remaining deviation is $w = 0.08$ (at most 0.05 for the
states of Table~\ref{tab:ood_results}).

\paragraph{Training and validation.}
To test generalization to unseen molecules rather than to unseen orbitals of
known molecules, the data were split by complex: all spin
states, charges, stretched geometries and repeated crystal-structure
determinations of a complex were placed together in the training, validation
or test set. The 3\,454 states belong to 1\,630 distinct complexes, which were
divided into 1\,103 complexes for training, 125 for validation and 402 for
testing, which contain 2\,329, 259 and 866 states, respectively. No complex
occurs in more than one set. Model~1 was trained on the
Box--Cox-transformed entropies with a mean-absolute-error loss, and the
checkpoint with the lowest validation error was retained. Five models were
trained from different initial weights on the same split, and their mean
prediction is used throughout. On the held-out test complexes the ensemble
reaches an MAE of 0.025 in $s^{(1)}$ (0.080 on the orbitals with
$s^{(1)}\ge0.10$) and classifies orbitals as active ($s^{(1)}\ge0.10$) with a
precision of 0.86, a recall of 0.80 and $F_1=0.83$, pooled over all test
orbitals (single models: $F_1=0.818\pm0.002$).

For comparison with the model of
Golub \emph{et al.}~\cite{golub2021}, whose code and trained models are not
available, we retrained its architecture (five hidden layers of 896 units,
scoring each orbital separately) with its 29 orbital features on the same
data, split and initial seeds. This model reaches $F_1=0.69$ on the test
complexes (MAE 0.036) and 0.47 on the benchmark of Sec.~\ref{sec:benchmark}
(mean over states, 0.54 pooled). With the 60 descriptors of this work the same
network reaches 0.80 and 0.62 (0.72), and \texttt{SpinAware} 0.83 and 0.65 (0.77).
Most of the gain therefore comes from the spin-resolved descriptors, and processing
the orbitals jointly adds a smaller further improvement (Supplementary
Table~S25).

Model~2 was trained on
the logarithm of the mutual information, on the same training, validation and test
complexes as Model~1, using the 3\,240 states for which mutual-information
labels are available (2\,174, 241 and 825 states). Five models were trained
from different initial weights. On the test complexes the predicted and DMRG
values of $\ln I_{ij}$ correlate with a coefficient of $0.865\pm0.001$, and
pairs with $I_{ij}>0.005$ are identified with an area under the receiver
operating characteristic curve of 0.918 (mean and standard deviation over the
five models). All scalers and transformation parameters
were fitted on the training data only. Architectures, optimizers and all
other hyperparameters of both models are listed in Supplementary
Table~S4.

\paragraph{Correlated reference calculations.}
To judge the predicted active spaces by the quantity of chemical
interest, the spin-state gap, we computed the energies of the selected
spaces with CASCI, in the
fixed orbitals of the lowest spin-clean UHF reference
(Sec.~\ref{ssec:noon}). Spaces of up to 16 orbitals were solved exactly
and larger spaces by DMRG-CASCI, using spin-adapted [SU(2)] DMRG in the UHF
natural orbitals (UNOs), so that each state has the intended total spin by
construction. Dynamic correlation outside the active
space was added with strongly contracted NEVPT2~\cite{Angeli2001}, using
reduced density matrices from DMRG where needed. Full-molecule
calculations with unrestricted coupled cluster with singles, doubles and
perturbative triples [UCCSD(T)]~\cite{Raghavachari1989} and frozen core, together
with the $T_1$ diagnostic~\cite{Lee1989}, serve as an independent reference
for the spin-state gaps. All calculations used the def2-SVP basis.

\paragraph{Software.}
Mean-field, CASCI, NEVPT2 and coupled-cluster calculations were performed
with PySCF~\cite{pyscf2018,pyscf2020}, DMRG calculations with
\textsc{block2}~\cite{Zhai2023}, and the models were implemented in
PyTorch~\cite{Paszke2019} and trained with AdamW~\cite{Loshchilov2019} and
cosine learning-rate annealing~\cite{Loshchilov2017}. Data processing and
figures used NumPy~\cite{Harris2020}, SciPy~\cite{Virtanen2020},
scikit-learn~\cite{Pedregosa2011} and Matplotlib~\cite{Hunter2007}. The quantum calculations used
Qiskit~\cite{Qiskit2024}, \texttt{ffsim}~\cite{ffsim} and
\texttt{qiskit-addon-sqd}~\cite{qiskitaddonsqd} on IBM quantum processors
(Sec.~\ref{sec:sqd}). Crystal structures were taken from the Cambridge
Structural Database~\cite{Groom2016}. Software versions are listed in
Supplementary Table~S5.

\subsection{Out-of-distribution benchmark}
\label{sec:benchmark}

\paragraph{Design principle.}
A model for active-space selection is useful only if it transfers beyond the
chemistry that generated its training labels. We therefore assessed
transferability on a benchmark of experimentally characterized complexes whose
difficulty was defined from chemical composition and connectivity alone. The
assignment uses neither the model nor the reference wavefunctions, so the
difficulty scale is independent of the quantity being evaluated. Descriptors of
electronic complexity obtained from the reference calculations (active-space
size, maximal single-orbital entropy and the entropy gap at the selection
threshold) were used only to interpret errors, not to assign difficulty.
Throughout, ``out of distribution'' refers to chemistry, that is, to metals,
ligands, binding modes and oxidation states absent from the training set, and not to
the descriptor space, in which the benchmark orbitals lie close to the training
data (Sec.~\ref{ssec:transfer}).

\paragraph{Reference chemical space.}
The training set (Sec.~\ref{sec:data}) comprises 3\,454 spin-pure
calculations on 1\,895 distinct mononuclear structures of Cr, Mn, Fe, Co
and Ni. The 2\,912 constructed complexes carry two to six monodentate
ligands on a metal in oxidation state $+2$ or $+3$, taken from a fixed
library of \ce{F-}, \ce{Cl-}, \ce{Br-}, \ce{OH-}, \ce{S^2-}, \ce{H2O},
\ce{NH3}, \ce{NCS-}, \ce{N3-} and CO. Their geometries are based on the idealized
structures of the SC1MC data set~\cite{Golub2022SC1MC}, which were
optimized at the Hartree--Fock level, and on complexes derived from these
templates by ligand exchange at standard metal--ligand distances. All spin
states of a complex share one geometry, at which both the UHF input and
the DMRG label of each state are computed. The 542 complexes taken from
experimental crystal structures add organic ligands and other binding
modes. No benchmark complex, or any structure sharing its crystallographic
refcode, is part of the training set. The benchmark instead introduces
motifs that the training geometries, checked through their metal--ligand
and ligand--ligand connectivity, do not contain or contain only in
isolated cases: the metals Ti and V, metal--methyl bonds, $\eta^3$-allyl
and $\eta^6$-arene ligands, a morpholine donor, a halide bridging to
aluminum, and zero-valent chromium carrying four or five carbonyl
ligands (\cpd{5} and \cpd{6}). Chelating ligands, by contrast, are common
in training and are not counted as novel. Because all training elements
lie at or below Kr and are described all-electron by the def2-SVP basis,
while heavier elements would require an effective core potential that
changes the orbital descriptors, the benchmark was restricted to
complexes of elements up to Kr (Supplementary Section~S1).

A test complex can depart from this space along four independent directions.
A direction counts as a departure when the corresponding motif is absent from
the training geometries or occurs there only in isolated cases:
\begin{itemize}
  \item[\textbf{M}] \emph{metal identity}: the metal is absent from training (here Ti and V);
  \item[\textbf{L}] \emph{ligand chemistry}: at least one ligand type is absent or rare;
  \item[\textbf{T}] \emph{binding topology}: at least one binding mode is absent or rare; chelation as such is common and does not count;
  \item[\textbf{O}] \emph{formal oxidation state}: the oxidation state lies outside $\{+2,+3\}$ of the constructed complexes and its characteristic coordination (here zero-valent chromium with four or five carbonyls) is absent.
\end{itemize}
We define the novelty index $\Nood \in \{0,\dots,4\}$ as the number of these
directions along which a complex departs from the training distribution.

\begin{figure}[tp]
  \centering
  \boxedgraphics{width=\dimexpr\linewidth-9pt\relax,height=0.55\textheight,keepaspectratio}{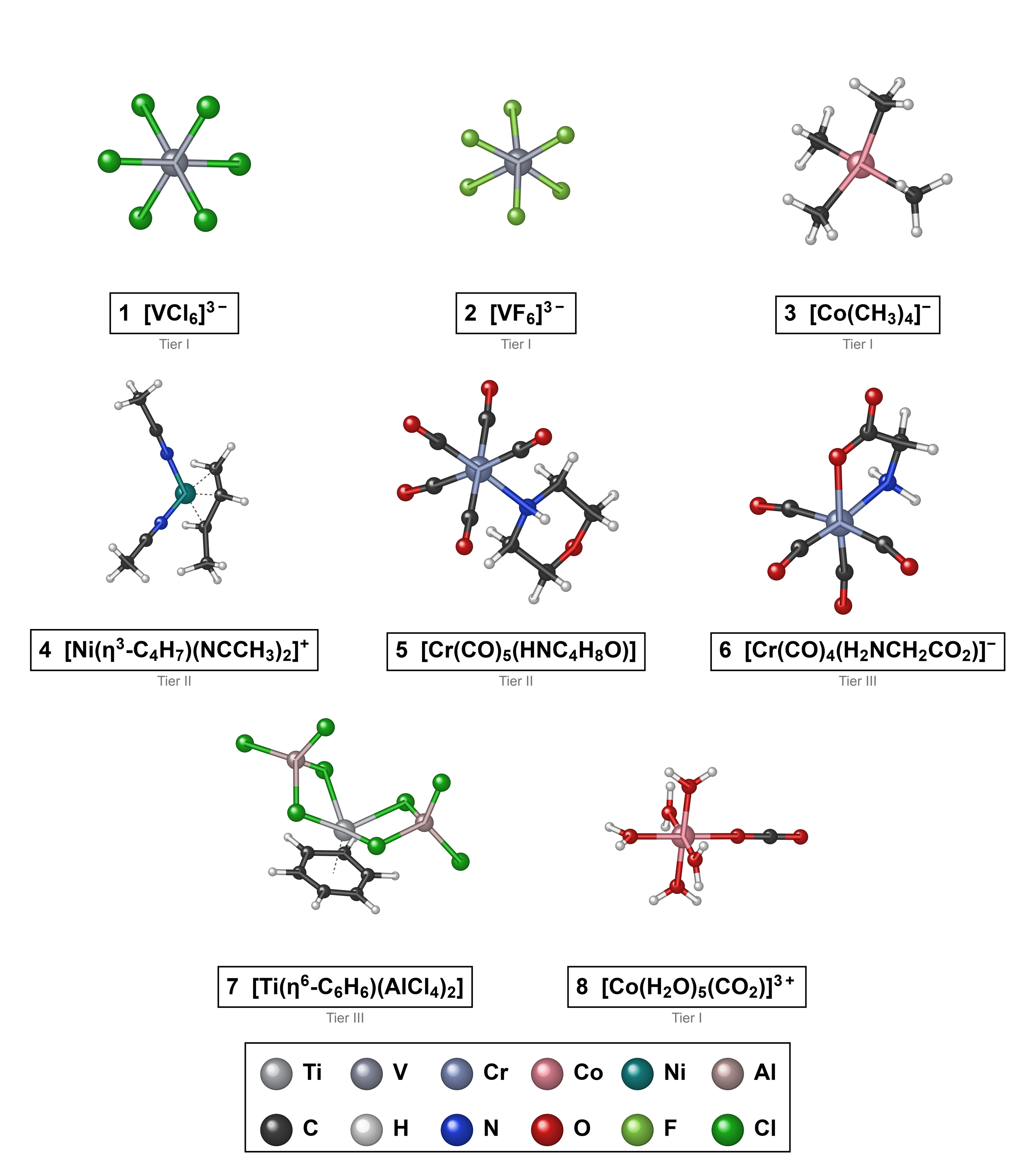}
  \caption{\textbf{Structures of the complexes studied in the main text.}
    \cpd{1}--\cpd{7}: out-of-distribution benchmark complexes, taken from
    experimental crystal structures (counter-ions and lattice solvent omitted).
    \cpd{8}: application system \CoCOthree\ at the equilibrium geometry of the
    published Co--O dissociation series (Co--O 1.97~\AA), with \ce{CO2} bound
    end-on through one oxygen atom. The tier of each complex (Table~\ref{tab:ood_set})
    is given below its formula. The complexes of Supplementary Section~S4 are
    shown in Supplementary Fig.~S1.}
  \label{fig:structures}
\end{figure}

\paragraph{Difficulty tiers.}
Complexes with $\Nood \le 1$ were classed as near-distribution (Tier~I), those
with $\Nood = 2$ as intermediate (Tier~II), and those with $\Nood \ge 3$ as
far-from-distribution (Tier~III). The seven complexes discussed in the main
text are listed in Table~\ref{tab:ood_set} and shown in
Fig.~\ref{fig:structures}. They span five metals (Ti, V, Cr,
Co and Ni), $d^2$ to $d^8$ electron counts, formal oxidation states from 0 to
$+3$, three to six coordination sites and 7 to 26 atoms. The two
vanadium(III) ions \cpd{1} and \cpd{2} probe the response of the active
space to ligand field strength and low-symmetry distortion at a fixed
metal. Five of the seven
crystal structures are salts. In every case the calculation was performed on the
isolated metal-containing species, and counter-ions and lattice solvent were
omitted. The source crystal structures and the omitted counter-ions are listed
in Supplementary Table~S1.

\begin{table*}[t]
\centering
\caption{\textbf{Out-of-distribution benchmark complexes: composition, novelty
and coordination geometry.} Novelty directions: M, metal; L, ligand;
T, binding topology; O, oxidation state. Multiplicities are given as $2S+1$;
those marked $^{\ast}$ have UHF references outside the spin-contamination
criterion and are reported separately. Sites count an $\eta^n$ ligand as one
site at its centroid (ctr). $\tau_4$ is the four-coordinate geometry index;
$\Sigma$ is the sum of deviations of the twelve \emph{cis} angles from
$90^\circ$; bite, chelate bite angle.}
\label{tab:ood_set}\label{tab:ood_geom}
\footnotesize
\setlength{\tabcolsep}{3.5pt}
\renewcommand{\arraystretch}{1.15}
\def\cell#1{\begin{tabular}[t]{@{}l@{}}#1\end{tabular}}
\begin{tabular}{@{}cl ccc ccc c l l l@{}}
\toprule
 & & \multicolumn{3}{c}{Electronic} & \multicolumn{3}{c}{Novelty} &
 \multicolumn{4}{c}{Coordination geometry} \\
\cmidrule(lr){3-5}\cmidrule(lr){6-8}\cmidrule(l){9-12}
No. & Complex & $d^n$ & $q$ & $2S+1$ & Directions & $\Nood$ & Tier &
Sites & Geometry & M--L (\AA) & Descriptors \\
\midrule
\cpd{1} & \ce{[VCl6]^3-}                        & $d^2$ & $-3$ & 1, 3             & M       & 1 & I   & 6 & octahedral & V--Cl 2.384--2.389 & $\Sigma = 2.5^\circ$ \\
\cpd{2} & \ce{[VF6]^3-}                         & $d^2$ & $-3$ & 1, 3             & M       & 1 & I   & 6 & \cell{distorted\\octahedral} & V--F 1.899--2.018 & $\Sigma = 29.0^\circ$ \\
\cpd{3} & \ce{[Co(CH3)4]-}                      & $d^6$ & $-1$ & 1, 3$^{\ast}$, 5 & L       & 1 & I   & 4 & distorted & Co--C 1.951--2.011 & $\tau_4 = 0.39$ \\
\addlinespace
\cpd{4} & \ce{[Ni({$\eta^3$}{-}C4H7)(NCCH3)2]+} & $d^8$ & $+1$ & 1, 3             & L, T    & 2 & II  & 3 & planar & \cell{Ni--N 1.882--1.890\\Ni--C 1.854--2.133} & $\sum\angle = 359.9^\circ$ \\
\cpd{5} & \ce{[Cr(CO)5(HNC4H8O)]}               & $d^6$ & $0$  & 1, 3$^{\ast}$, 5, 7$^{\dagger}$ & L, O    & 2 & II  & 6 & octahedral & \cell{Cr--N 2.196\\Cr--C 1.820--1.907} & $\Sigma = 23.3^\circ$ \\
\addlinespace
\cpd{6} & \ce{[Cr(CO)4(H2NCH2CO2)]-}            & $d^6$ & $-1$ & 1, 3$^{\ast}$    & L, O    & 2 & II  & 6 & \cell{distorted\\octahedral} & \cell{Cr--N 2.150\\Cr--O 2.218\\Cr--C 1.731--1.891} & \cell{$\Sigma = 55.2^\circ$\\bite $75.1^\circ$} \\
\cpd{7} & \ce{[Ti({$\eta^6$}{-}C6H6)(AlCl4)2]}  & $d^2$ & $0$  & 1, 3             & M, L, T & 3 & III & 5 & piano stool & \cell{Ti--ctr 2.069\\Ti--Cl 2.589--2.611} & \cell{Cl--Ti--Cl\\$77.7$--$78.1^\circ$} \\
\bottomrule
\end{tabular}
\end{table*}

\paragraph{Structural features.}
Coordination geometries were characterized from the experimental coordinates
(Table~\ref{tab:ood_geom}), taken from the Cambridge Structural
Database~\cite{Groom2016}. An $\eta^n$ ligand is counted as one coordination
site located at the centroid of its bound carbon atoms. Four-coordinate
geometries are described by $\tau_4 = [360^\circ - (\alpha + \beta)]/141^\circ$,
where $\alpha$ and $\beta$ are the two largest L--M--L angles ($\tau_4 = 1$ for
a tetrahedron and 0 for a square plane)~\cite{Yang2007tau4}, and octahedral distortion by
$\Sigma = \sum_{i=1}^{12} |90^\circ - \theta_i|$ over the twelve \emph{cis}
angles.
The structures of the individual complexes are described in
Supplementary Section~S1.

\paragraph{Spin states and reference wavefunctions.}
For each complex we considered the spin multiplicities accessible for its
$d$-electron count (Table~\ref{tab:ood_set}). UHF
references were obtained from multiple initial guesses followed by internal
stability analysis, and the lowest-energy solution satisfying
$|\langle \hat{S}^2\rangle - S(S+1)| \le \max[0.1,\,0.1\,S(S+1)]$ was retained.
States for which no such solution was found are marked in
Table~\ref{tab:ood_set} and excluded from aggregate statistics. Screening for
the stable solution is essential: in a control calculation on
\ce{[Co(H2O)5(CO2)]^3+}, an unscreened quintet reference lay
149.8~m$E_\mathrm{h}$ above the stable solution and changed the DMRG active
space from three to ten orbitals. For \CoMe\ the lowest computed state is the
triplet, whose UHF reference does not meet this criterion. The retained singlet
and quintet are therefore not the lowest computed states and serve as test states.
For \CrMorph\ the triplet likewise fails the criterion. Its singlet, the expected
ground state of an 18-electron Cr(0) carbonyl, lies 50\kcal\ below the triplet
at the UHF level and 72\kcal\ below it in the window DMRG (each state in its own
orbitals) and is included in the benchmark.

\paragraph{Reference labels.}
Single-orbital entropies $s_1$ and mutual information $I_{ij}$ were computed
with DMRG as for the training labels (Sec.~\ref{sec:data}).
Orbitals with
$s_1 \ge 0.10$ were classed as active. The total spin of every benchmark label
was verified in the same way (Sec.~\ref{sec:data}), and the closed-shell singlets were
computed with the $\hat S^2$ penalty ($\langle\hat S^2\rangle \approx 10^{-5}$).
Because the window position and the
orbital ordering change with spin multiplicity, active orbitals were compared
between spin states through the molecular-orbital overlap
$|\mathbf{C}_1^{\mathsf{T}}\mathbf{S}\mathbf{C}_2|$ rather than by orbital index.

\section{Results and discussion}
\label{sec:results}

\subsection{Transferability to out-of-distribution complexes}
\label{ssec:transfer}

We evaluated \texttt{SpinAware} on six benchmark complexes that differ from the training
data in metal, ligand, binding topology or oxidation state: the
vanadium(III) ions \VClsix\ (\cpd{1}) and \VFsix\ (\cpd{2}), the
tetramethylcobaltate \CoMe\ (\cpd{3}), the chromium(0) carbonyls \CrMorph\
(\cpd{5}) and \CrGly\ (\cpd{6}), and the titanium arene complex \TiArene\
(\cpd{7}). The nickel allyl complex \NiAllyl\ (\cpd{4}) and
\CoCOtwo\ (\cpd{8}) are analyzed separately in
Secs.~\ref{ssec:app}--\ref{sec:sqd}, where the predicted active spaces are
carried through to multireference and quantum-hardware calculations. We analyze
the 11 spin states of these six complexes for which a spin-clean UHF
reference exists, because the descriptors, the DMRG labels and any subsequent
multireference calculation all rest on this reference. The remaining states are
listed in Supplementary Section~S4, together with the septet of \CrMorph, which
requires an electron outside the $d$ shell and is kept there as a high-spin test
case ($^{\dagger}$ in Table~\ref{tab:ood_set}). For every state, the predicted
single-orbital entropy profile is compared with the DMRG reference over the
full 40-orbital window.

An orbital is classed as active when its entropy reaches a threshold $\tau$.
Let $\mathcal{A}_{\mathrm{DMRG}}$ and $\mathcal{A}_{\mathrm{ML}}$ be the active
sets obtained from the reference and predicted entropies. Treating
$\mathcal{A}_{\mathrm{DMRG}}$ as the ground truth, precision
$P = |\mathcal{A}_{\mathrm{ML}} \cap \mathcal{A}_{\mathrm{DMRG}}| / |\mathcal{A}_{\mathrm{ML}}|$
is the fraction of predicted active orbitals that are correct, and recall
$R = |\mathcal{A}_{\mathrm{ML}} \cap \mathcal{A}_{\mathrm{DMRG}}| / |\mathcal{A}_{\mathrm{DMRG}}|$
is the fraction of true active orbitals that are recovered. Their harmonic mean,
\begin{equation}
  F_1 \;=\; \frac{2PR}{P+R}
      \;=\; \frac{2\,\mathrm{TP}}{2\,\mathrm{TP} + \mathrm{FP} + \mathrm{FN}},
\end{equation}
where TP, FP and FN count true-positive, false-positive and false-negative
orbitals, rewards active spaces that are both complete and compact: an active
space that includes every orbital ($R = 1$, low $P$) or only one correct orbital
($P = 1$, low $R$) both score poorly. When both sets are empty, $F_1 = 1$ by
convention. Alongside $F_1$ we report the orbital-wise MAE
of $s_1$. The two metrics are complementary: the MAE measures how faithfully
the entropy profile is reproduced, whereas $F_1$ measures whether the resulting
active-space decision is correct.

A two-configuration model illustrates the chemical scale of the thresholds.
Consider a doubly occupied
orbital that loses a fraction $w$ of its electron pair to a correlating
partner, the elementary event of static correlation. Its single-orbital entropy
is then $s_1 = -(1-w)\ln(1-w) - w\ln w$, and its natural occupation is
$n = 2(1-w)$. The threshold $\tau = 0.10$, the value used in earlier
machine-learned active-space selection~\cite{golub2021}, corresponds to
$w \approx 0.02$, that is, to an orbital that carries about 2\% of a
doubly excited configuration and has a natural occupation of about 1.96 (or
0.04 for the receiving virtual orbital). The permissive threshold
$\tau = 0.07$ corresponds to $w \approx 0.013$ and occupations of about 1.97
and 0.03. It therefore lies close to the conventional natural-occupation criterion
for an active orbital, occupations outside 0.02--1.98~\cite{Pulay1988,Keller2015}, which
we also use in the NOON scans of
Sec.~\ref{ssec:noon}. These thresholds provide practical screening
criteria, whose consequences for the selected spaces are assessed by the
NOON scans and energy comparisons.

\begin{table*}[t]
\centering
\caption{\textbf{Prediction quality on the out-of-distribution benchmark.}
$n_{\mathrm{DMRG}}$ and $n_{\mathrm{ML}}$ are the numbers of orbitals with
$s_1 \ge \tau$ in the reference and predicted profiles; MAE is the orbital-wise
mean absolute error of $s_1$ over the 40-orbital window. Predictions:
mean of the five Model~1 runs; references: spin-checked DMRG labels
(Sec.~\ref{sec:benchmark}).}
\label{tab:ood_results}
\footnotesize
\setlength{\tabcolsep}{4pt}
\begin{tabular}{@{}cllccccccc@{}}
\toprule
 & & & \multicolumn{3}{c}{$\tau = 0.10$} & \multicolumn{3}{c}{$\tau = 0.07$} & \\
\cmidrule(lr){4-6}\cmidrule(lr){7-9}
Tier & Complex & $2S+1$ & $n_{\mathrm{DMRG}}$ & $n_{\mathrm{ML}}$ & $F_1$ & $n_{\mathrm{DMRG}}$ & $n_{\mathrm{ML}}$ & $F_1$ & MAE \\
\midrule
I   & \VClsix\ (\cpd{1}) & 1 & 8  & 4  & 0.67 & 10 & 9  & 0.95 & 0.047 \\
    &                    & 3 & 4  & 6  & 0.80 & 4  & 9  & 0.62 & 0.017 \\
    & \VFsix\ (\cpd{2})  & 1 & 6  & 5  & 0.73 & 7  & 16 & 0.61 & 0.044 \\
    &                    & 3 & 0  & 5  & 0.00 & 2  & 13 & 0.27 & 0.022 \\
    & \CoMe\ (\cpd{3})   & 1 & 9  & 9  & 1.00 & 14 & 13 & 0.81 & 0.021 \\
    &                    & 5 & 4  & 0  & 0.00 & 8  & 6  & 0.71 & 0.014 \\
\addlinespace
II  & \CrMorph\ (\cpd{5}) & 1 & 11 & 14 & 0.88 & 19 & 19 & 0.95 & 0.024 \\
    &                     & 5 & 11 & 11 & 0.82 & 22 & 16 & 0.79 & 0.031 \\
    & \CrGly\ (\cpd{6})   & 1 & 18 & 15 & 0.79 & 26 & 25 & 0.98 & 0.024 \\
\addlinespace
III & \TiArene\ (\cpd{7}) & 1 & 8  & 11 & 0.84 & 9  & 11 & 0.90 & 0.029 \\
    &                     & 3 & 4  & 8  & 0.67 & 6  & 11 & 0.71 & 0.020 \\
\midrule
\multicolumn{3}{@{}l}{Tier I mean (6 states)}   & & & 0.53 & & & 0.66 & 0.028 \\
\multicolumn{3}{@{}l}{Tier II mean (3 states)}  & & & 0.83 & & & 0.91 & 0.026 \\
\multicolumn{3}{@{}l}{Tier III mean (2 states)} & & & 0.75 & & & 0.80 & 0.025 \\
\multicolumn{3}{@{}l}{All (11 states)}          & & & 0.65 & & & 0.75 & 0.027 \\
\bottomrule
\end{tabular}
\end{table*}

Averaged over the 11 spin states, \texttt{SpinAware} reproduces the DMRG entropy
profiles with an MAE of 0.027 and recovers the active space with $F_1 = $ 0.65
at $\tau = 0.10$ (Table~\ref{tab:ood_results}). Pooled over all orbitals of the
11 states, as for the test set, $F_1 = 0.77$ at $\tau = 0.10$ and 0.81 at $\tau = 0.07$,
and the MAE is 0.077 over the 83 DMRG-active orbitals and 0.015 over the remaining
357 orbitals.
We see no sign of degradation with
the novelty index: the mean $F_1$ is 0.53 for Tier~I, 0.83 for Tier~II and
0.75 for Tier~III, although Tiers~II and III contain only two complexes
and one complex. The chromium glycinate \CrGly, which departs from the
training set in ligand and oxidation state, and the titanium arene
\TiArene, which departs in metal, ligand and binding topology, are among the best-predicted systems ($F_1 = $ $0.67$--$0.84$ at
$\tau = 0.10$ and $0.71$--$0.98$ at $\tau = 0.07$), whereas the most difficult
cases are the Tier~I vanadium and cobalt ions. The tier-averaged MAE is
likewise flat (0.028, 0.026 and 0.025 for Tiers~I--III). Chemical novelty as
such is therefore not what limits transfer. The model's input is a set of
orbital-resolved electronic quantities (energies, exchange couplings,
metal--ligand composition and spin polarization) rather than an identity of
the ligands, so what it has to transfer is how a correlated metal--ligand
orbital looks electronically, not which ligands it has seen. The
comparison with the architecture of Golub \emph{et al.} in
Sec.~\ref{sec:data} points the same way: on this benchmark it reaches
$F_1 = 0.47$ with its 29 orbital features and 0.62 with our 60
descriptors, against 0.65 for \texttt{SpinAware}, so most of the transfer
comes from the spin-resolved descriptors, and processing the orbitals
jointly adds a smaller improvement. This is also
seen directly in the space of the model's inputs. For every orbital we computed
the distance to its nearest training orbital in the standardized
60-dimensional descriptor space and compared it with the same distance for
complexes held out from training (10\% of the training complexes, all their
spin states and stretched geometries kept together). The benchmark orbitals lie
at a median distance only 1.09 times that of the held-out complexes, and 1.8\%
of them fall beyond the 95th percentile of the held-out distribution, compared
with 5\% by construction for the held-out complexes themselves. Of the
correlated orbitals ($s_1 \ge 0.10$), only one lies beyond it. The distance does
not follow the novelty tiers: the Tier~III titanium arene \TiArene\ lies at the
held-out median, whereas the Tier~I complexes \VFsix\ and \CoMe\ are the
farthest from training. The benchmark is therefore new in its chemistry but not
in the electronic patterns of its orbitals, which the descriptors map onto
those of the training set. The prediction error depends only weakly on this
distance (rank correlation 0.04 over all orbitals and 0.15 over the correlated
ones). Only for the correlated orbitals farthest from the training data (top
quarter of the distance) does the MAE rise, to 0.11 against 0.06--0.07 for the
others. The residual
errors trace instead to two electronic regimes, described below, that can
occur in any tier.

\begin{figure}[tp]
  \centering
  \boxedgraphics{width=\dimexpr\linewidth-9pt\relax,height=0.75\textheight,keepaspectratio}{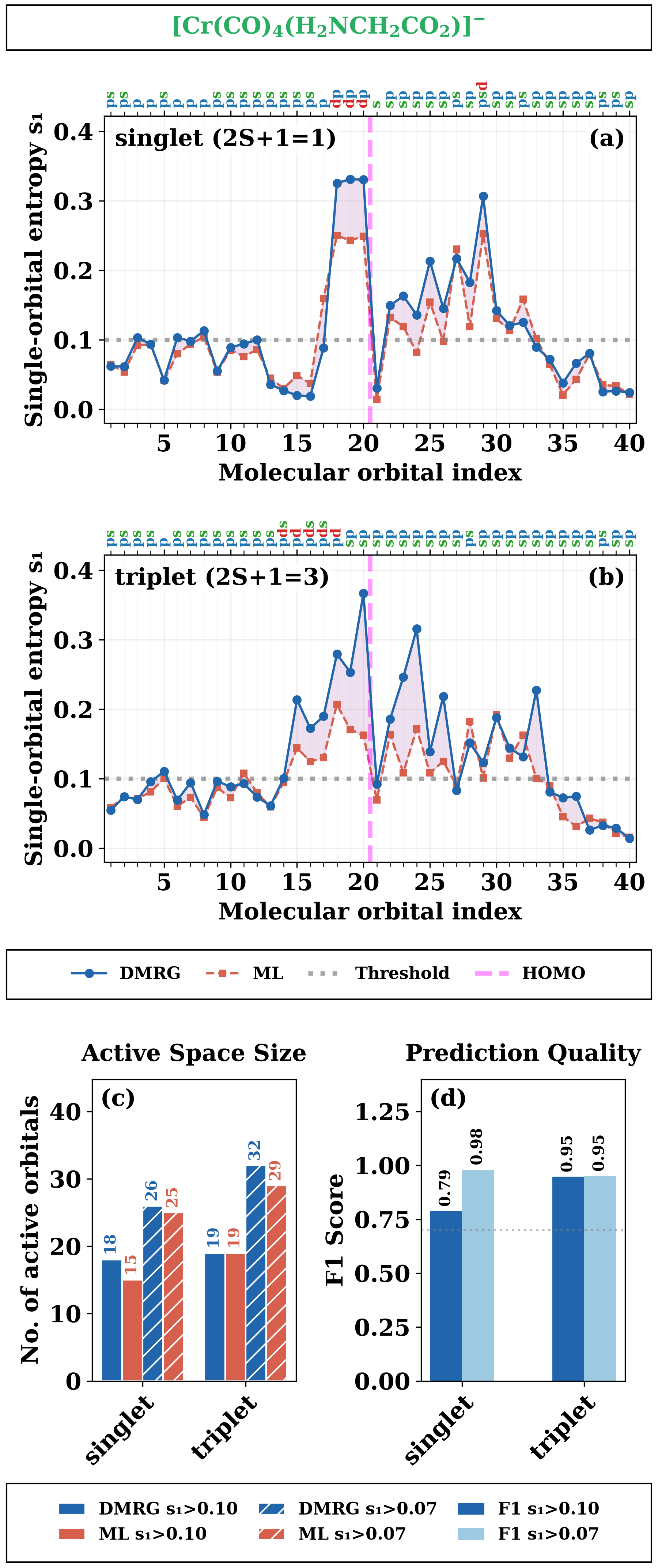}
  \caption{\textbf{\CrGly\ (\cpd{6}, Tier~II): single-orbital entropies.} An N,O-chelating amino acid on zero-valent chromium, a rare ligand and an oxidation state absent from training, is transferred accurately. Panels: singlet (a) and triplet (b) profiles; number of active orbitals (c) and $F_1$ (d). Upper panels: $s_1$ over the 40-orbital window from DMRG (blue) and \texttt{SpinAware} (red); dotted line, $\tau = 0.10$; dashed line, HOMO--LUMO boundary; letters, dominant orbital character. Lower panels: solid and hatched bars, $\tau = 0.10$ and 0.07; dotted line, $F_1 = 0.70$.}
  \label{fig:ood_c8}
\end{figure}

The chromium carbonyls and the titanium arene complex contain ligands, binding
modes and oxidation states that are absent from or rare in training: zero-valent chromium carrying four or five
$\pi$-acidic carbonyls in \CrMorph\ and \CrGly, an N,O-chelating amino acid in
\CrGly, and an $\eta^6$-arene on titanium(II) in \TiArene. In the chromium(0) carbonyls,
the $d^6$ metal donates electron density from its filled $t_{2g}$-type
orbitals into the empty $\pi^\ast$ orbitals of the carbonyls, and the
carbonyls donate into the metal through $\sigma$ bonds. The strongly correlated
orbitals are therefore the metal-d--ligand bonding and antibonding
combinations on either side of the frontier gap, labeled d/p in the entropy
plots. For the glycinate \CrGly\ (Fig.~\ref{fig:ood_c8}), the predicted profile
follows the reference across the whole window and places its maxima on these
orbitals, and at $\tau = 0.07$ it recovers the 26-orbital DMRG active space with
$F_1 = $ 0.98. For the morpholine complex \CrMorph\ (Supplementary Fig.~S4),
\texttt{SpinAware} selects the same manifold with $F_1 = $ 0.88 and 0.82 in the singlet and quintet. These active spaces are large (11--18 orbitals at $\tau = 0.10$),
so the agreement cannot be attributed to a few dominant orbitals.

\begin{figure}[tp]
  \centering
  \boxedgraphics{width=\dimexpr\linewidth-9pt\relax,height=0.75\textheight,keepaspectratio}{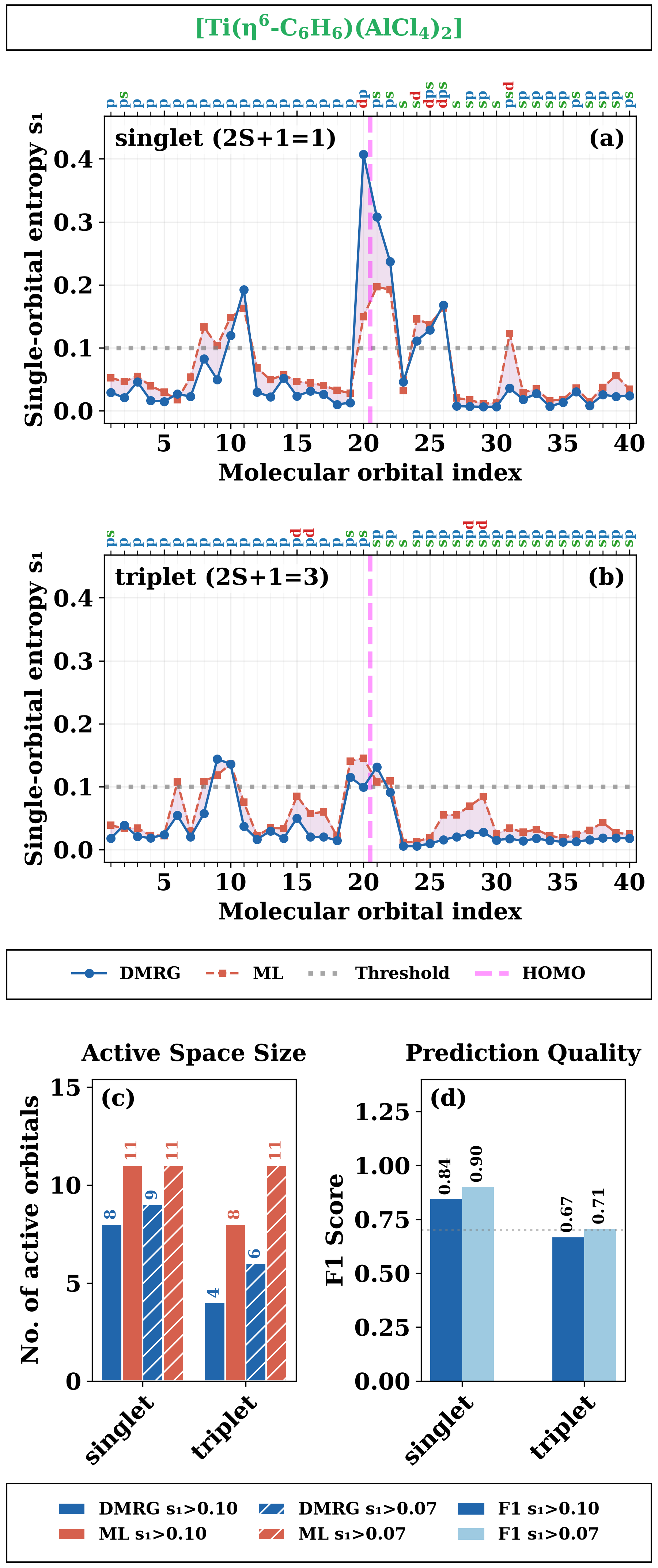}
  \caption{\textbf{\TiArene\ (\cpd{7}, Tier~III): single-orbital entropies.} An $\eta^6$-arene complex of titanium, a metal absent from training, is transferred accurately. Panels: singlet (a) and triplet (b) profiles; number of active orbitals (c) and $F_1$ (d). Upper panels: $s_1$ over the 40-orbital window from DMRG (blue) and \texttt{SpinAware} (red); dotted line, $\tau = 0.10$; dashed line, HOMO--LUMO boundary; letters, dominant orbital character. Lower panels: solid and hatched bars, $\tau = 0.10$ and 0.07; dotted line, $F_1 = 0.70$.}
  \label{fig:ood_c9}
\end{figure}

\begin{figure}[tp]
  \centering
  \includegraphics[width=\linewidth,height=0.6\textheight,keepaspectratio]{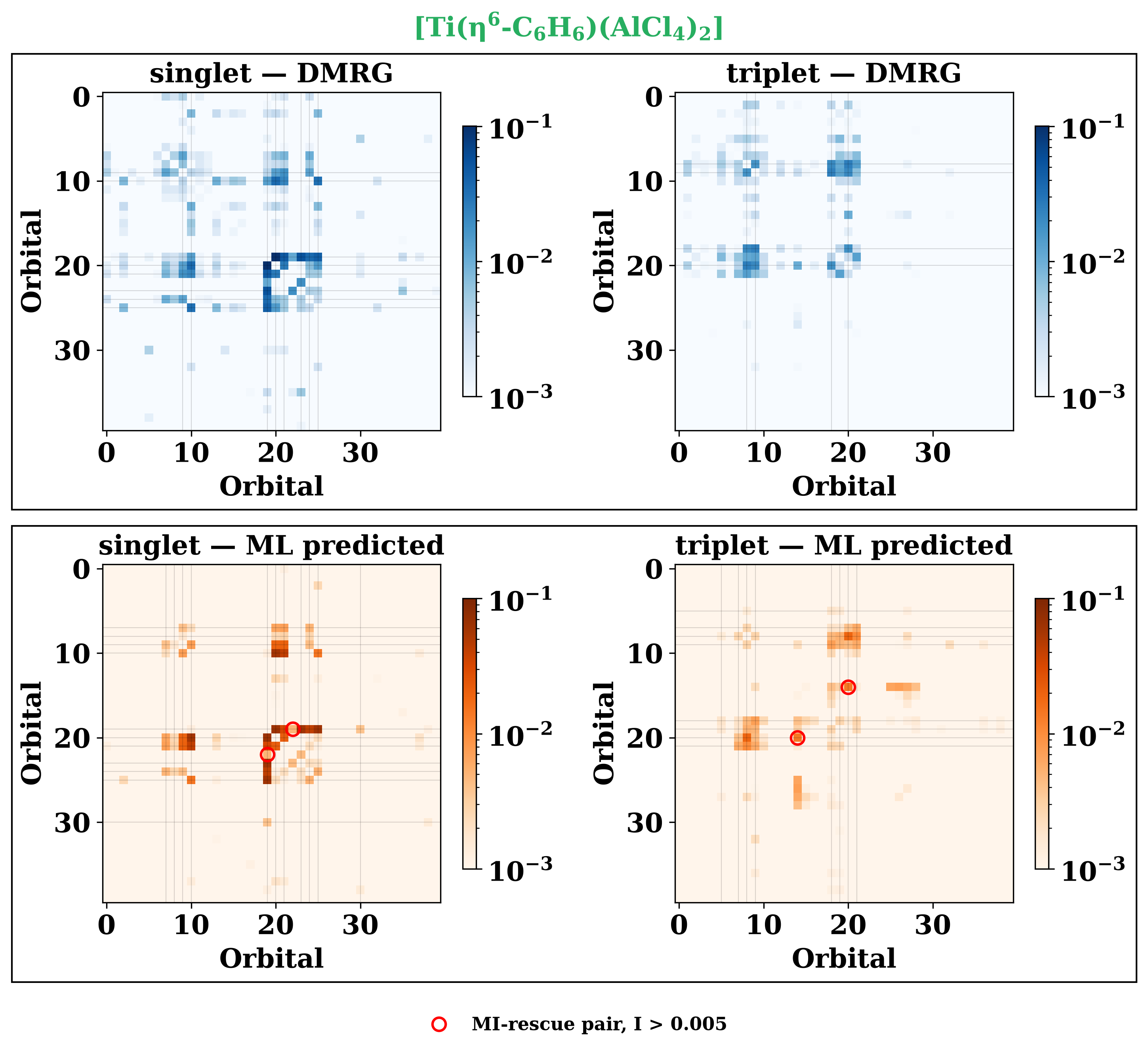}
  \caption{\textbf{\TiArene\ (\cpd{7}, Tier~III): mutual information.} The two strongly coupled orbital blocks of the reference appear in the prediction for both spin states. Panels: singlet (left) and triplet (right); DMRG (top row) and \texttt{SpinAware} (bottom row). Logarithmic color scale over the 40-orbital window; red circles, rescue pairs (predicted $s_1 < 0.10$ and $I_{ij} > 0.005$ with an orbital of predicted $s_1 \ge 0.10$).}
  \label{fig:mi_c9}
\end{figure}

The titanium arene complex \TiArene\ shows this transfer for a metal that is
absent from training (Figs.~\ref{fig:ood_c9} and \ref{fig:mi_c9}). Titanium(II)
is a $d^2$ ion, and its two d electrons are partly delocalized onto the
benzene ring by metal-to-arene back-donation. The shortened and lengthened
C--C bonds of the coordinated ring in the crystal structure are the
structural signature of this interaction. Accordingly, the dominant
correlation in the singlet resides in a frontier orbital of mixed Ti~d and
arene $\pi^\ast$ character (DMRG $s_1 = 0.41$). The prediction under-estimates
this orbital but reproduces the correlated orbitals immediately above the gap
($F_1 = $ 0.84 at $\tau = 0.10$ and 0.90 at $\tau = 0.07$, MAE 0.029). In the
triplet, all four DMRG-active
orbitals are recovered, together with four further orbitals near the threshold
($F_1 = 0.67$ and 0.71, MAE 0.020).
The predicted mutual information reproduces both strongly coupled blocks of
the reference in each spin state (Fig.~\ref{fig:mi_c9}): the metal--arene
block around the frontier gap and a second block of Ti--Cl orbitals of the
tetrachloroaluminate ligands. Titanium(II) shares the $d^2$ configuration of
the vanadium(III) ions, which are not well predicted (see below), yet it is
predicted reliably, presumably because back-donation into the arene spreads its
correlation over several orbitals of moderate entropy rather than
concentrating it in one strongly correlated d pair.

\begin{figure}[tp]
  \centering
  \boxedgraphics{width=\dimexpr\linewidth-9pt\relax,height=0.80\textheight,keepaspectratio}{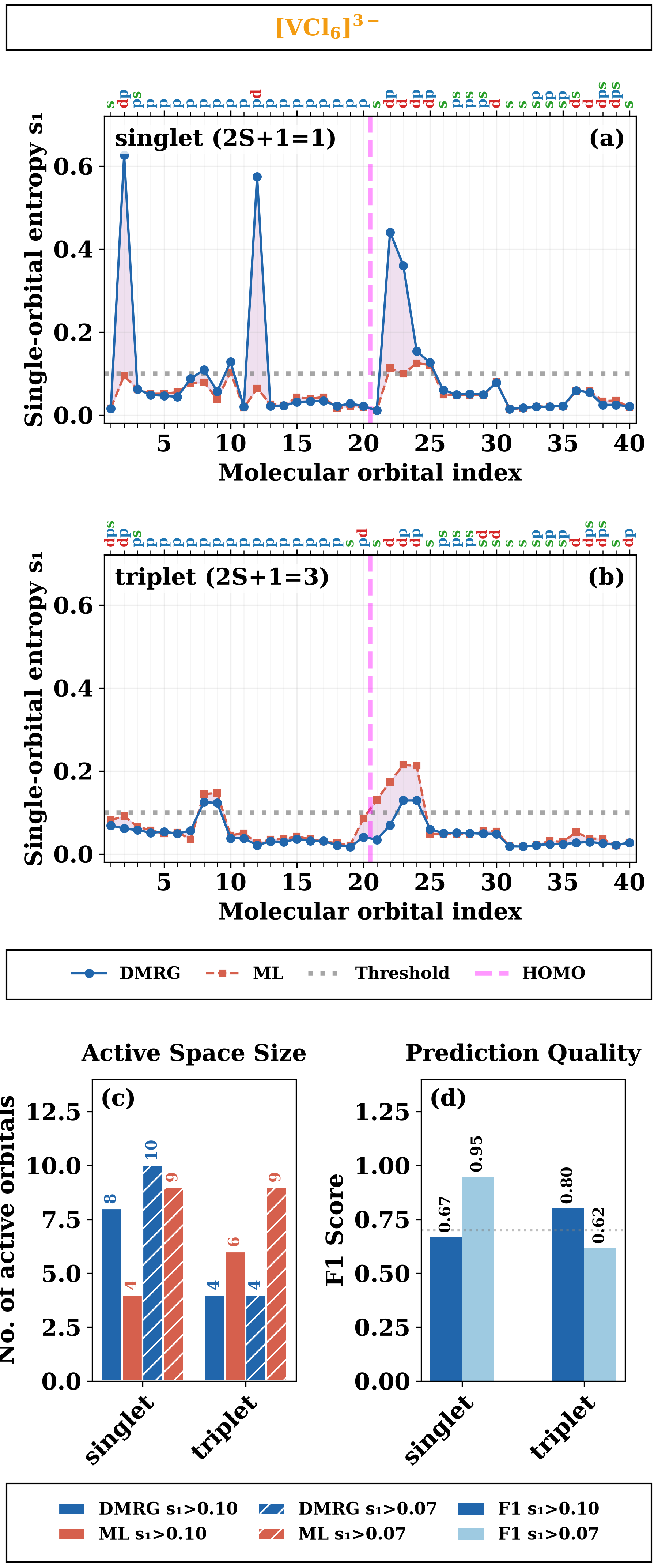}
  \caption{\textbf{\VClsix\ (\cpd{1}, Tier~I): single-orbital entropies.} The strongly correlated V~d orbitals of the singlet are located, but their entropies are compressed. Panels: singlet (a) and triplet (b) profiles; number of active orbitals (c) and $F_1$ (d). Upper panels: $s_1$ over the 40-orbital window from DMRG (blue) and \texttt{SpinAware} (red); dotted line, $\tau = 0.10$; dashed line, HOMO--LUMO boundary; letters, dominant orbital character. Lower panels: solid and hatched bars, $\tau = 0.10$ and 0.07; dotted line, $F_1 = 0.70$.}
  \label{fig:ood_c1}
\end{figure}

The residual errors arise in two electronic regimes. The first is
\emph{strong static correlation}. In the singlet states of
the vanadium(III) ions \VClsix\ and \VFsix\ (Fig.~\ref{fig:ood_c1}a and
Supplementary Fig.~S2a), DMRG assigns individual orbital entropies of about
0.6 (maximal $s_1$ = 0.63 and 0.66), close to the value of $\ln 2 \approx 0.69$ per orbital expected for a
pair of orbitals that share two electrons equally. These large entropies
identify open-shell singlet character: in an octahedral $d^2$ ion the two
electrons occupy two of three degenerate $t_{2g}$ orbitals, and pairing them
in a single orbital, as a closed-shell determinant requires, is energetically
unfavorable. The singlet is therefore an equal mixture of several
configurations, and a single determinant is a poor reference. \texttt{SpinAware} places
elevated entropy on the correct V~d orbitals but compresses its magnitude
(maximal predicted $s_1 \approx$ 0.15 for \VFsix, against 0.66 from DMRG),
which accounts for the largest MAEs in the set (0.044--0.047 for the vanadium
singlets). Such values lie in the tail of the training distribution, where the
median per-state maximal entropy is 0.22 and only 19\% of the states reach a
maximal entropy above 0.6, so the regression shrinks toward typical values. The error affects the
magnitude of the entropy, not the position of the correlated orbitals.

\begin{figure}[tp]
  \centering
  \boxedgraphics{width=\dimexpr\linewidth-9pt\relax,height=0.80\textheight,keepaspectratio}{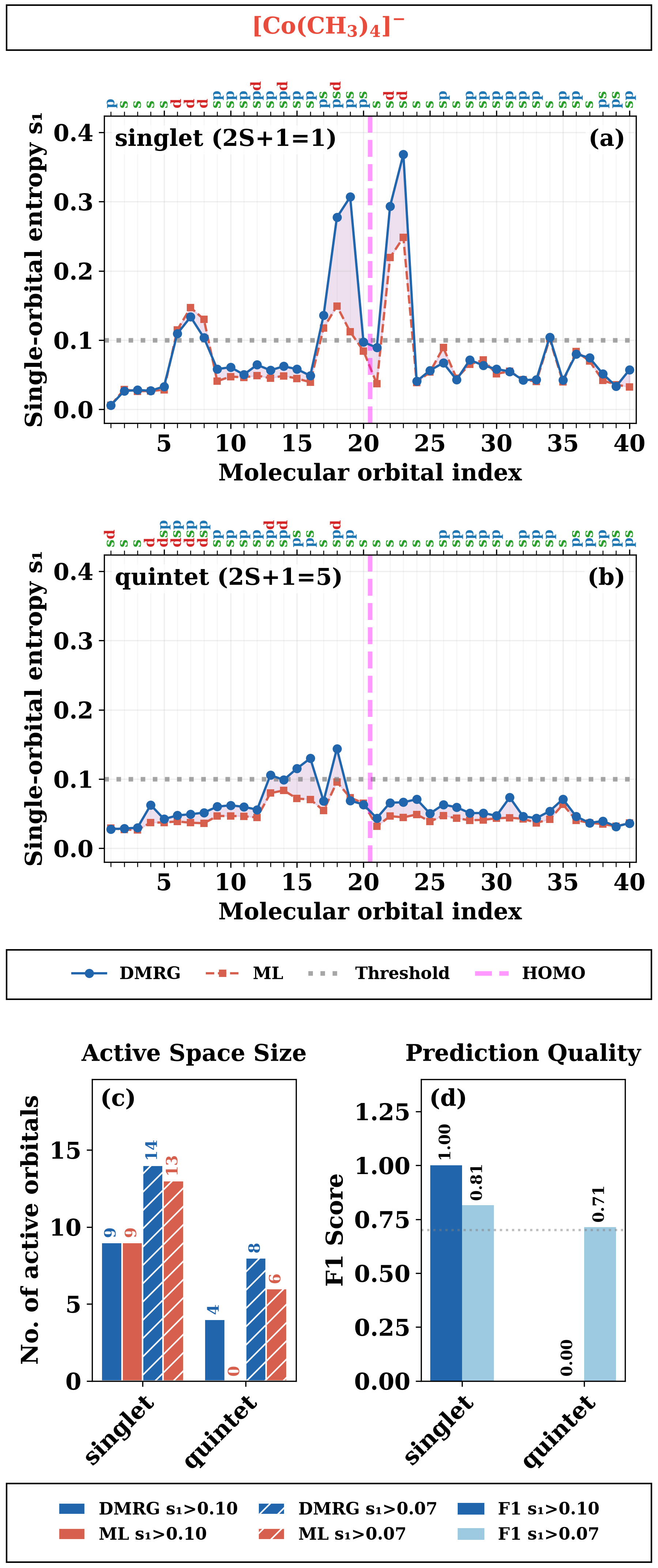}
  \caption{\textbf{\CoMe\ (\cpd{3}, Tier~I): single-orbital entropies.} Weak correlation near the threshold gives small errors but unstable active-space decisions in the high-spin states. Panels: singlet (a) and quintet (b) profiles; number of active orbitals (c) and $F_1$ (d). Upper panels: $s_1$ over the 40-orbital window from DMRG (blue) and \texttt{SpinAware} (red); dotted line, $\tau = 0.10$; dashed line, HOMO--LUMO boundary; letters, dominant orbital character. Lower panels: solid and hatched bars, $\tau = 0.10$ and 0.07; dotted line, $F_1 = 0.70$.}
  \label{fig:ood_c3}
\end{figure}

The second regime is \emph{weak correlation concentrated near the threshold}.
In the high-spin quintet of \CoMe\ and the triplet of \VFsix\
(Fig.~\ref{fig:ood_c3}c and Supplementary Fig.~S2b), the reference profiles are
almost flat, with all entropies below 0.15. In these states the d electrons
occupy separate orbitals with parallel spins. Exchange between parallel spins
stabilizes a single high-spin determinant, and the Pauli principle forbids
the pair excitations that create static correlation within the half-filled
orbitals, so little static correlation remains to be resolved. \texttt{SpinAware}
reproduces both profiles with small absolute errors (MAE 0.014--0.022), yet $F_1$ at $\tau = 0.10$
falls to zero, because a few orbitals lying just above $\tau$ in the reference
fall just below it in the prediction, or vice versa. Small differences near the entropy threshold can therefore change
the classification of an orbital. The consequences for the energies are
assessed separately for the application systems. A hard
threshold applied to a flat profile is an ill-conditioned decision, which is
why profile-level accuracy and set-level agreement must be assessed together.
The low entropies identify weakly correlated states within the
reference window.

The failure cases share chemical features that are under-represented in the
training data, which suggests where the learned representation is incomplete.
(i)~\emph{Electron count and metal size.} The training metals (Cr--Ni in
oxidation states $+2$ and $+3$) span $d^3$--$d^8$, so the $d^2$ configurations
of the vanadium(III) and titanium(II) complexes are unseen. Early transition
metals also have more diffuse 3d orbitals and smaller intra-atomic exchange
integrals, which shift the energetic and exchange descriptors on which the
model relies. That \TiArene\ is nonetheless well predicted, while the
vanadium(III) singlets are not, indicates that the bottleneck is the strongly
correlated $t_{2g}^2$ singlet manifold rather than the $d^2$ count itself.
(ii)~\emph{$\sigma$-alkyl donors.} \CoMe\ binds through four strongly covalent
Co--C $\sigma$ bonds with no $\pi$ component, a bonding motif absent from
training, where all carbon donors are $\pi$-acceptors. Its predicted profiles
are uniformly flatter than the reference in all three spin states.
(iii)~\emph{Spin-state discrimination is learned per metal.} For the metals
represented in training, the model reproduces the dependence of correlation on
spin state: for \CoMe, both the predicted and the DMRG maximal entropies
decrease from the singlet to the quintet, as expected when exchange
stabilization grows with the number of unpaired electrons. For vanadium, which
is absent from training, the predicted singlet is no more correlated than the triplet
(maximal predicted $s_1 \approx 0.12$ and 0.21 for \VClsix), whereas DMRG shows a strongly correlated singlet and a
weakly correlated triplet. Spin awareness therefore transfers across ligands
but not, at present, to unseen metals. This identifies early transition
metals, and strongly correlated low-spin states in general, as the most
valuable additions to future training sets. The states at risk cannot be
recognized from the UHF reference alone: the closed-shell solutions of both
vanadium singlets show no significant broken-symmetry instability (lowering
below 0.1~m$E_\mathrm{h}$). They are, however, recognized from their
chemistry (a low-spin state of a metal absent from training) before any
correlated calculation.
Lowering the threshold from 0.10 to 0.07 raises the mean $F_1$ from 0.65 to
0.75 (Table~\ref{tab:ood_results}), and it does so in every tier, most strongly
in Tier~I (0.53 to 0.66). In the back-bonding complexes \CrGly\ and
\TiArene, the model places the entropy of several correlated metal--ligand
orbitals just below 0.10, and the permissive threshold recovers them. The
average, however, conceals the opposite effect in weakly correlated states. In
the triplet of \VClsix, $F_1$ falls from 0.80 to 0.62, and in the triplet of
\VFsix\ the lower threshold admits 13 predicted orbitals against two in the
reference, because in flat profiles a lower threshold admits orbitals that
carry no static correlation. A fixed entropy threshold therefore cannot serve
every state equally well.

The choice also carries an asymmetric cost. Omitting a strongly correlated
orbital causes a qualitative failure of the subsequent multireference
calculation, whereas including a weakly correlated one only increases its cost.
A permissive entropy threshold, which favors recall, is therefore the safer
screening criterion, provided the resulting candidate set can be refined by a
criterion that adapts to the individual system.

Single-orbital entropy alone cannot provide this refinement: it measures how
correlated each orbital is, not which orbitals are correlated with each other.
Chemically, static correlation always involves at least two orbitals (the
orbital that loses electron density and the one that receives it, such as the
bonding and antibonding partners of a metal--ligand bond), and an orbital of
modest $s_1$ can be indispensable if it is the partner of a strongly
correlated one. Pairwise mutual information $I_{ij}$ supplies this missing
information. We therefore do not use any single threshold to fix the active
space. Instead, we combine the predicted entropies and mutual information into
a compact pool of candidate orbitals, from which the final active space is
selected by a NOON scan.

\begin{figure}[tp]
  \centering
  \includegraphics[width=\linewidth,height=0.6\textheight,keepaspectratio]{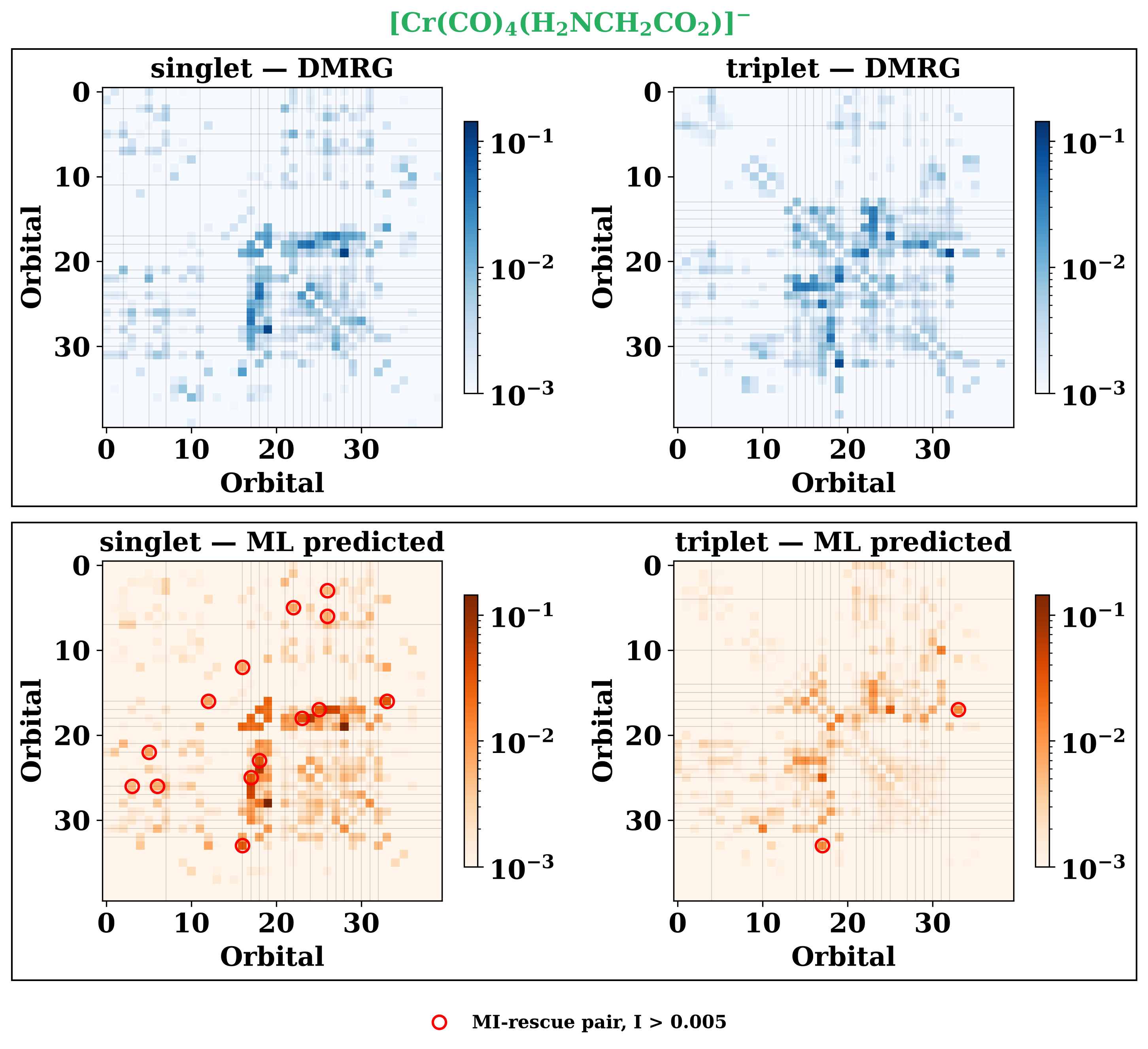}
  \caption{\textbf{\CrGly\ (\cpd{6}, Tier~II): mutual information.} The coupled metal--ligand block is recovered; in the singlet the pool holds all 18 DMRG-active orbitals, against 13 from the entropy alone. Panels: singlet (left) and triplet (right); DMRG (top row) and \texttt{SpinAware} (bottom row). Logarithmic color scale over the 40-orbital window; red circles, rescue pairs (predicted $s_1 < 0.10$ and $I_{ij} > 0.005$ with an orbital of predicted $s_1 \ge 0.10$).}
  \label{fig:mi_c8}
\end{figure}

\subsection{Mutual information and the candidate orbital pool}
\label{ssec:pool}

\begin{figure}[tp]
  \centering
  \includegraphics[width=\linewidth,height=0.6\textheight,keepaspectratio]{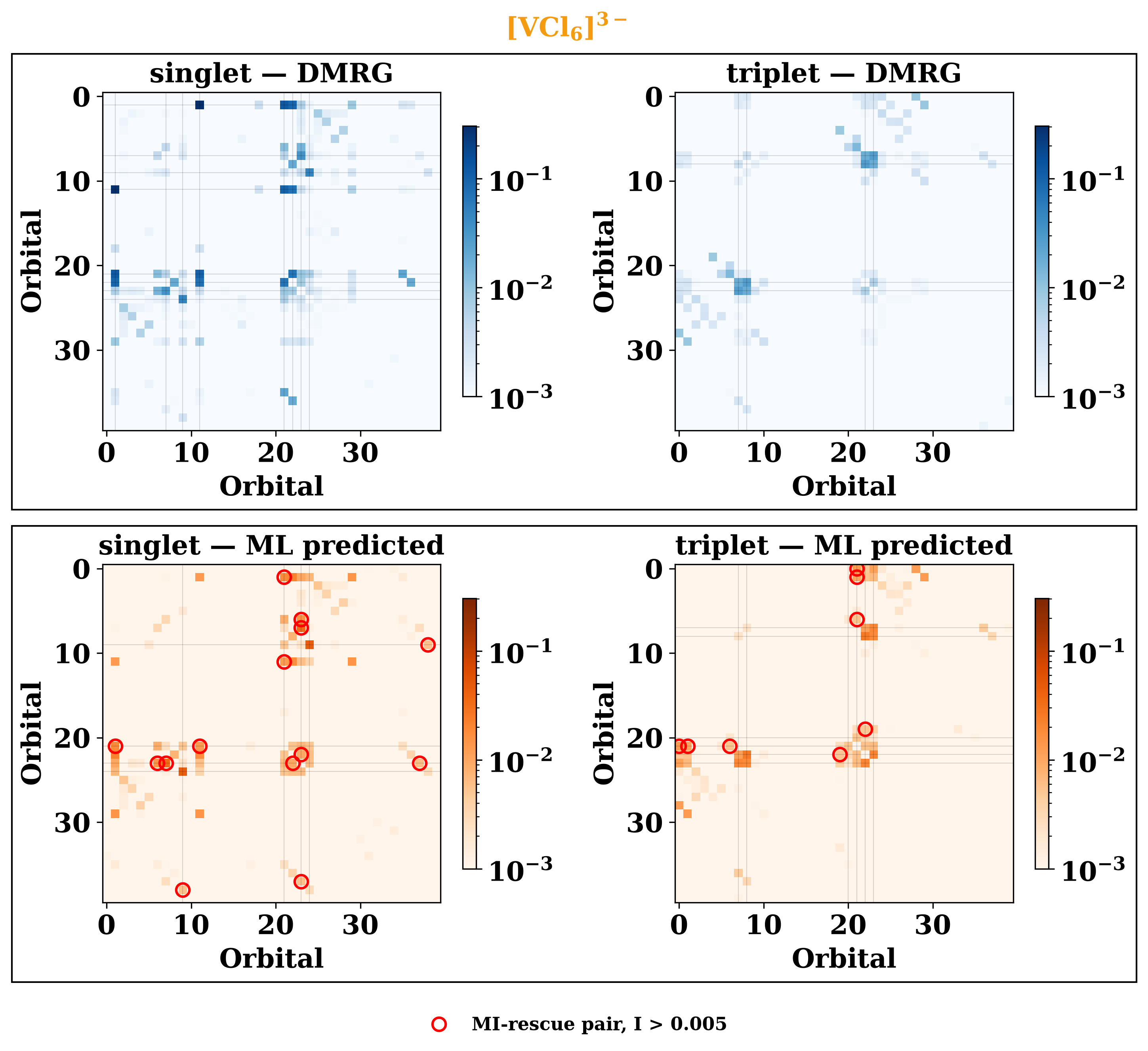}
  \caption{\textbf{\VClsix\ (\cpd{1}, Tier~I): mutual information.} The isolated, strongly correlated pairs of the singlet are placed correctly but with much smaller values. Panels: singlet (left) and triplet (right); DMRG (top row) and \texttt{SpinAware} (bottom row). Logarithmic color scale over the 40-orbital window; red circles, rescue pairs (predicted $s_1 < 0.10$ and $I_{ij} > 0.005$ with an orbital of predicted $s_1 \ge 0.10$).}
  \label{fig:mi_c1}
\end{figure}

The mutual information $I_{ij}$ (Eq.~\ref{eq:mi}) measures how strongly two
orbitals are correlated with each other. The pairwise \texttt{SpinAware} network
(Sec.~\ref{ssec:model2}), trained on the DMRG pair labels, predicts $I_{ij}$
for all 780 orbital pairs of the 40-orbital window. Its predictions reproduce
the pattern of pairwise correlation even for chemistry absent from training.
For \TiArene, the metal--arene and Ti--Cl blocks of the reference, around
window orbitals 19--25 and 8--10, appear in the prediction for both spin
states (Fig.~\ref{fig:mi_c9}). For \CrGly, the coupled block of metal--ligand
orbitals between window orbitals 17 and 30, which describes $\sigma$-donation
and $\pi$-back-donation between chromium and its carbonyl, amine and
carboxylate ligands, is recovered (Fig.~\ref{fig:mi_c8}). The model therefore
learns not only how correlated each orbital is but also which orbitals are
correlated with each other. As for the single-orbital entropies, the magnitude
of the strongest correlations is underestimated. In the vanadium(III) singlets,
DMRG shows a few isolated pairs of very large mutual information (the
$t_{2g}$ orbitals that share the two d electrons), which the prediction places
correctly but with much smaller values (Fig.~\ref{fig:mi_c1} and
Supplementary Fig.~S3). The prediction still identifies the correct partners
of the most correlated orbitals, which is the information needed to assemble
an active space.

We build the candidate pool $\mathcal{P}$ from two complementary criteria. The
first is the permissive entropy threshold, which recovers orbitals whose
predicted entropy falls just below the selection threshold. The second is a
mutual-information rescue, which adds orbitals that are only weakly correlated
on their own but strongly correlated with an active orbital, such as the
partner that completes a correlated pair. With
$\mathcal{A} = \{i : s^{\mathrm{ML}}_{1,i} \ge 0.10\}$,
\begin{equation}
  \mathcal{P} \;=\; \{\, i : s^{\mathrm{ML}}_{1,i} \ge 0.07 \,\}
  \;\cup\; \{\, i \notin \mathcal{A} : \max_{j \in \mathcal{A}} I^{\mathrm{ML}}_{ij} > 0.005 \,\}.
\end{equation}
The rescue threshold reflects the scale of mutual information: values of
order $10^{-1}$, $10^{-2}$ and $10^{-3}$ are characteristic of nondynamic,
static and dynamic correlation, respectively~\cite{boguslawski2012}, so
$I_{ij} > 0.005$ selects couplings that stand above the dynamic-correlation
background. The pool is designed for completeness rather than precision: its
purpose is to contain every orbital that the final active space may need, while
remaining much smaller than the full window. All quantities in the pool are
predicted, and no correlated calculation is required to construct it.

\begin{figure}[tp]
  \centering
  \includegraphics[width=\linewidth,height=0.6\textheight,keepaspectratio]{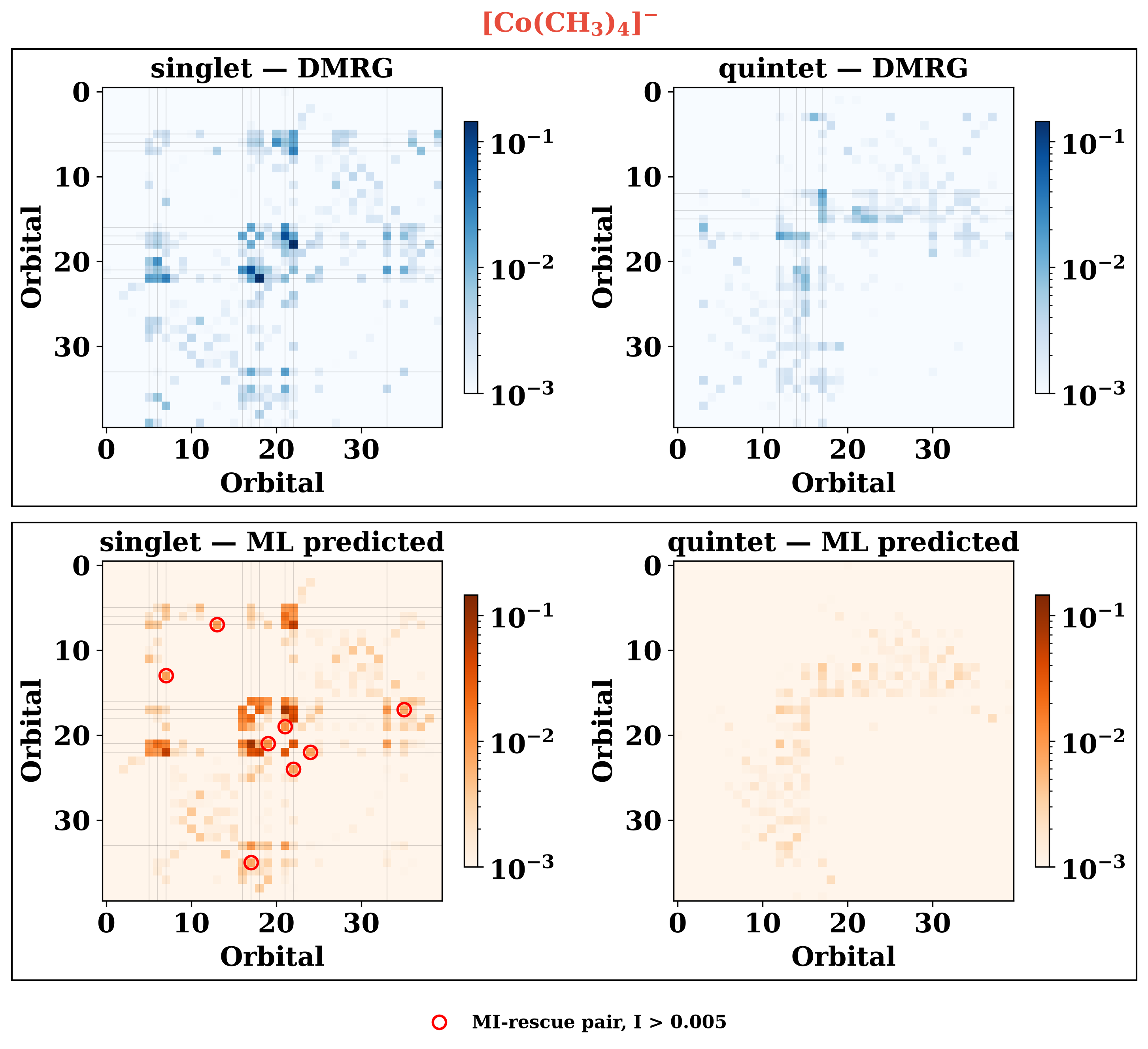}
  \caption{\textbf{\CoMe\ (\cpd{3}, Tier~I): mutual information.} The coupled blocks of the reference appear in the prediction for both spin states.} Panels: singlet (left) and quintet (right); DMRG (top row) and \texttt{SpinAware} (bottom row). Logarithmic color scale over the 40-orbital window; red circles, rescue pairs (predicted $s_1 < 0.10$ and $I_{ij} > 0.005$ with an orbital of predicted $s_1 \ge 0.10$).
  \label{fig:mi_c3}
\end{figure}

A fixed-budget control compares the candidate pool with entropy
ranking at the same state-specific number of orbitals. The archived controls
for the earlier checkpoints are reported in the Supplementary Information.
With the final models and the spin-checked references, over the 11 states of
Table~\ref{tab:pool}, the pool and the entropy ranking at the same budget both
recover 82 of the 83 DMRG-active orbitals, and for the application systems the pool
recovers 28 of 29 against 25 for the entropy ranking (Sec.~\ref{ssec:app}).

\begin{table*}[t]
\centering
\caption{\textbf{Coverage of the DMRG active space by the candidate pool.}
For each spin state, $n_{\mathrm{DMRG}}$ is the number of window orbitals with
DMRG $s_1 \ge 0.10$ (the DMRG active space, at most 40 orbitals per state). The
next columns give how many of these orbitals are found by the predicted entropy
alone ($s^{\mathrm{ML}}_1 \ge 0.10$, and $\ge 0.07$ without the mutual-information rescue) and by the candidate pool $\mathcal{P}$,
and the size of the pool out of the 40 orbitals of the window. The last row
sums the active orbitals over the 11 spin states and averages the pool size.
In the column for $s_1^{\mathrm{ML}} \ge 0.07$, the number of orbitals kept is given
in parentheses (mean in the last row).}
\label{tab:pool}
\footnotesize
\setlength{\tabcolsep}{4pt}
\def\hd#1{\begin{tabular}[b]{@{}c@{}}#1\end{tabular}}
\begin{tabular}{@{}cllccccc@{}}
\toprule
Tier & Complex & State & $n_{\mathrm{DMRG}}$ & \hd{Found by\\$s_1^{\mathrm{ML}} \ge 0.10$} & \hd{Found by\\$s_1^{\mathrm{ML}} \ge 0.07$ (kept)} & \hd{Found by\\pool} & \hd{Pool size\\(of 40)} \\
\midrule
I   & \VClsix\ (\cpd{1}) & singlet & 8  & 4 (50\%)   & 7 (9)   & 8 (100\%)  & 12 \\
    &                    & triplet & 4  & 4 (100\%)  & 4 (9)   & 4 (100\%)  & 10 \\
    & \VFsix\ (\cpd{2})  & singlet & 6  & 4 (67\%)   & 6 (16)  & 6 (100\%)  & 19 \\
    &                    & triplet & 0  & --         & -- (13) & --         & 16 \\
    & \CoMe\ (\cpd{3})   & singlet & 9  & 9 (100\%)  & 9 (13)  & 9 (100\%)  & 15 \\
    &                    & quintet & 4  & 0 (0\%)    & 4 (6)   & 4 (100\%)  & 6  \\
\addlinespace
II  & \CrMorph\ (\cpd{5}) & singlet & 11 & 11 (100\%) & 11 (19) & 11 (100\%) & 20 \\
    &                     & quintet & 11 & 9 (82\%)   & 10 (16) & 10 (91\%)  & 19 \\
    & \CrGly\ (\cpd{6})   & singlet & 18 & 13 (72\%)  & 18 (25) & 18 (100\%) & 27 \\
\addlinespace
III & \TiArene\ (\cpd{7}) & singlet & 8  & 8 (100\%)  & 8 (11)  & 8 (100\%)  & 12 \\
    &                     & triplet & 4  & 4 (100\%)  & 4 (11)  & 4 (100\%)  & 11 \\
\midrule
\multicolumn{3}{@{}l}{Sum over the 11 spin states} & 83 & 66 (80\%) & 81 (98\%); 13.5 & 82 (99\%) & 15.2 (mean) \\
\bottomrule
\end{tabular}
\end{table*}
Summed over the 11 spin states, DMRG identifies 83 active orbitals: between 0
and 18 per state, out of the 40 window orbitals of each state
(Table~\ref{tab:pool}). The predicted entropy alone, at $\tau = 0.10$,
recovers 66 of these 83 orbitals (80\%). The candidate pool recovers 82 of them
(99\%), while containing on average only 15.2 of the 40 window orbitals of a
state. The gains are concentrated where the entropy analysis located the main
weakness of the model: the strongly correlated singlets, in which the predicted
entropies are compressed below the threshold. For the singlet of \CrGly, the
pool raises the number of recovered orbitals from 13 to all 18, and for the
singlets of \VClsix\ and \VFsix\ from 4 and 4 to all 8 and 6,
respectively. On the benchmark, however, most of this gain is
already obtained by the lower entropy threshold alone: orbitals with
$s_1^{\mathrm{ML}} \ge 0.07$ recover 81 of the 83 orbitals with 13.5 orbitals per
state, and ranking the orbitals by predicted entropy alone, at the same budget as
the pool, recovers 82, as many as the pool. The mutual information is obtained
from the same DMRG calculations at no additional cost and acts as a safety net:
it recovers weakly correlated partners of strongly correlated orbitals where
they occur, which is the case for the application systems
(Sec.~\ref{ssec:app}). The reason is chemical: an orbital whose
own entropy is underestimated is still the partner of a correlated pair, and
the predicted mutual information with that partner remains large.

The size of the pool adapts to the electronic structure without any
system-specific adjustment. It contains 10--19 orbitals for the vanadium ions,
the singlet of \CoMe\ and the titanium complex, whose correlation is confined
to the metal d shell and its nearest ligand orbitals, and 19--27 orbitals for
the chromium carbonyls, whose correlation extends over the full metal--ligand
$\sigma$- and $\pi$-manifold of six ligands. In the weakly correlated triplet
of \VFsix, the rescue adds six orbitals although DMRG places no orbital
above the threshold. The pool then over-covers, which costs efficiency but not
completeness. No orbital is rescued in the high-spin quintet of \CoMe,
consistent with the absence of static correlation when all d electrons occupy
separate orbitals with parallel spins. The four orbitals that
DMRG places marginally above the threshold are nevertheless in the pool, because
the model predicts them above 0.07 (Fig.~\ref{fig:mi_c3}). The single orbital missed
by the pool belongs to the quintet of \CrMorph. It is ranked 23rd by the predicted
entropy and is not coupled to an orbital of the core.

The pool is not itself the active space: it is a compact, nearly complete set
of candidates. The final active space is selected from the pool by a NOON scan,
in which correlated calculations are performed for active spaces of increasing
size drawn from the pool, and orbitals are retained only while their natural
occupation numbers deviate significantly from 0 and 2 (Sec.~\ref{ssec:noon}).
Because the pool contains on average about 38\% of the window, the scan
explores a small, chemically targeted space. In the following section we apply
the complete procedure (prediction of entropies and mutual information,
construction of the pool and NOON scan) to \NiAllyl\ (\cpd{4}) and \CoCOtwo\
(\cpd{8}), and use the resulting active spaces in SQD.

\subsection{Predicted active spaces for two application systems}
\label{ssec:app}

We next apply the complete prediction to two systems that were withheld from
the analysis above and are used in the remainder of this work for the NOON scan
and for SQD. The first is the benchmark complex
\cpd{4} of Tier~II (Table~\ref{tab:ood_set}), the cationic nickel(II) allyl
complex \ce{[Ni($\eta^3${-}C4H7)(NCCH3)2]+}, a planar $d^8$ complex with an
$\eta^3$-crotyl ligand. The second, \ce{[Co(H2O)5(CO2)]^{2+/3+}} (\cpd{8}), is
a model of \ce{CO2} binding to cobalt that has recently served as a benchmark
for SQD on quantum hardware~\cite{David2026}.
\ce{CO2} binds end-on through one oxygen atom ($\eta^1$-O, $\kappa^1O$,
Fig.~\ref{fig:structures}), with a
linear Co--O--C unit (179$^\circ$) and an essentially undistorted, linear
\ce{CO2} (O--C--O 180$^\circ$, C--O 1.18~\AA\ to the bound and 1.10~\AA\ to the
free oxygen). Monodentate binding through a single donor atom is the dominant
topology of the training data, and water is part of the constructed ligand
library. Only the \ce{CO2} ligand is new, so $\Nood = 1$ (Tier~I). We used the equilibrium
geometry of the published Co--O dissociation series (Co--O 1.97~\AA) and
considered the singlet and quintet of cobalt(III) ($q = +3$) and the quartet
of cobalt(II) ($q = +2$). Together, the two systems cover nickel(II) $d^8$,
cobalt(III) $d^6$ and cobalt(II) $d^7$ centers and multiplicities from singlet
to quintet. For every state, the reference is the lowest-energy UHF solution
that satisfies the spin-contamination criterion, and the prediction-quality metrics
are summarized in Table~\ref{tab:app}. For the singlets of \cpd{4} and
\cpd{8}, the stability search finds a lower UHF solution only by breaking spin
symmetry ($\langle\hat S^2\rangle = 0.74$ and 1.15, 10.3\kcal\ below the
closed-shell solution for \cpd{4}). Such a solution mixes singlet and triplet
character and fails the spin criterion, so the closed-shell solution is the
lowest spin-clean reference and is used, as for every other state. The
broken-symmetry solution signals open-shell (diradical) character of these
singlets, consistent with the large $T_1$ diagnostic of \cpd{4} (0.062). This
static correlation is recovered by the multireference treatment starting from
the spin-pure orbitals (Supplementary Table~S8).

\begin{table*}[t]
\centering
\caption{\textbf{Predicted active spaces for the application systems.}
Metrics as in Table~\ref{tab:ood_results}; pool coverage and pool size as in
Table~\ref{tab:pool}. $q$ is the total charge. $^{\ddagger}$The missing orbital is
the singly occupied Co~3d orbital, which enters every active space through the
chemical seed of the NOON scan (Sec.~\ref{ssec:noon}).}
\label{tab:app}
\small
\begin{tabular}{@{}ccl ccc ccc c cc@{}}
\toprule
 & & & \multicolumn{3}{c}{$\tau = 0.10$} & \multicolumn{3}{c}{$\tau = 0.07$} & & & \\
\cmidrule(lr){4-6}\cmidrule(lr){7-9}
No. & $q$ & State & $n_{\mathrm{DMRG}}$ & $n_{\mathrm{ML}}$ & $F_1$ & $n_{\mathrm{DMRG}}$ & $n_{\mathrm{ML}}$ & $F_1$ & MAE & Found by pool & Pool size \\
\midrule
\cpd{4}  & $+1$ & singlet & 10 & 8  & 0.67 & 20 & 13 & 0.73 & 0.031 & 10 (100\%) & 20 \\
         & $+1$ & triplet & 7  & 5  & 0.67 & 12 & 11 & 0.70 & 0.016 & 7 (100\%)  & 14 \\
\addlinespace
\cpd{8}  & $+3$ & singlet & 6  & 4  & 0.60 & 8  & 6  & 0.86 & 0.018 & 6 (100\%)  & 14 \\
         & $+2$ & quartet & 3  & 1  & 0.50 & 7  & 4  & 0.73 & 0.013 & 3 (100\%)  & 6 \\
         & $+3$ & quintet & 3  & 1  & 0.50 & 7  & 4  & 0.55 & 0.011 & 2 (67\%)$^{\ddagger}$ & 7 \\
\midrule
\multicolumn{3}{@{}l}{All states} & 29 & & & & & & & 28 (97\%) & 12.2 (mean) \\
\bottomrule
\end{tabular}
\end{table*}
\begin{figure}[!tbp]
  \centering
  \boxedgraphics{width=\dimexpr\linewidth-9pt\relax,height=0.75\textheight,keepaspectratio}{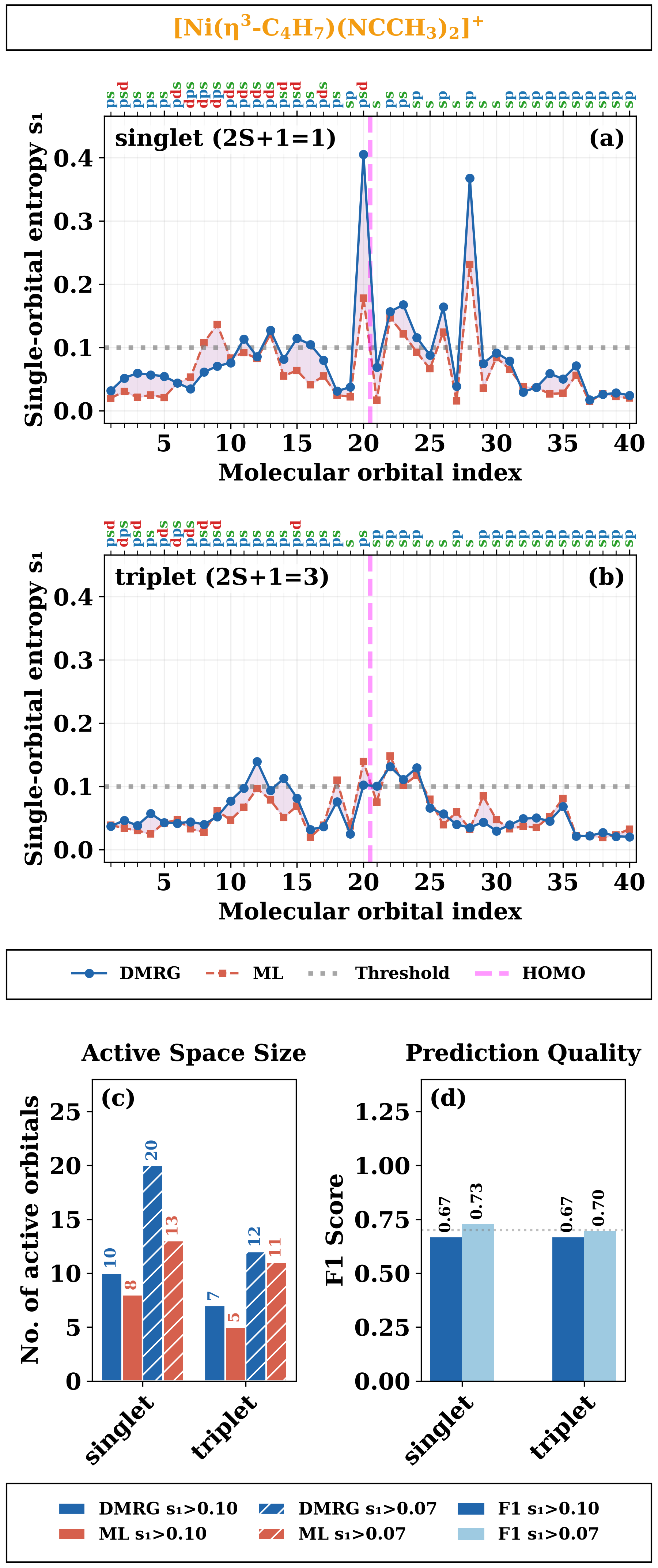}
  \caption{\textbf{\NiAllyl\ (\cpd{4}, Tier~II): single-orbital entropies.} The two dominant Ni~d/allyl orbitals of the singlet are located but under-predicted. Panels: singlet (a) and triplet (b) profiles; number of active orbitals (c) and $F_1$ (d). Upper panels: $s_1$ over the 40-orbital window from DMRG (blue) and \texttt{SpinAware} (red); dotted line, $\tau = 0.10$; dashed line, HOMO--LUMO boundary; letters, dominant orbital character. Lower panels: solid and hatched bars, $\tau = 0.10$ and 0.07; dotted line, $F_1 = 0.70$.}
  \label{fig:app_s1_c5}
\end{figure}

\begin{figure}[!tbp]
  \centering
  \boxedgraphics{width=\dimexpr\linewidth-9pt\relax,height=0.80\textheight,keepaspectratio}{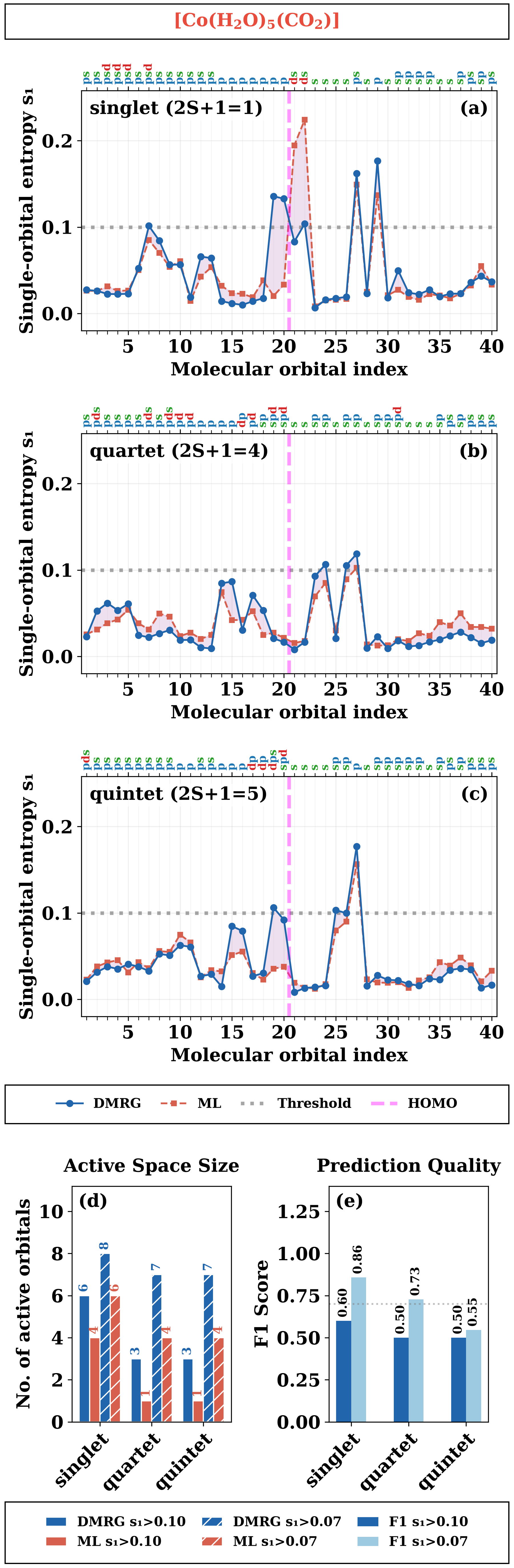}
  \caption{\textbf{\CoCOtwo\ (\cpd{8}, Tier~I): single-orbital entropies.} The flat profiles of the cobalt(II) quartet and the cobalt(III) quintet are reproduced closely, but only their strongest orbital is placed above the threshold. Panels: cobalt(III) singlet (a), cobalt(II) quartet (b) and cobalt(III) quintet (c) profiles; number of active orbitals (d) and $F_1$ (e). Upper panels: $s_1$ over the 40-orbital window from DMRG (blue) and \texttt{SpinAware} (red); dotted line, $\tau = 0.10$; dashed line, HOMO--LUMO boundary; letters, dominant orbital character. Lower panels: solid and hatched bars, $\tau = 0.10$ and 0.07; dotted line, $F_1 = 0.70$.}
  \label{fig:app_s1_c10}
\end{figure}

In the nickel(II) allyl complex \cpd{4} (Fig.~\ref{fig:app_s1_c5}), the singlet
is dominated by two orbitals of mixed Ni~d/allyl~$\pi$ character, the occupied
frontier orbital ($s_1 = 0.41$) and a virtual partner ($s_1 = 0.37$), as
expected for the partially diradical character of an $\eta^3$-allyl--metal
interaction. The prediction locates both orbitals but compresses their
entropies (0.18 and 0.23), and it places several moderately
correlated orbitals just below the threshold. The active set is therefore
incomplete (six of ten orbitals at $\tau = 0.10$, $F_1 = 0.67$), and the
permissive threshold recovers 12 of 20 ($F_1 = 0.73$). The triplet is
reproduced closely (MAE 0.016, $F_1 = 0.67$ and 0.70).

For \cpd{8} (Fig.~\ref{fig:app_s1_c10}), the cobalt(III) singlet is recovered
with $F_1 = 0.60$ and 0.86, but the prediction over-estimates the entropies of
the two lowest virtual orbitals, which DMRG places at or below the threshold,
and under-estimates those of the two highest occupied orbitals (DMRG
$s_1 = 0.13$, predicted below 0.04). The cobalt(II) quartet and the
cobalt(III) quintet are examples of the near-threshold regime identified for
the benchmark: DMRG assigns three orbitals entropies above 0.10 in each, with
maxima of 0.12 and 0.18. The prediction follows these flat profiles closely
(MAE 0.013 and 0.011) but places only one orbital above the threshold
($F_1 = 0.50$ at $\tau = 0.10$, 0.73 and 0.55 at $\tau = 0.07$).

\begin{figure}[!tbp]
  \centering
  \includegraphics[width=\linewidth,height=0.6\textheight,keepaspectratio]{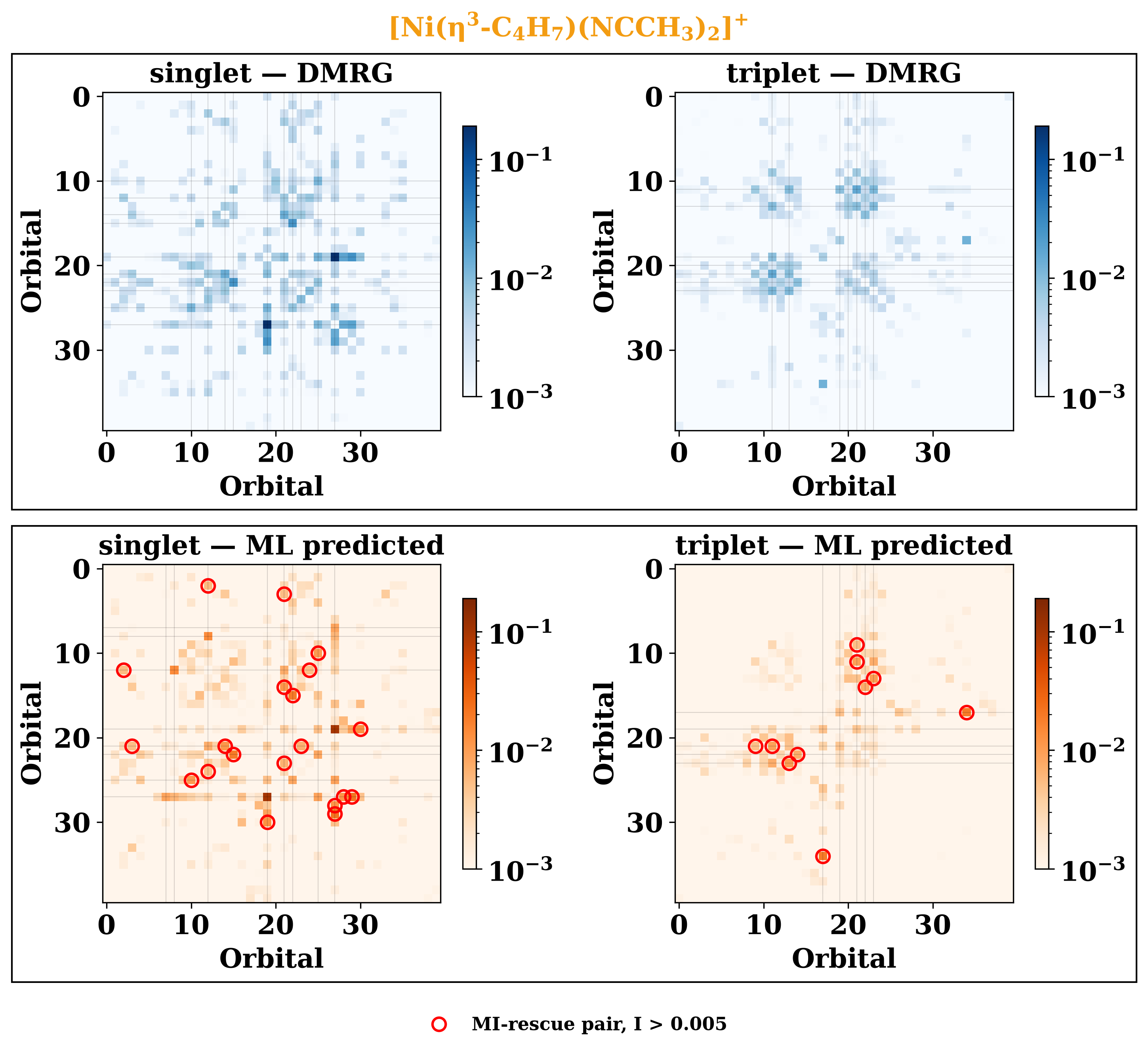}
  \caption{\textbf{\NiAllyl\ (\cpd{4}, Tier~II): mutual information.} The strongest pair of the singlet is reproduced, and the rescues follow the rows of its two dominant orbitals. Panels: singlet (left) and triplet (right); DMRG (top row) and \texttt{SpinAware} (bottom row). Logarithmic color scale over the 40-orbital window; red circles, rescue pairs (predicted $s_1 < 0.10$ and $I_{ij} > 0.005$ with an orbital of predicted $s_1 \ge 0.10$).}
  \label{fig:app_mi_c5}
\end{figure}

\begin{figure*}[!tp]
  \centering
  \includegraphics[width=\linewidth,height=0.6\textheight,keepaspectratio]{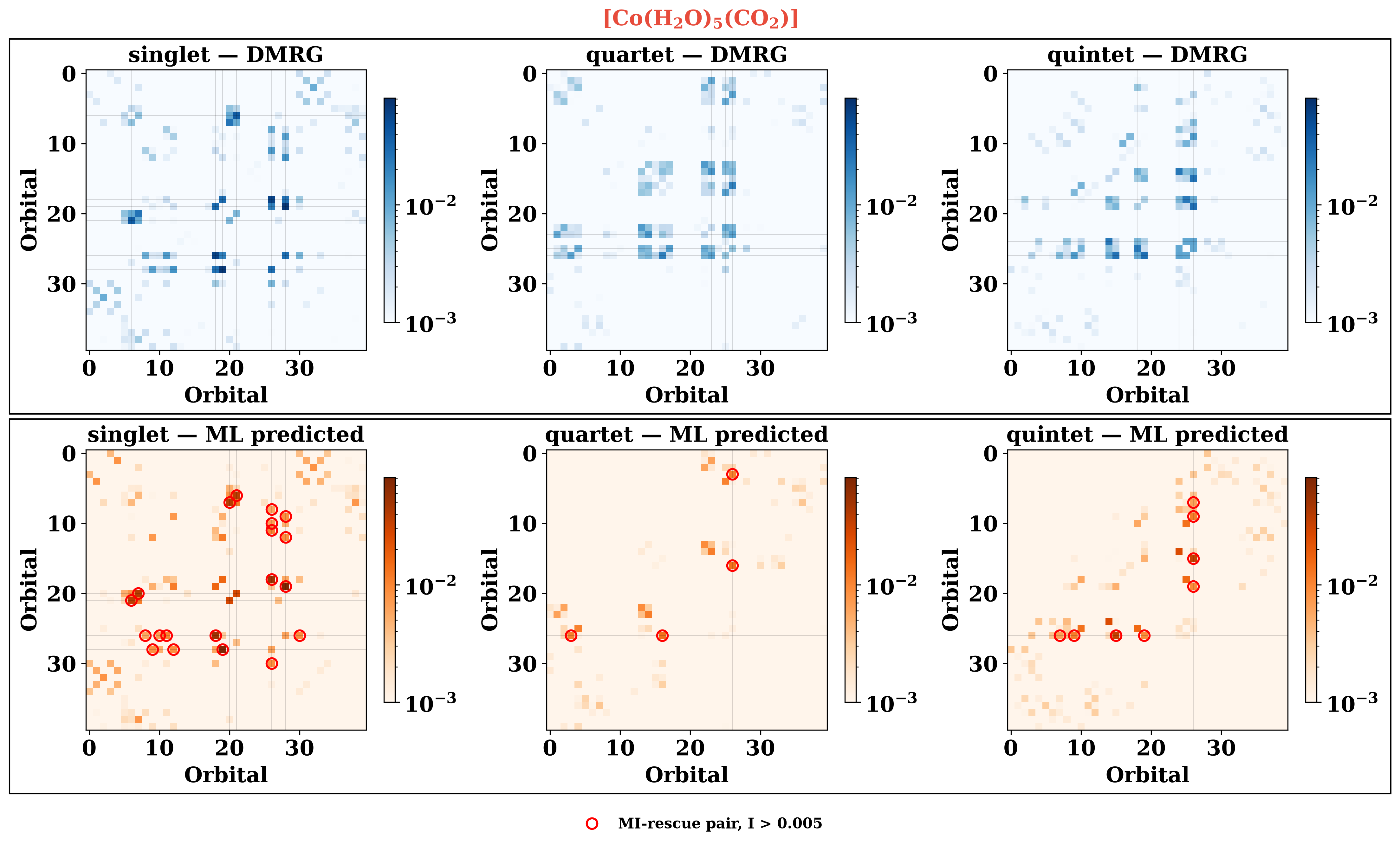}
  \caption{\textbf{\CoCOtwo\ (\cpd{8}, Tier~I): mutual information.} In the cobalt(II) quartet, the rescued pairs coincide with the coupled blocks of the reference. Panels: cobalt(III) singlet (left), cobalt(II) quartet (center) and cobalt(III) quintet (right); DMRG (top row) and \texttt{SpinAware} (bottom row). Logarithmic color scale over the 40-orbital window; red circles, rescue pairs (predicted $s_1 < 0.10$ and $I_{ij} > 0.005$ with an orbital of predicted $s_1 \ge 0.10$).}
  \label{fig:app_mi_c10}
\end{figure*}

The predicted mutual-information matrices reproduce the coupling patterns of the
reference in both systems (Figs.~\ref{fig:app_mi_c5} and \ref{fig:app_mi_c10}).
For the singlet of \cpd{4}, the strongest pair of the reference, which couples
the two dominant Ni~d/allyl orbitals, appears at the same position in the
prediction, and the rescue pairs run along the rows and columns of these two
orbitals (Fig.~\ref{fig:app_mi_c5}). The rescue thus recovers the partners of
the orbitals whose entropies are compressed, the case in which the mutual
information is most needed. For the quartet of \cpd{8}, the rescued pairs
coincide with the coupled blocks of the reference matrix
(Fig.~\ref{fig:app_mi_c10}).

For the five states, the candidate pool recovers 28 of the 29 DMRG-active orbitals
with 12.2 orbitals on average (Table~\ref{tab:app}), compared with 15 (52\%)
from the entropy alone. The remaining one, the singly occupied Co~3d orbital of
the quintet of \cpd{8}, is included by the chemical seed. The largest gain occurs for the singlet of \cpd{4}, in
which the pool recovers all ten active orbitals against six from the entropy
alone, because the rescue adds the partners of the two dominant orbitals whose
entropies are compressed. Here the mutual information adds information that the
entropies do not contain: at the same budget, ranking by predicted entropy
alone recovers 25 of the 29 orbitals and the lower threshold alone 23, whereas
the pool recovers 28, including the Ni--allyl bond orbital of the singlet of
\cpd{4} (ranked 22nd by entropy). In an application, however, no DMRG reference is
available to tell which pool orbitals are required. The pool must therefore be
reduced to a final active space by a criterion that relies only on the
correlated calculation itself. This reduction is also a practical necessity for
quantum hardware, where every spatial orbital in the active space requires two
qubits in the Jordan--Wigner mapping, so that the pools of 6--20 orbitals
would require 12--40 qubits. The following subsection describes the
NOON scan that selects the final active space from the
pool.
\subsection{Natural-orbital occupation scans}
\label{ssec:noon}

The candidate pool tells us which orbitals may need to be active, but in an
application there is no DMRG reference to decide which of them actually are.
That decision must come from the correlated wavefunction itself. The most
direct measure is provided by the NOONs,
the eigenvalues $n_k$ of the one-particle reduced density matrix
\begin{equation}
  \gamma_{pq} = \sum_{\sigma} \langle \Psi | \hat a^\dagger_{p\sigma} \hat a_{q\sigma} | \Psi \rangle ,
\end{equation}
with $\gamma\,\mathbf{u}_k = n_k\,\mathbf{u}_k$ and $0 \le n_k \le 2$. For a
single determinant every orbital is either doubly occupied ($n_k = 2$) or empty
($n_k = 0$), apart from the singly occupied orbitals of an open-shell state
($n_k = 1$). Static correlation moves electron density from occupied to virtual
natural orbitals, so that occupations depart from these integer values. The size
of the departure measures how strongly an orbital participates in the
correlation, and occupations outside the range 0.02--1.98 are the classic
signature of an orbital that must be treated as active~\cite{Pulay1988,Keller2015,stein2016}. Unlike the single-orbital entropy, which is
computed in a fixed window, the NOONs are properties of the wavefunction in a
given active space. They therefore test directly whether adding an orbital to
that space changes the correlated description.

A single calculation in the full pool would show the final occupations, but not
which orbitals are responsible for them. Correlation also involves pairs
of orbitals: an occupied orbital can lose density only if a suitable virtual
partner is present to receive it. An orbital added without its partner
therefore leaves the occupations unchanged, however important it is. We
therefore grow the active space one orbital at a time, in the order given by
the \texttt{SpinAware} prediction, and follow the NOONs after each step. A sudden change
in the occupations marks the step at which a correlated pair is completed, a
flat step marks an orbital that is only a passenger, and a plateau indicates that the occupation-based diagnostics have stabilized over the sampled range. The scan turns the predicted ranking into a testable statement: a
good ranking produces its jumps early and reaches the plateau with few
orbitals.

Each scan starts from the lowest spin-clean UHF reference on which the
predictions were made (def2-SVP) and works in the fixed basis of its UNOs~\cite{Pulay1988}. The initial active space is fixed by
a chemical rule rather than by the model. It contains (i) every UNO with a
fractional occupation $0.10 < n < 1.90$, which includes the singly occupied
orbitals of open-shell states, and (ii) the five orbitals of largest metal-3d
character.
A singly occupied orbital has a definite occupation in the reference
determinant and therefore zero single-orbital entropy there in the
$M_s = S$ component used for the labels (in other $M_s$ components of the same
state it carries up to $\ln 2$). In the correlated state its entropy is usually
small (below about 0.03) but can reach the threshold, as for one singly
occupied site of the cobalt(III) quintet of \cpd{8} ($s_1 = 0.106$), so no
entropy-based criterion, predicted or DMRG, can select it reliably, yet it
must be active to describe the spin coupling.

The 3d orbitals are identified by projecting the doubly occupied and the
empty UNOs, separately, onto the metal 3d atomic orbitals, in the spirit of
AVAS~\cite{Sayfutyarova2017}. This rotation within each block leaves the UHF
density unchanged. Every space must contain five orbitals with more than 50\%
metal-3d projection, otherwise it is rejected. Beyond this seed, all orbitals
are taken from the \texttt{SpinAware} prediction, in the order defined by the pool:
first orbitals with predicted $s_1 \ge 0.07$ in decreasing order of $s_1$,
then the mutual-information-rescued orbitals, and finally the remaining window
orbitals. Each predicted window orbital is mapped onto the UNO of maximum
overlap ($|S| \ge 0.3$), and orbitals already in the space, such as the singly
occupied ones, are skipped. After each addition, the wavefunction is obtained
by exact CASCI up to 16
orbitals and by DMRG-CASCI from 17 to 20 orbitals (\textsc{block2}, spin
adapted, Fiedler orbital ordering, 20 sweeps with bond dimension increased from
250 to a final $M = 1000$). For the 17-orbital spaces of all five
states, the first DMRG step, DMRG reproduces the exact CASCI energy to
$10^{-7}$~$E_\mathrm{h}$ and the NOONs to $3\times10^{-6}$. The scripts and results
of all NOON scans, CASCI, NEVPT2 and DMRG calculations are available from the corresponding authors upon reasonable request.

To follow the scan, we monitor the NOONs $n$
($0 \le n \le 2$) of the active space. In a single-determinant description
every occupation is an integer: $n = 2$ for a doubly occupied orbital, $n = 0$
for an empty (virtual) one and $n = 1$ for a singly occupied one. Static
correlation mixes in further configurations and moves electron density from
occupied into virtual orbitals, so the occupations deviate from these integers.
We quantify the deviation in three ways: the depletion $2 - n$ of an occupied
natural orbital (the density it has lost), the population $n$ of a virtual
natural orbital (the density it has gained), and, for open-shell states, the
deviation $|n - 1|$ of a singly occupied orbital from exactly one electron.
For each step we report the strongest correlated pair, i.e.\ the largest
depletion among the occupied and the largest population among the virtual
natural orbitals, together with the second-strongest pair and the largest
$|n - 1|$. A deviation above 0.02 (occupations below 1.98 or above 0.02), the
conventional criterion for an active orbital and close to the occupation
equivalent of $\tau = 0.07$ (Sec.~\ref{ssec:transfer}), indicates clear static correlation.
The two members of a pair need not have equal values, because density can flow
from several occupied orbitals into one virtual orbital, or from one occupied
orbital into several virtual ones. For example, in the 16-orbital space of the
singlet of \NiAllyl\ (\cpd{4}), the most depleted occupied orbital has $n = 1.955$
(depletion 0.045), whereas the most populated virtual orbital has $n = 0.064$.

\begin{figure}[!tbp]
  \centering
  \boxedgraphics{width=\dimexpr\linewidth-9pt\relax,height=0.7\textheight,keepaspectratio}{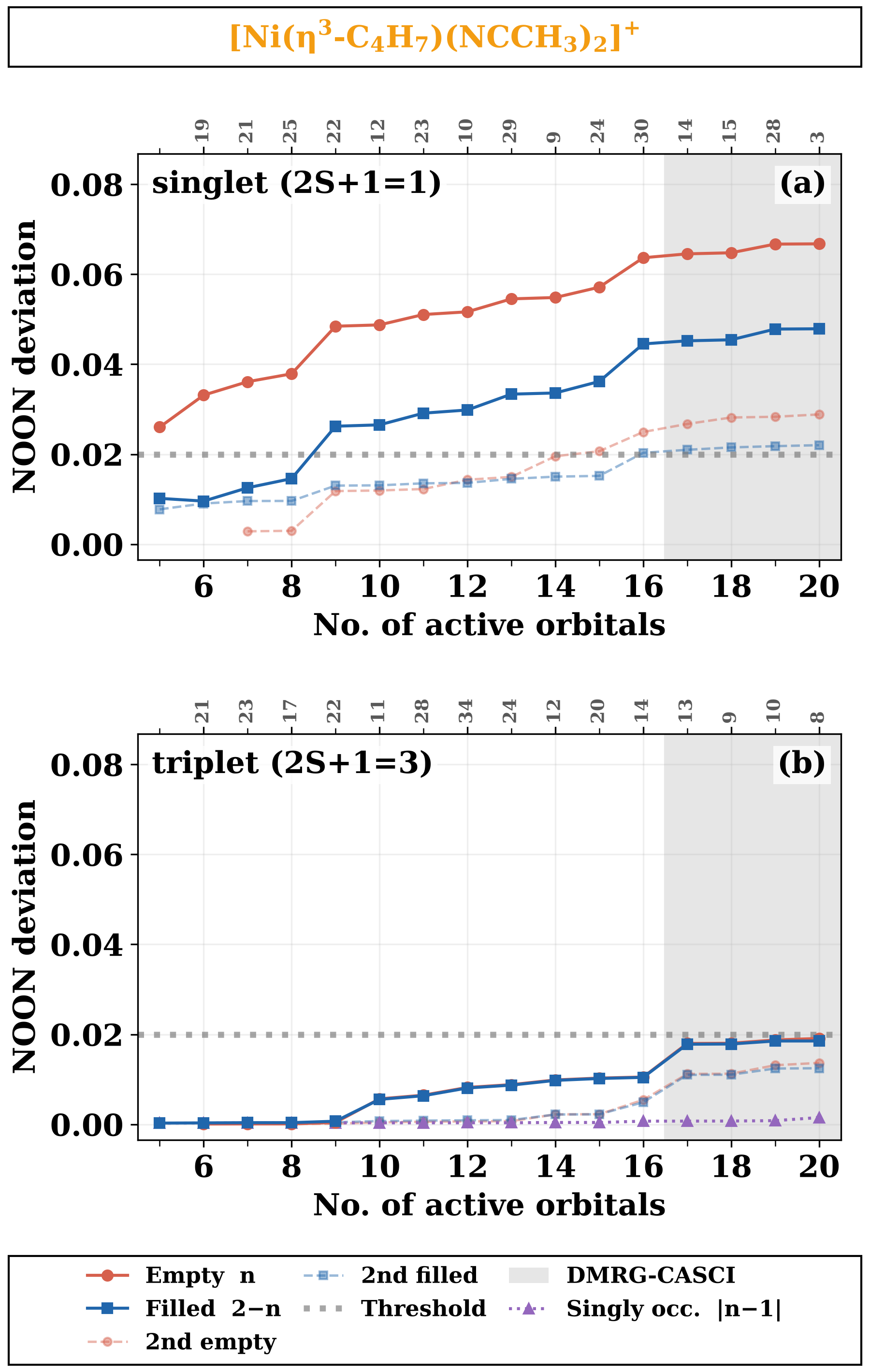}
  \caption{\textbf{\NiAllyl\ (\cpd{4}): NOON scans.}
  Singlet (a) and triplet (b). The active space grows from the five Ni~d
  orbitals by one predicted orbital per step (window index of each added orbital
  on top; numbered from 0, HOMO = 19). Solid lines: strongest correlated pair,
  shown as $2 - n$ of the occupied (blue) and $n$ of the virtual (red) natural
  orbital; dashed lines: second pair; purple: $|n - 1|$ of singly occupied
  orbitals. Dotted line: 0.02. Gray region: DMRG-CASCI; otherwise exact CASCI.}
  \label{fig:noon_c5}
\end{figure}

In the closed-shell singlet of \NiAllyl\ (\cpd{4}), the correlation resides in the bond
between nickel and the allyl ligand, and a rule based on the metal d shell alone
misses most of it (Fig.~\ref{fig:noon_c5}a). The five Ni~d orbitals already
show moderate correlation (pair strength 0.026), from a filled Ni~d orbital to
an empty one. The first three predicted orbitals, windows 19, 21 and 25, raise
it gradually to 0.038. The fourth, window 22, completes the allyl$\rightarrow$Ni
pair at nine orbitals and raises the strongest pair from 0.038 to 0.049 (empty
orbital) and from 0.015 to 0.026 (filled orbital). This sequence illustrates
the pair principle: the orbitals with the highest entropy are the right ones,
but the correlation appears in full only once both partners of the pair are
present. The two values differ because the empty Ni~d orbital receives density
from several filled orbitals, Ni~d and allyl, at once. Further orbitals add
small contributions, and a final step at 16 orbitals (window 30, a
mutual-information-rescued orbital) brings the pair strength to 0.064, within
0.003 of its 20-orbital value. In this closed-shell state, the reference
wavefunction offers no spin-polarized pair from which the correlated ligand
orbitals could have been identified. The predicted ranking places them among
its first four picks.

\begin{figure}[!tbp]
  \centering
  \boxedgraphics{width=\dimexpr\linewidth-9pt\relax,height=0.82\textheight,keepaspectratio}{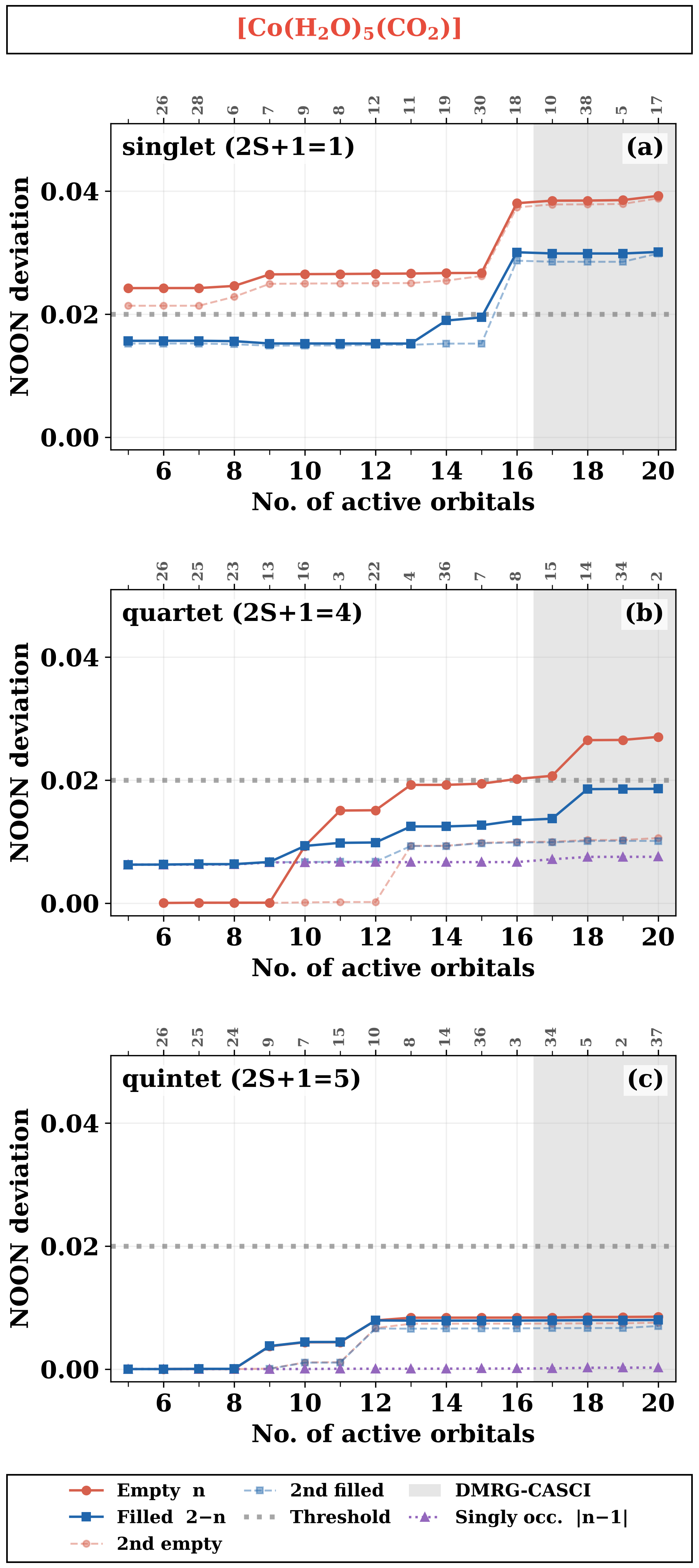}
  \caption{\textbf{\CoCOtwo\ (\cpd{8}): NOON scans.}
  Cobalt(III) singlet (a), cobalt(II) quartet (b) and cobalt(III) quintet (c).
  In the singlet the strongest pair
  grows when the two highest occupied orbitals are added (14 and 16 orbitals); in the quartet the weak ligand
  pair forms when the mutual-information-rescued occupied orbitals are added.
  Lines and shading as in Fig.~\ref{fig:noon_c5}.}
  \label{fig:noon_c10}
\end{figure}

\CoCOtwo\ (\cpd{8}) is a control in which the correlation is weaker
(Fig.~\ref{fig:noon_c10}). In the cobalt(III) singlet, the d shell alone gives
a pair strength of 0.024 (t$_{2g}$$\rightarrow$e$_g$), and the first eight
predicted orbitals raise it only to 0.027. The two highest occupied orbitals,
windows 19 and 18, whose entropies the prediction under-estimates
(Sec.~\ref{ssec:app}), enter only at 14 and 16 orbitals, and the second raises the
strongest pair to 0.038, within 0.001 of its 20-orbital value. In the
cobalt(III) quintet all values stay below 0.01, and the singly occupied
orbitals remain at one electron. The high-spin triplet of \NiAllyl\ (\cpd{4})
behaves similarly: all pair strengths stay below 0.02, and the two largest
steps, at 10 and 17 orbitals, add only weak correlation between a filled and
an empty orbital (Fig.~\ref{fig:noon_c5}b). In these states the predicted
orbitals do not create spurious correlation but only enlarge the space.

The cobalt(II) quartet of \CoCOtwo\ (\cpd{8}) provides a direct test of the
mutual-information rescue (Fig.~\ref{fig:noon_c10}b). The scan starts with a
weak coupling (0.006) between a filled Co~d orbital and the singly occupied
orbitals. The orbitals ranked highest by entropy, windows 26, 25, 23 and 13,
leave the occupations almost unchanged (0.007). The weak ligand pair forms
only when the rescued occupied partners, windows 16, 3 and 4, are added,
raising the strongest pair to 0.019 at 13 orbitals, and two further rescued
orbitals, windows 15 and 14, bring it to 0.027 at 18 orbitals. Without the
rescue, these partners would have been ranked among the remaining window
orbitals, and the pair would have appeared much later in the scan. This is the
pair principle seen from the other side: the entropy identifies the virtual
half of the pair, and the mutual information supplies its occupied partner.

Extending every scan to 20 orbitals with DMRG-CASCI tests the stability of the occupation-based correlation diagnostics (Table~\ref{tab:noon}). We characterize this stability by the
change in the strongest pair strength between a given size and the largest
scanned space. This diagnostic tracks the dominant static correlation, rather than providing a bound on the spin-gap error. For both complexes the occupations reach a plateau. The singlet
of \NiAllyl\ (\cpd{4}) reaches it at 16 orbitals, when window orbital 30 is added (0.064 at
16 and 0.067 at 20 orbitals), and the triplet takes a last step at 17 orbitals,
from 0.011 to 0.018. For \CoCOtwo\ (\cpd{8}), the singlet reaches its plateau at 16
orbitals (0.038, against 0.039 at 20), the quintet already at 12, and the
quartet takes a last step at 18 orbitals (from 0.021 to 0.027). At 16 orbitals
the pair strength of every state lies within 0.009 of its 20-orbital value,
below the 0.02 that marks clear static correlation.
The size of the energy calculations is set by NEVPT2: 16 orbitals is the
largest active space for which the strongly contracted NEVPT2 correction
remains tractable with the full four-particle reduced density matrix, so the
Ni triplet and the Co quartet are used at 16 orbitals although their
occupations take a last small step at 17 and 18 orbitals. We
therefore use 16 orbitals for all spin states of a complex in the energy
calculations: CAS(16,16) for the singlet and triplet of
\NiAllyl{}, and CAS(22,16), CAS(19,16) and CAS(20,16) for the singlet, quartet and
quintet of \CoCOtwo{}. %
Energetic convergence is assessed separately below and in the Supplementary Information.

Converged occupations are a necessary condition for an adequate active space,
but not a sufficient one. We therefore test the selected spaces directly on the
quantity of interest, the spin-state gap. Table~\ref{tab:gaps} reports the gaps
obtained by CASCI and by CASCI with strongly contracted NEVPT2 for active
spaces of equal size in both spin states, built from the same predicted
ranking, and compares them with UCCSD(T) for the full molecule. The CASCI energies are exact at all sizes, and we use NEVPT2 evaluated from the DMRG one- to four-particle
reduced density matrices (bond dimension 1000), which reproduces conventional NEVPT2 to
$2\times10^{-7}\,E_\mathrm{h}$ at 13 orbitals, and at 16 orbitals the DMRG energy matches exact
CASCI to $3\times10^{-7}\,E_\mathrm{h}$. The CASCI gaps lie far from the
reference, by about 36--45\,kcal\,mol$^{-1}$ for \NiAllyl\ (\cpd{4}) at all sizes and
21--24\,kcal\,mol$^{-1}$ for \CoCOtwo\ (\cpd{8}) up to 13 orbitals, because the dynamic correlation outside the active
space, which CASCI omits, differs strongly between the two spin states. With
NEVPT2 the gaps change by at most 4.5\,kcal\,mol$^{-1}$ between 9 and
16 orbitals but do not approach the coupled-cluster reference
systematically: at 16 orbitals they differ from UCCSD(T) by 6.3\,kcal\,mol$^{-1}$
for \NiAllyl{} and 5.3\,kcal\,mol$^{-1}$ for \CoCOtwo{}, and at 9--13 orbitals by
4.2--5.0 and 0.9--3.6\,kcal\,mol$^{-1}$. For \NiAllyl{} the NEVPT2 gap changes by 1.8\mbox{\kcal} between 13 and 16 orbitals, and for \CoCOtwo{} it still increases by 2.8\,kcal\,mol$^{-1}$ over this range. Comparison between multireference perturbation theory in the predicted
spaces and full-molecule coupled cluster quantifies the cross-method
differences at the present basis and fixed geometries. %
In particular, the remaining change for
\CoCOtwo{} warrants caution.The largest
changes occur at the last step. For \CoCOtwo{}, the CASCI gap drops by
21.7\,kcal\,mol$^{-1}$ from 13 to 16 orbitals, when the two highest occupied
orbitals enter the space. NEVPT2 absorbs most of this change, but the corrected
gap still rises by 2.8\,kcal\,mol$^{-1}$. For \NiAllyl{}, adding the rescued
orbital 30 at 16 orbitals changes the corrected gap by 1.8\,kcal\,mol$^{-1}$. At
16 orbitals the corrected gap therefore still depends on which ligand orbitals
lie at the boundary of the active space. The T1 diagnostic of the singlet of
\NiAllyl{} (0.062, above the usual 0.05 guideline) is consistent with
the partly multiconfigurational character identified in the scan and also
limits the status of UCCSD(T) as a definitive reference.

\begin{table*}[tp]
  \caption{\textbf{Summary of the NOON scans.} Pair strength is the larger of
  $2-n$ (occupied) and $n$ (virtual) for the strongest correlated pair. Window
  orbitals are numbered from 0 (HOMO = 19). The last column gives the active
  space used for the energy calculations (same number of orbitals for all spin
  states of a complex) and, in parentheses, the size at which the strongest
  pair takes its last step.}
  \label{tab:noon}
  \centering
  \begin{tabular}{lllccl}
    \toprule
    State & Correlation resides in & Key orbitals in the ranking
        & \multicolumn{2}{c}{Pair strength} & Active space used \\
    \cmidrule(lr){4-5}
            &                        &
        & 16 orb. & 20 orb. & (plateau) \\
    \midrule
    \multicolumn{6}{@{}l}{\NiAllyl\ (\cpd{4})} \\
    singlet & Ni--allyl bond      & 4th pick (window 22) & 0.064 & 0.067 & CAS(16,16) (16) \\
    triplet & Ni d shell          & none (weak)          & 0.011 & 0.019 & CAS(16,16) (17, weak) \\
    \addlinespace
    \multicolumn{6}{@{}l}{\CoCOtwo\ (\cpd{8})} \\
    singlet & Co d + highest occupied & windows 19, 18 (late)         & 0.038 & 0.039 & CAS(22,16) (16) \\
    quartet & Co d + ligand pair      & rescued by mutual information & 0.020 & 0.027 & CAS(19,16) (18, weak) \\
    quintet & Co d shell              & none (passengers)             & 0.008 & 0.009 & CAS(20,16) (12) \\
    \bottomrule
  \end{tabular}
\end{table*}

An archived fixed-budget comparison of the earlier model spaces with
metal-$d$ plus frontier completion and a specified ligand-aware AVAS selection
is reported in the Supplementary Information. All three selections match the
active-electron allocation of the earlier ML spaces. The final-model spaces
use different active-electron allocations, so the interpretation of those
controls is restricted to the earlier spaces.

The scans support a simple protocol that combines chemical rules with the
prediction. The active space is started from the metal d shell and the singly
occupied orbitals, and it is completed in the order of the predicted pool until
the strongest pair strength lies within 0.01 of its value in the largest
scanned space, with the same number of orbitals for all spin states of a
complex. When the correlation extends into the metal--ligand bonds, as in the
singlet of \NiAllyl\ (\cpd{4}), the prediction supplies a ligand-orbital ranking: a metal-$d$-shell rule alone does not identify the correlated ligand orbitals, whereas the predicted
ranking places them among its first picks. When the correlation is confined to
the d shell, as in the quartet and quintet of \CoCOtwo\ (\cpd{8}) and the triplet of \NiAllyl{}, the rule alone
suffices, and the predicted orbitals remain passengers without distorting the
result. In the cobalt(III) singlet, two occupied orbitals whose entropies are
under-predicted enter only at 14 and 16 orbitals, so the scan detects their
contribution only at the end of the range used for the energies. Throughout, the scans confirm that high entropy alone does not make an
orbital important in a given space: correlation appears only when both partners
of a pair are present, which is why the pairwise information in the pool is
essential. For every scanned space we retain the active-space Hamiltonian,
constructed exactly as in the CASCI calculations, so that quantum and classical
solvers address the same problem: 80 Hamiltonians for \NiAllyl{} and \CoCOtwo{},
from 5 to 20 spatial orbitals (10--40 qubits).

Every spatial orbital requires two qubits in the Jordan--Wigner mapping, so
the 9-, 11-, 13- and 16-orbital spaces of the scans correspond to 18, 22, 26
and 32 qubits. We carry all four sizes to quantum hardware, with the same
number of orbitals for both states of a complex
(Sec.~\ref{sec:sqd}). Because the quantum calculation and CASCI solve the same
active-space Hamiltonian, the comparison between them isolates the error of the
quantum solver, while the change of the gap with the size of the space is the
active-space truncation, quantified classically in Table~\ref{tab:gaps}.

\subsection{Sample-based quantum diagonalization on quantum hardware}
\label{sec:sqd}

The active space selected by \texttt{SpinAware} determines the size of the quantum
problem: an active space of $n_\mathrm{act}$ orbitals requires
$2n_\mathrm{act}$ qubits in the Jordan--Wigner mapping. A compact but complete
active space is therefore what makes a transition-metal complex accessible to
current quantum processors, whose circuits are limited by gate errors and
decoherence. The aim of this section is to show that the predicted active
spaces can be carried through to such a processor and still reproduce the
spin-state gap of the exact classical calculation in the same space. We use SQD~\cite{Kanno2023,RobledoMoreno2025}, currently the main
approach for chemistry on pre-fault-tolerant hardware beyond 30 qubits,
because it needs no variational optimization on the device and is robust to
noise. Variational algorithms~\cite{Peruzzo2014,Kandala2017,Tilly2022}, by
contrast, must optimize circuit parameters on the device, where flat energy
landscapes slow the optimization~\cite{McClean2018,Fitzek2024qBang}, and they
need error mitigation and verification to reach chemical
accuracy~\cite{Temme2017,Cai2023,Lolur2023REM,Skogh2024,Zou2025MREM}. All active spaces studied here (18--32 qubits) can still be solved
exactly by CASCI. This is deliberate: the exact energy of the same
Hamiltonian is available as a reference, so every hardware result can be
checked. The calculations are a proof of concept for the workflow and make no
claim of quantum advantage.
\begin{table}[tb]
  \caption{\textbf{Spin-state gaps for active spaces of equal size.}
  $\Delta E = E(\text{low spin}) - E(\text{high spin})$ in kcal\,mol$^{-1}$;
  negative values mean that the low-spin state is lower. CASCI gaps from exact CASCI; NEVPT2 is strongly
  contracted, from DMRG reduced density matrices (bond dimension 1000). CC: UCCSD(T) for the full molecule (frozen core).
  S, T, Q: singlet, triplet, quintet.
  def2-SVP throughout.}
  \label{tab:gaps}
  \centering
  \begin{tabular}{llrrrrr}
    \toprule
    & & \multicolumn{4}{c}{Active orbitals} & \\
    \cmidrule(lr){3-6}
    Gap & Method & 9 & 11 & 13 & 16 & CC \\
    \midrule
    \multicolumn{7}{@{}l}{\NiAllyl\ (\cpd{4})} \\
    S$-$T & CASCI         &  2.8    &  3.0    &  0.3    & $-5.4$  & \\
          & CASCI+NEVPT2  & $-37.7$ & $-36.9$ & $-37.3$ & $-35.6$ & $-41.9$ \\
    \addlinespace
    \multicolumn{7}{@{}l}{\CoCOthree\ (\cpd{8})} \\
    S$-$Q & CASCI         & 38.9 & 37.7 & 35.6 & 14.0 & \\
          & CASCI+NEVPT2  & 18.4 & 15.7 & 17.4 & 20.2 & 14.8 \\
    \bottomrule
  \end{tabular}
\end{table}
SQD is a hybrid procedure in which the quantum computer only supplies
electronic configurations and a classical computer does the diagonalization.
Each measurement of the $2n_\mathrm{act}$ qubits yields a bitstring
$x\in\{0,1\}^{2n_\mathrm{act}}$ that encodes one occupation-number vector, i.e.\
one Slater determinant, with one bit per spin orbital. Bitstrings with the
wrong numbers of $\alpha$ and $\beta$ electrons, which arise mainly from
hardware noise, are repaired by self-consistent configuration recovery, in
which bits are flipped according to the current average orbital occupations.
The distinct $\alpha$ and $\beta$ strings of a batch of samples are combined
as a tensor product, and the active-space Hamiltonian is diagonalized in the
resulting subspace with a constraint on $\langle\hat S^2\rangle$. Spin symmetry can alternatively be imposed on the quantum side, by
spin-preserving circuits~\cite{Gard2020,Anselmetti2021} or by working in a
spin-adapted representation~\cite{Gandon2025}. Because the
diagonalization is variational, the SQD energy is an upper bound to the exact
energy $E_\mathrm{CASCI}$ of the same Hamiltonian, which serves as the
reference throughout. It approaches $E_\mathrm{CASCI}$ when the subspace
contains the determinants that carry the ground-state wavefunction. The
classical part is identical for all calculations below. What differs is the
source of the bitstrings, and the accuracy of SQD depends entirely on how well
these sources find the important determinants. The Hamiltonians are those of
the NOON scans (Sec.~\ref{ssec:noon}), and both spin states of a complex are
treated in active spaces of equal size.

We compare three sources of samples. (i)~\emph{LUCJ.} The local unitary cluster Jastrow (LUCJ)
ansatz~\cite{Motta2023} prepares an approximation to the ground state on the
quantum computer, and measuring it gives determinants with probabilities
close to their weights in that state. Even on ideal hardware, such
ground-state sampling finds new determinants inefficiently~\cite{Reinholdt2025}.
The circuit consists of orbital
rotations and diagonal (number--number) Jastrow terms, and in the local variant
the Jastrow terms act only between neighboring qubits, which matches the
hardware connectivity. We use one repetition with parameters taken from the
CCSD amplitudes of the active space, without variational optimization
(\texttt{ffsim}~\cite{ffsim}), and draw 40\,000 shots. (ii)~\emph{SqDRIFT.}
SqDRIFT~\cite{Piccinelli2025} uses no ansatz and no parameters. It builds on
sample-based Krylov quantum diagonalization~\cite{Yu2025SKQD}: the
time-evolved states $e^{-i\hat H t}|\mathrm{HF}\rangle$ of the Hartree--Fock
determinant spread over the determinants that couple to it through the
Hamiltonian, and therefore sample the determinants that matter for the ground
state. The time evolution is approximated by the qDRIFT
protocol~\cite{Campbell2019}: each circuit applies a random sequence of $L=10$
exponentials of individual Hamiltonian terms, drawn with probabilities
proportional to their coefficients. Many such random circuits are run, and
their samples are pooled. Under mild assumptions this gives a convergence
guarantee for the SQD energy~\cite{Piccinelli2025}. We use 200 random circuits
with 100 shots each at two evolution times, $t=1$ and 10~$\hbar/E_\mathrm{h}$,
giving 40\,000 shots per state. (iii)~\emph{Uniformly random bitstrings.}
Every bit is set to 0 or 1 with probability one half, with the same number of
samples and the same SQD settings, and without any quantum computer. This is a
control experiment: configuration recovery and the tensor-product subspace can
by themselves produce a useful subspace from unstructured input, so a
hardware result is only meaningful if it is better than random input.

The hardware calculations used \texttt{ibm\_phoenix} (IBM Nighthawk r2, a
square lattice in which each qubit couples to four neighbors) and, for one
LUCJ comparison, \texttt{ibm\_fez} (IBM Heron r2, a heavy-hexagon lattice with
at most three neighbors per qubit), without error mitigation. The active-space
Hamiltonians, the sampling and SQD scripts and the raw measurement counts of all
hardware and simulator runs are available from the corresponding authors upon reasonable request. The cost of a
circuit on hardware is characterized by three quantities, given in
Table~\ref{tab:circ_main} for the largest case, the singlet of \NiAllyl\ (\cpd{4}) at 32
qubits: the number of two-qubit gates, the dominant source of error; the
depth, i.e.\ the number of sequential layers of native gates after
transpilation, which sets the run time and the exposure to decoherence; and
the estimated success probability (ESP), the product of all gate and readout
fidelities, which approximates the fraction of shots that are free of any
error. The two circuit types differ characteristically. Each SqDRIFT
exponential acts on a string of qubits set by the Jordan--Wigner mapping and
is applied one after the other, so the circuit is deep (about 1170 layers on average)
but contains few two-qubit gates (518). The LUCJ circuit is built from layers
of gates that act in parallel, so it is shallower (430--570 layers) but
contains about twice as many two-qubit gates (928 on \texttt{ibm\_phoenix} and
1077 on \texttt{ibm\_fez}, where the sparser heavy-hexagon lattice requires
additional swap gates). Because the gate count dominates the error, the ESP of
SqDRIFT (0.38) is 3 and 30 times higher than that of LUCJ (0.12 and 0.012). Even for SqDRIFT most shots therefore contain at least one error. SQD
tolerates this because configuration recovery and the variational
diagonalization do not require error-free samples. Each source used 5--13~s of
processor time. The processors, the qubit layout and further circuit details
are given in Supplementary Section~S7. All sources were
processed with the same SQD settings (\texttt{qiskit-addon-sqd}~\cite{qiskitaddonsqd},
three batches of 300 samples), and every value is the mean and standard
deviation over six SQD seeds or six independent random sets.

\begin{table}[t]
\centering
\caption{\textbf{Circuits for the singlet of \NiAllyl\ (\cpd{4}) at 32 qubits,
CAS(16,16).} Two-qubit gates (2q), depth after transpilation to the native
gates of the processor, estimated success probability (ESP) and processor
time per source (40\,000 shots). SqDRIFT values: mean over the 400 random
circuits (Supplementary Table~S13).}
\label{tab:circ_main}
\footnotesize
\setlength{\tabcolsep}{4pt}
\begin{tabular}{@{}llcccc@{}}
\toprule
Source & Processor & 2q & Depth & ESP & Time (s) \\
\midrule
SqDRIFT & \texttt{ibm\_phoenix} & 518  & 1171 & 0.38  & 6  \\
LUCJ    & \texttt{ibm\_phoenix} & 928  & 427  & 0.12  & 5  \\
LUCJ    & \texttt{ibm\_fez}     & 1077 & 566  & 0.012 & 13 \\
Random  & none (classical)      & --   & --   & --    & -- \\
\bottomrule
\end{tabular}
\end{table}

\begin{figure*}[!tp]
  \centering
  \setlength{\fboxsep}{4pt}\setlength{\fboxrule}{0.5pt}%
  \fbox{\begin{minipage}{0.97\textwidth}
    \centering
    \textbf{(a)}~\NiAllyl\ (\cpd{4}), singlet and triplet, 16 orbitals\\[2pt]
    \includegraphics[width=\linewidth]{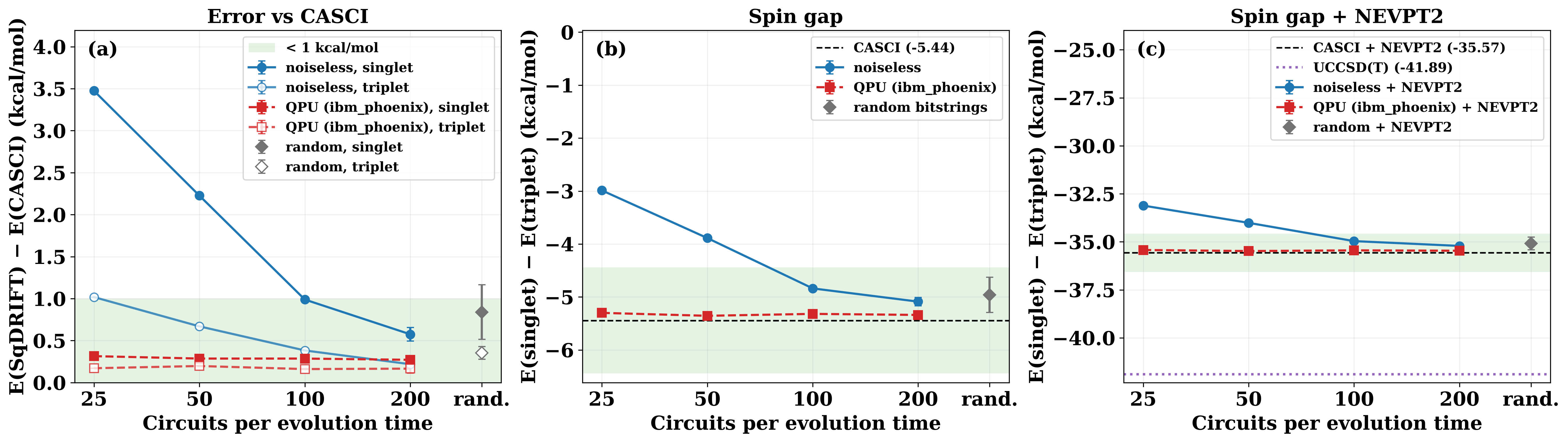}
  \end{minipage}}\\[6pt]
  \fbox{\begin{minipage}{0.97\textwidth}
    \centering
    \textbf{(b)}~\CoCOthree\ (\cpd{8}), singlet and quintet, 16 orbitals\\[2pt]
    \includegraphics[width=\linewidth]{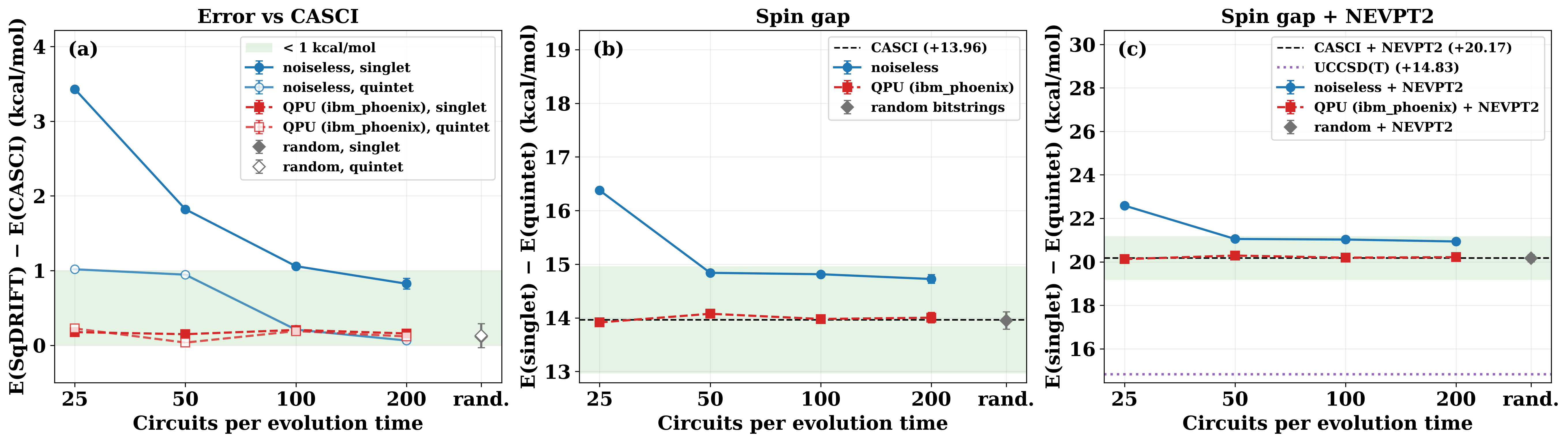}
  \end{minipage}}
  \caption{\textbf{SqDRIFT on 32 qubits: noiseless simulation and
  \texttt{ibm\_phoenix}.} 16-orbital active spaces of \NiAllyl\ (\cpd{4};
  \textbf{a}) and \CoCOthree\ (\cpd{8}; \textbf{b}). Left: SQD error relative to CASCI for
  the low-spin (filled symbols) and high-spin (open symbols) state as a function
  of the number of random circuits per evolution time. Center: spin-state gap
  compared with CASCI. Right: gap with the NEVPT2 correction of the same active
  space, compared with CASCI+NEVPT2 and full-molecule UCCSD(T). Gray diamonds:
  uniformly random bitstrings with the same total number of samples. Green
  band: within 1\kcal\ of the reference. Error bars: standard deviation over
  six SQD seeds or six random sets.}
  \label{fig:sqd_conv}
\end{figure*}

At 32 qubits (16 orbitals), the largest spaces selected by the NOON scans,
SqDRIFT on \texttt{ibm\_phoenix} reaches chemical accuracy for both states of
both complexes already with 25 circuits per evolution time
(Fig.~\ref{fig:sqd_conv}). For the Ni singlet the error falls from 0.31 to
0.27\kcal\ between 25 and 200 circuits, whereas noiseless SqDRIFT starts at
3.5\kcal\ and reaches $0.58\pm0.08$\kcal\ (six SQD seeds) only at 200
circuits. The Co singlet behaves in the same way (3.4\kcal\ at 25 and
$0.82\pm0.07$\kcal\ at 200 noiseless circuits, against 0.16\kcal\ on the
hardware). The reason is sample diversity: noiseless randomized evolutions from
the Hartree--Fock state remain concentrated on few configurations, while
hardware noise spreads the samples over many more determinants, which the
tensor-product subspace and configuration recovery turn into a larger and
better subspace. For the weakly correlated high-spin states the noiseless
samples suffice as well (0.22 and 0.06\kcal\ at 200 circuits). With 200
circuits, the hardware spin gaps are $-5.34\pm0.06$\kcal\ for \NiAllyl\ (\cpd{4},
CASCI $-5.44$) and $14.00\pm0.10$\kcal\ for \CoCOthree\ (\cpd{8}, CASCI 13.96). With
the NEVPT2 correction they become $-35.46$ and 20.21\kcal, against CASCI+NEVPT2
values of $-35.57$ and 20.17\kcal\ and UCCSD(T) values of $-41.89$ and
14.83\kcal.

\begin{figure*}[!tp]
  \centering
  \setlength{\fboxsep}{4pt}\setlength{\fboxrule}{0.5pt}%
  \fbox{\begin{minipage}{0.97\textwidth}
    \centering
    \textbf{(a)}~\NiAllyl\ (\cpd{4}), singlet and triplet\\[2pt]
    \includegraphics[width=\linewidth]{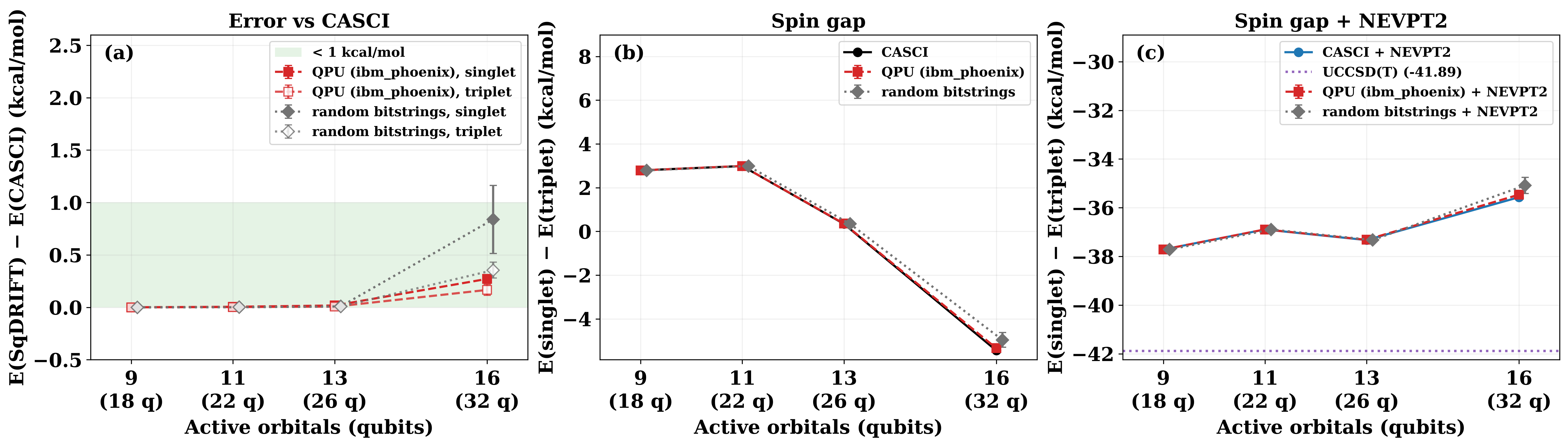}
  \end{minipage}}\\[6pt]
  \fbox{\begin{minipage}{0.97\textwidth}
    \centering
    \textbf{(b)}~\CoCOthree\ (\cpd{8}), singlet and quintet\\[2pt]
    \includegraphics[width=\linewidth]{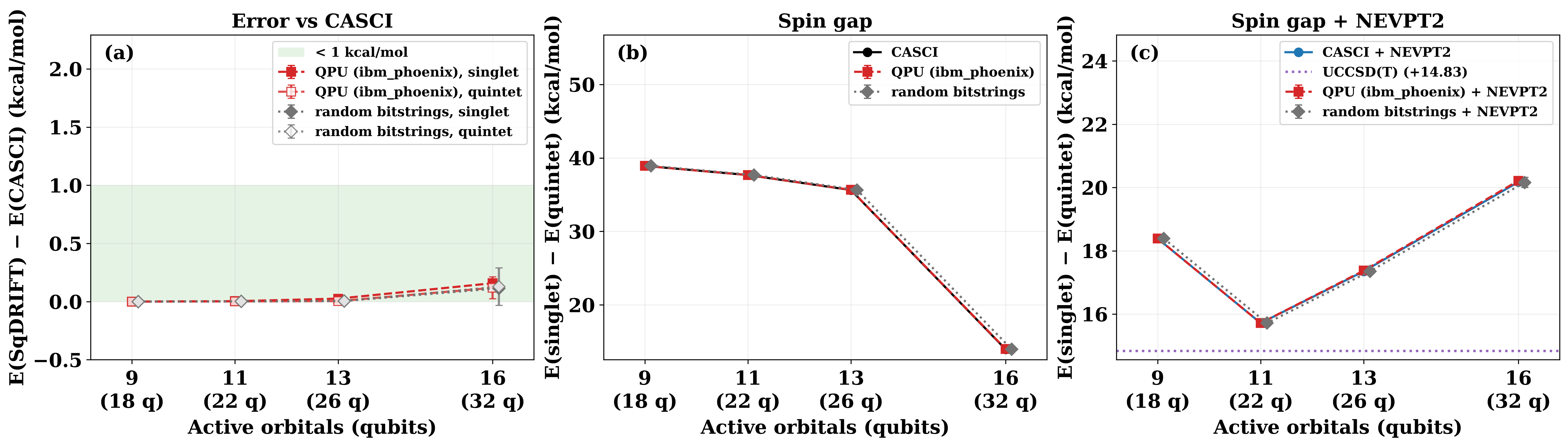}
  \end{minipage}}
  \caption{\textbf{Spin-state gaps from \texttt{ibm\_phoenix} for active spaces
  of 9 to 16 orbitals (18--32 qubits).} \NiAllyl\ (\cpd{4}; \textbf{a}) and \CoCOthree\ (\cpd{8};
  \textbf{b}); both states of a complex use active spaces of the same size from
  the predicted ranking. Left: SQD error relative to CASCI for each state.
  Center: spin-state gap compared with CASCI. Right: gap with the NEVPT2
  correction, compared with CASCI+NEVPT2 and UCCSD(T). Red: SqDRIFT on
  \texttt{ibm\_phoenix} ($L=10$, 200 circuits per evolution time, 100 shots
  per circuit); gray: uniformly random bitstrings. Error bars: standard
  deviation over six SQD seeds or six random sets.}
  \label{fig:sqd_size}
\end{figure*}

We then repeated the hardware calculation for the 9-, 11-, 13- and
16-orbital spaces of the predicted ranking (Fig.~\ref{fig:sqd_size}). At every
size and for every state, the SqDRIFT error stays below 0.3\kcal. It is below
0.025\kcal\ up to 26 qubits and grows to 0.12--0.27\kcal\ at 32 qubits.
The hardware gaps follow the CASCI gaps to within 0.10\kcal\ at all sizes,
so that the hardware reproduces the size dependence of the gap analyzed in
Table~\ref{tab:gaps}: with the NEVPT2 correction, the gap of \NiAllyl\ (\cpd{4})
varies between $-37.7$ and $-35.5$\kcal\ and that of \CoCOthree\ (\cpd{8})
between 15.7 and 20.2\kcal\ as the space grows from 9 to 16 orbitals. Up to
26 qubits, uniformly random bitstrings are as accurate as the hardware samples
(errors below 0.01\kcal), because the subspace they span covers most of the
configuration space. At 32 qubits this is no longer the case for \NiAllyl{}:
random sampling gives errors of $0.84\pm0.33$ and $0.35\pm0.08$\kcal\ for the
singlet and triplet, against $0.27\pm0.03$ and $0.17\pm0.05$\kcal\ on the
hardware, and the random gap ($-4.96\pm0.33$\kcal) deviates from CASCI by
0.5\kcal. For the more weakly correlated \CoCOthree{}, random bitstrings remain
comparable to the hardware at 32 qubits ($0.11\pm0.02$ and $0.13\pm0.16$\kcal,
against $0.16\pm0.02$ and $0.12\pm0.09$\kcal). The hardware samples are also
more often physically valid: at 32 qubits 2.4--5.4\% of them carry the correct
particle numbers, against 0.4--3.9\% of random bitstrings.

\begin{figure*}[!tp]
  \centering
  \setlength{\fboxsep}{4pt}\setlength{\fboxrule}{0.5pt}%
  \fbox{\begin{minipage}[t]{0.475\textwidth}
    \centering
    \includegraphics[width=\linewidth]{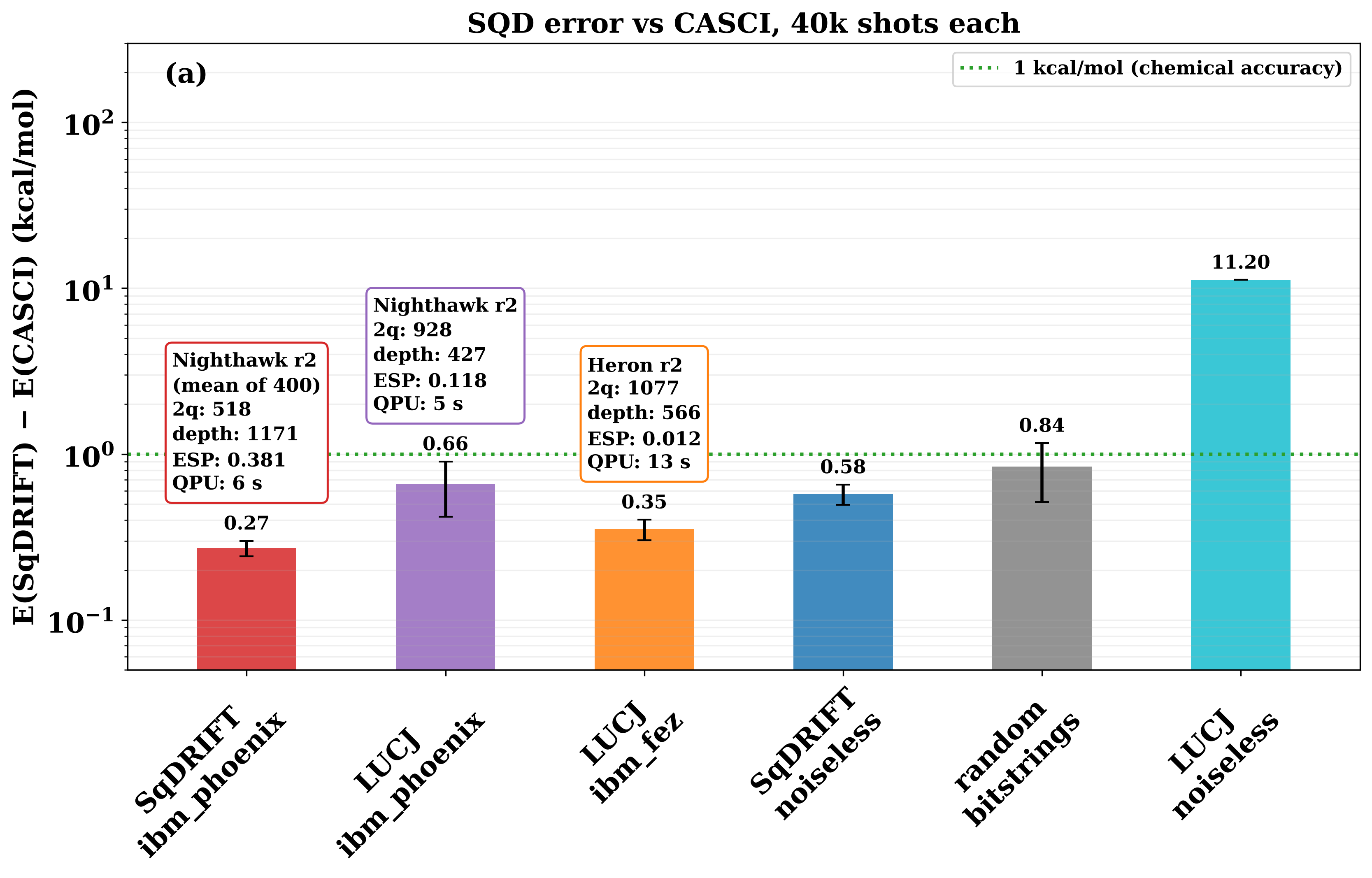}
  \end{minipage}}\hfill
  \fbox{\begin{minipage}[t]{0.475\textwidth}
    \centering
    \includegraphics[width=\linewidth]{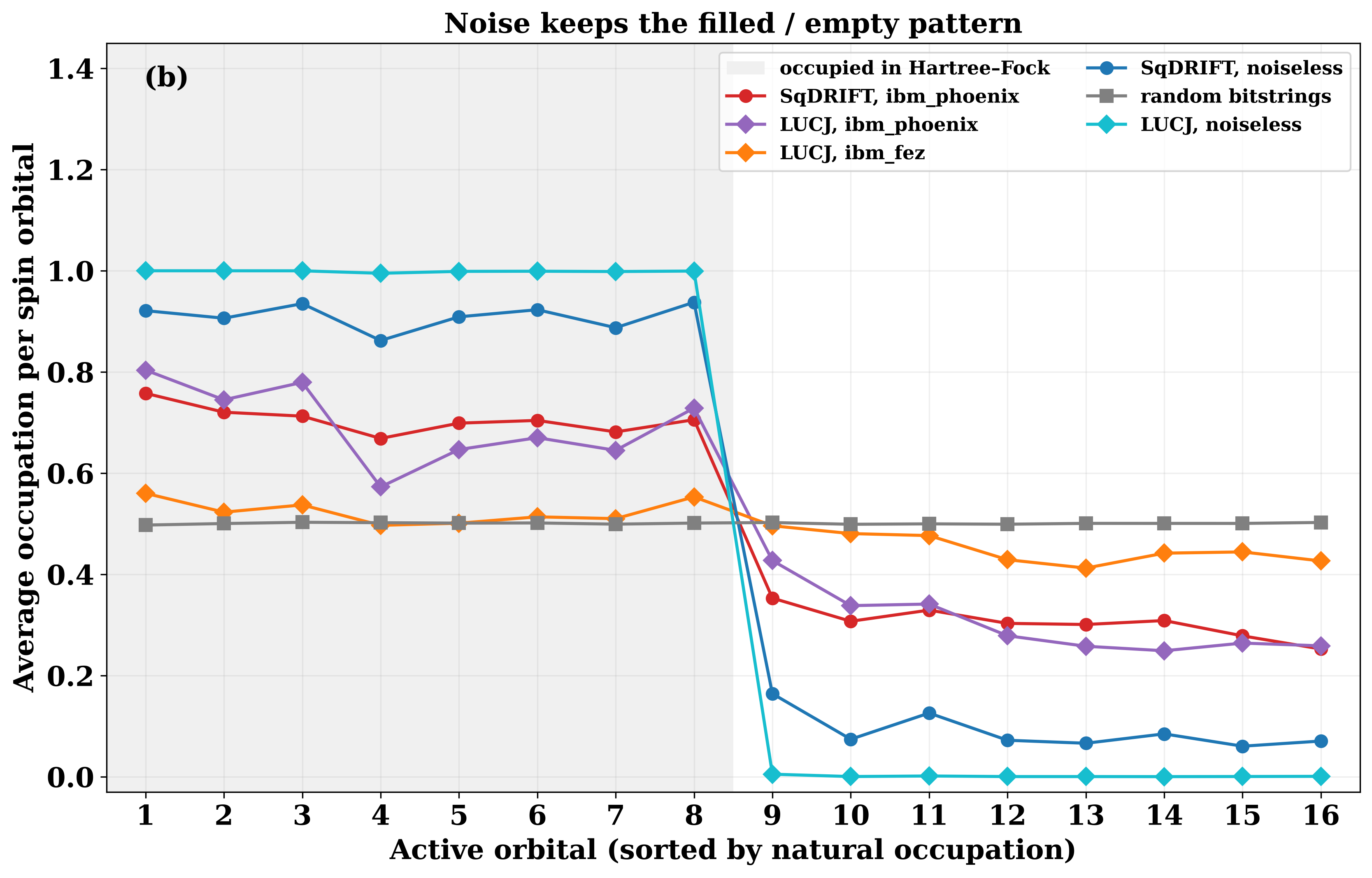}
  \end{minipage}}
  \caption{\textbf{Sample sources for the singlet of \NiAllyl\ (\cpd{4}) at 32 qubits,
  CAS(16,16).} \textbf{a},~SQD error relative to CASCI with 40\,000 samples
  from each source and identical SQD settings (logarithmic scale; mean and
  standard deviation over six SQD seeds or random sets). Boxes give the chip,
  the number of two-qubit gates (2q), the circuit depth, the estimated success
  probability (ESP) and the processor time; for SqDRIFT the mean over the
400 circuits. Dotted line: 1\kcal.
  \textbf{b},~Average occupation of each active spin orbital in the samples,
  with the orbitals sorted by natural occupation; the shaded region marks the
  orbitals occupied in the Hartree--Fock determinant.}
  \label{fig:sqd_samplers}
\end{figure*}

Figure~\ref{fig:sqd_samplers} compares the sample sources for the most
demanding case, the singlet of \NiAllyl\ (\cpd{4}) at 32 qubits, whose correlation extends
into the Ni--allyl bond (Sec.~\ref{ssec:noon}). Noiseless LUCJ is the least
accurate source (11.2\kcal): its CCSD-parametrized state is dominated by the
Hartree--Fock determinant, so the samples fill the occupied orbitals and leave
the virtual orbitals empty (Fig.~\ref{fig:sqd_samplers}b), and the 40\,000
samples contain only 74 distinct configurations. Uniformly random bitstrings
have the opposite problem: every orbital is occupied with probability one
half, most samples have the wrong particle numbers, and the error is
$0.84\pm0.33$\kcal. The hardware samples lie between these limits. Noise
spreads them over many configurations but preserves the filled/empty pattern
of the reference, with average occupations of about 0.5--0.8 for the occupied
and 0.25--0.5 for the virtual orbitals. As a result, all three hardware sources
reach chemical accuracy, and SqDRIFT gives the lowest error,
$0.27\pm0.03$\kcal, with the fewest two-qubit gates and the highest success
probability (Table~\ref{tab:circ_main}). LUCJ gives $0.35\pm0.05$\kcal\ on
\texttt{ibm\_fez} but $0.66\pm0.24$\kcal\ on \texttt{ibm\_phoenix}, where its circuit
runs with a ten times higher ESP: more of the phoenix samples reproduce the
narrow noiseless distribution (35\,642 distinct configurations among the 40\,000
shots, against 39\,999 on \texttt{ibm\_fez}), and their occupations keep more of
the Hartree--Fock pattern. For LUCJ in this case, the lower error rate of
\texttt{ibm\_phoenix} therefore did not give a better SQD energy.

This benefit of hardware noise refers to the
systems and circuits studied here, and whether it persists for larger active
spaces or other noise levels remains open.

These results carry the complete workflow, from a single UHF calculation
through the predicted active space to a quantum processor, up to 32 qubits
for two open-shell transition-metal complexes, and recover spin-state gaps
that agree with exact CASCI for the same active-space Hamiltonians to
0.10\kcal. The separate classical CASCI+NEVPT2 calculations give spin-state gaps
that differ from full-molecule UCCSD(T) by 5.3--6.3\kcal\ at 16 orbitals.
This cross-method comparison remains subject to active-space and
reference-method uncertainties. The role of the hardware is to provide samples that are diverse but
structured. At 32 qubits this is the first size at which uniformly random
sampling becomes clearly worse for the more strongly correlated nickel complex. All spaces studied here remain within the
reach of exact diagonalization, so the calculations validate the workflow and
do not indicate a quantum advantage. A 22-qubit LUCJ calculation for the three states of \CoCOtwo\ (\cpd{8}) on
\texttt{ibm\_phoenix} reproduces CASCI to within 0.02\kcal\ (Supplementary
Section~S7).

\subsection{Scope and limitations}
\label{ssec:limits}

Five limitations follow from the results. First, the model compresses strong
static correlation: entropies of about 0.6, as in the open-shell singlets of the
vanadium(III) ions, are predicted at the correct orbitals but with much
smaller magnitude. These states are rare in the training distribution, and
they are recognized from their chemistry, low-spin states of metals absent
from training, rather than from the UHF reference, which shows no
significant instability. Second, in weakly correlated high-spin states
the entropy profile is flat near any threshold, so the absolute predicted
values are less reliable than the predicted ranking. In the quintet of
\cpd{3}, no orbital is predicted above 0.10, yet the four DMRG-active orbitals appear at
ranks 1, 3, 5 and 6 of the prediction, and
a rank-based or relative criterion would recover them. Third, the singly
occupied $d$ orbitals of high-spin states carry little entanglement in any
entropy-based measure, DMRG or predicted, and are included by a chemical rule
rather than by the model. Fourth, the workflow requires a spin-clean
UHF reference. Several intermediate-spin states of the wider benchmark do not
meet this criterion, so the most strongly spin-contaminated open-shell states
are at present outside its scope. Fifth, the model works in a fixed window
of 40 orbitals around the frontier gap. The boundary descriptors indicate
coupling to orbitals outside the window but cannot recover them, so complexes
with extended ligand systems may need a larger window, which we have not
tested. All results refer to the def2-SVP basis,
gas-phase species at fixed geometries, and elements up to Kr. The basis-set error could be reduced without enlarging the active
space, and hence the qubit count, with the transcorrelated method, which builds
short-range electron correlation into an effective Hamiltonian through a
Jastrow factor~\cite{BoysHandy1969,Motta2020TC,McArdle2020TC,Sokolov2023TC,Dobrautz2024TC,Magnusson2024TC}.

\section{Conclusions}
\label{sec:conclusions}

We have presented \texttt{SpinAware}, a model that predicts the orbital entanglement of
transition-metal complexes (DMRG single-orbital entropies and orbital-pair
mutual information) from a single UHF calculation, and we have used it to
build active spaces for classical and quantum multireference calculations.
The model combines spin-resolved descriptors that encode the spin state
through $\alpha$--$\beta$ occupation and energy differences, continuous
metal--ligand composition in place of binary shell flags, and a
transformer that scores each orbital in the context of all other
orbitals of the window. Most of the gain over earlier models comes from
the descriptors. Processing the orbitals jointly adds a smaller
improvement, and the exchange information enters through the
descriptors rather than through the attention bias.

On a benchmark of experimentally characterized complexes defined
independently of the model, no systematic decrease in accuracy is observed
across the compositional novelty tiers considered here:
complexes that depart from the training data in metal, ligand, binding
topology and oxidation state at once are among the best predicted. The
residual errors are electronic rather than chemical in origin, and occur in
two regimes: strong static correlation, whose magnitude is compressed, and
weak correlation near the selection threshold, where a hard cut is
ill-conditioned. A candidate pool of predicted entropies and mutual-information partners
contains 99\% of the DMRG-active orbitals of the eleven benchmark spin
states (82 of 83) while holding, per state, 38\% of the window, and 28 of the 29 active orbitals for the two application systems.

A NOON scan selects the final active space from this
pool. Where the correlation extends into the metal--ligand bonds, the
predicted ranking places the decisive ligand orbitals among its first picks.
Where it is confined to the $d$ shell, the predicted orbitals remain
passengers and do not distort the result. From 13 to 16 orbitals, the corrected gaps change by
1.8\kcal\ for nickel and 2.8\kcal\ for cobalt. The selected spaces give
spin-state gaps that, after classically evaluated NEVPT2 corrections, differ from full-molecule
UCCSD(T) by 0.9--6.3\kcal\ for active spaces of 9--16 orbitals, and at 16 orbitals the
corrected gap still depends on the ligand orbitals at the boundary of the space This comparison remains subject to residual active-space and
reference-method uncertainties. On IBM quantum processors, SQD with samples
from randomized time evolutions (SqDRIFT) reproduces exact CASCI energies
for the same active-space Hamiltonians to within 0.3\kcal\ up to 32 qubits, where noisy hardware samples
converge faster than noiseless ones and, for the nickel complex, outperform uniformly random
bitstrings. Extending the training data to early transition metals and
strongly correlated low-spin states, using the predicted ranking rather than
absolute thresholds, and applying the workflow to active spaces beyond the
reach of exact diagonalization are the natural next steps.

Beyond these steps, the prediction is cheap enough to serve
as a component of larger automated workflows. High-throughput explorations of
reaction mechanisms already launch multiconfigurational calculations
automatically where coupled-cluster diagnostics indicate strong
correlation~\cite{Unsleber2023}, and an end-to-end demonstration combining
high-performance computing, entanglement-based active-space selection and
logical qubits has shown how such a pipeline can deliver the active site of a
catalytic reaction to a quantum computer~\cite{MicrosoftQuantum2024}. In both
cases the strongly correlated configurations are identified with correlated
calculations. \texttt{SpinAware} could provide the same orbital-level information from
the UHF calculation of each spin state, and thereby carry these pipelines to the
spin-state chemistry of open-shell catalysts. Spin-forbidden reactions are a
natural target: free-energy simulations along a spin-gap collective variable
now reach crossing regions between surfaces of different
multiplicity~\cite{Mullukkandy2026}, and spin-state-consistent active spaces
could supply multireference reference points along such paths.
A second direction is embedding. Active-space embedding
treats a correlated fragment of a molecule or solid with a quantum solver in a
classical environment~\cite{Battaglia2024}, and the orbital-embedding method eCAS
assembles localized active spaces that can be solved independently, adding
environment orbitals according to a mutual-correlation
criterion~\cite{Brunisholz2026}. Predicted orbital entropies and mutual
information could supply such criteria at mean-field cost, once the training
data are extended from isolated complexes to metal sites in larger
environments, such as metalloenzymes and defects in solids. Finally, agentic
frameworks for multireference calculations reduce errors relative to fixed
protocols but vary widely in strategy and cost from run to
run~\cite{PerezSanchez2026}. A fast, quantitative predictor of orbital
entanglement could serve these agents as a tool that supplies a reproducible
starting point for the active space, while AI-driven discovery of quantum
algorithms~\cite{Finger2026} could then be applied to the Hamiltonians of the
selected spaces.

\begin{acknowledgments}
 S.K. acknowledges funding from the Quantum Science Center, a National Quantum Information Science Research Center of the U.S. Department of Energy (DOE), operated at Oak Ridge National Laboratory (ORNL).  W.D. acknowledges funding from the German Federal Ministry of Research, Technology and Space (BMFTR) under the research program Quantensysteme and funding measure Quantum Futur 3 for project No. 13N17229, as well as from the Helmholtz Association via the Initiative and Networking Fund for project No. VH-NG-21-08 under the Helmholtz Investigator program and for projects KA-QUS-02 (qFLOW) and KA-QUS-03 (QT-Batt) under the Helmholtz Quantum Use case call. We thank the Scultetus Center at CASUS for funding the research stay of M.A.A.A., which enabled this work.

We gratefully acknowledge the computing time made available to us on the high-performance computers at the NHR Center Paderborn Center for Parallel Computing (PC2) (projects hpc-prf-qehpc and hpc-prf-nniap). PC2 is jointly supported by the Federal Ministry of Research, Technology and Space of Germany and the state governments participating in the NHR (www.nhr-verein.de/unsere-partner). We also acknowledge computing time on the Hazel cluster at North Carolina State University and access to the IBM Quantum systems \texttt{ibm\_phoenix} and \texttt{ibm\_fez}. The views expressed are those of the
authors and do not reflect the official policy or position of IBM or the IBM
Quantum team.

\end{acknowledgments}

\paragraph*{Author contributions.}
M.A.A.A.\ conceived the method, generated the training and benchmark data,
developed and trained the models, performed the DMRG, NOON-scan, CASCI, NEVPT2 and
quantum-hardware calculations, analyzed the results and wrote the original draft.
W.D.\ and S.K.\ contributed equally: both supervised the project, acquired funding
and contributed to writing the manuscript. W.D.\ additionally provided access to
high-performance computing resources and S.K.\ to quantum-computing resources.
T.D.K.\ acquired funding and contributed to writing the manuscript.
All authors reviewed and approved the final manuscript.

\paragraph*{Data availability.}
The trained models, the benchmark structures with their DMRG reference labels,
the results of the NOON scans and of the CASCI, NEVPT2, DMRG and UCCSD(T)
calculations, the active-space Hamiltonians,  the raw measurement counts of
all quantum-hardware runs and the training data set are available from the corresponding
authors upon reasonable request.

\paragraph*{Code availability.}
The SpinAware code for feature generation and prediction, together with the
scripts for the NOON scans, the classical reference calculations and the SQD,
SqDRIFT and LUCJ workflows, the five trained Model~1 and Model~2 pairs,
their training scripts, the split by complex and a script that recomputes the
benchmark tables from the released models,is available from the corresponding authors
upon reasonable request.

\bibliographystyle{apsrev4-2}
\bibliography{refs}

\end{document}


\sloppy
\maketitle
\tableofcontents
\clearpage

\section{Benchmark complexes and reference calculations}

\paragraph{Composition of the benchmark.}
The out-of-distribution benchmark comprises thirteen mononuclear complexes taken
from experimental crystal structures in the Cambridge Structural Database (CSD).
Seven of them (\cpd{1}--\cpd{7}), chosen to span the three difficulty tiers,
are analyzed in detail in the main text; complex \cpd{4} is also used there as an
application system. The remaining six (\cpd{9}--\cpd{14}) complement this set
with further metals and coordination environments and are analyzed in
Sections~\ref{sec:si_quality} and \ref{sec:si_figures}; their composition is
given in Table~\ref{tab:si_comp}. Complex \cpd{8},
\ce{[Co(H2O)5(CO2)]^{2+/3+}}, is not a crystal structure and is described in the
main text (Section~III\,C). The source crystal structures are
listed in Table~\ref{tab:si_csd}.
In every case the calculation was performed on the isolated metal-containing
species; counter-ions and lattice solvent were omitted.

\paragraph{Scope.}
The training set contains only elements up to Kr, for which the def2-SVP basis
is all-electron. Elements beyond Kr are described with an effective core
potential, which replaces the core electrons and thereby changes the one- and
two-electron descriptors of the valence orbitals in a way that is not
represented in the training data. Complexes containing such elements lie
outside the scope of the present model, and all benchmark complexes were
therefore restricted to elements up to Kr.

\paragraph{Reference wavefunctions.}
Because UHF solutions depend on the initial guess, the reference of every spin
state was obtained from five starting points (restricted-open-shell or
restricted starts from minimal-basis and H\"uckel guesses, and unrestricted
minimal-basis, H\"uckel and atomic guesses), each followed through internal
stability analysis. The lowest solution satisfying
$|\langle \hat S^2\rangle - S(S+1)| \le \max[0.1,\,0.1\,S(S+1)]$ was retained,
and all DMRG labels and predictions refer to this solution. States for which no
such solution exists are marked $^\ast$ and excluded from all averages.
Closed-shell singlet solutions that satisfy this criterion but are internally
unstable towards a broken-symmetry solution with larger spin contamination are
retained; this is the case for the singlets of \cpd{4} and \cpd{8}
(Section~\ref{sec:si_app}).
Table~\ref{tab:si_uhf} lists, for every benchmark state, the UHF energy of the
reference used for the labels and predictions and the outcome of this
screening. Where the screening found a lower spin-clean solution than the
first calculation, the descriptors and the DMRG label were recomputed on it.

\begin{table}[h!]
\centering
\caption{\textbf{Source crystal structures of the benchmark complexes.}
Compound names are given as deposited in the CSD. Calculations used only the
metal-containing species; the counter-ions listed here were omitted.}
\label{tab:si_csd}
\footnotesize
\setlength{\tabcolsep}{4pt}
\begin{tabular}{@{}clll@{}}
\toprule
No. & CSD refcode & CSD compound name & Counter-ion(s) omitted \\
\midrule
\cpd{1} & YUDFUY & tetrakis(Methylammonium) hexachloro-vanadium(iii) chloride
        & 4 \ce{CH3NH3+}, \ce{Cl-} \\
\cpd{2} & LIDYUT & Diethylenetriammonium hexafluoro-vanadium(iii)
        & \ce{[H3N(CH2)2NH2(CH2)2NH3]^3+} \\
\cpd{3} & ZONCUD & \parbox[t]{6.6cm}{(acetylacetonato)-tetrakis(1-methylpyrrolidin-2-one)-magnesium(ii) tetramethyl-cobalt(iii)}
        & \parbox[t]{4.2cm}{\raggedright (acetylacetonato)tetrakis(1-methylpyrrolidin-2-one)magnesium(II)} \\
\cpd{4} & HAVTIH & \parbox[t]{6.6cm}{($\eta^3$-syn-Crotyl)-bis(acetonitrile)-nickel(ii) hexafluorophosphate} & \ce{PF6-} \\
\cpd{5} & VILFUR & Pentacarbonyl-morpholine-chromium(0) & none \\
\cpd{6} & RIQPUC & Tetraethylammonium tetracarbonyl-glycinato-chromium(0) & \ce{(C2H5)4N+} \\
\cpd{7} & BELYIA & not given in the entry$^{a}$ & none \\
\addlinespace
\cpd{9}  & ADUGIT & \parbox[t]{6.6cm}{\raggedright bis(3,3-difluorocyclobutan-1-aminium) tetrachloro-chromium(ii)} & 2 \ce{C4H8F2N+} \\
\cpd{10} & HEGNUE & tris(imidazolium) hexafluoro-vanadium(iii) & 3 \ce{C3H5N2+} \\
\cpd{11} & QOVWIH & \parbox[t]{6.6cm}{\raggedright Tetraphenylphosphonium (acetonitrile)-dibromo-chloro-cobalt(ii)} & \ce{(C6H5)4P+} \\
\cpd{12} & SEDGAN & \parbox[t]{6.6cm}{\raggedright bromo-tricarbonyl-(2,7-dimethyl-1-oxido-1,8-naphthyridine)-manganese(i)} & none \\
\cpd{13} & SESDUP & \parbox[t]{6.6cm}{\raggedright ($\eta^6$-Benzene)-bis(bis($\mu_2$-chloro)-dichloro-aluminum)-titanium(ii) benzene solvate (second structure of \cpd{7})} & \ce{C6H6} (lattice solvent) \\
\cpd{14} & VEBPUN & \parbox[t]{6.6cm}{\raggedright bis(3-Diethylamino-5-phenyl-1,2,4-dithiazolium) tetrachloro-nickel(ii)} & 2 \ce{C12H15N2S2+} \\
\bottomrule
\end{tabular}
\vspace{2pt}
\raggedright\footnotesize
$^{a}$Assigned as ($\eta^6$-benzene)bis(tetrachloroaluminato)titanium(II) from
composition and connectivity (CSD deposition number 1108465).
\end{table}

\paragraph{Compound names.}
\begingroup\raggedright
\begin{list}{}{%
  \setlength{\leftmargin}{1.6em}\setlength{\labelwidth}{1.2em}%
  \setlength{\labelsep}{0.4em}\setlength{\itemsep}{1pt}%
  \setlength{\parsep}{0pt}\setlength{\topsep}{2pt}}
\item[\cpd{1}] hexachloridovanadate(III)
\item[\cpd{2}] hexafluoridovanadate(III)
\item[\cpd{3}] tetramethylcobaltate(III)
\item[\cpd{4}] bis(acetonitrile-$\kappa N$)($\eta^3$-\textit{syn}-but-2-en-1-yl)nickel(II)
\item[\cpd{5}] pentacarbonyl(morpholine-$\kappa N$)chromium(0)
\item[\cpd{6}] tetracarbonyl(glycinato-$\kappa^2 N,O$)chromate(0)
\item[\cpd{7}] ($\eta^6$-benzene)bis(tetrachloridoaluminato-$\kappa^2 \mathit{Cl},\mathit{Cl}'$)titanium(II)
\item[\cpd{8}] pentaaqua(carbon dioxide-$\kappa O$)cobalt(III) and cobalt(II)
\end{list}
\endgroup

\paragraph{Structures of complexes \cpd{1}--\cpd{7}.}
\emph{Tier I.} Complex \cpd{1} is a near-ideal octahedron
(V--Cl 2.384--2.389~\AA, $\Sigma = 2.5^\circ$) and differs from the training
set only in its metal. Complex \cpd{2} contains the same $d^2$ vanadium(III)
center in a strongly distorted fluoride environment (V--F 1.899--2.018~\AA,
$\Sigma = 29.0^\circ$, \emph{trans} angles 173.3--175.4$^\circ$), imposed in
the crystal by N--H$\cdots$F hydrogen bonding to the triammonium counter-ion.
A second, more symmetric crystal structure of \ce{[VF6]^3-} (complex
\cpd{10}, Section~\ref{sec:si_quality}) gives the same picture.
Complex \cpd{3} introduces $\sigma$-methyl carbon donors, which occur in only two
training calculations (Co--C 1.951--2.011~\AA). Its geometry lies far from a tetrahedron
($\tau_4 = 0.39$; largest angles 147.8 and 156.5$^\circ$).

\emph{Tier II.} Complex \cpd{4} is planar (sum of angles at Ni $359.9^\circ$)
and contains an asymmetrically bound $\eta^3$-\textit{syn}-crotyl ligand
(Ni--C 1.854--2.133~\AA, Ni--centroid 1.751~\AA) \emph{trans} to two
acetonitrile donors (Ni--N 1.882--1.890~\AA, N--Ni--N 95.6$^\circ$).
Complex \cpd{5} is a zero-valent chromium carbonyl bearing a single
N-bound morpholine (Cr--N 2.196~\AA, $\Sigma = 23.3^\circ$); the carbonyl
\emph{trans} to nitrogen is markedly shorter than the four \emph{cis} carbonyls
(Cr--C 1.820 versus 1.886--1.907~\AA), consistent with enhanced back-donation
to the CO opposite the weak $\pi$-acceptor.
Complex \cpd{6} is an anionic chromium(0) carbonylate bearing
an N,O-chelating glycinate. Its small bite angle (75.1$^\circ$;
Cr--N 2.150~\AA, Cr--O 2.218~\AA) makes it the most distorted octahedron in the
set ($\Sigma = 55.2^\circ$), and the Cr--C distances span 1.731--1.891~\AA.
Because chelating ligands are common in training, its chelation does not count
as a departure; its amino-acidate ligand (rare) and its zero-valent carbonyl
chemistry (absent) do.

\emph{Tier III.} Complex \cpd{7} departs from training in metal, ligand and topology
simultaneously: an $\eta^6$-benzene titanium(II) four-legged piano stool
(Ti--centroid 2.069~\AA, Ti--C 2.466--2.547~\AA) with two $\kappa^2$-bound
tetrachloridoaluminate ligands (Ti--Cl 2.589--2.611~\AA,
Cl--Ti--Cl 77.7--78.1$^\circ$, Ti$\cdots$Al 3.40--3.44~\AA). The coordinated
ring shows pronounced C--C bond-length alternation (1.330--1.473~\AA),
consistent with substantial metal-to-arene back-donation.

\paragraph{Complementary complexes.}
Table~\ref{tab:si_comp} lists the composition of complexes \cpd{9}--\cpd{14}
as used in the calculations. Complexes \cpd{10} and \cpd{13} are second
crystal structures of the ions in \cpd{2} and \cpd{7}, respectively. The
novelty directions follow the definitions of the main text.

\begin{table}[h!]
\centering
\caption{\textbf{Complementary benchmark complexes \cpd{9}--\cpd{14}.} Atom
counts refer to the isolated metal-containing species; $q$ is the total
charge. Novelty directions as in the main text (M, metal; L, ligand; T,
binding topology; O, oxidation state).}
\label{tab:si_comp}
\footnotesize
\begin{tabular}{@{}cllcccll@{}}
\toprule
No. & Refcode & Species & Atoms & $q$ & $2S+1$ & Directions & Tier \\
\midrule
\cpd{9}  & ADUGIT & \ce{[CrCl4]^2-} & 5 & $-2$ & 5 & none & I \\
\cpd{10} & HEGNUE & \ce{[VF6]^3-} (second structure of \cpd{2}) & 7 & $-3$ & 1, 3 & M & I \\
\cpd{11} & QOVWIH & \ce{[CoBr2Cl(NCCH3)]-} & 10 & $-1$ & 4 & L & I \\
\cpd{12} & SEDGAN & \ce{[MnBr(C13H10N2O4)]} & 31 & 0 & 1, 3$^\ast$ & O & I \\
\cpd{13} & SESDUP & \ce{[Ti($\eta^6${-}C6H6)(AlCl4)2]} (second structure of \cpd{7}) & 23 & 0 & 1 & M, L, T & III \\
\cpd{14} & VEBPUN & \ce{[NiCl4]^2-} & 5 & $-2$ & 1, 3 & none & I \\
\bottomrule
\end{tabular}

\vspace{2pt}
\raggedright\footnotesize
$^\ast$No spin-clean UHF reference; excluded from all averages.
\end{table}

\section{Training data}
\label{sec:si_training}

The training set comprises 3\,454 spin-pure calculations (UHF/def2-SVP
reference, DMRG with \textsc{block2}, bond dimension $M=1024$, 16 sweeps,
40-orbital window of 20 occupied and 20 virtual orbitals). Its composition is
given in Table~\ref{tab:si_train}. States whose UHF reference deviates from the
exact $\langle\hat S^2\rangle$ by more than 0.5 were excluded.The total spin of every
DMRG label was verified with the exact $\langle\hat S^2\rangle$ (including the
overlaps between $\alpha$ and $\beta$ orbitals and the UHF core); 787 labels were
obtained with an energy penalty on $\hat S^2-S(S+1)$ (main text,
Sec.~II\,C). The training references were not stability-screened:
the model learns the map from a given UHF reference to the DMRG entropies in its
orbitals, and descriptors and labels always refer to the same reference. All calculations
use the ionic charge implied by the metal oxidation state and the formal
ligand charges. None of the benchmark complexes of the main text or of
Table~\ref{tab:si_comp} is part of the training data.

Figure~\ref{fig:si_trainset} summarizes the training set and the
in-distribution test. The 3\,454 states are distributed over Cr (524), Mn
(747), Fe (875), Co (727) and Ni (581) and over multiplicities 1--6 (595, 387,
813, 488, 808 and 363 states). The maximal single-orbital entropy per state has
a median of 0.22; 87.5\% of the states contain at least one orbital with
$s^{(1)}>0.10$, and 18.9\%, 8.0\% and 4.0\% reach 0.6, 0.8 and 1.0, up to the
limit of $\ln 4$. In the crystal-structure subset the metal is bound to N, C,
O, Cl, S, F, Br and P donors, with a few B, I, Se, Si and Sb contacts. On the
test set (866 states of 402 complexes), the five-model ensemble of Model~1
reproduces the DMRG entropies with an MAE of 0.025 (0.080 on orbitals with
$s^{(1)}>0.10$), and the five-model ensemble of Model~2 reproduces
$\log I_{ij}$ with a correlation coefficient of 0.88 over 643\,500 orbital
pairs.

\begin{table}[h!]
\centering
\caption{\textbf{Composition of the training set.} Metals Cr, Mn, Fe, Co and
Ni; spin multiplicities 1--6; total charges $-9$ to $+3$, with all but nine
calculations between $-5$ and $+3$. Ligands: \ce{F-},
\ce{Cl-}, \ce{Br-}, \ce{OH-}, \ce{S^2-}, \ce{NCS-}, \ce{N3-}, \ce{H2O},
\ce{NH3}, CO, and the ligands of the crystal-structure subset.}
\label{tab:si_train}
\small
\begin{tabular}{@{}lrl@{}}
\toprule
Subset & Calculations & Description \\
\midrule
Production I       & 681     & constructed complexes, equilibrium geometries \\
Production II      & 1\,321  & constructed complexes, equilibrium geometries \\
Pilot              & 90      & constructed complexes, equilibrium geometries \\
Charge test        & 13      & validation of ionic charges \\
Stretched          & 807     & metal--ligand bonds stretched by factors 1.5 and 2.0 \\
Crystal structures & 542     & experimental CSD structures; N, C, O, Cl, S, F, Br and P donors \\
\midrule
Total              & 3\,454  & \\
\bottomrule
\end{tabular}
\end{table}
\section{Model, training and software settings}
\label{sec:si_settings}

\begin{table}[h!]
\centering
\caption{\textbf{Architectures and training settings of the two models.}
Both models use the same split by complex; five models per task were trained
from different initial weights, and their mean prediction is used. Test metrics
refer to the held-out test complexes.}
\label{tab:si_hyper}
\footnotesize
\begin{tabular}{@{}p{2.5cm}p{5.8cm}p{5.8cm}@{}}
\toprule
 & Model 1: single-orbital entropy & Model 2: mutual information \\
\midrule
Input & 40 orbitals $\times$ 60 descriptors and $40\times40$ exchange matrix
      & orbital pair: 401 features (Eq.~21) \\
Architecture & embedding to $d=256$; 4 pre-norm transformer layers, 8 heads,
      FFN dimension 1024; exchange bias $\alpha^{(l)}\tilde K_{ij}$ per layer,
      initialized at 0; head $256\to64\to1$, linear output during training,
      clamped at zero in evaluation and inference
      & input projection to 384; 3 gated residual blocks; head $384\to192\to1$; GELU \\
Parameters & 3\,189\,125 & 1\,565\,185 \\
Target & Box--Cox ($\lambda=0.1377934$ for the final models, $\epsilon=10^{-8}$) and min--max scaled $s^{(1)}$
       & $\ln(I_{ij}+10^{-7})$ \\
Loss & mean absolute error over valid orbitals
     & Huber ($\delta=2$), weights $\propto 1+100\,I_{ij}$ \\
Optimizer & AdamW, lr $3\times10^{-4}$, weight decay $10^{-4}$
          & AdamW, lr $3\times10^{-4}$, weight decay $10^{-2}$ \\
Schedule & cosine annealing to $10^{-6}$, 2000 epochs; best validation checkpoint
         & up to 500 epochs; early stopping, patience 80 \\
Batch size & 128 molecules & 4096 pairs \\
Regularization & dropout 0.1; gradient clipping at norm 1.0 & dropout 0.15 \\
Data split & by complex: 1\,103 / 125 / 402 complexes, 2\,329 / 259 / 866 states (train / validation / test)
           & same complexes; 2\,174 / 241 / 825 states with mutual-information labels \\
Test & ensemble: MAE 0.025, $F_1 = 0.83$ ($s^{(1)}\ge0.10$); single models $F_1 = 0.818\pm0.002$
     & $r(\ln\hat I,\ln I) = 0.865\pm0.001$; AUC($I>0.005$) $=0.918$ \\
Hardware & one NVIDIA A100, $\approx$20 min per model
         & one NVIDIA A100, $\approx$1~h per model \\
\bottomrule
\end{tabular}
\end{table}

\begin{table}[h!]
\centering
\caption{\textbf{Software, quantum hardware and computing resources.}}
\label{tab:si_software}
\footnotesize
\begin{tabular}{@{}llp{8.4cm}@{}}
\toprule
Package or resource & Version & Used for \\
\midrule
PySCF & 2.13.1 & UHF and stability analysis, integrals, CASCI, sc-NEVPT2, UCCSD(T) \\
\textsc{block2} & 0.5.3 & DMRG labels, DMRG-CASCI, RDMs for DMRG-NEVPT2 \\
PyTorch & 2.5.1 & Models 1 and 2 \\
NumPy / SciPy & 2.4.6 / 1.17.1 & data handling; Box--Cox transformation \\
scikit-learn & 1.9.0 & min--max scaling, metrics \\
Matplotlib & 3.11.2 & figures \\
Qiskit & 2.5.2 & circuits, transpilation \\
\texttt{ffsim} & 0.0.84 & LUCJ circuits from CCSD amplitudes \\
\texttt{qiskit-addon-sqd} & 0.13.1 & configuration recovery and SQD \\
Qiskit Aer & 0.17.2 & noiseless and noisy (\texttt{FakeNighthawk}) simulation \\
Qiskit Aer GPU & 0.18.0 & noiseless 32-qubit SqDRIFT (cuStateVec 1.6.0) \\
\texttt{qiskit-fermions} & 0.2.0 & fermionic operators for SqDRIFT \\
Qiskit IBM Runtime & 0.50.0 & hardware execution \\
\midrule
\texttt{ibm\_phoenix} & IBM Nighthawk r2 & SqDRIFT and LUCJ, 18--32 qubits; LUCJ for \cpd{8}, 22 qubits \\
\texttt{ibm\_fez} & IBM Heron r2 & LUCJ, 32 qubits \\
Noctua~2 (PC$^2$, Paderborn) & -- & DMRG labels, model training and inference \\
Hazel (NC State University) & -- & CASCI, NEVPT2, UCCSD(T), SQD preparation and analysis \\
Otus (PC$^2$, Paderborn) & -- & noiseless 32-qubit SqDRIFT simulation (NVIDIA H100) \\
ACES (Texas A\&M University) & -- & DMRG \\
\bottomrule
\end{tabular}
\end{table}

\subsection{Exchange statistics in an idealized model}
\label{sec:si_kstats}

The way the exchange statistics of Eqs.~(7) and
(8) of the main text resolve the three archetypes of
Fig.~2a--c of the main text can be seen in an
idealized model. Let orbital A couple with strength $K_0$ to
$m$ partners, let orbital B couple with the same strength $K_0$ to a
single partner, and let orbital C couple with a small strength
$\delta\ll K_0$ to all $N-1$ orbitals; all remaining couplings are
neglected. The leading coupling is $K_0$ for both A and B, so the sorted
list sees no difference. The statistics of Eqs.~(7) and
(8) do ($n_i$ for $K_0>\kappa>\delta$):
\begin{equation*}
\setlength{\arraycolsep}{4pt}
\begin{array}{l|ccccc}
 & \max_j K_{ij} & S_i & n_i & C_i & \mathrm{PR}_i \\ \hline
\text{A} & K_0 & mK_0 & m & 1/m & m \\
\text{B} & K_0 & K_0  & 1 & 1   & 1 \\
\text{C} & \delta & (N-1)\delta & 0 & 1/(N-1) & N-1
\end{array}
\end{equation*}
For a $d$ orbital coupled to the four other $d$ orbitals of its center,
$m=4$: the total coupling, the partner count and the participation
ratio are all four times larger than for the pair orbital, although
their strongest couplings are equal. The relative spread $s_i/\mu_i$
tells the same story from the other side. It is large for B, whose
coupling is a single spike, and smaller for A, whose coupling is spread
over several partners. The spectator C is separated from both by its
small $\max_j K_{ij}$ and $S_i$ and by $n_i=0$. Its participation ratio
is nevertheless the largest of the three, because weak coupling spread
evenly is also ``delocalized''. The shape descriptors
$C_i$ and $\mathrm{PR}_i$ are therefore meaningful only together with
the magnitudes $S_i$ and $\max_j K_{ij}$, which is why both kinds of
statistic are supplied to the model.

\section{Prediction quality of complexes \cpd{9}--\cpd{14} and of spin-contaminated states}
\label{sec:si_quality}

\begin{figure}[h!]
\centering
\includegraphics[width=0.8\linewidth]{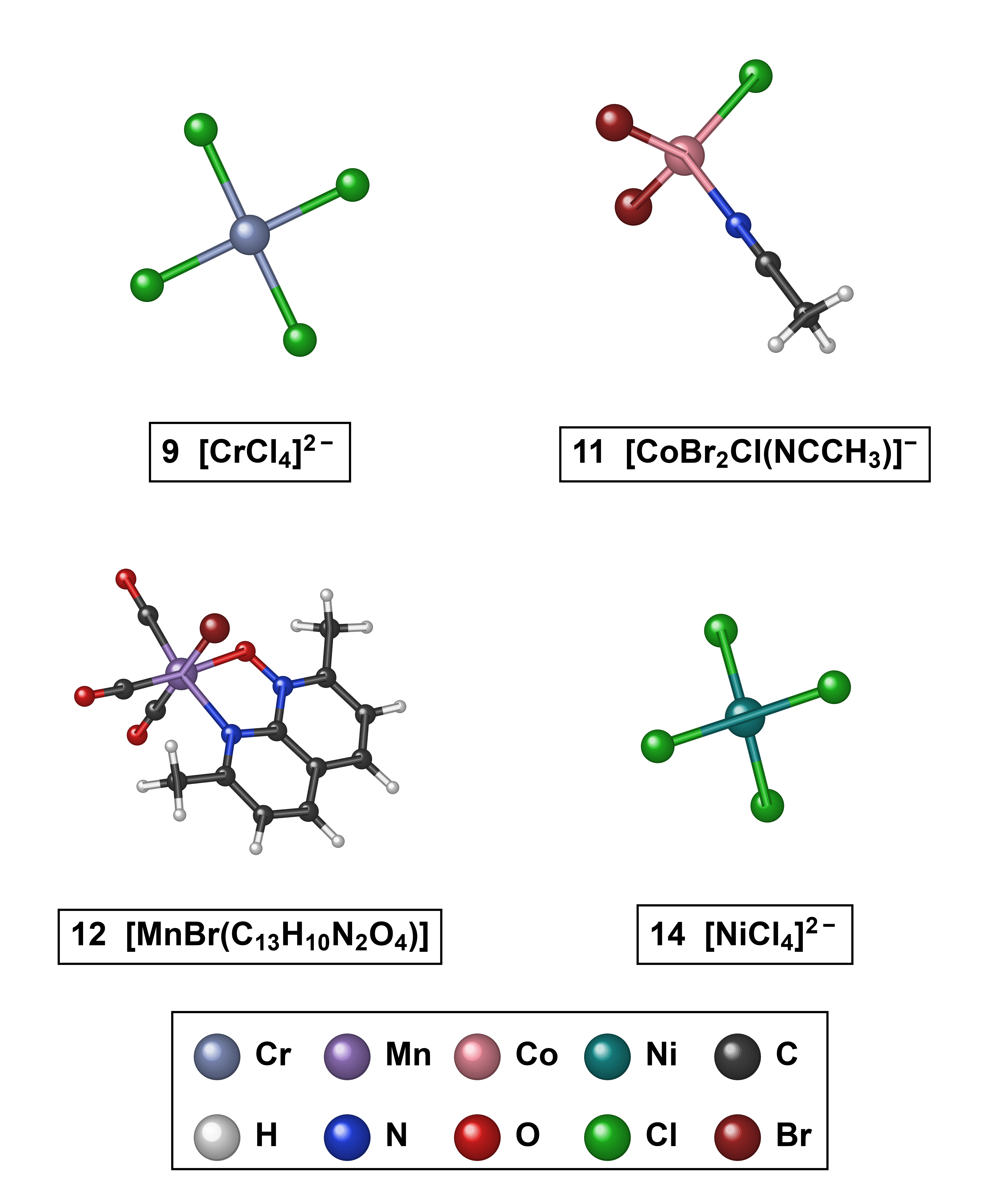}
\caption{\textbf{Structures of the additional benchmark complexes} \cpd{9},
\cpd{11}, \cpd{12} and \cpd{14}, taken from experimental crystal structures
(counter-ions and lattice solvent omitted). Complexes \cpd{10} and \cpd{13} are
second crystal structures of \cpd{2} and \cpd{7} and are shown in
Fig.~4 of the main text.}
\label{fig:si_struct}
\end{figure}

The prediction quality for complexes \cpd{1}--\cpd{7} is reported in
Table~III of the main text; Table~\ref{tab:si_ext} reports
the same quantities for complexes \cpd{9}--\cpd{14}, whose structures are
shown in Fig.~\ref{fig:si_struct}. Over the eight spin states
of these complexes, the predicted active space agrees with DMRG with
$F_1 = $ 0.61 at $\tau = 0.10$ and 0.83 at $\tau = 0.07$ (MAE 0.019), close to
the values for the eleven spin states of complexes \cpd{1}--\cpd{7} (0.65, 0.75 and 0.027). The
behavior of the two groups is therefore consistent, and the two electronic
regimes described in the main text appear in both.

\begin{table}[h!]
\centering
\caption{\textbf{Prediction quality for complexes \cpd{9}--\cpd{14}.}
$n_{\mathrm{DMRG}}$ and $n_{\mathrm{ML}}$ are the numbers of orbitals with
$s_1 \ge \tau$ in the reference and predicted profiles. The state marked
$^\ast$ has no spin-clean UHF reference and is excluded from the mean.
The last row gives the septet of \cpd{5}, which requires an electron outside the
$d$ shell and is reported here as a high-spin test case; it is not included in the mean.}
\label{tab:si_ext}
\small
\begin{tabular}{@{}cllcccccccc@{}}
\toprule
 & & & & \multicolumn{3}{c}{$\tau = 0.10$} & \multicolumn{3}{c}{$\tau = 0.07$} & \\
\cmidrule(lr){5-7}\cmidrule(lr){8-10}
No. & Refcode & Complex & $2S+1$ & $n_{\mathrm{DMRG}}$ & $n_{\mathrm{ML}}$ & $F_1$ & $n_{\mathrm{DMRG}}$ & $n_{\mathrm{ML}}$ & $F_1$ & MAE \\
\midrule
\cpd{9}  & ADUGIT & \ce{[CrCl4]^2-}      & 5        & 2  & 1  & 0.67 & 6  & 6  & 1.00 & 0.006 \\
\cpd{10} & HEGNUE & \ce{[VF6]^3-}$^{a}$  & 1        & 5  & 6  & 0.91 & 7  & 9  & 0.88 & 0.039 \\
         &        &                      & 3        & 0  & 4  & 0.00 & 2  & 8  & 0.40 & 0.019 \\
\cpd{11} & QOVWIH & \ce{[CoBr2Cl(NCCH3)]-} & 4      & 4  & 7  & 0.73 & 13 & 18 & 0.71 & 0.018 \\
\cpd{12} & SEDGAN & \ce{[MnBr(C13H10N2O4)]} & 1     & 14 & 13 & 0.89 & 16 & 18 & 0.94 & 0.021 \\
         &        &                      & 3$^\ast$ & -- & -- & --   & -- & -- & --   & --    \\
\cpd{13} & SESDUP & \ce{[Ti($\eta^6${-}C6H6)(AlCl4)2]}$^{b}$ & 1 & 9 & 10 & 0.95 & 9 & 12 & 0.86 & 0.029 \\
\cpd{14} & VEBPUN & \ce{[NiCl4]^2-}      & 1        & 7  & 6  & 0.77 & 14 & 15 & 0.90 & 0.016 \\
         &        &                      & 3        & 0  & 1  & 0.00 & 6  & 7  & 0.92 & 0.005 \\
\midrule
\multicolumn{4}{@{}l}{Mean (8 states)} & & & 0.61 & & & 0.83 & 0.019 \\
\addlinespace
\cpd{5}  & VILFUR & \ce{[Cr(CO)5(HNC4H8O)]} & 7     & 6  & 10 & 0.62 & 20 & 18 & 0.79 & 0.022 \\
\bottomrule
\end{tabular}

\vspace{2pt}
\raggedright\footnotesize
$^{a}$A second, more symmetric crystal structure of the ion in \cpd{2}.
$^{b}$A second crystal structure of \cpd{7}.
\end{table}

\paragraph{Spin states without a spin-clean reference.}
For three spin states of the main benchmark, the triplets of
\ce{[Co(CH3)4]-} (\cpd{3}), \ce{[Cr(CO)5(HNC4H8O)]} (\cpd{5}) and
\ce{[Cr(CO)4(H2NCH2CO2)]-} (\cpd{6}), no UHF solution satisfying the
spin-contamination criterion of the main text was found. A spin-contaminated
reference mixes in higher spin states, so its orbitals, its exchange matrix
and the DMRG entropies computed on them describe a mixture rather than the
target state; neither the descriptors nor the reference labels are then
meaningful, and these states are left out of every average in the main text.
For completeness, Table~\ref{tab:si_excl} lists the results obtained on the
lowest-energy contaminated solutions. The numbers are similar to those of the
spin-clean states, but they are not used in any conclusion. For the triplet of
\cpd{3} ($\langle\hat S^2\rangle_\mathrm{UHF} = 2.47$, above the limit of 2.2) a DMRG
label exists and is included in the released data, but the state was excluded
before the predictions were made and is therefore not evaluated.

\begin{table}[h!]
\centering
\caption{\textbf{Spin states excluded from the main-text statistics.}
Results on the lowest-energy, spin-contaminated UHF solutions. Columns as in
Tables~III and IV of the main text. The
triplet of \cpd{3} was not evaluated(see text).}
\label{tab:si_excl}
\footnotesize
\setlength{\tabcolsep}{3pt}
\begin{tabular}{@{}llccccccccc@{}}
\toprule
 & & \multicolumn{3}{c}{$\tau = 0.10$} & \multicolumn{3}{c}{$\tau = 0.07$} & & & \\
\cmidrule(lr){3-5}\cmidrule(lr){6-8}
Complex & $2S+1$ & $n_{\mathrm{DMRG}}$ & $n_{\mathrm{ML}}$ & $F_1$ & $n_{\mathrm{DMRG}}$ & $n_{\mathrm{ML}}$ & $F_1$ & MAE & pool cov. & $|P|$ \\
\midrule
\cpd{3} \ce{[Co(CH3)4]-}          & 3 & -- & -- & --   & -- & -- & --   & --    & --          & -- \\
\cpd{5} \ce{[Cr(CO)5(HNC4H8O)]}   & 3 & 11 & 12 & 0.96 & 29 & 22 & 0.82 & 0.021 & 11 (100\%) & 23 \\
\cpd{6} \ce{[Cr(CO)4(H2NCH2CO2)]-} & 3 & 19 & 19 & 0.95 & 32 & 29 & 0.95 & 0.036 & 19 (100\%) & 29 \\
\bottomrule
\end{tabular}
\end{table}

\paragraph{Weak correlation near the threshold.}
Every state with $F_1 = 0$ is a weakly correlated high-spin state, the triplets
of \cpd{10} and \cpd{14}. DMRG places no orbital above the threshold, and the
prediction reproduces the flat profile with a small error (MAE 0.005--0.019)
but places one to four orbitals marginally above 0.10. The quintet of \cpd{9}
and the quartet of \cpd{11} behave in the same way, with errors of 0.006 and
0.018.

\paragraph{Strong correlation and metal identity.}
The singlets of the vanadium(III) ion \cpd{10} and of the nickel(II) ion
\cpd{14} are strongly correlated (maximal DMRG $s_1 = $ 0.65 and 0.46), and in
both the correlated orbitals are located correctly ($F_1 = $ 0.91 and 0.77).
Their entropies, however, are compressed to different degrees: the largest
predicted entropy is 0.18 for vanadium, which is absent from training, but 0.34
for nickel, which is part of the training set. Together with the vanadium
singlets of \cpd{1} and \cpd{2}, this shows that the compression of strong
correlation discussed in the main text is mainly a property of metals absent
from training.

\paragraph{Geometry and further metals.}
The two crystal structures of \ce{[VF6]^3-}, the distorted \cpd{2} and the more
symmetric \cpd{10}, give the same picture: a strongly correlated singlet and a
weakly correlated triplet, with similar $F_1$ values, so the results do not
depend on a particular geometry. The manganese complex \cpd{12}, the only
manganese complex of the benchmark, and the titanium complex \cpd{13} are
predicted with $F_1 = $ 0.89 and 0.95. Complex \cpd{13}, a second crystal
structure of the titanium arene complex \cpd{7}, confirms the accurate
transfer to titanium found for \cpd{7} in the main text. The novelty
directions and tiers of \cpd{9}--\cpd{14} are given in
Table~\ref{tab:si_comp}.

\section{Entropy profiles and mutual information of individual complexes}
\label{sec:si_figures}

The main text shows the entropy profiles and mutual information of \cpd{1},
\cpd{3}, \cpd{6} and \cpd{7} (Figs.~8--10)
and of the application systems \cpd{4} and \cpd{8}. Figures~\ref{fig:si_s1_c2}--\ref{fig:si_mi_c14} show
the same analysis for the remaining benchmark complexes, one complex per page.
In the entropy profiles, the gray dotted line marks the selection threshold
$\tau = 0.10$, the magenta dashed line the HOMO--LUMO boundary, and the letters
above each panel the dominant atomic-orbital character (s, p, d) of each
orbital.

\paragraph{\cpd{2}, \ce{[VF6]^3-} (Figs.~\ref{fig:si_s1_c2} and \ref{fig:si_mi_c2}).}
This ion contains the same $d^2$ vanadium(III) center as \cpd{1} in a distorted
fluoride environment. As for \cpd{1}, the singlet is strongly correlated: DMRG
assigns entropies of up to 0.66 to frontier orbitals with V~d character, whereas the largest predicted entropy is 0.15. The
correlated orbitals are nevertheless partly recovered (five predicted against
six in DMRG; $F_1 = $ 0.73 at $\tau = 0.10$ and 0.61 at $\tau = 0.07$;
MAE 0.044), and the mutual information places the strongly correlated pairs
correctly but with much smaller values. The triplet is weakly correlated: no
DMRG entropy reaches 0.10, and the prediction reproduces the flat
profile (MAE 0.022) but places five orbitals marginally above the threshold
($F_1 = 0$ at $\tau = 0.10$ and 0.27 at $\tau = 0.07$). Here the
mutual-information rescue adds six orbitals, so that the candidate pool
(16 orbitals) over-covers, which costs efficiency but not completeness.

\paragraph{\cpd{5}, \ce{[Cr(CO)5(HNC4H8O)]} (Figs.~\ref{fig:si_s1_c5} and \ref{fig:si_mi_c5}).}
The singlet is the
ground state of this 18-electron chromium(0) carbonyl: it lies 50\,kcal\,mol$^{-1}$
below the triplet at the UHF level and 72\,kcal\,mol$^{-1}$ in the window DMRG
(each state in its own orbitals). Its active space (11 orbitals at $\tau = 0.10$,
19 at $\tau = 0.07$) is formed by the metal-d and carbonyl $\pi^\ast$ orbitals on
either side of the frontier gap and is reproduced closely ($F_1 = 0.88$ and 0.95;
MAE 0.024); the candidate pool (20 orbitals) contains all eleven active orbitals.
The quintet is predicted with $F_1 = 0.82$ and 0.79 (MAE 0.031); its pool
(19 orbitals) contains ten of the eleven active orbitals. The septet requires an
electron outside the $d$ shell and is reported only here, as a high-spin test case
($F_1 = 0.62$ and 0.79; MAE 0.022; Table~\ref{tab:si_ext}). The triplet, whose
reference is spin-contaminated, is shown for completeness but excluded from all
averages.

\paragraph{\cpd{9}, \ce{[CrCl4]^2-} (Figs.~\ref{fig:si_s1_c9} and \ref{fig:si_mi_c9}).}
The tetrahedral chromium(II) quintet is weakly correlated (maximal DMRG
$s_1 = 0.12$). Two orbitals lie marginally above the threshold in DMRG and one
in the prediction ($F_1 = 0.67$), and the agreement improves at the permissive
threshold ($F_1 = $ 1.00), with the smallest error of the benchmark (MAE 0.006).

\paragraph{\cpd{10}, \ce{[VF6]^3-}, second structure (Figs.~\ref{fig:si_s1_c10} and \ref{fig:si_mi_c10}).}
The more symmetric structure of the vanadium(III) ion reproduces the behavior
of \cpd{2}. The singlet is strongly correlated (maximal DMRG $s_1 = $ 0.65); all five active orbitals are recovered ($F_1 = $ 0.91 and 0.88), with entropies
compressed to at most 0.18. The triplet is weakly correlated (maximum 0.07) and
belongs to the near-threshold regime ($F_1 = 0$ and 0.40; MAE 0.019).

\paragraph{\cpd{11}, \ce{[CoBr2Cl(NCCH3)]-} (Figs.~\ref{fig:si_s1_c11} and \ref{fig:si_mi_c11}).}
The cobalt(II) quartet is weakly correlated (maximal DMRG $s_1 = 0.13$). The
prediction follows the flat profile (MAE 0.018) but places seven orbitals above
the threshold against four in DMRG ($F_1 = $ 0.73 and 0.71).

\paragraph{\cpd{12}, \ce{[MnBr(C13H10N2O4)]} (Figs.~\ref{fig:si_s1_c12} and \ref{fig:si_mi_c12}).}
The manganese singlet has a large active space of 14 orbitals (maximal DMRG
$s_1 = 0.27$), which the prediction reproduces closely ($F_1 = $ 0.89 and 0.94;
MAE 0.021). The triplet has no spin-clean UHF solution and is not shown.

\paragraph{\cpd{13}, \ce{[Ti($\eta^6${-}C6H6)(AlCl4)2]}, second structure (Figs.~\ref{fig:si_s1_c13} and \ref{fig:si_mi_c13}).}
The titanium singlet (maximal DMRG $s_1 = 0.38$) is the most accurately
predicted state of the benchmark at $\tau = 0.10$: all nine DMRG-active orbitals
are recovered with one additional orbital ($F_1 = 0.95$; MAE 0.029).

\paragraph{\cpd{14}, \ce{[NiCl4]^2-} (Figs.~\ref{fig:si_s1_c14} and \ref{fig:si_mi_c14}).}
The tetrahedral nickel(II) singlet is strongly correlated (maximal DMRG $s_1 = 0.46$). Because nickel is part of the training
set, its entropies are compressed far less than those of the vanadium singlets
(maximal predicted $s_1 = $ 0.34); five of the seven active orbitals are
recovered ($F_1 = $ 0.77 and 0.90; MAE 0.016), and the mutual-information rescue
adds further partner orbitals. The triplet is weakly correlated (maximum 0.08) and
belongs to the near-threshold regime ($F_1 = 0$ and 0.92; MAE 0.005).

\complexfigure{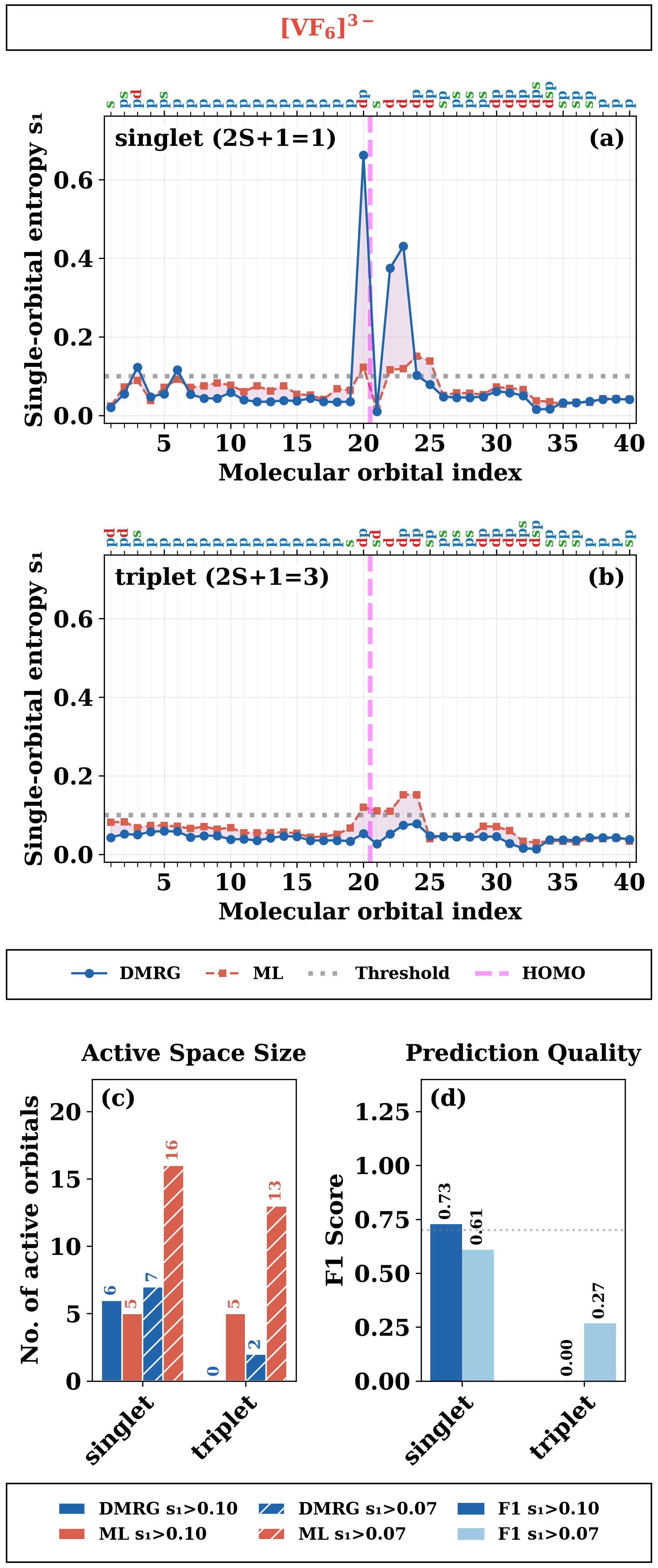}{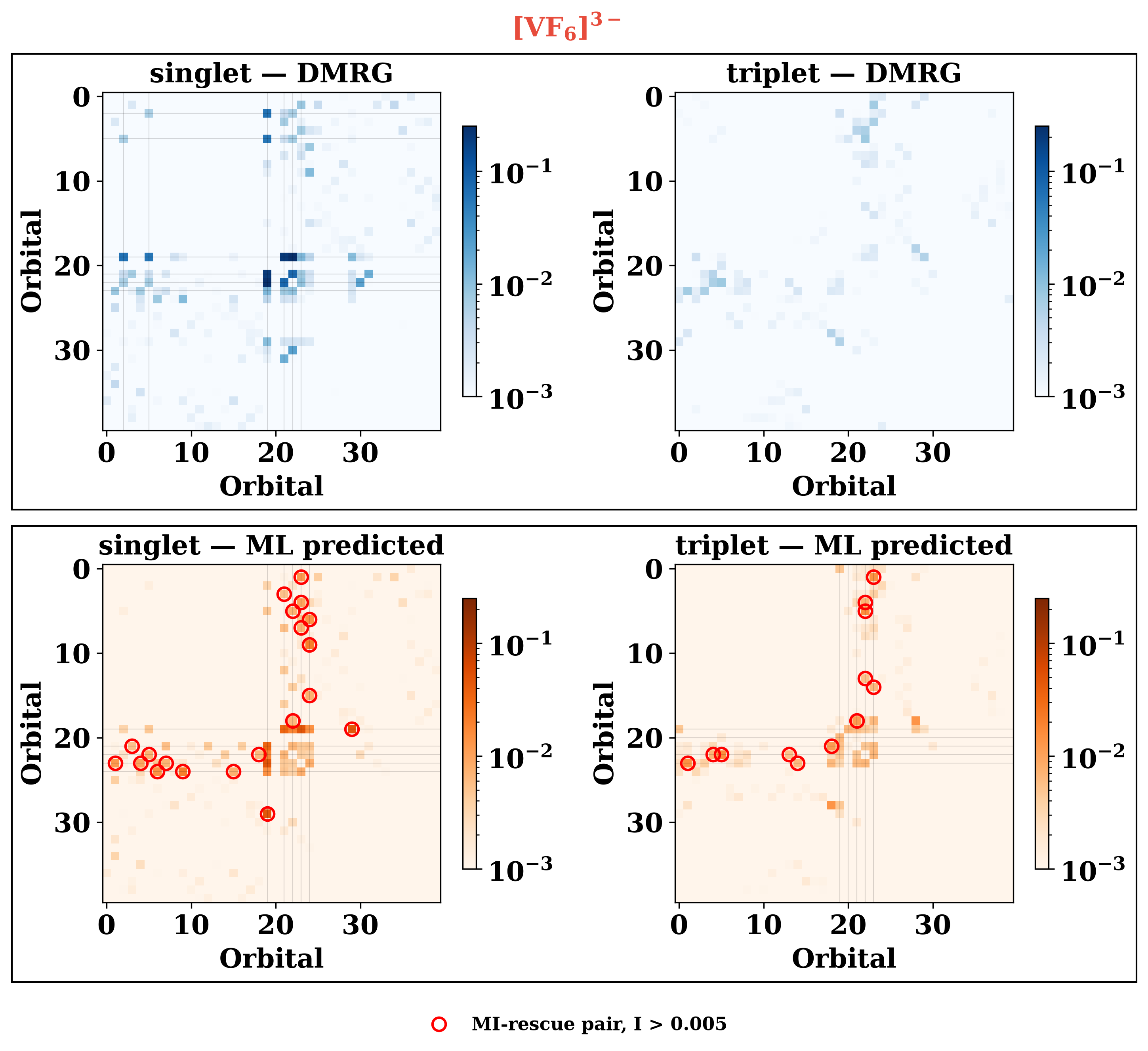}
  {\cpd{2}, \ce{[VF6]^3-} (Tier~I).}
  {Singlet and triplet, the triplet on its lowest spin-clean UHF solution.}
  {c2}

\complexfigure{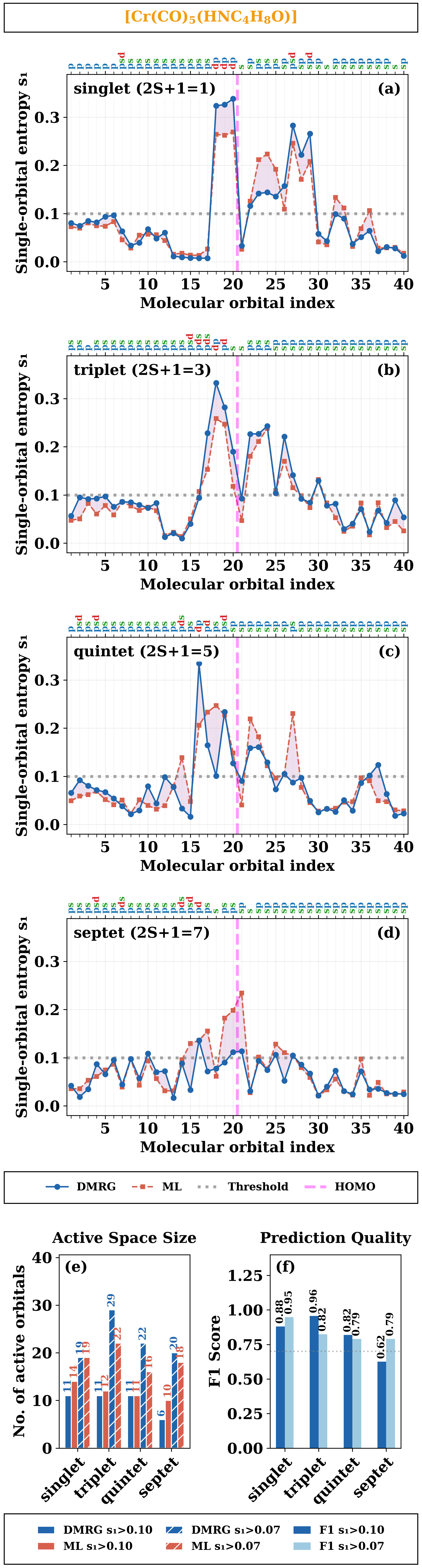}{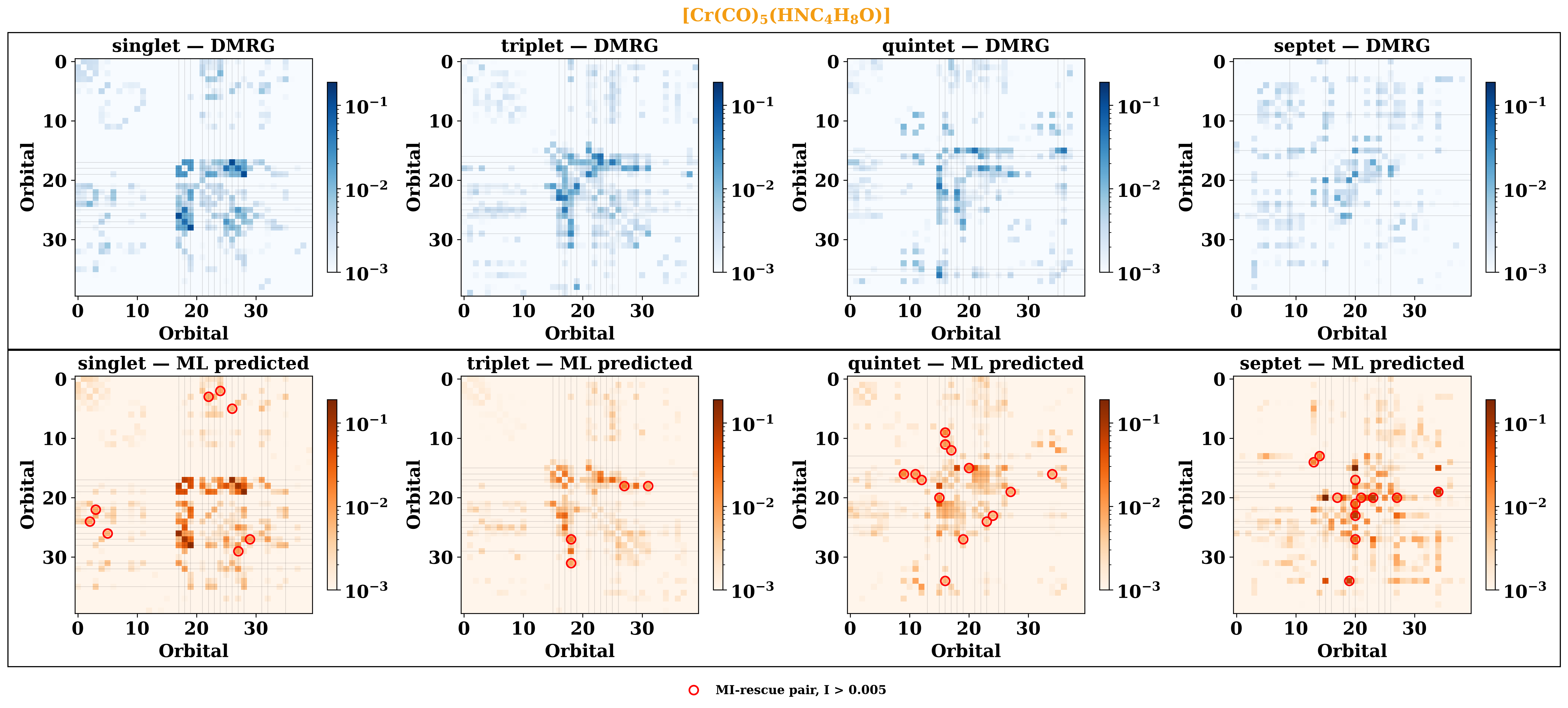}
  {\cpd{5}, \ce{[Cr(CO)5(HNC4H8O)]} (Tier~II).}
  {Triplet (spin-contaminated reference, excluded from averages), quintet and septet.}
  {c5}

\complexfigure{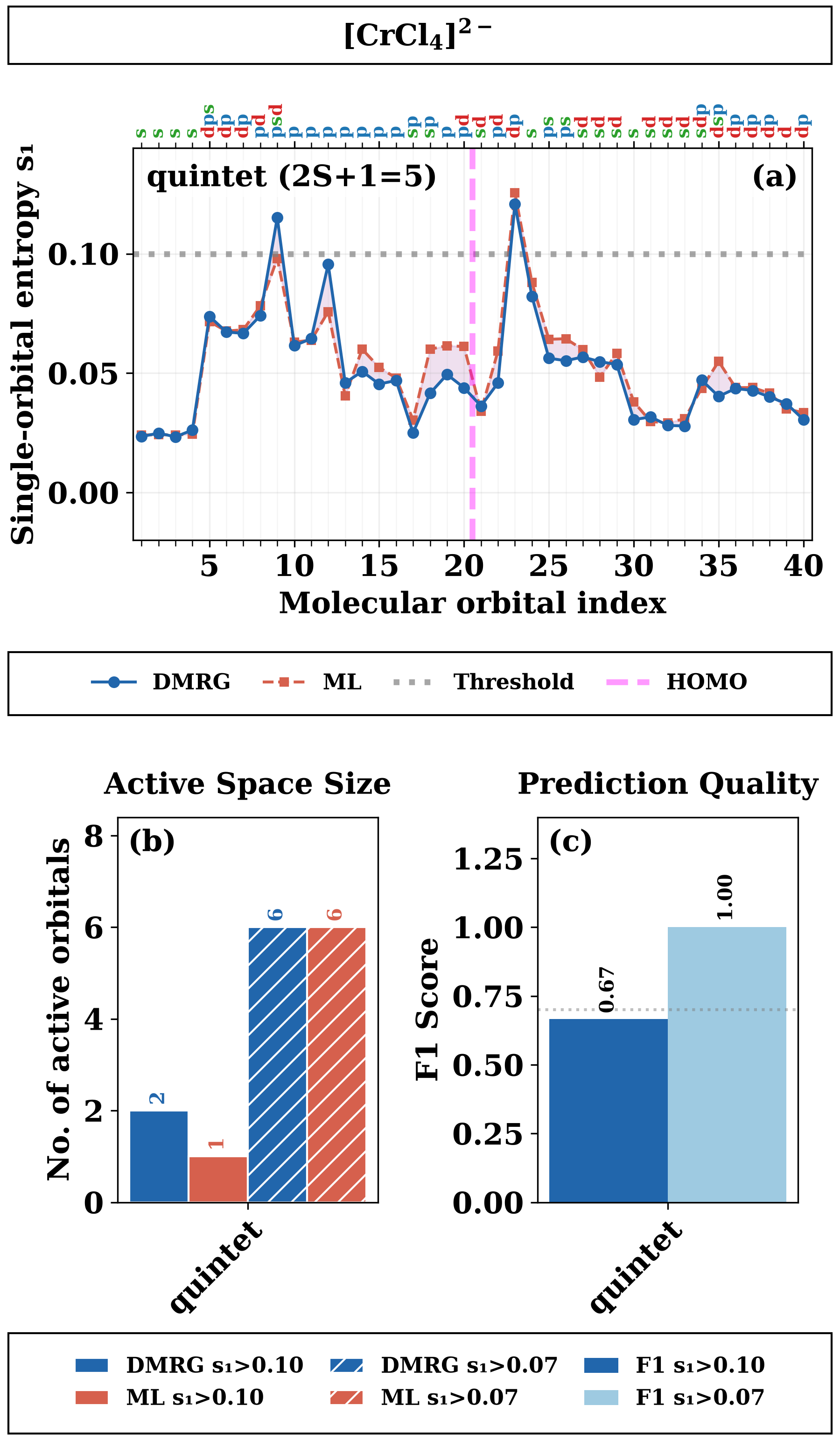}{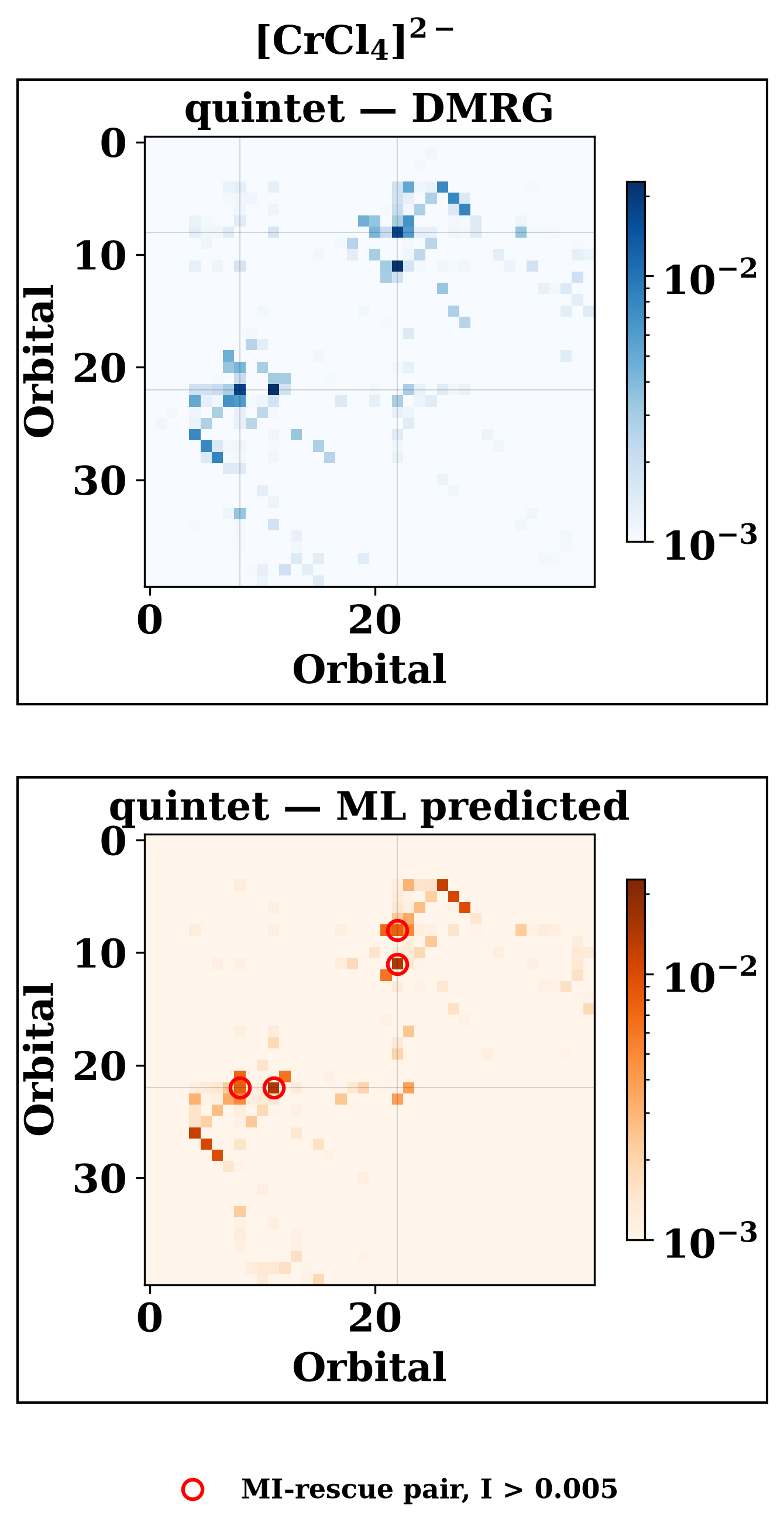}
  {\cpd{9}, \ce{[CrCl4]^2-} (ADUGIT).}
  {Quintet.}
  {c9}

\complexfigure{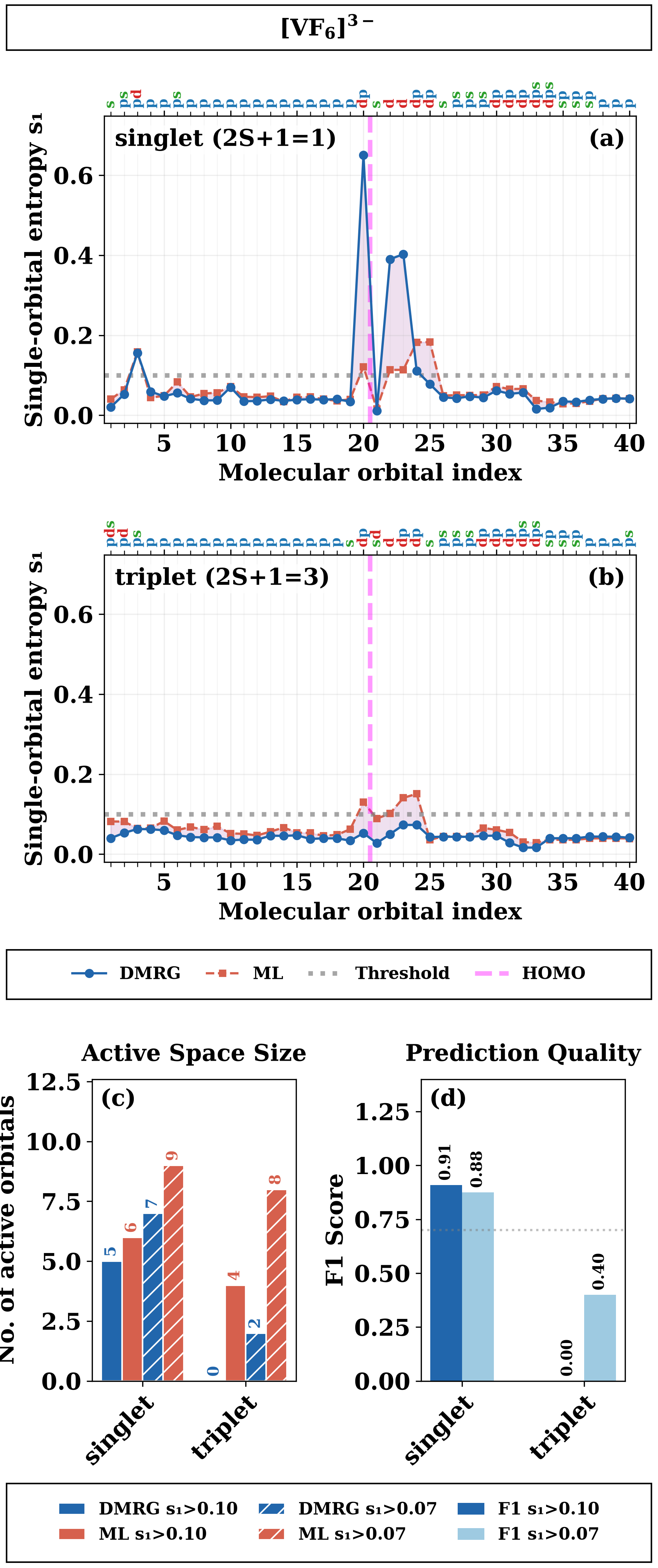}{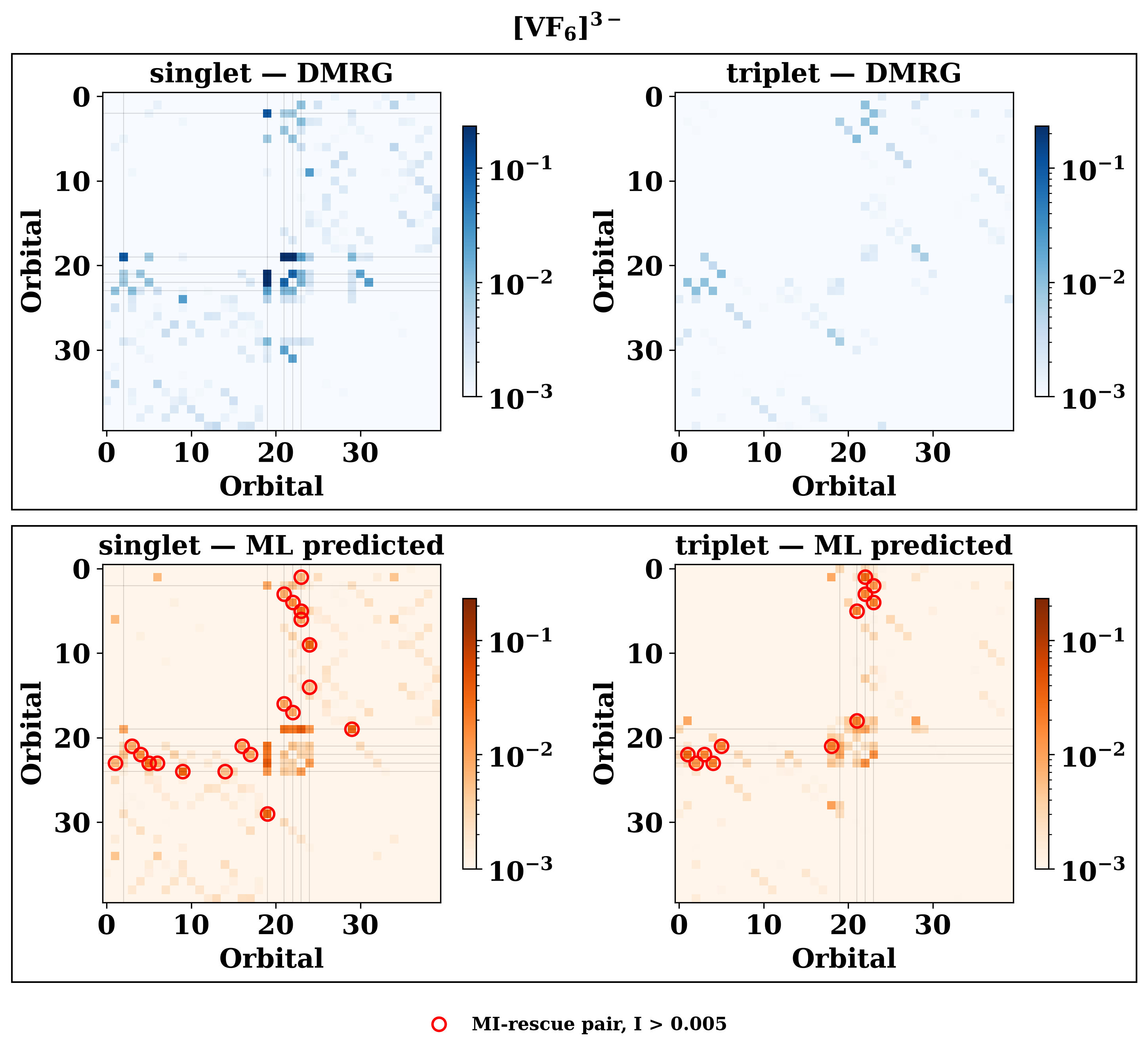}
  {\cpd{10}, \ce{[VF6]^3-}, second crystal structure (HEGNUE).}
  {Singlet and triplet.}
  {c10}

\complexfigure{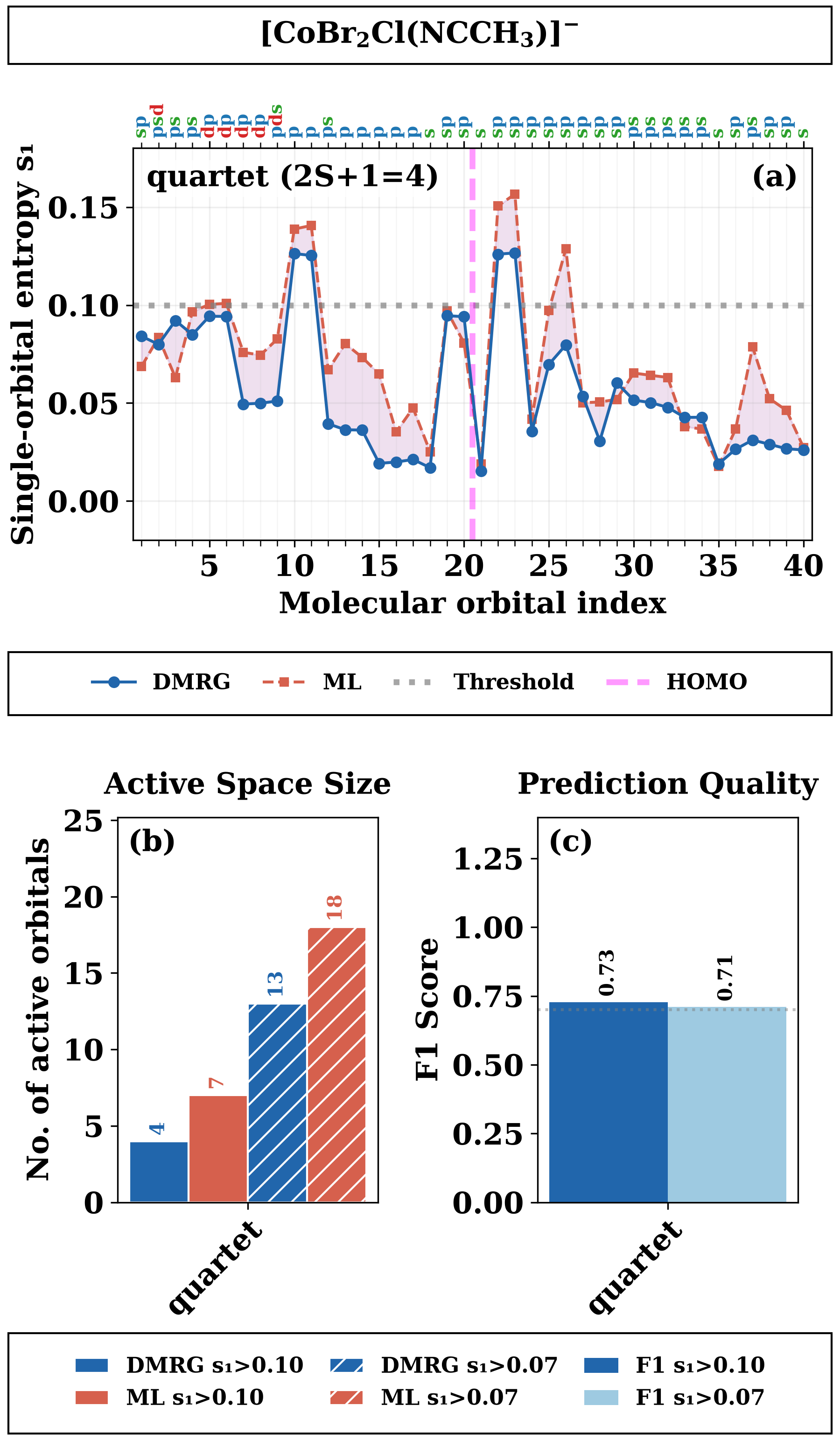}{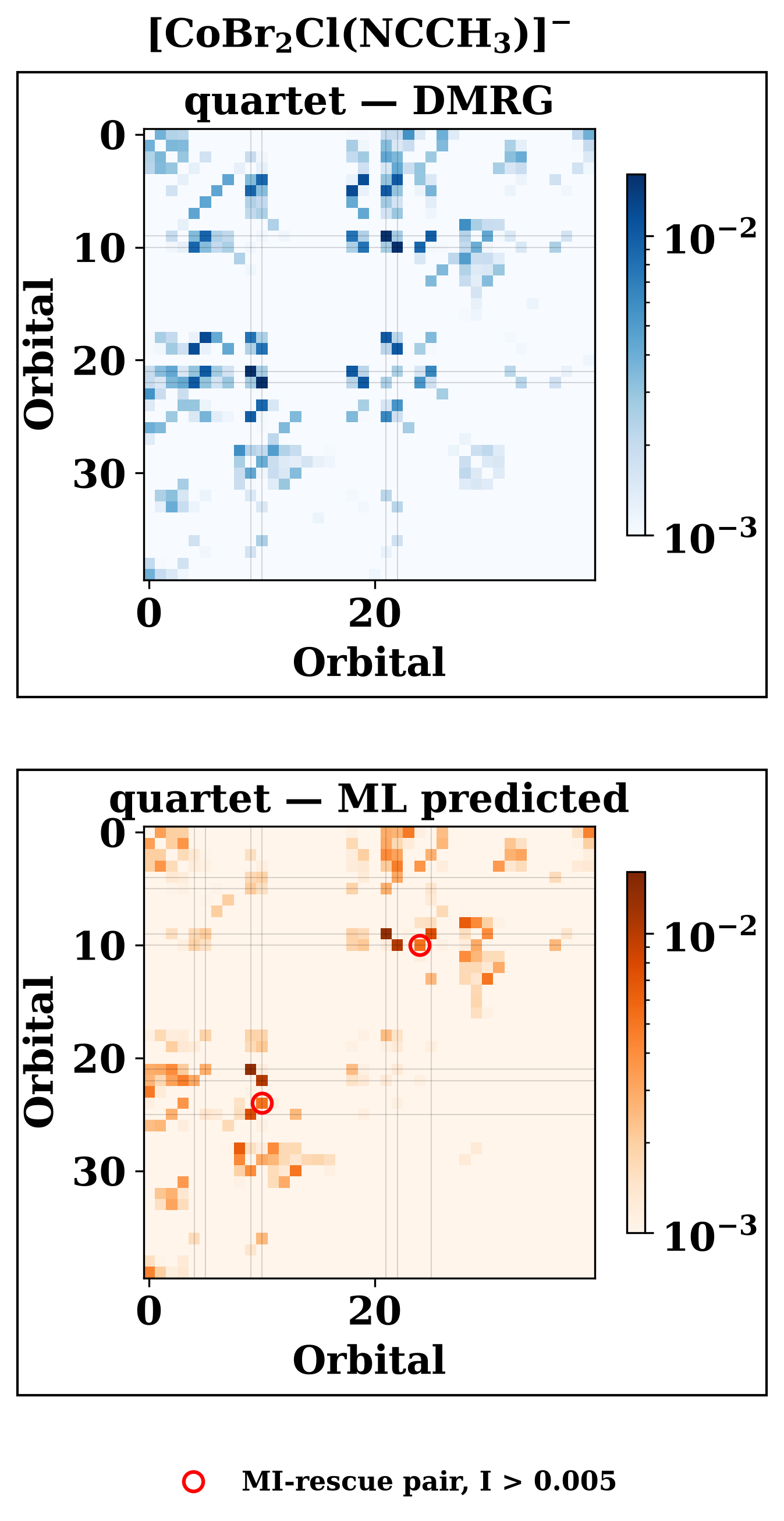}
  {\cpd{11}, \ce{[CoBr2Cl(NCCH3)]-} (QOVWIH).}
  {Quartet.}
  {c11}

\complexfigure{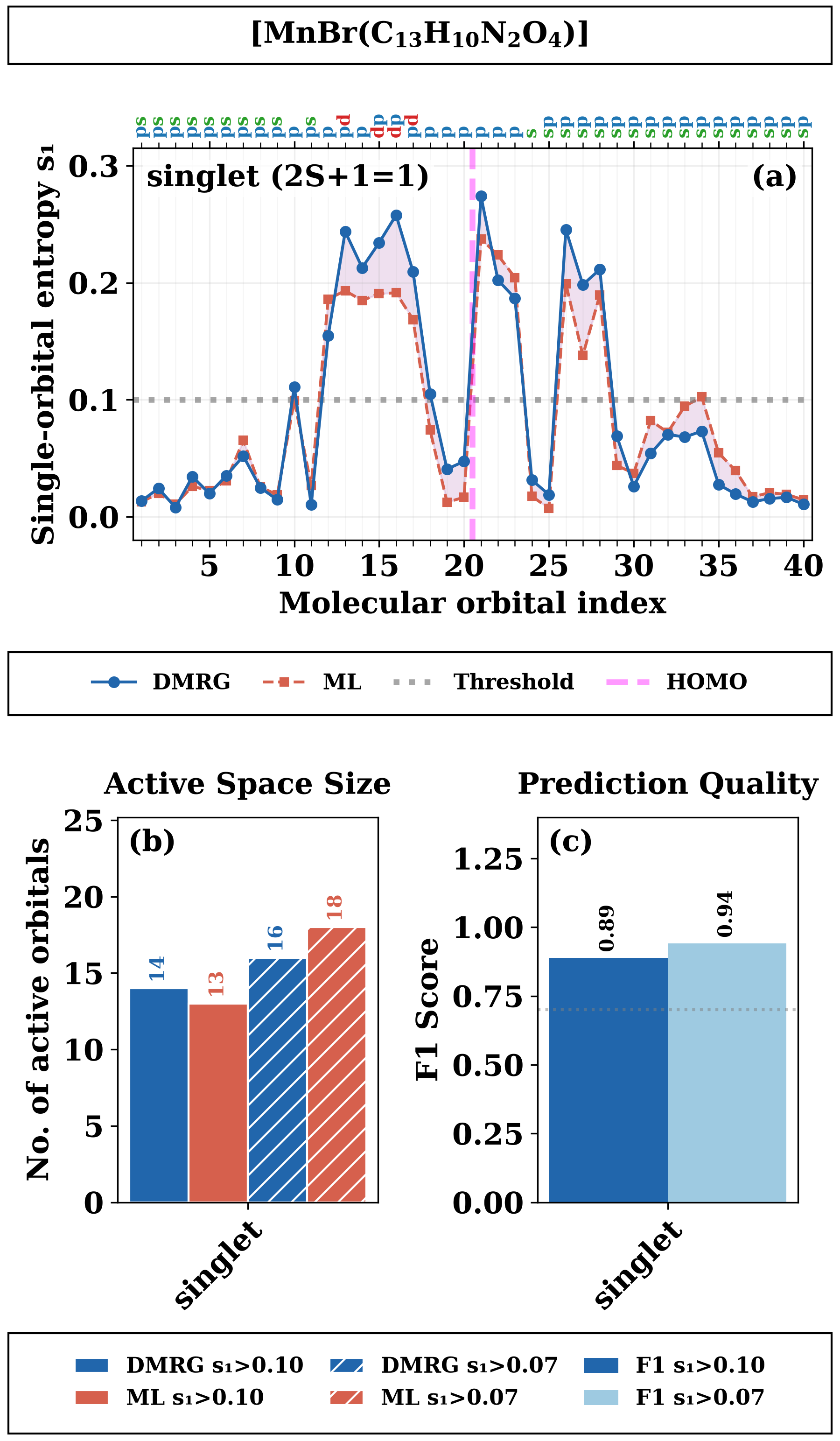}{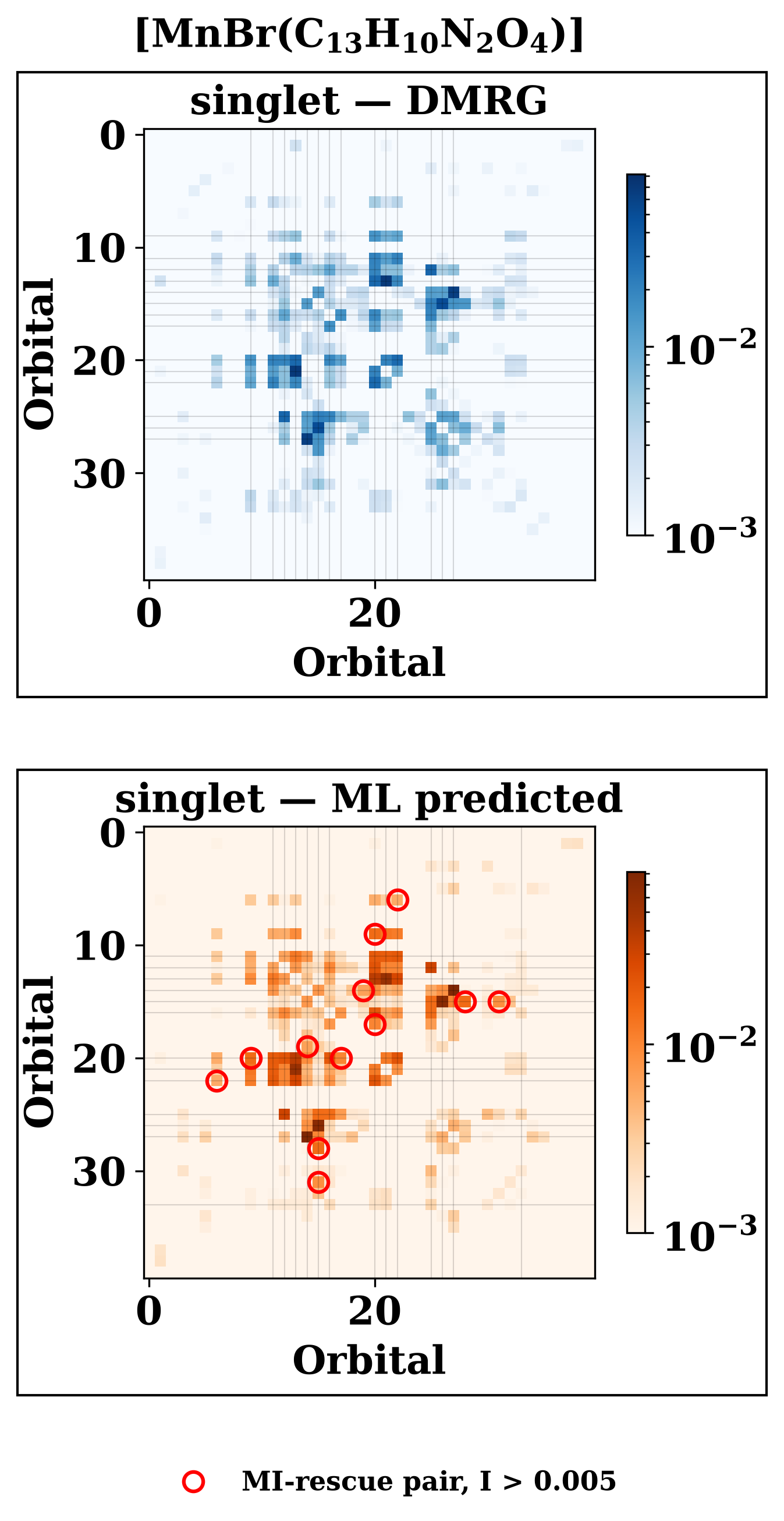}
  {\cpd{12}, \ce{[MnBr(C13H10N2O4)]} (SEDGAN).}
  {Singlet; the triplet has no spin-clean reference.}
  {c12}

\complexfigure{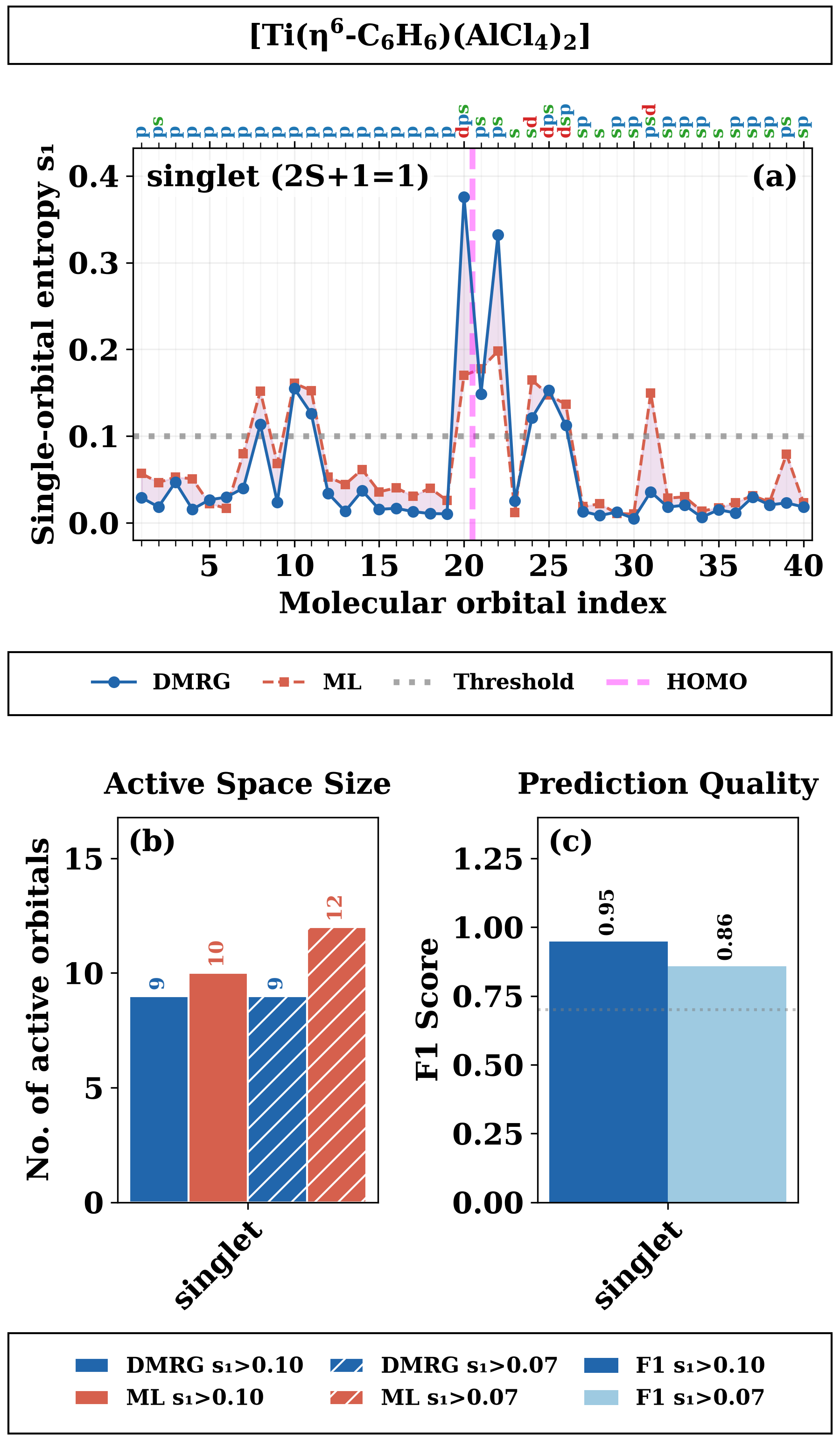}{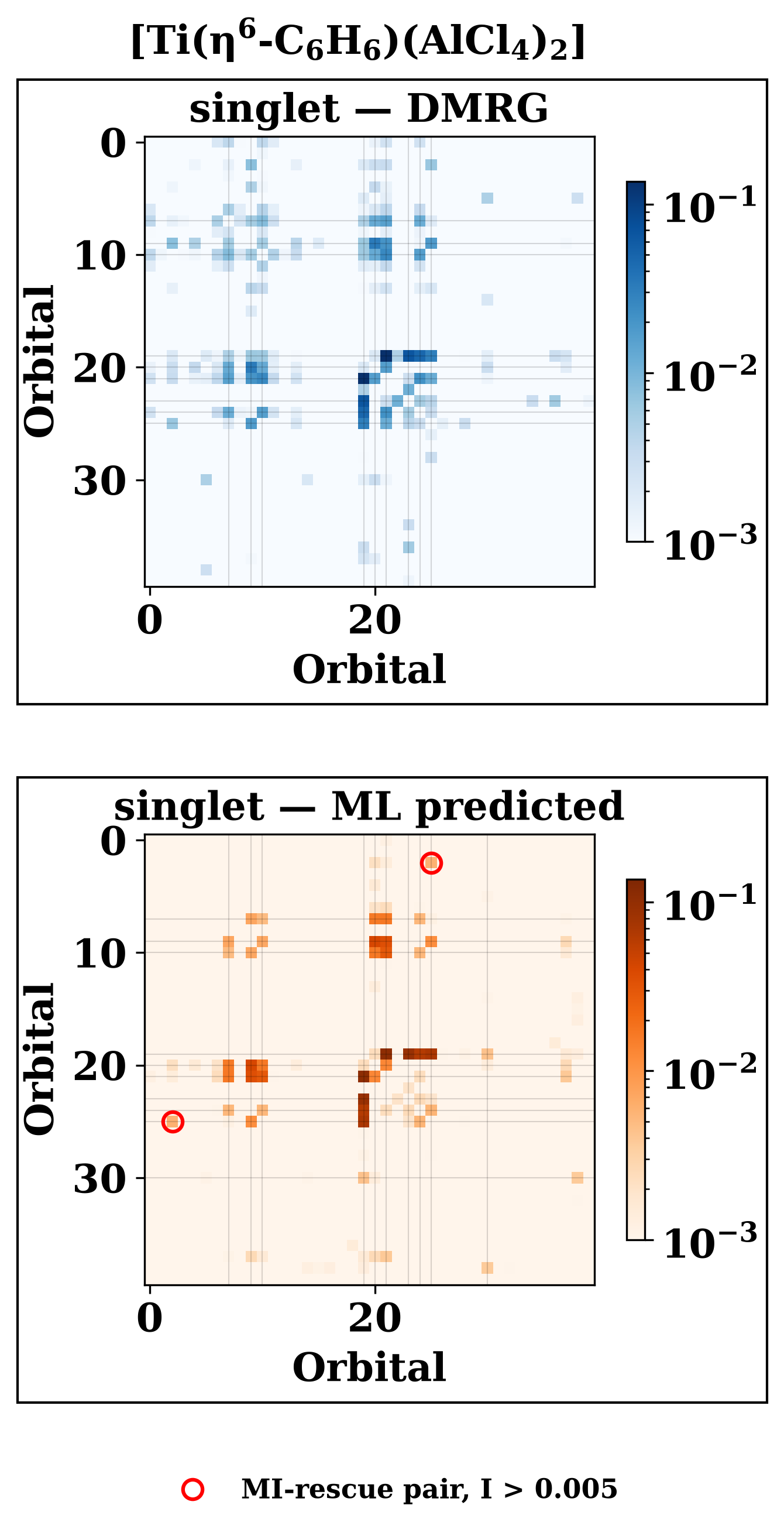}
  {\cpd{13}, \ce{[Ti($\eta^6${-}C6H6)(AlCl4)2]}, second structure (SESDUP).}
  {Singlet.}
  {c13}

\complexfigure{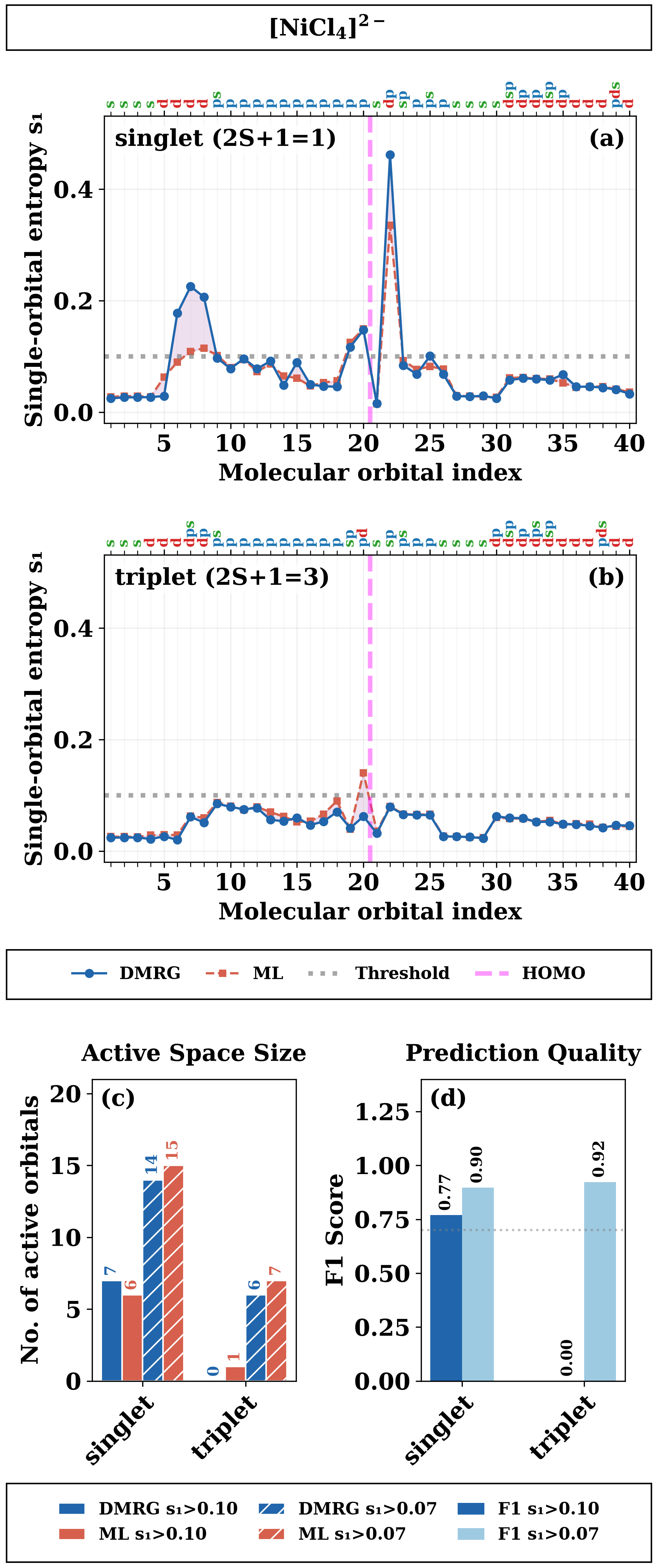}{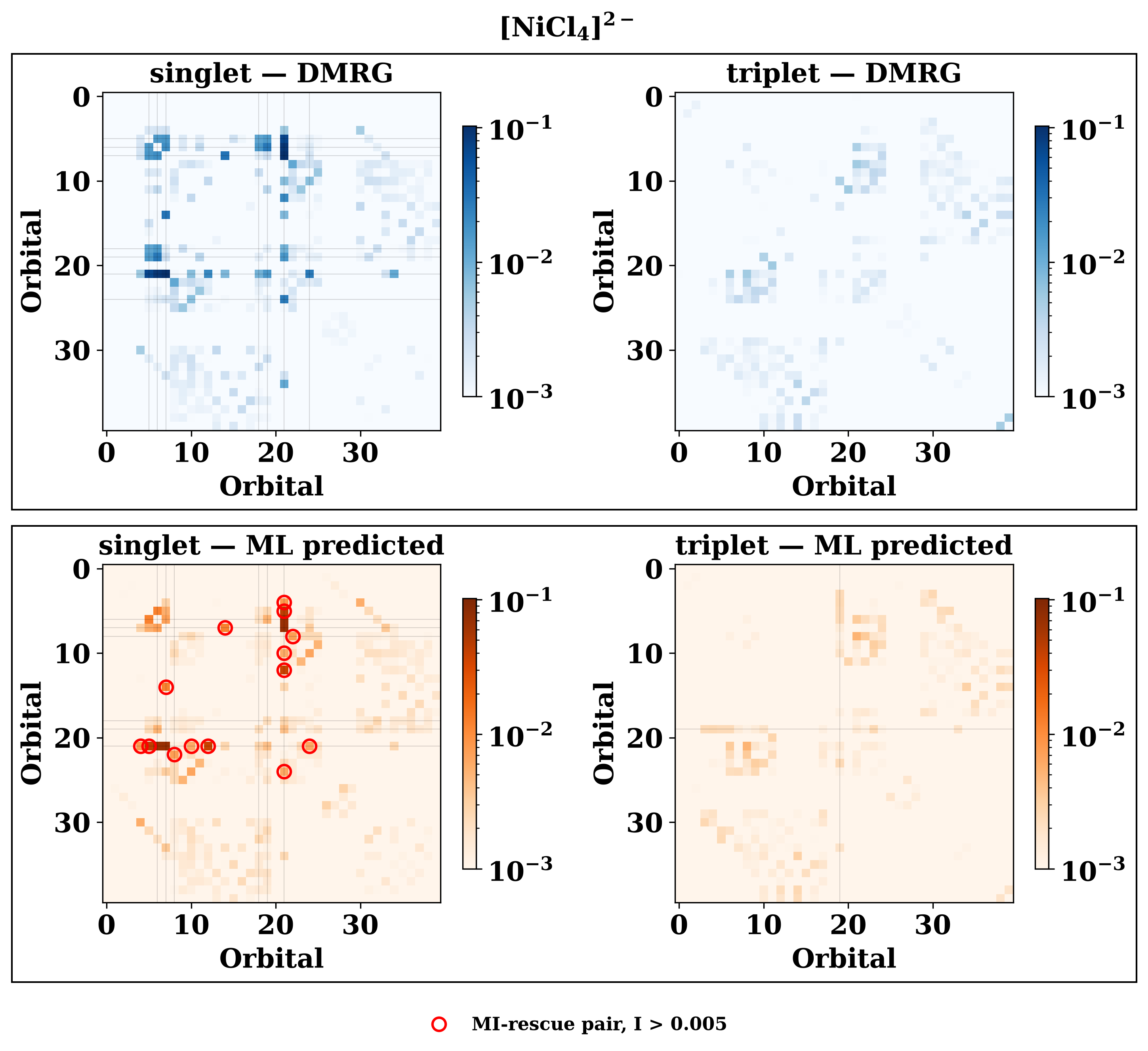}
  {\cpd{14}, \ce{[NiCl4]^2-} (VEBPUN).}
  {Singlet and triplet.}
  {c14}

\section{Application systems: reference wavefunctions and energies}
\label{sec:si_app}

Table~\ref{tab:si_app} lists the UHF references and the energies of the
11-orbital active spaces of the earlier scans (for the final-model spaces see
Table~\ref{tab:si_app_final}). For both
complexes, the closed-shell singlet UHF solution is internally unstable
towards a broken-symmetry solution with large spin contamination, which the
spin criterion of the main text rejects. For \cpd{4} the broken-symmetry
solution lies 16.4~m$E_\mathrm{h}$ (10.3\,kcal\,mol$^{-1}$) below the
closed-shell solution, with $\langle\hat S^2\rangle = 0.737$, which indicates
substantial open-shell character of the singlet; the triplet nevertheless
remains 12.4\,kcal\,mol$^{-1}$ below even the broken-symmetry singlet at the
UHF level. For \cpd{8} the lower solution has $\langle\hat S^2\rangle=1.15$.
All singlet results refer to the closed-shell solution. The UCCSD(T) $T_1$
diagnostic is 0.062 for the singlet of \cpd{4} and 0.033/0.030 for the
singlet/quintet of \cpd{8}.

\paragraph{Distinct accuracy comparisons.}
The SQD errors in Section~\ref{sec:si_sqd} compare SQD with exact CASCI
for the same active-space Hamiltonian and therefore quantify the solver
error within a fixed space. The comparison of CASCI+NEVPT2 gaps with
full-molecule UCCSD(T) in the main text instead tests agreement between
different correlation treatments. The NEVPT2 corrections are evaluated
classically. This agreement neither measures the SQD error nor establishes
absolute gap accuracy or complete active-space convergence.

\begin{table}[h!]
\centering
\caption{\textbf{Reference energies of the earlier application spaces} (def2-SVP,
$E_\mathrm{h}$). CASCI and NEVPT2 for the 11-orbital spaces built from UHF
natural orbitals (the Hamiltonians used for SQD); ``corr.'' is the
strongly contracted NEVPT2 correlation energy. Complex \cpd{8} at
$r(\mathrm{Co{-}O_{CO_2}})=1.97$~\AA.}
\label{tab:si_app}
\footnotesize
\begin{tabular}{@{}llcrccccc@{}}
\toprule
No. & State & $q$ & $E_\mathrm{UHF}$ & $\langle\hat S^2\rangle$ & Stable & Space & $E_\mathrm{CASCI}$ & NEVPT2 corr. \\
\midrule
\cpd{4} & singlet & $+1$ & $-1925.292985$ & 0.000 & no  & CAS(14,11) & $-1925.31790365$ & $-1.84457722$ \\
        & triplet & $+1$ & $-1925.329155$ & 2.032 & yes & CAS(16,11) & $-1925.32197079$ & $-1.77718400$ \\
\addlinespace
\cpd{8} & singlet & $+3$ & $-1947.429359$ & 0.000 & no  & CAS(14,11) & $-1947.448896$ & $-1.791714$ \\
        & quartet & $+2$ & $-1948.133725$ & 3.755 & yes & CAS(11,11) & $-1948.132758$ & $-1.787308$ \\
        & quintet & $+3$ & $-1947.515173$ & 6.007 & yes & CAS(12,11) & $-1947.516294$ & $-1.754639$ \\
\bottomrule
\end{tabular}
\end{table}

The archived UCCSD(T) total energies of \cpd{4} are
$-1927.29704203\,E_\mathrm{h}$ (singlet) and
$-1927.23028785\,E_\mathrm{h}$ (triplet). Expanded active-space energies
and fixed-budget selection controls are given in Section~\ref{sec:si_controls}.
The full-molecule UCCSD(T) total energies of \cpd{8} (cobalt(III),
17 frozen core orbitals) are $-1949.18164270\,E_\mathrm{h}$ (singlet) and
$-1949.20528028\,E_\mathrm{h}$ (quintet), with $T_1$ diagnostics of 0.033 and
0.030.
The spaces above were built from UHF natural orbitals before the final models; they are kept for
reference in the archived controls. All energies of the main text and the hardware calculations use the spaces of the
final models, whose 11- and 16-orbital energies are listed in Table~\ref{tab:si_app_final}; all sizes from
9 to 20 orbitals are in Table~\ref{tab:si_expanded_final}.

\begin{table}[h!]
\centering
\caption{\textbf{Energies of the final-model active spaces used in the main text}
($E_\mathrm{h}$, def2-SVP). 11 orbitals: spaces of the 22-qubit hardware calculations; 16 orbitals: spaces of
the 32-qubit calculations and of Table~VII of the main text. CASCI exact; ``corr.'' is the strongly contracted
NEVPT2 correlation energy from DMRG reduced density matrices.}
\label{tab:si_app_final}
\footnotesize
\begin{tabular}{@{}llcrrcrr@{}}
\toprule
 & & \multicolumn{3}{c}{11 orbitals} & \multicolumn{3}{c}{16 orbitals} \\
\cmidrule(lr){3-5}\cmidrule(l){6-8}
No. & State & Space & $E_\mathrm{CASCI}$ & corr. & Space & $E_\mathrm{CASCI}$ & corr. \\
\midrule
\cpd{4} & singlet & CAS(12,11) & $-1925.318131$ & $-1.841127$ & CAS(16,16) & $-1925.337872$ & $-1.823021$ \\
        & triplet & CAS(12,11) & $-1925.322892$ & $-1.777568$ & CAS(16,16) & $-1925.329195$ & $-1.775014$ \\
\addlinespace
\cpd{8} & singlet & CAS(14,11) & $-1947.452378$ & $-1.792978$ & CAS(22,16) & $-1947.498423$ & $-1.736745$ \\
        & quartet & CAS(13,11) & $-1948.137431$ & $-1.787823$ & CAS(19,16) & $-1948.149494$ & $-1.770458$ \\
        & quintet & CAS(12,11) & $-1947.512408$ & $-1.757989$ & CAS(20,16) & $-1947.520676$ & $-1.746642$ \\
\bottomrule
\end{tabular}
\end{table}

\section{Sample-based quantum diagonalization: details}
\label{sec:si_sqd}

\begin{figure}[h!]
\centering
\includegraphics[width=\linewidth]{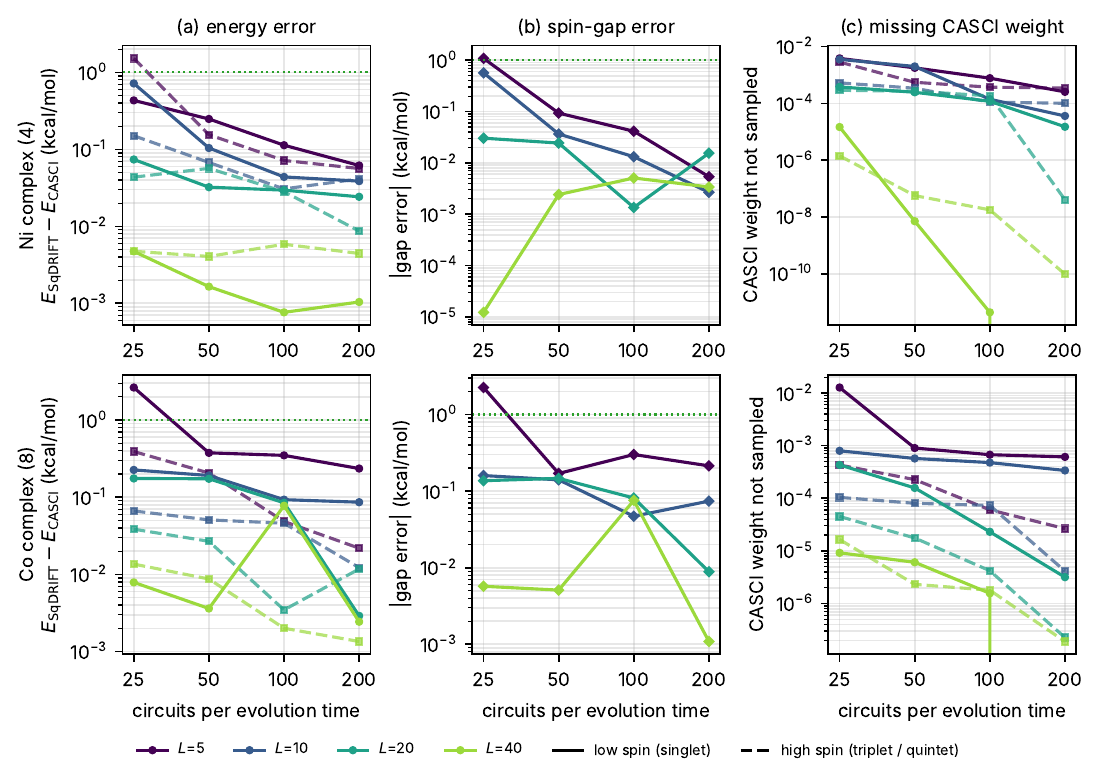}
\caption{\textbf{Noiseless SqDRIFT at 22 qubits (11 orbitals): sequence
length and number of circuits.} Rows: \cpd{4} (singlet and triplet) and
\cpd{8} (singlet and quintet). (a)~Energy error relative to CASCI;
(b)~error of the spin gap; (c)~CASCI weight of the configurations that were not
sampled, versus the number of circuits per evolution time (100 shots each), for
sequence lengths $L = 5$, 10, 20 and 40. Dotted lines: 1\,kcal\,mol$^{-1}$.}
\label{fig:si_sweep}
\end{figure}

\paragraph{SqDRIFT and LUCJ on \texttt{ibm\_phoenix} and \texttt{ibm\_fez}.}
Table~\ref{tab:si_sqdrift} lists the SQD errors and spin gaps of
Fig.~20 for every active-space size. SqDRIFT circuits used
random sequences of $L=10$ Hamiltonian exponentials, 200 circuits per evolution
time and 100 shots per circuit; SQD used three batches of 300 samples, with
six SQD seeds (19--24) for the hardware samples and six independent sets of
uniformly random bitstrings. A sweep of the sequence length ($L=5$, 10, 20 and
40) at 22 qubits is shown in Fig.~\ref{fig:si_sweep}. Each state was
sampled at two evolution times, $t = 1$ and $t = 10$ (atomic units), with 200
circuits per time; the qDRIFT sequences were drawn with seed 19, so every
hardware circuit has a noiseless twin. Circuits were compiled with the Qiskit
preset pass manager (optimization level 3, transpiler seed 42) against the
backend calibration at the time of preparation.

\begin{figure}[h!]
\centering
\includegraphics[width=0.62\linewidth]{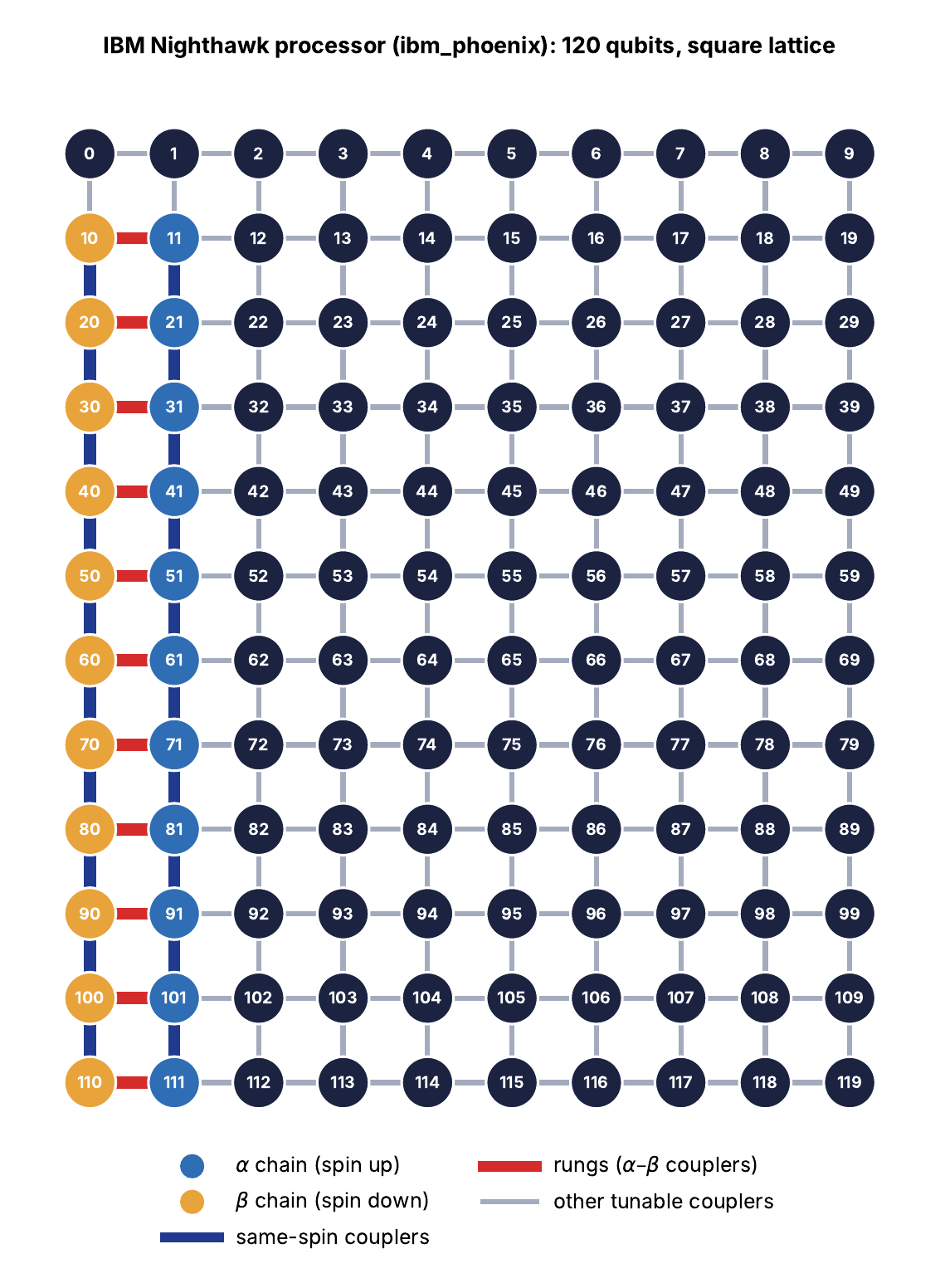}
\caption{\textbf{Qubit layout of the IBM Nighthawk processors.} 120 qubits on
a square lattice with nearest-neighbor tunable couplers. Coloured: the
22-qubit ladder of the LUCJ circuits for \cpd{8} on \texttt{ibm\_phoenix}, with
the $\alpha$ spin orbitals on qubits 11, 21, \ldots, 111 and the $\beta$ spin
orbitals on qubits 10, 20, \ldots, 110; red rungs carry the opposite-spin
interactions.}
\label{fig:si_nighthawk}
\end{figure}

\begin{table}[h!]
\centering
\caption{\textbf{SQD with SqDRIFT samples from \texttt{ibm\_phoenix} and with
uniformly random bitstrings} (mean over six seeds or random sets; final-model spaces). Errors
$E_\mathrm{SQD}-E_\mathrm{CASCI}$ and gaps $E(\text{low})-E(\text{high})$ in
kcal\,mol$^{-1}$; high spin is the triplet for \cpd{4} and the quintet for
\cpd{8}. ``Valid'' is the percentage of singlet samples with the correct
particle numbers (hardware) and the corresponding fraction for uniformly
random bitstrings.}
\label{tab:si_sqdrift}
\footnotesize
\setlength{\tabcolsep}{3.5pt}
\begin{tabular}{@{}llrrrrrrrrr@{}}
\toprule
 & Orbitals & \multicolumn{2}{c}{Error, QPU} & \multicolumn{2}{c}{Error, random} & \multicolumn{3}{c}{Gap} & \multicolumn{2}{c}{Valid (\%)} \\
\cmidrule(lr){3-4}\cmidrule(lr){5-6}\cmidrule(lr){7-9}\cmidrule(l){10-11}
No. & (qubits) & low & high & low & high & CASCI & QPU & random & QPU & random \\
\midrule
\cpd{4} & 9 (18)  & 0.000 & 0.000 & 0.000 & 0.000 & 2.79    & 2.79    & 2.79    & 17.1 & 6.1 \\
\cpd{4} & 11 (22) & 0.005 & 0.002 & 0.001 & 0.001 & 2.99    & 2.99    & 2.99    & 12.2 & 5.1 \\
\cpd{4} & 13 (26) & 0.017 & 0.006 & 0.009 & 0.008 & 0.34    & 0.36    & 0.35    & 9.6  & 4.4 \\
\cpd{4} & 16 (32) & 0.271 & 0.166 & 0.839 & 0.354 & $-5.44$ & $-5.34$ & $-4.96$ & 5.4  & 3.9 \\
\addlinespace
\cpd{8} & 9 (18)  & 0.000 & 0.000 & 0.000 & 0.000 & 38.91   & 38.91   & 38.91   & 15.2 & 6.1 \\
\cpd{8} & 11 (22) & 0.003 & 0.002 & 0.001 & 0.001 & 37.67   & 37.67   & 37.67   & 6.5  & 2.6 \\
\cpd{8} & 13 (26) & 0.025 & 0.004 & 0.005 & 0.005 & 35.63   & 35.65   & 35.63   & 3.6  & 0.8 \\
\cpd{8} & 16 (32) & 0.158 & 0.118 & 0.114 & 0.128 & 13.96   & 14.00   & 13.95   & 2.4  & 0.4 \\
\bottomrule
\end{tabular}
\end{table}
\paragraph{Quantum processors.}
All hardware calculations used IBM superconducting processors with
fixed-frequency transmon qubits and tunable couplers (Table~\ref{tab:si_qpu}).
\texttt{ibm\_phoenix} belongs to the Nighthawk family:
120 qubits on a $10\times12$ square lattice in which each qubit couples to up
to four neighbors (218 couplers; Fig.~\ref{fig:si_nighthawk}).
\texttt{ibm\_fez} is a Heron~r2 processor with 156 qubits on a heavy-hexagon
lattice, in which each qubit couples to at most three neighbors. The square
lattice accommodates the two spin chains of the Jordan--Wigner ordering as a
ladder---the $\alpha$ and $\beta$ spin orbitals of the same spatial orbital on
neighboring qubits---so that the opposite-spin interactions of LUCJ act on
the rungs without swap gates. The heavy-hexagon lattice lacks such ladders,
which explains the larger two-qubit gate count of the same LUCJ circuit on
The SqDRIFT circuits were compiled one by one, and the compiler chose a different placement for
each, so the 400 circuits of one state together touch most of the qubits of \texttt{ibm\_phoenix} (95 and
99 of the 120 qubits for the 32-qubit quintet and singlet of \cpd{8}). A single layout therefore does not
exist for these runs. The 32-qubit LUCJ circuit of \cpd{4} used physical qubits 64--66, 70--72, 74--76,
80--86, 90--96, 100--104 and 110--114 of \texttt{ibm\_phoenix}, and qubits 87--93, 97, 98, 105--111,
117, 118, 123--129, 136, 137 and 143--147 of \texttt{ibm\_fez}.

\begin{table}[h!]
\centering
\caption{\textbf{Quantum processors.} Median error rates and coherence times over all qubits and couplers,
from the calibration stored with the compiled circuits.}
\label{tab:si_qpu}
\footnotesize
\setlength{\tabcolsep}{4pt}
\begin{tabular}{@{}lcc@{}}
\toprule
 & \texttt{ibm\_phoenix} & \texttt{ibm\_fez} \\
\midrule
Processor family & Nighthawk r2 & Heron r2 \\
Qubits & 120 & 156 \\
Topology (max.\ degree) & square lattice (4) & heavy hexagon (3) \\
Native gates & CZ, $\sqrt{X}$, $X$, $R_Z$ & CZ, $\sqrt{X}$, $X$, $R_Z$ \\
Median two-qubit (CZ) error & $1.76\times10^{-3}$ & $2.45\times10^{-3}$ \\
Median readout error & $4.4\times10^{-3}$ & $1.22\times10^{-2}$ \\
Median $T_1$ / $T_2$ ($\mu$s) & 180 / 149 & 134 / 100 \\
Calibration used & 10 Nov 2026 & 10 Nov 2026 \\
Used for & SqDRIFT 18--32 q; LUCJ 32 q (\cpd{4}) and 22 q (\cpd{8}) & LUCJ 32 q (\cpd{4}) \\
\bottomrule
\end{tabular}
\end{table}

\begin{table}[h!]
\centering
\caption{\textbf{SqDRIFT circuits on \texttt{ibm\_phoenix} for all active spaces.} Mean two-qubit
gate count, depth and ESP over the 400 circuits of each state (two evolution times $\times$ 200 circuits,
$L=10$), with the calibration of Table~\ref{tab:si_qpu}; processor time for 40\,000 shots. Low spin: singlet;
high spin: triplet (\cpd{4}) or quintet (\cpd{8}).}
\label{tab:si_sqdrift_circ}
\footnotesize
\setlength{\tabcolsep}{3.5pt}
\begin{tabular}{@{}llcccccccc@{}}
\toprule
 & Orbitals & \multicolumn{2}{c}{2q} & \multicolumn{2}{c}{Depth} & \multicolumn{2}{c}{ESP} & \multicolumn{2}{c}{Time (s)} \\
\cmidrule(lr){3-4}\cmidrule(lr){5-6}\cmidrule(lr){7-8}\cmidrule(l){9-10}
No. & (qubits) & low & high & low & high & low & high & low & high \\
\midrule
\cpd{4} & 9 (18)  & 334 & 293 & 786  & 700  & 0.568 & 0.606 & 5 & 5 \\
\cpd{4} & 11 (22) & 369 & 356 & 848  & 822  & 0.528 & 0.534 & 5 & 5 \\
\cpd{4} & 13 (26) & 413 & 396 & 936  & 901  & 0.477 & 0.488 & 5 & 5 \\
\cpd{4} & 16 (32) & 518 & 519 & 1171 & 1178 & 0.381 & 0.384 & 6 & 6 \\
\addlinespace
\cpd{8} & 9 (18)  & 354 & 317 & 857  & 760  & 0.556 & 0.580 & 5 & 5 \\
\cpd{8} & 11 (22) & 482 & 353 & 1162 & 827  & 0.448 & 0.535 & 6 & 5 \\
\cpd{8} & 13 (26) & 542 & 431 & 1285 & 961  & 0.388 & 0.457 & 6 & 5 \\
\cpd{8} & 16 (32) & 608 & 534 & 1422 & 1188 & 0.326 & 0.370 & 6 & 6 \\
\bottomrule
\end{tabular}
\end{table}

\begin{table}[h!]
\centering
\caption{\textbf{Circuits for the singlet of \cpd{4} at 32 qubits,
CAS(16,16).} Two-qubit gate count (2q), depth and estimated success
probability (ESP, the product of the gate and readout fidelities of the physical qubits used) after
transpilation for the listed processor. SqDRIFT: two evolution times
($t=1$ and 10) $\times$ 200 random circuits $\times$ 100 shots; mean over all 400 circuits (the means
for $t=1$ and $t=10$ differ by at most 1 two-qubit gate, 14 layers and 0.002 in ESP).}
\label{tab:si_circ}
\footnotesize
\setlength{\tabcolsep}{3pt}
\begin{tabular}{@{}llccccc@{}}
\toprule
Sample source & Processor & Physical qubits & 2q & Depth & ESP & QPU time (s) \\
\midrule
SqDRIFT & \texttt{ibm\_phoenix} & $\ge 32$$^{a}$ & 518  & 1171 & 0.381 & 6 \\
LUCJ    & \texttt{ibm\_phoenix} & 33$^{b}$       & 928  & 427  & 0.118 & 5 \\
LUCJ    & \texttt{ibm\_fez}     & 32             & 1077 & 566  & 0.012 & 13 \\
\bottomrule
\end{tabular}

\vspace{2pt}
\raggedright\footnotesize
$^{a}$Per circuit; each circuit was placed separately by the compiler. $^{b}$One qubit in addition
to the 32 data qubits is used for routing.
\end{table}

\paragraph{LUCJ calculation for \cpd{8} at 22 qubits.} The three states of \cpd{8} were
also computed with LUCJ circuits on \texttt{ibm\_phoenix}, in the 11-orbital spaces of the predicted
ranking, CAS(14,11), CAS(13,11) and CAS(12,11) for the singlet, quartet and quintet, with 100\,000 shots per
state. An earlier run on \texttt{ibm\_miami} in spaces built from UHF natural orbitals is superseded by
this calculation.

\paragraph{Circuits.} The LUCJ circuits (one repetition) were parametrized
from the CCSD $t_1$ and $t_2$ amplitudes of each active space by double
factorization of $t_2$, without variational optimization (\texttt{ffsim}
\texttt{UCJOpSpinBalanced} for the singlet, \texttt{UCJOpSpinUnbalanced} for
the quartet and quintet). In the local variant, same-spin Jastrow terms act
between nearest neighbors along each spin chain and opposite-spin terms only
on the ladder rungs; a full LUCJ variant with all pairs was run on the
simulator as a test. The circuits were transpiled with the Qiskit preset pass
manager at optimization level~3 together with the \texttt{ffsim}
pre-initialization passes, on the ladder $\alpha:[11,21,\ldots,111]$, $\beta:[10,20,\ldots,110]$ of \texttt{ibm\_phoenix}
(summed CZ error of the layout 0.060), without dynamical decoupling or error suppression
(Table~\ref{tab:si_jobs}). The transpilation added no swap gates and kept every qubit on the
ladder.

\begin{table}[h!]
\centering
\caption{\textbf{Executed LUCJ circuits for \cpd{8}} (\texttt{ibm\_phoenix}, 11 November 2026).
ESP estimated with the calibration of Table~\ref{tab:si_qpu}.}
\label{tab:si_jobs}
\footnotesize
\begin{tabular}{@{}lcccccc@{}}
\toprule
State & Qubits & Depth & Two-qubit gates & ESP & Shots & QPU time \\
\midrule
Singlet & 22 & 195 & 344 & 0.39 & 100\,000 & 7 s \\
Quartet & 22 & 196 & 338 & 0.40 & 100\,000 & 7 s \\
Quintet & 22 & 200 & 334 & 0.40 & 100\,000 & 7 s \\
\bottomrule
\end{tabular}
\end{table}

\paragraph{SQD settings.} Configuration recovery, subspace construction and
diagonalization used \texttt{qiskit-addon-sqd} with the settings of
Table~\ref{tab:si_sqdset}. The self-consistency loop was stopped when the
energy changed by less than $10^{-5}$~$E_\mathrm{h}$ and the average
occupations by less than $10^{-7}$. Table~\ref{tab:si_sqdres} compares the
different sources of samples.

\begin{table}[h!]
\centering
\caption{\textbf{SQD settings for the sources of samples of \cpd{8}.}}
\label{tab:si_sqdset}
\footnotesize
\begin{tabular}{@{}lcc@{}}
\toprule
 & Random bitstrings & QPU \\
\midrule
Bitstrings & 100\,000 uniform & 100\,000 measured \\
Samples per batch & 1000 & 1000 \\
Batches / max.\ iterations & 3 / 30 & 3 / 30 \\
\bottomrule
\end{tabular}
\end{table}

\begin{table}[h!]
\centering
\caption{\textbf{SQD error $E_\mathrm{SQD}-E_\mathrm{CAS}$ (kcal\,mol$^{-1}$) and sample
statistics} for the three states of \cpd{8} (final-model 11-orbital spaces, $10^5$ samples per source).
Chemical accuracy is 1\,kcal\,mol$^{-1}$ = 1.6~m$E_\mathrm{h}$. Noiseless LUCJ was not repeated for these
spaces.}
\label{tab:si_sqdres}
\footnotesize
\begin{tabular}{@{}lllcccc@{}}
\toprule
State & Space & Source & Distinct bitstrings & HF fraction & Valid fraction & Error \\
\midrule
Singlet & CAS(14,11) & Random & 98\,769 & 0.000 & 0.026 & $1\times10^{-8}$ \\
 & & QPU & 21\,524 & 0.273 & 0.379 & $1.0\times10^{-4}$ \\
\addlinespace
Quartet & CAS(13,11) & Random & 98\,769 & 0.000 & 0.018 & $3\times10^{-5}$ \\
 & & QPU & 25\,689 & 0.218 & 0.312 & $1.0\times10^{-4}$ \\
\addlinespace
Quintet & CAS(12,11) & Random & 98\,769 & 0.000 & 0.013 & $5\times10^{-6}$ \\
 & & QPU & 21\,150 & 0.247 & 0.346 & $1.2\times10^{-2}$ \\
\bottomrule
\end{tabular}
\end{table}

\begin{figure}[h!]
\centering
\includegraphics[width=\linewidth]{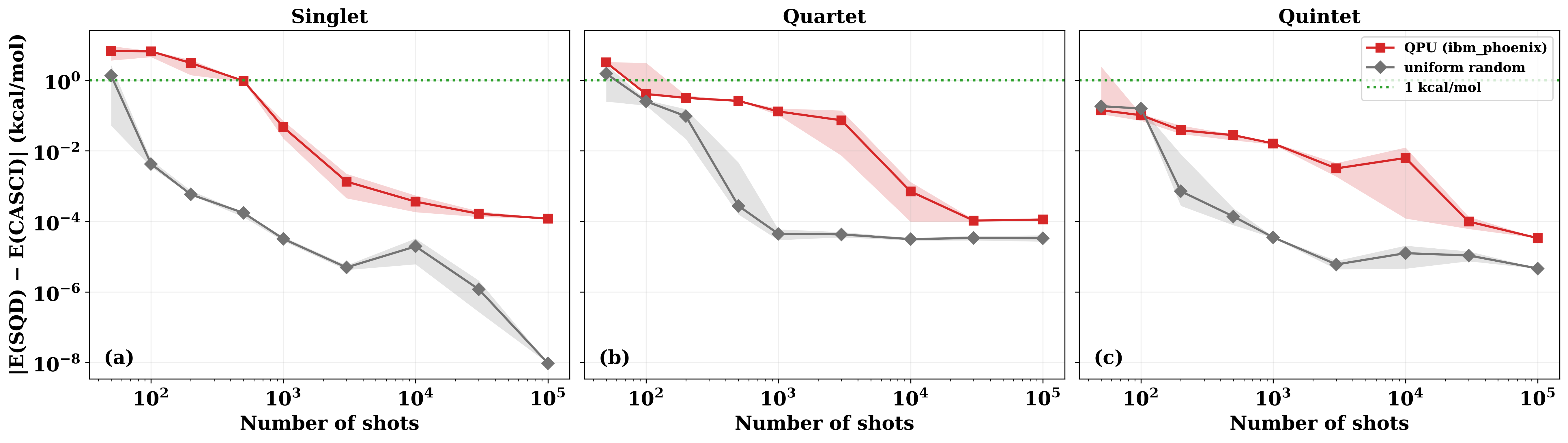}
\caption{\textbf{Shots needed for SQD of \cpd{8} (22 qubits, final-model 11-orbital spaces).}
Error of SQD relative to exact diagonalization versus the number of shots, for subsamples of the
\texttt{ibm\_phoenix} LUCJ samples (red) and for uniformly random bitstrings (gray). Symbols: median over the
seeds (three below $10^3$ shots, two above; one QPU set at $10^5$); shaded: range. Dotted: 1\,kcal\,mol$^{-1}$.
Runs in which no valid configuration was sampled are omitted (Table~\ref{tab:si_lowshot}).}
\label{fig:si_shots}
\end{figure}

\paragraph{Subsampling.} To determine how many shots are needed, and whether the hardware samples
differ from uniformly random bitstrings, random subsets of the measured bitstrings and the same numbers of
random bitstrings were processed with the settings of Table~\ref{tab:si_sqdset}
(Tables~\ref{tab:si_subsample} and \ref{tab:si_lowshot}, Fig.~\ref{fig:si_shots}); no further quantum
time was used. All runs with $10^3$ to $10^5$ shots are far within chemical accuracy (largest error
0.16\,kcal\,mol$^{-1}$, quartet at $10^3$ shots). The hardware samples are 15--27 times more often physically
valid than random bitstrings, but random bitstrings give larger subspaces and equal or smaller errors: in
these 22-qubit spaces, configuration recovery turns $10^5$ random bitstrings into the complete singlet
space (108\,900 configurations), so that SQD is exact. The phoenix samples are concentrated on few
configurations (22--27\% of all shots are the Hartree--Fock determinant), and small subsets therefore give
small subspaces. At 50--100 shots, random samples sometimes contain no valid configuration (one of three
runs for the singlet and quartet and two of three for the quintet at 50 shots; one of three for the quintet
at 100 shots), so configuration recovery cannot start, whereas the hardware samples always can. Chemical
accuracy in all seeds is reached from 500 (hardware) and 100 (random) shots for the singlet, from 200 and
100 shots for the quartet, and from 100 and 200 shots for the quintet; chemical accuracy is taken here as
1\,kcal\,mol$^{-1}$. With the full data the error of the quintet depends on the SQD seed
($1.2\times10^{-2}$ with the seed of Table~\ref{tab:si_sqdres}, $3\times10^{-5}$ with another), which shows
the scatter of SQD at this level.

\begin{table}[h!]
\centering
\caption{\textbf{Subsampling of the measured data} (\texttt{ibm\_phoenix}, final-model
11-orbital spaces; range over two seeds, one set at $10^5$ QPU shots). Error in kcal\,mol$^{-1}$;
subspace dimension out of 108\,900 (singlet), 76\,230 (quartet) and 54\,450 (quintet).}
\label{tab:si_subsample}
\footnotesize
\begin{tabular}{@{}llccccc@{}}
\toprule
State & Shots & Source & Valid fraction & Distinct valid & Subspace dimension & Error \\
\midrule
Singlet & 1\,000 & QPU & 0.39--0.40 & 72--77 & 36\,864--46\,225 & 0.022--0.071 \\
 & & random & 0.03 & 29--32 & 102\,400--104\,329 & $3$--$4\times10^{-5}$ \\
 & 10\,000 & QPU & 0.37--0.39 & 312--320 & 86\,436--90\,000 & $2$--$5\times10^{-4}$ \\
 & & random & 0.03 & 261--274 & 103\,684--105\,625 & $6\times10^{-6}$--$3\times10^{-5}$ \\
 & 100\,000 & QPU & 0.38 & 1\,272 & 89\,401 & $1\times10^{-4}$ \\
 & & random & 0.03 & 2\,589--2\,602 & 108\,900 & $1\times10^{-8}$ \\
\addlinespace
Quartet & 1\,000 & QPU & 0.29--0.31 & 55--72 & 16\,059--19\,012 & 0.10--0.16 \\
 & & random & 0.01 & 15 & 61\,180--61\,777 & $3$--$6\times10^{-5}$ \\
 & 10\,000 & QPU & 0.31 & 330--331 & 47\,382--47\,804 & $1\times10^{-4}$--$1\times10^{-3}$ \\
 & & random & 0.02 & 192--194 & 62\,856--62\,918 & $3\times10^{-5}$ \\
 & 100\,000 & QPU & 0.31 & 1\,390 & 51\,025 & $1\times10^{-4}$ \\
 & & random & 0.02 & 1\,807--1\,811 & 66\,584--67\,485 & $3$--$4\times10^{-5}$ \\
\addlinespace
Quintet & 1\,000 & QPU & 0.33--0.38 & 62--80 & 7\,760--8\,547 & 0.015--0.017 \\
 & & random & 0.01 & 12--14 & 45\,188--46\,008 & $3$--$4\times10^{-5}$ \\
 & 10\,000 & QPU & 0.34 & 260--276 & 30\,084--31\,161 & $1\times10^{-4}$--0.012 \\
 & & random & 0.01 & 128--149 & 45\,920--46\,001 & $5\times10^{-6}$--$2\times10^{-5}$ \\
 & 100\,000 & QPU & 0.35 & 1\,085 & 32\,400 & $3\times10^{-5}$ \\
 & & random & 0.01 & 1\,296--1\,304 & 51\,645--52\,635 & $5\times10^{-6}$ \\
\bottomrule
\end{tabular}
\end{table}

\begingroup

\section{Archived fixed-reference and fixed-budget controls}
\label{sec:si_controls}

\subsection{Released-model inference and reference matching}
These archived controls use the earlier released entropy and mutual-information
checkpoints, without retraining, threshold optimization or fitting to benchmark
labels. All eleven states of the benchmark at the time of these controls were reevaluated. Nine reproduce
the printed active-orbital counts, $F_1$ scores and mean absolute errors to
the stated precision. The released triplet references for complexes
\cpd{1} and \cpd{2} differ from those underlying the main tables and are
excluded from these paired controls rather than combined across references.
The control cohort therefore contains nine states of six complexes and
83 DMRG-active orbitals. All selection methods within a comparison use the
same 40-orbital reference window and its DMRG labels.

The stored Box--Cox exponent is $0.10910732265471183$, with
$\epsilon=10^{-8}$. For Model~1, the stored affine min--max transformation
acts on zero-based descriptor columns 0--3, 30, 31, 44--49, 58 and 59.
The remaining columns are unchanged. The exchange matrix is divided by
$2.6162941455841064\,E_\mathrm{h}$. These are the fitted transformations
distributed with the checkpoints, not transformations refitted for this control.
Inference used Python 3.13.5, PyTorch 2.5.1 (CPU), NumPy 2.4.6,
SciPy 1.17.1 and scikit-learn 1.9.0. Figures used Matplotlib 3.11.2.
These versions describe the independent inference control, not an inferred
training environment.

\subsection{Information gain at equal orbital budget}
The entropy-only comparator selects the $|\mathcal P|$ orbitals with the
largest predicted single-orbital entropy in each state. It therefore has
exactly the same state-specific orbital budget as the entropy-plus-MI pool.
The pool recovers 79 of 83 DMRG-active orbitals (95.2\%), compared with
77 (92.8\%) for equal-budget entropy ranking and 76 (91.6\%) for the
fixed permissive entropy threshold of 0.07. The mean pool size is 16.9
spatial orbitals. The two equal-budget gains occur in the quintet of
\cpd{5} (13 versus 12 recovered orbitals) and the singlet of \cpd{6}
(18 versus 17). The quintet of \cpd{3} remains a failure case, with
zero of four active orbitals recovered by either selection.

The paired increase in pooled recall is 2.4 percentage points. A cluster
bootstrap resampling the six complexes, retaining all their spin states
together, gives a 95\% percentile interval of 0.0--4.1 percentage points
(20\,000 resamples, random seed 73291). This small cohort does not establish
a universal or statistically resolved advantage. It does establish that
the two additional recoveries are not caused by a larger orbital budget.
They are additional recovered reference orbitals at unchanged pool size,
not additional selected orbitals or a demonstrated reduction in qubit count.

For budget curves, orbitals above the permissive entropy threshold are
ranked by predicted entropy, followed by MI-rescued orbitals ranked by
their largest predicted coupling to a predicted active orbital. The remaining
orbitals follow in entropy order. The frontier-energy comparator ranks
the same window by distance from the midpoint of its central occupied and
virtual orbital energies. Random selection has expected recall $m/40$
at budget $m$. Orbital recall tests screening completeness, not spin-gap accuracy.

\emph{Final models.} The numbers above refer to the released (paper)
checkpoints and the original labels. With the final five-model ensembles and
the spin-checked labels, over the eleven states of Table~IV
(83 DMRG-active orbitals), the pool recovers 82 orbitals with 15.2 orbitals per
state, equal-budget entropy ranking also 82, and the fixed threshold of 0.07
alone 81 with 13.5 orbitals per state. For the application systems
(29 active orbitals) the pool recovers 28, equal-budget entropy ranking 25 and
the threshold of 0.07 alone 23. The quintet of \cpd{3} is no longer a failure
case: all four of its active orbitals are in the pool. The released benchmark data
have been regenerated from the final label files, so the triplet references of
\cpd{1} and \cpd{2} now agree with the tables.

\begin{figure}[h!]
\centering
\includegraphics[width=\linewidth]{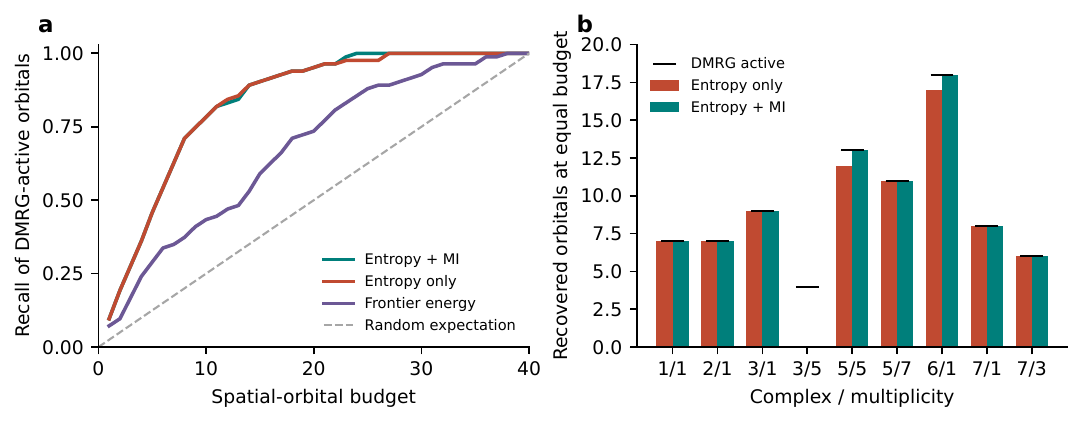}
\caption{Fixed-reference screening controls for nine benchmark states.
\textbf{a}, Pooled recall at the same spatial-orbital budget for entropy-plus-MI
ranking, entropy ranking and frontier-energy ranking. The dashed line is the
expectation for random selection. \textbf{b}, Recovered DMRG-active orbitals
at the state-specific pool budget, identical for both selections. Black marks
give the reference counts. Labels denote complex number and multiplicity.
The empty bars for \cpd{3}, quintet, retain the negative case.}
\label{fig:si_screening_controls}
\end{figure}

\subsection{Attention sensitivity is not a retraining ablation}
The four fitted exchange-bias coefficients are $-0.1333$, $-2.1200$,
$-2.9219$ and $-3.0672$. In the five final models all four coefficients
are negative as well, from $-0.37$ to $-0.81$ in the first and from $-2.0$
to $-2.8$ in the last layer. Thus, exchange bias does not necessarily increase
attention between strongly exchange-coupled orbitals. Setting only the
attention-bias matrix to zero at inference, while retaining all exchange-derived
descriptors and the trained weights, changes the mean entropy error of the
nine-state control cohort from 0.03589 to 0.03593. This small response is
a sensitivity result for the fitted checkpoint. It does not remove exchange
information from the input, and it does not establish the result of training
a model without exchange bias. Retraining controls with and without the exchange bias are given in
Section~\ref{sec:si_ablation}.

\subsection{Equal-budget energetic controls for the nickel complex}
For \cpd{4}, three selection methods were compared for the earlier model spaces
at 9, 11, 13 and 16 spatial orbitals (Fig.~\ref{fig:si_nickel_budget_controls}). The singlet/triplet active-electron counts are 10/12,
14/16, 14/18 and 18/22, respectively, inherited from the released scans and
identical for every method within a spin state. Geometry, def2-SVP basis,
UHF spin state, inactive-electron count and strongly contracted NEVPT2
treatment were also matched. The comparison tests orbital selection at
these fixed electron allocations, rather than optimization over all allocations.
All calculations used PySCF 2.13.1. The nine- and eleven-orbital CASCI
problems were solved exactly. The nine-orbital ML spaces reconstruct the
archived CASCI energies within $3.9\times10^{-7}\,E_\mathrm{h}$, with spin
expectation values within $2\times10^{-9}$ of their targets.

The chemical comparator retains the same metal-$d$ and fractionally occupied
orbitals, then fills the remaining occupied and virtual slots in Fock-energy
order towards the frontier. The AVAS comparator uses Ni $3d$ and ligand C/N
$2p$ minimal-basis target functions. Occupied and virtual blocks are rotated
separately to diagonalize the target projector, and the highest-overlap
orbitals are retained with the same occupied, fractional and virtual counts
as the ML space. This is one specified ligand-aware AVAS choice, not an
optimization over AVAS targets.

The thirteen- and sixteen-orbital controls use spin-adapted DMRG with block2
0.5.4 and full one- through four-particle density matrices for strongly
contracted NEVPT2. A fixed random seed of 4382 and a sweep schedule of
4 sweeps at bond dimension 250, 4 at 500 and up to 12 at 1000 were followed
by a 6-sweep check at 2000. The allowed energy change was
$10^{-6}\,E_\mathrm{h}$. If the one-particle density changed by more than
$10^{-4}$ in Frobenius norm, the higher-bond state was refined until its
energy and density changes fell below $10^{-8}\,E_\mathrm{h}$ and $10^{-5}$,
respectively. Orbital reordering permutes the same selected space.
Density-matrix electron traces were checked, and the active orbitals were
kept unchanged during inactive/virtual semicanonicalization.
Tests against exact eleven-orbital CASCI+NEVPT2 agree within
$6.9\times10^{-7}\,E_\mathrm{h}$ for both spin states. The sixteen-orbital
ML corrected totals reproduce the archived values within
$1.2\times10^{-6}\,E_\mathrm{h}$. Full state energies are listed below.

\begin{table}[h!]
\centering
\caption{Archived nickel spin gaps for equal-budget selection controls.
$\Delta E=E(\mathrm S)-E(\mathrm T)$ in kcal\,mol$^{-1}$.
The full-molecule UCCSD(T) comparison is $-41.8889$\,kcal\,mol$^{-1}$.
The last column is the absolute difference between the corrected gap and
this comparison, not an error relative to an exact reference.}
\label{tab:si_nickel_controls}
\small
\begin{tabular}{@{}rlrrr@{}}
\toprule
Orbitals & Selection & CASCI & CASCI+NEVPT2 & $|\Delta E-\Delta E_{\rm CC}|$ \\
\midrule
9 & ML-guided & 4.1759 & $-34.8443$ & 7.0446 \\
 & Metal-$d$ plus frontier & 1.8632 & $-42.4388$ & 0.5500 \\
 & AVAS (Ni $3d$, C/N $2p$) & 24.3650 & $-21.6208$ & 20.2681 \\
\addlinespace
11 & ML-guided & 2.5521 & $-39.7389$ & 2.1499 \\
 & Metal-$d$ plus frontier & 1.3033 & $-43.2951$ & 1.4062 \\
 & AVAS (Ni $3d$, C/N $2p$) & $-9.7248$ & $-27.0993$ & 14.7896 \\
\addlinespace
13 & ML-guided & $-0.2413$ & $-41.5449$ & 0.3440 \\
 & Metal-$d$ plus frontier & 1.4589 & $-44.4596$ & 2.5707 \\
 & AVAS (Ni $3d$, C/N $2p$) & 7.0067 & $-43.0915$ & 1.2026 \\
\addlinespace
16 & ML-guided & $-0.3073$ & $-41.4937$ & 0.3952 \\
 & Metal-$d$ plus frontier & 6.8794 & $-41.3467$ & 0.5422 \\
 & AVAS (Ni $3d$, C/N $2p$) & 6.0453 & $-48.2833$ & 6.3944 \\
\bottomrule
\end{tabular}
\end{table}

For these earlier spaces, chemical frontier completion is closer to UCCSD(T) at nine and eleven
orbitals. ML is closest at thirteen orbitals, whereas ML and chemical
selection are similarly close at sixteen. From thirteen to sixteen orbitals,
the corrected gaps change by $+0.0512$, $+3.1129$ and $-5.1918$\,kcal\,mol$^{-1}$
for ML, chemical and AVAS selection, respectively. Within this archived comparison, the evidence supports
local budget stability of the ML ranking, not universal superiority or a
unique energetic optimum. Screening recovery, spin-gap stability and
agreement with a comparison method remain distinct tests.

Occupation-based stability does not establish complete energetic convergence.
The archived ML CASCI gaps at eighteen and twenty orbitals are $-3.5259$
and $-9.5058$\,kcal\,mol$^{-1}$, compared with approximately $-0.24$ and
$-0.31$\,kcal\,mol$^{-1}$ at thirteen and sixteen orbitals. Their NEVPT2
corrections are not available. Thus the thirteen-to-sixteen-orbital plateau
is local to the reported budget range. UCCSD(T) is a cross-method comparison
with its own uncertainty, particularly for the nickel singlet with a T1
diagnostic of 0.062. These controls concern one complex, fixed geometries
and basis, one chemical rule and one specified AVAS target choice. They do
not isolate the energetic contribution of MI or constitute retraining ablations.
The ML rows of Tables~\ref{tab:si_nickel_controls} and
\ref{tab:si_nickel_control_energies} use the earlier ML spaces, with 18 and 22
active electrons at 16 orbitals. The final-model spaces use 16 active electrons
in both spin states at that size (Table~\ref{tab:si_expanded_final}). Their
CASCI+NEVPT2 gaps are $-37.72$, $-36.90$, $-37.33$ and $-35.57$\,kcal\,mol$^{-1}$
at 9, 11, 13 and 16 orbitals, respectively, with absolute differences from
UCCSD(T) of 4.17, 4.99, 4.56 and 6.32\,kcal\,mol$^{-1}$. The change from 13 to
16 orbitals is $+1.76$\,kcal\,mol$^{-1}$. The chemical and AVAS controls retain
the earlier electron allocations, so their comparison applies to the
earlier model spaces.

\begin{figure}[htbp]
  \centering
  \includegraphics[width=\textwidth]{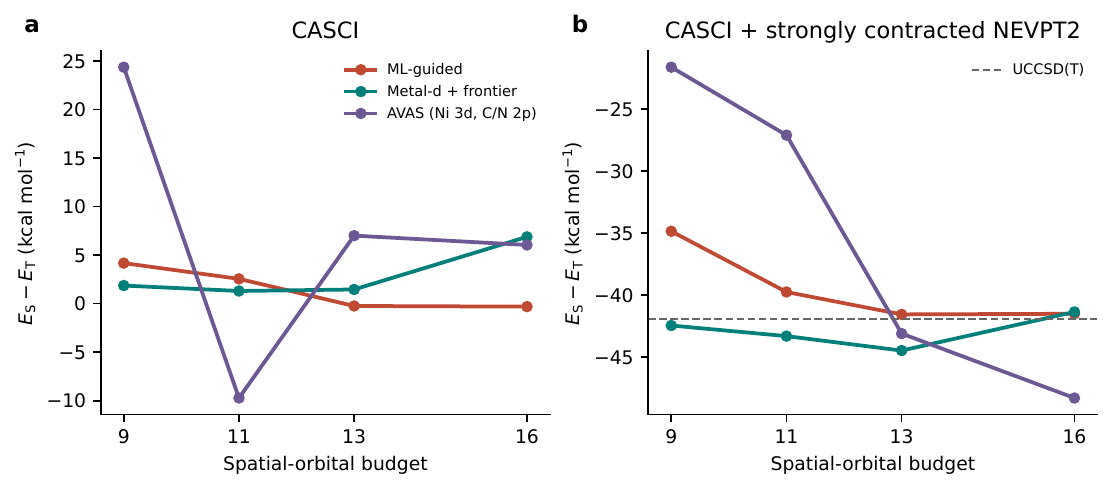}
  \caption{\textbf{Archived fixed-budget nickel selection controls.}
  Singlet--triplet gaps for the earlier model spaces using ML-guided selection, metal-$d$ plus frontier
  completion and the specified Ni $3d$/ligand C,N $2p$ AVAS choice.
  \textbf{a}, CASCI. \textbf{b}, CASCI with strongly contracted NEVPT2.
  The dashed line is the full-molecule UCCSD(T) comparison, not an exact
  reference. All selections use the same orbital and active-electron budgets
  within each spin state.}
  \label{fig:si_nickel_budget_controls}
\end{figure}

\begin{table}[h!]
\centering
\caption{Archived state energies for the fixed-budget nickel controls in $E_\mathrm{h}$.
The total energy is the sum of CASCI and the strongly contracted NEVPT2
correction. Chemical denotes metal-$d$ plus frontier completion, and AVAS
uses the Ni $3d$/ligand C,N $2p$ target defined above.}
\label{tab:si_nickel_control_energies}
\small
\begin{tabular}{@{}rllrrr@{}}
\toprule
Orbitals & Selection & State & $E_{\rm CASCI}$ & $E_{\rm corr}^{(2)}$ & $E_{\rm total}$ \\
\midrule
9 & ML-guided & singlet & $-1925.315151584$ & $-1.841852541$ & $-1927.157004125$ \\
9 & ML-guided & triplet & $-1925.321806365$ & $-1.779669872$ & $-1927.101476237$ \\
9 & Chemical & singlet & $-1925.317496288$ & $-1.841977481$ & $-1927.159473769$ \\
9 & Chemical & triplet & $-1925.320465491$ & $-1.771377690$ & $-1927.091843181$ \\
9 & AVAS & singlet & $-1925.314997999$ & $-1.813487276$ & $-1927.128485275$ \\
9 & AVAS & triplet & $-1925.353826055$ & $-1.740204240$ & $-1927.094030295$ \\
\addlinespace
11 & ML-guided & singlet & $-1925.317903773$ & $-1.844579011$ & $-1927.162482784$ \\
11 & ML-guided & triplet & $-1925.321970794$ & $-1.777183972$ & $-1927.099154766$ \\
11 & Chemical & singlet & $-1925.318465539$ & $-1.835667245$ & $-1927.154132784$ \\
11 & Chemical & triplet & $-1925.320542502$ & $-1.764595169$ & $-1927.085137671$ \\
11 & AVAS & singlet & $-1925.372669865$ & $-1.754464199$ & $-1927.127134064$ \\
11 & AVAS & triplet & $-1925.357172473$ & $-1.726776121$ & $-1927.083948593$ \\
\addlinespace
13 & ML-guided & singlet & $-1925.322619408$ & $-1.841358500$ & $-1927.163977907$ \\
13 & ML-guided & triplet & $-1925.322234841$ & $-1.775537063$ & $-1927.097771904$ \\
13 & Chemical & singlet & $-1925.321140461$ & $-1.830816982$ & $-1927.151957443$ \\
13 & Chemical & triplet & $-1925.323465363$ & $-1.757641188$ & $-1927.081106551$ \\
13 & AVAS & singlet & $-1925.383660835$ & $-1.762374869$ & $-1927.146035703$ \\
13 & AVAS & triplet & $-1925.394826649$ & $-1.682538435$ & $-1927.077365085$ \\
\addlinespace
16 & ML-guided & singlet & $-1925.337938549$ & $-1.824807917$ & $-1927.162746467$ \\
16 & ML-guided & triplet & $-1925.337448834$ & $-1.759173194$ & $-1927.096622027$ \\
16 & Chemical & singlet & $-1925.324040341$ & $-1.822395365$ & $-1927.146435706$ \\
16 & Chemical & triplet & $-1925.335003371$ & $-1.745542203$ & $-1927.080545574$ \\
16 & AVAS & singlet & $-1925.429692231$ & $-1.701785963$ & $-1927.131478194$ \\
16 & AVAS & triplet & $-1925.439325992$ & $-1.615207892$ & $-1927.054533884$ \\
\bottomrule
\end{tabular}
\end{table}

\subsection{Expanded active-space energy record}
\begin{table}[h!]
\centering

\caption{Archived active-space energies for the two application complexes
in $E_\mathrm{h}$. $n$ and $N_e$ are spatial orbitals and active electrons.
CASCI is exact except at 16 orbitals, where the listed energy is DMRG-CASCI
with bond dimension 1000. Corrections are from the archived DMRG-based
strongly contracted NEVPT2 calculations. Dashes mean that no correction
was available and must not be interpreted as zero. S, T and Q denote
singlet, triplet and quintet.}
\label{tab:si_expanded_energies}
\footnotesize
\begin{tabular}{@{}llrrrr@{}}
\toprule
Complex & State & $n$ & $N_e$ & $E_\mathrm{CASCI}$ & NEVPT2 correction \\
\midrule
4 & S & 9 & 10 & $-1925.31515120$ & $-1.84185024$ \\
4 & T & 9 & 12 & $-1925.32180636$ & $-1.77966975$ \\
4 & S & 11 & 14 & $-1925.31790365$ & $-1.84457722$ \\
4 & T & 11 & 16 & $-1925.32197079$ & $-1.77718400$ \\
4 & S & 13 & 14 & $-1925.32262013$ & $-1.84135682$ \\
4 & T & 13 & 18 & $-1925.32223484$ & $-1.77553706$ \\
4 & S & 16 & 18 & $-1925.33793892$ & $-1.82480639$ \\
4 & T & 16 & 22 & $-1925.33744881$ & $-1.75917321$ \\
4 & S & 18 & 20 & $-1925.34554050$ & -- \\
4 & T & 18 & 24 & $-1925.33992165$ & -- \\
4 & S & 20 & 24 & $-1925.35811007$ & -- \\
4 & T & 20 & 24 & $-1925.34296155$ & -- \\
8 & S & 9 & 10 & $-1947.44880348$ & $-1.79396949$ \\
8 & Q & 9 & 10 & $-1947.51591372$ & $-1.75611795$ \\
8 & S & 11 & 14 & $-1947.44889642$ & $-1.79171263$ \\
8 & Q & 11 & 12 & $-1947.51629393$ & $-1.75464034$ \\
8 & S & 13 & 18 & $-1947.45432395$ & $-1.78833967$ \\
8 & Q & 13 & 16 & $-1947.51695917$ & $-1.75318705$ \\
8 & S & 16 & 22 & $-1947.46613539$ & $-1.77310736$ \\
8 & Q & 16 & 22 & $-1947.52063699$ & $-1.74220366$ \\
8 & S & 18 & 26 & $-1947.46839519$ & -- \\
8 & Q & 18 & 26 & $-1947.52177049$ & -- \\
8 & S & 20 & 26 & $-1947.47512933$ & -- \\
8 & Q & 20 & 28 & $-1947.52436599$ & -- \\
\bottomrule
\end{tabular}
\end{table}

The table above records the earlier active spaces. The final-model
energies used in the main text are listed in Table~\ref{tab:si_expanded_final},
including NEVPT2 corrections at 9, 11, 13 and 16 orbitals. Corrections beyond
16 orbitals are unavailable for both sets of spaces. For the final-model
spaces, the cobalt corrected gap rises by 2.8\,kcal\,mol$^{-1}$ from 13 to
16 orbitals and lies 5.3\,kcal\,mol$^{-1}$ above UCCSD(T) at 16 orbitals.

\begin{table}[h!]
\centering
\caption{\textbf{Active-space energies of the final-model spaces} ($E_\mathrm{h}$). $n$ and $N_e$
are spatial orbitals and active electrons. CASCI is exact up to 16 orbitals and DMRG-CASCI (bond dimension
1000) from 17 orbitals. NEVPT2 corrections are strongly contracted, from DMRG reduced density matrices;
dashes: not evaluated. S, T, Q4 and Q: singlet, triplet, quartet and quintet.}
\label{tab:si_expanded_final}
\footnotesize
\begin{tabular}{@{}llrrrr@{}}
\toprule
Complex & State & $n$ & $N_e$ & $E_\mathrm{CASCI}$ & NEVPT2 correction \\
\midrule
4 & S & 9 & 10 & $-1925.31515120$ & $-1.84185024$ \\
4 & S & 11 & 12 & $-1925.31813061$ & $-1.84112738$ \\
4 & S & 13 & 14 & $-1925.32366004$ & $-1.83594785$ \\
4 & S & 16 & 16 & $-1925.33787196$ & $-1.82302056$ \\
4 & S & 17 & 18 & $-1925.34278749$ & -- \\
4 & S & 18 & 20 & $-1925.34554050$ & -- \\
4 & S & 20 & 22 & $-1925.35045651$ & -- \\
4 & T & 9 & 10 & $-1925.31959669$ & $-1.77729502$ \\
4 & T & 11 & 12 & $-1925.32289190$ & $-1.77756848$ \\
4 & T & 13 & 12 & $-1925.32420941$ & $-1.77590959$ \\
4 & T & 16 & 16 & $-1925.32919522$ & $-1.77501361$ \\
4 & T & 17 & 18 & $-1925.34067393$ & -- \\
4 & T & 18 & 20 & $-1925.34086699$ & -- \\
4 & T & 20 & 24 & $-1925.34699467$ & -- \\
\addlinespace
8 & S & 9 & 10 & $-1947.44880348$ & $-1.79396949$ \\
8 & S & 11 & 14 & $-1947.45237817$ & $-1.79297840$ \\
8 & S & 13 & 18 & $-1947.46349050$ & $-1.77837368$ \\
8 & S & 16 & 22 & $-1947.49842301$ & $-1.73674459$ \\
8 & S & 17 & 24 & $-1947.49936941$ & -- \\
8 & S & 18 & 24 & $-1947.50148733$ & -- \\
8 & S & 20 & 28 & $-1947.50486510$ & -- \\
8 & Q4 & 9 & 9 & $-1948.12764351$ & $-1.78657429$ \\
8 & Q4 & 11 & 13 & $-1948.13743118$ & $-1.78782302$ \\
8 & Q4 & 13 & 15 & $-1948.14778801$ & $-1.77211932$ \\
8 & Q4 & 16 & 19 & $-1948.14949384$ & $-1.77045834$ \\
8 & Q4 & 17 & 21 & $-1948.14988595$ & -- \\
8 & Q4 & 18 & 23 & $-1948.15306552$ & -- \\
8 & Q4 & 20 & 25 & $-1948.15491853$ & -- \\
8 & Q & 9 & 8 & $-1947.51081608$ & $-1.76126055$ \\
8 & Q & 11 & 12 & $-1947.51240789$ & $-1.75798921$ \\
8 & Q & 13 & 16 & $-1947.52027130$ & $-1.74924580$ \\
8 & Q & 16 & 20 & $-1947.52067566$ & $-1.74664227$ \\
8 & Q & 17 & 20 & $-1947.52092390$ & -- \\
8 & Q & 18 & 22 & $-1947.52167430$ & -- \\
8 & Q & 20 & 24 & $-1947.52292520$ & -- \\
\bottomrule
\end{tabular}
\end{table}

\endgroup

\section{Stability screening of the UHF references}
\label{sec:si_uhf}
\begin{table}[h!]
\centering
\caption{\textbf{UHF references of the benchmark states.} $E_\mathrm{UHF}$ of the
reference used (in $E_\mathrm{h}$) and the result of the stability screening:
``confirmed'' means that the screening returned the stored solution
(within 0.1~m$E_\mathrm{h}$); otherwise the lowering relative to the first
calculation is given, and the labels refer to the lower solution.}
\label{tab:si_uhf}
\footnotesize
\begin{tabular}{@{}llll@{}}
\toprule
Complex & State & $E_\mathrm{UHF}$ used & Screening \\
\midrule
\cpd{1}  & singlet & $-3698.904480$ & confirmed (closed shell) \\
         & triplet & $-3699.000000$ & lower by 6.1~m$E_\mathrm{h}$ \\
\cpd{2}  & singlet & $-1538.709354$ & confirmed (closed shell) \\
         & triplet & $-1538.805934$ & lower by 5.2~m$E_\mathrm{h}$ \\
\cpd{3}  & singlet & $-1539.288110$ & confirmed \\
         & triplet & --             & no spin-clean solution$^\ast$ \\
         & quintet & $-1539.376178$ & confirmed \\
\cpd{5}  & singlet & $-1892.070711$ & new calculation, closed shell \\
         & triplet & --             & no spin-clean solution$^\ast$ \\
         & quintet & $-1891.945629$ & lower by 34.3~m$E_\mathrm{h}$ \\
         & septet  & $-1891.871014$ & lower by 19.4~m$E_\mathrm{h}$ \\
\cpd{6}  & singlet & $-1775.684255$ & confirmed \\
         & triplet & --             & no spin-clean solution$^\ast$ \\
\cpd{7}  & singlet & $-5238.034090$ & confirmed \\
         & triplet & $-5238.101363$ & lower by 61.5~m$E_\mathrm{h}$ \\
\cpd{9}  & quintet & $-2880.846959$ & confirmed \\
\cpd{10} & singlet & $-1538.704175$ & lower by 1.3~m$E_\mathrm{h}$ \\
         & triplet & $-1538.802165$ & lower by 0.8~m$E_\mathrm{h}$ \\
\cpd{11} & quartet & $-7116.882956$ & lower by 15.2~m$E_\mathrm{h}$ \\
\cpd{12} & singlet & $-4627.372240$ & confirmed \\
         & triplet & --             & no spin-clean solution$^\ast$ \\
\cpd{13} & singlet & $-5238.044962$ & confirmed \\
\cpd{14} & singlet & $-3344.166809$ & confirmed \\
         & triplet & $-3344.290676$ & lower by 13.1~m$E_\mathrm{h}$ \\
\bottomrule
\end{tabular}
\end{table}

\section{Retraining controls of the exchange bias}
\label{sec:si_ablation}
Model~1 was retrained, on the same data, split and settings as the final models
and with the same five initial-weight seeds, (a) with the exchange bias held at
zero in all layers and (b) with the exchange matrix of every molecule permuted by
a fixed random permutation of its orbitals, so that the couplings are assigned to
the wrong orbital pairs while all descriptors are unchanged.
Table~\ref{tab:si_ablation} compares the five-model ensembles. Neither control
reduces the accuracy on the test complexes, and the model without exchange bias
also reaches the same accuracy on the benchmark. The exchange bias
in attention therefore adds no measurable accuracy in the trained model; the
exchange information reaches the model through the exchange descriptors of each
orbital.

\begin{table}[h!]
\centering
\caption{\textbf{Retraining controls of the exchange bias} (five-model ensembles).
Test: 866 states of 402 held-out complexes. Benchmark: eleven states as in
Table~III, evaluated before the singlet of \cpd{5} was added
(septet of \cpd{5} included); $F_1$ is the mean over states. The permuted-exchange
control was evaluated on the test set only.}
\label{tab:si_ablation}
\footnotesize
\begin{tabular}{@{}lcccccc@{}}
\toprule
 & \multicolumn{3}{c}{Test} & \multicolumn{3}{c}{Benchmark} \\
\cmidrule(lr){2-4}\cmidrule(lr){5-7}
Model~1 & MAE & MAE ($s_1\ge0.10$) & $F_1$ & MAE & $F_1$ (0.10 / 0.07) & Pool \\
\midrule
final (exchange bias) & 0.025 & 0.080 & 0.83 & 0.027 & 0.63 / 0.74 & 77/78 \\
no exchange bias      & 0.025 & 0.080 & 0.83 & 0.026 & 0.70 / 0.77 & 77/78 \\
permuted exchange     & 0.025 & 0.079 & 0.83 & -- & -- & -- \\
\bottomrule
\end{tabular}
\end{table}

\begin{table}[h!]
\centering
\caption{\textbf{SQD of \cpd{8} with few shots} (22 qubits, \texttt{ibm\_phoenix}
LUCJ samples and uniformly random bitstrings, final-model 11-orbital spaces). Error relative to exact
diagonalization in kcal\,mol$^{-1}$ for three subsampling seeds; n.v.: no
valid configuration in the sample, so configuration recovery cannot start.}
\label{tab:si_lowshot}
\footnotesize
\begin{tabular}{@{}llll@{}}
\toprule
State & Shots & QPU samples & Random bitstrings \\
\midrule
Singlet & 50 & 3.66, 6.78, 9.37 & n.v., 0.05, 2.66 \\
 & 100 & 6.65, 4.58, 6.82 & $3\times10^{-3}$, $6\times10^{-3}$, $4\times10^{-3}$ \\
 & 200 & 1.40, 3.10, 3.82 & $6\times10^{-4}$, $6\times10^{-4}$, $7\times10^{-4}$ \\
 & 500 & 0.98, 0.91, 0.97 & $1\times10^{-4}$, $2\times10^{-4}$, $2\times10^{-4}$ \\
\addlinespace
Quartet & 50 & 3.29, 3.22, 3.11 & 0.25, n.v., 2.85 \\
 & 100 & 3.16, 0.41, 0.38 & 0.25, 0.28, 0.19 \\
 & 200 & 0.32, 0.30, 0.37 & 0.02, 0.10, 0.15 \\
 & 500 & 0.26, 0.26, 0.26 & $3\times10^{-4}$, $2\times10^{-4}$, $5\times10^{-3}$ \\
\addlinespace
Quintet & 50 & 0.14, 0.11, 2.46 & n.v., 0.19, n.v. \\
 & 100 & 0.07, 0.11, 0.10 & 0.17, 0.15, n.v. \\
 & 200 & 0.03, 0.04, 0.05 & $3\times10^{-4}$, $7\times10^{-4}$, $8\times10^{-3}$ \\
 & 500 & 0.02, 0.03, 0.03 & $1\times10^{-4}$, $2\times10^{-4}$, $8\times10^{-5}$ \\
\bottomrule
\end{tabular}
\end{table}

\begin{figure}[h!]
\centering
\includegraphics[width=\linewidth]{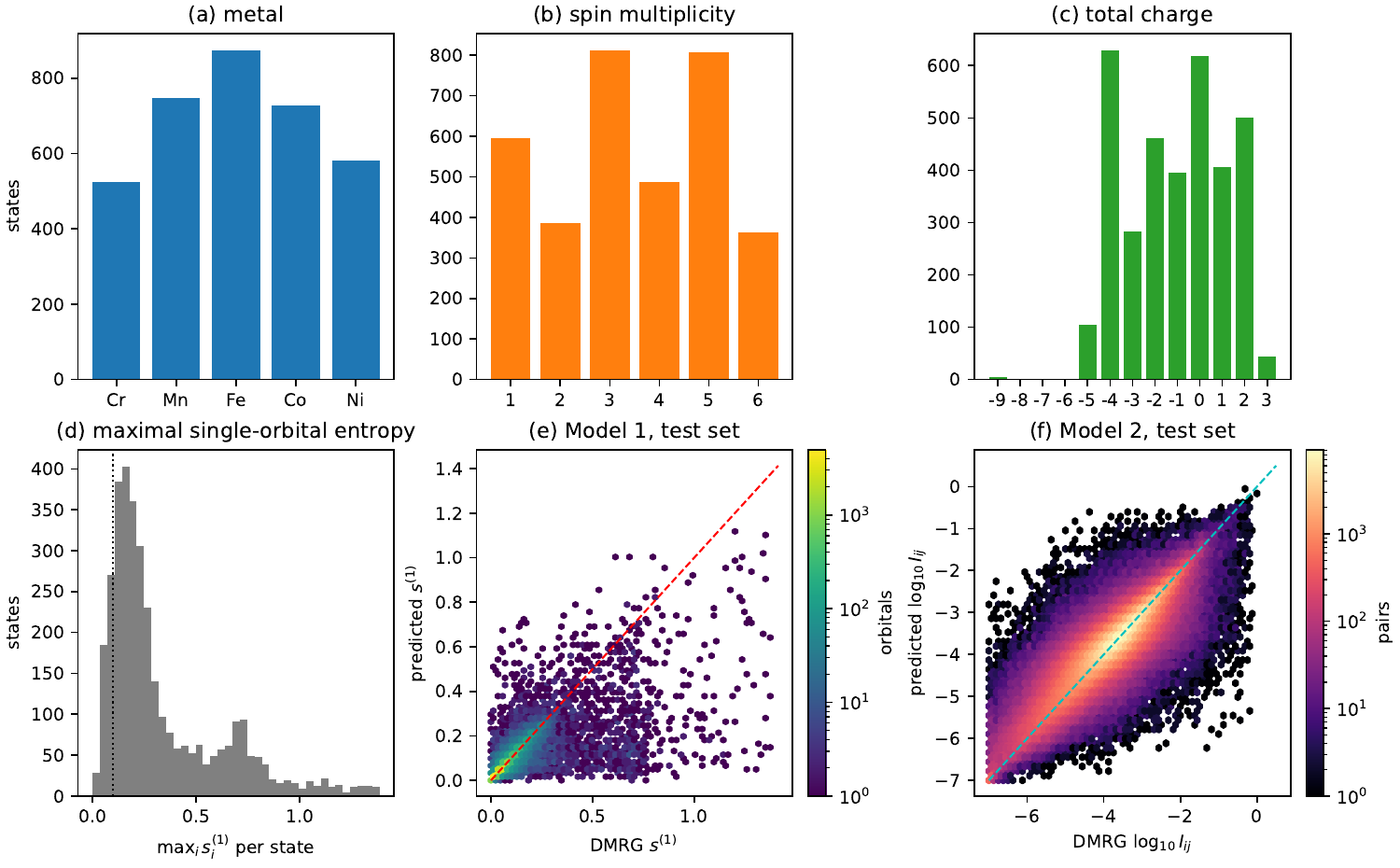}
\caption{\textbf{Training set and in-distribution test.} (a)--(c)~Numbers of
states by metal, spin multiplicity and total charge (3\,454 states).
(d)~Maximal single-orbital entropy per state; dotted line: selection threshold
0.10. (e)~Predicted versus DMRG single-orbital entropies on the test set
(866 states of 402 held-out complexes; mean of the five Model~1 runs).
(f)~Predicted versus DMRG mutual information on the same test complexes
(643\,500 orbital pairs; mean of the five Model~2 runs), on a logarithmic scale;
values below $10^{-7}$ are shown at $10^{-7}$. Large entropies are underestimated in (e), the
compression of strong correlation discussed in the main text.
Color: number of orbitals or pairs per bin (logarithmic).}
\label{fig:si_trainset}
\end{figure}

\section{Comparison with the model of Golub \emph{et al.}}
\label{sec:si_golub}
The code and trained models of Golub \emph{et al.}\ (J.\ Chem.\ Theory Comput.\ \textbf{17}, 6053 (2021)) are not
available, and their model was trained on spin-restricted orbitals of the
SC1MC set, with one spin state per complex. We therefore retrained its
architecture on the data of this work: a feed-forward network with five
hidden layers of 896 units (ReLU, dropout 0.1) that scores each orbital from
its own features, trained with an L1 loss on Box--Cox and min--max scaled
$s^{(1)}$ (Adam, learning rate $10^{-3}$ with cosine decay, batch size 1024,
2000 epochs, best validation epoch kept). As for Model~1, the output layer is
linear during training and clamped at zero afterwards. Two feature sets were
used: the 29 features of Golub \emph{et al.} (descriptors 0--18 and 40--49 of
Table~I, with identical definitions) and all 60
descriptors of this work. Data, split and the five initial seeds are those of
Model~1, and the five-model ensembles are compared in
Table~\ref{tab:si_golub}. With Golub's features, recall on the test complexes
drops to 0.63 (\texttt{SpinAware} 0.80), so about a third of the active orbitals are
missed; the largest benchmark losses occur for open-shell states, e.g.\ the
triplet of \cpd{7} ($F_1=0.20$, against 0.67 with the 60 descriptors). Replacing
the features by the 60 descriptors recovers most of the difference, and the
transformer of \texttt{SpinAware} adds a smaller further improvement, mainly at the
threshold of 0.07.

\begin{table}[h!]
\centering
\caption{\textbf{Comparison with the model of Golub \emph{et al.}} Five-model
ensembles trained on the same data and split. Test: 866 states of 402
held-out complexes, metrics pooled over all orbitals. Benchmark: the eleven
states of Table~III; $F_1$ at $\tau=0.10$ / 0.07.}
\label{tab:si_golub}
\footnotesize
\setlength{\tabcolsep}{3pt}
\begin{tabular}{@{}llcccccc@{}}
\toprule
 & & \multicolumn{3}{c}{Test} & \multicolumn{3}{c}{Benchmark} \\
\cmidrule(lr){3-5}\cmidrule(lr){6-8}
Network & Features & MAE ($s^{(1)}\ge0.10$) & Recall / precision & $F_1$ & $F_1$, mean over states & $F_1$, pooled & MAE \\
\midrule
Golub \emph{et al.} & 29 (Golub) & 0.036 (0.105) & 0.63 / 0.76 & 0.69 & 0.47 / 0.53 & 0.54 / 0.57 & 0.044 \\
Golub \emph{et al.} & 60 (this work) & 0.028 (0.084) & 0.77 / 0.83 & 0.80 & 0.62 / 0.67 & 0.72 / 0.74 & 0.031 \\
\texttt{SpinAware} & 60 (this work) & 0.025 (0.080) & 0.80 / 0.86 & 0.83 & 0.65 / 0.75 & 0.77 / 0.81 & 0.027 \\
\bottomrule
\end{tabular}
\end{table}

\section{Orbital-resolved single-orbital entropies}

For each complex and spin state, the tables below list every orbital of the
40-orbital window with its dominant atomic-orbital character, orbital energy,
occupation in the UHF reference, and the DMRG and predicted single-orbital
entropies. Atomic-orbital character is given for the $\alpha$ component of each
orbital from L\"owdin populations. These tables are numbered O1--O23.

\newcounter{savedtable}\setcounter{savedtable}{\value{table}}%
\setcounter{table}{0}\renewcommand{\thetable}{O\arabic{table}}%
\input{si_tables/si_tables.tex}\setcounter{table}{\value{savedtable}}\renewcommand{\thetable}{S\arabic{table}}%

%% file: si_tables/si_tables.tex
\clearpage

\begin{longtable}{@{}rp{6.2cm}rcrr@{}}
\caption{Single-orbital entropies for complex \textbf{1}, \ce{[VCl6]^3-}, singlet ($2S+1 = 1$): DMRG reference and SpinAware prediction. $E_\mathrm{UHF} = -3698.904480$~$E_\mathrm{h}$; $\langle \hat S^2\rangle = 0.000$ (exact 0.000). Values of $s_1 \ge 0.10$ are in bold. Atomic-orbital character is given for the $\alpha$ component of each orbital (L\"owdin weights $\ge 0.08$).}\label{tab:si_s1_1_singlet}\\
\toprule
No. & Atomic-orbital character ($\alpha$) & $\varepsilon$ ($E_\mathrm{h}$) & Occ. & $s_1$ DMRG & $s_1$ SpinAware \\
\midrule\endfirsthead
\multicolumn{6}{l}{\small\itshape continued from previous page}\\\toprule
No. & Atomic-orbital character ($\alpha$) & $\varepsilon$ ($E_\mathrm{h}$) & Occ. & $s_1$ DMRG & $s_1$ SpinAware \\
\midrule\endhead
\bottomrule\endlastfoot
1 & $s$(Cl) & -0.533 & 2 & 0.015 & 0.017 \\
2 & $d_{x^2-y^2}$(V) $d_{yz}$(V) $p_x$(Cl) $d_{xz}$(V) & -0.067 & 2 & \textbf{0.626} & 0.095 \\
3 & $p_z$(Cl) $p_y$(Cl) $p_x$(Cl) $s$(V) & -0.049 & 2 & 0.063 & 0.062 \\
4 & $p_x$(Cl) $p_z$(Cl) $p_y$(Cl) & -0.007 & 2 & 0.048 & 0.051 \\
5 & $p_z$(Cl) $p_y$(Cl) $p_x$(Cl) & -0.002 & 2 & 0.046 & 0.051 \\
6 & $p_y$(Cl) $p_x$(Cl) & 0.001 & 2 & 0.044 & 0.055 \\
7 & $p_z$(Cl) $p_y$(Cl) $p_x$(Cl) & 0.002 & 2 & 0.087 & 0.076 \\
8 & $p_z$(Cl) $p_x$(Cl) $p_y$(Cl) $d_{z^2}$(V) & 0.002 & 2 & \textbf{0.109} & 0.079 \\
\end{longtable}

\clearpage

\begin{longtable}{@{}rp{6.2cm}rcrr@{}}
\caption{Single-orbital entropies for complex \textbf{1}, \ce{[VCl6]^3-}, triplet ($2S+1 = 3$): DMRG reference and SpinAware prediction. $E_\mathrm{UHF} = -3699.000000$~$E_\mathrm{h}$; $\langle \hat S^2\rangle = 2.010$ (exact 2.000). Values of $s_1 \ge 0.10$ are in bold. Atomic-orbital character is given for the $\alpha$ component of each orbital (L\"owdin weights $\ge 0.08$).}\label{tab:si_s1_1_triplet}\\
\toprule
No. & Atomic-orbital character ($\alpha$) & $\varepsilon$ ($E_\mathrm{h}$) & Occ. & $s_1$ DMRG & $s_1$ SpinAware \\
\midrule\endfirsthead
\multicolumn{6}{l}{\small\itshape continued from previous page}\\\toprule
No. & Atomic-orbital character ($\alpha$) & $\varepsilon$ ($E_\mathrm{h}$) & Occ. & $s_1$ DMRG & $s_1$ SpinAware \\
\midrule\endhead
\bottomrule\endlastfoot
1 & $d_{yz}$(V) $d_{xz}$(V) $d_{xy}$(V) $d_{x^2-y^2}$(V) & -0.156 & 2 & 0.069 & 0.081 \\
2 & $d_{xz}$(V) $d_{x^2-y^2}$(V) $d_{yz}$(V) & -0.156 & 2 & 0.061 & 0.091 \\
3 & $p_z$(Cl) $p_y$(Cl) $p_x$(Cl) $s$(V) & -0.050 & 2 & 0.058 & 0.066 \\
4 & $p_z$(Cl) $p_y$(Cl) $p_x$(V) & -0.003 & 2 & 0.051 & 0.057 \\
5 & $p_x$(Cl) $p_y$(Cl) $p_z$(Cl) & -0.003 & 2 & 0.053 & 0.049 \\
6 & $p_x$(Cl) $p_z$(Cl) $p_y$(Cl) & -0.003 & 2 & 0.049 & 0.052 \\
7 & $p_x$(Cl) $p_z$(Cl) $p_y$(Cl) & 0.005 & 2 & 0.056 & 0.035 \\
8 & $p_x$(Cl) $p_z$(Cl) $p_y$(Cl) $d_{z^2}$(V) & 0.006 & 2 & \textbf{0.125} & \textbf{0.144} \\
\end{longtable}

\clearpage

\begin{longtable}{@{}rp{6.2cm}rcrr@{}}
\caption{Single-orbital entropies for complex \textbf{2}, \ce{[VF6]^3-}, singlet ($2S+1 = 1$): DMRG reference and SpinAware prediction. $E_\mathrm{UHF} = -1538.709354$~$E_\mathrm{h}$; $\langle \hat S^2\rangle = 0.000$ (exact 0.000). Values of $s_1 \ge 0.10$ are in bold. Atomic-orbital character is given for the $\alpha$ component of each orbital (L\"owdin weights $\ge 0.08$).}\label{tab:si_s1_2_singlet}\\
\toprule
No. & Atomic-orbital character ($\alpha$) & $\varepsilon$ ($E_\mathrm{h}$) & Occ. & $s_1$ DMRG & $s_1$ SpinAware \\
\midrule\endfirsthead
\multicolumn{6}{l}{\small\itshape continued from previous page}\\\toprule
No. & Atomic-orbital character ($\alpha$) & $\varepsilon$ ($E_\mathrm{h}$) & Occ. & $s_1$ DMRG & $s_1$ SpinAware \\
\midrule\endhead
\bottomrule\endlastfoot
1 & $s$(F) & -0.817 & 2 & 0.021 & 0.024 \\
2 & $p_z$(F) $p_x$(F) $p_y$(F) & -0.045 & 2 & 0.055 & 0.073 \\
3 & $p_y$(F) $p_z$(F) $p_x$(F) & -0.025 & 2 & \textbf{0.122} & 0.089 \\
4 & $p_y$(F) $p_x$(F) $p_z$(F) & -0.021 & 2 & 0.047 & 0.038 \\
5 & $p_y$(F) $p_x$(F) $p_z$(F) & -0.015 & 2 & 0.055 & 0.071 \\
6 & $p_x$(F) $p_y$(F) $p_z$(F) & -0.010 & 2 & \textbf{0.116} & 0.092 \\
7 & $p_z$(F) $p_x$(F) $p_y$(F) & 0.004 & 2 & 0.054 & 0.071 \\
8 & $p_y$(F) $p_z$(F) $p_x$(F) & 0.005 & 2 & 0.044 & 0.075 \\
\end{longtable}

\clearpage

\begin{longtable}{@{}rp{6.2cm}rcrr@{}}
\caption{Single-orbital entropies for complex \textbf{2}, \ce{[VF6]^3-}, triplet ($2S+1 = 3$): DMRG reference and SpinAware prediction. $E_\mathrm{UHF} = -1538.805934$~$E_\mathrm{h}$; $\langle \hat S^2\rangle = 2.004$ (exact 2.000). Values of $s_1 \ge 0.10$ are in bold. Atomic-orbital character is given for the $\alpha$ component of each orbital (L\"owdin weights $\ge 0.08$).}\label{tab:si_s1_2_triplet}\\
\toprule
No. & Atomic-orbital character ($\alpha$) & $\varepsilon$ ($E_\mathrm{h}$) & Occ. & $s_1$ DMRG & $s_1$ SpinAware \\
\midrule\endfirsthead
\multicolumn{6}{l}{\small\itshape continued from previous page}\\\toprule
No. & Atomic-orbital character ($\alpha$) & $\varepsilon$ ($E_\mathrm{h}$) & Occ. & $s_1$ DMRG & $s_1$ SpinAware \\
\midrule\endhead
\bottomrule\endlastfoot
1 & $p_x$(F) $d_{z^2}$(V) $p_z$(F) $d_{x^2-y^2}$(V) & -0.060 & 2 & 0.043 & 0.082 \\
2 & $p_x$(F) $p_y$(F) $d_{z^2}$(V) $d_{x^2-y^2}$(V) & -0.056 & 2 & 0.052 & 0.082 \\
3 & $p_x$(F) $p_z$(F) $p_y$(F) & -0.032 & 2 & 0.050 & 0.068 \\
4 & $p_y$(F) $p_z$(F) $p_x$(F) & -0.015 & 2 & 0.058 & 0.074 \\
5 & $p_y$(F) $p_z$(F) $p_x$(F) & -0.011 & 2 & 0.059 & 0.073 \\
6 & $p_z$(F) $p_x$(F) $p_y$(F) & -0.002 & 2 & 0.058 & 0.071 \\
7 & $p_y$(F) $p_x$(F) $p_z$(F) & 0.001 & 2 & 0.043 & 0.065 \\
8 & $p_x$(F) $p_y$(F) $p_z$(F) & 0.005 & 2 & 0.048 & 0.070 \\
\end{longtable}

\clearpage

\begin{longtable}{@{}rp{6.2cm}rcrr@{}}
\caption{Single-orbital entropies for complex \textbf{3}, \ce{[Co(CH3)4]-}, singlet ($2S+1 = 1$): DMRG reference and SpinAware prediction. $E_\mathrm{UHF} = -1539.288110$~$E_\mathrm{h}$; $\langle \hat S^2\rangle = 0.000$ (exact 0.000). Values of $s_1 \ge 0.10$ are in bold. Atomic-orbital character is given for the $\alpha$ component of each orbital (L\"owdin weights $\ge 0.08$).}\label{tab:si_s1_3_singlet}\\
\toprule
No. & Atomic-orbital character ($\alpha$) & $\varepsilon$ ($E_\mathrm{h}$) & Occ. & $s_1$ DMRG & $s_1$ SpinAware \\
\midrule\endfirsthead
\multicolumn{6}{l}{\small\itshape continued from previous page}\\\toprule
No. & Atomic-orbital character ($\alpha$) & $\varepsilon$ ($E_\mathrm{h}$) & Occ. & $s_1$ DMRG & $s_1$ SpinAware \\
\midrule\endhead
\bottomrule\endlastfoot
1 & $p_x$(Co) $p_z$(Co) & -2.822 & 2 & 0.006 & 0.006 \\
2 & $s$(C) $s$(H) $s$(Co) & -0.736 & 2 & 0.026 & 0.029 \\
3 & $s$(H) $s$(C) & -0.709 & 2 & 0.028 & 0.026 \\
4 & $s$(H) $s$(C) & -0.702 & 2 & 0.027 & 0.027 \\
5 & $s$(H) $s$(C) & -0.690 & 2 & 0.033 & 0.028 \\
6 & $d_{xy}$(Co) $d_{yz}$(Co) $d_{xz}$(Co) & -0.480 & 2 & \textbf{0.109} & \textbf{0.115} \\
7 & $d_{z^2}$(Co) $d_{xz}$(Co) $d_{x^2-y^2}$(Co) $d_{xy}$(Co) & -0.477 & 2 & \textbf{0.134} & \textbf{0.147} \\
8 & $d_{z^2}$(Co) $d_{x^2-y^2}$(Co) $d_{xz}$(Co) & -0.444 & 2 & \textbf{0.104} & \textbf{0.130} \\
\end{longtable}

\clearpage

\begin{longtable}{@{}rp{6.2cm}rcrr@{}}
\caption{Single-orbital entropies for complex \textbf{3}, \ce{[Co(CH3)4]-}, quintet ($2S+1 = 5$): DMRG reference and SpinAware prediction. $E_\mathrm{UHF} = -1539.376178$~$E_\mathrm{h}$; $\langle \hat S^2\rangle = 6.017$ (exact 6.000). Values of $s_1 \ge 0.10$ are in bold. Atomic-orbital character is given for the $\alpha$ component of each orbital (L\"owdin weights $\ge 0.08$).}\label{tab:si_s1_3_quintet}\\
\toprule
No. & Atomic-orbital character ($\alpha$) & $\varepsilon$ ($E_\mathrm{h}$) & Occ. & $s_1$ DMRG & $s_1$ SpinAware \\
\midrule\endfirsthead
\multicolumn{6}{l}{\small\itshape continued from previous page}\\\toprule
No. & Atomic-orbital character ($\alpha$) & $\varepsilon$ ($E_\mathrm{h}$) & Occ. & $s_1$ DMRG & $s_1$ SpinAware \\
\midrule\endhead
\bottomrule\endlastfoot
1 & $s$(C) $s$(H) $d_{x^2-y^2}$(Co) $d_{xz}$(Co) & -0.729 & 2 & 0.028 & 0.029 \\
2 & $s$(H) $s$(C) & -0.717 & 2 & 0.029 & 0.027 \\
3 & $s$(H) $s$(C) & -0.715 & 2 & 0.030 & 0.027 \\
4 & $d_{xy}$(Co) $d_{yz}$(Co) & -0.659 & 2 & 0.063 & 0.037 \\
5 & $d_{yz}$(Co) $d_{xy}$(Co) & -0.647 & 2 & 0.042 & 0.037 \\
6 & $d_{z^2}$(Co) $d_{x^2-y^2}$(Co) $d_{xz}$(Co) & -0.639 & 2 & 0.048 & 0.039 \\
7 & $d_{x^2-y^2}$(Co) $s$(H) $d_{xz}$(Co) $s$(C) & -0.626 & 2 & 0.049 & 0.037 \\
8 & $d_{xz}$(Co) $d_{z^2}$(Co) $d_{x^2-y^2}$(Co) & -0.582 & 2 & 0.052 & 0.036 \\
\end{longtable}

\clearpage

\begin{longtable}{@{}rp{6.2cm}rcrr@{}}
\caption{Single-orbital entropies for complex \textbf{4}, \ce{[Ni($\eta^3$-C4H7)(NCCH3)2]+}, singlet ($2S+1 = 1$): DMRG reference and SpinAware prediction. $E_\mathrm{UHF} = -1925.292985$~$E_\mathrm{h}$; $\langle \hat S^2\rangle = 0.000$ (exact 0.000). Values of $s_1 \ge 0.10$ are in bold. Atomic-orbital character is given for the $\alpha$ component of each orbital (L\"owdin weights $\ge 0.08$).}\label{tab:si_s1_4_singlet}\\
\toprule
No. & Atomic-orbital character ($\alpha$) & $\varepsilon$ ($E_\mathrm{h}$) & Occ. & $s_1$ DMRG & $s_1$ SpinAware \\
\midrule\endfirsthead
\multicolumn{6}{l}{\small\itshape continued from previous page}\\\toprule
No. & Atomic-orbital character ($\alpha$) & $\varepsilon$ ($E_\mathrm{h}$) & Occ. & $s_1$ DMRG & $s_1$ SpinAware \\
\midrule\endhead
\bottomrule\endlastfoot
1 & $p_z$(C) $p_z$(N) $p_y$(C) $p_y$(N) & -0.796 & 2 & 0.032 & 0.020 \\
2 & $p_z$(C) $s$(H) $p_y$(C) & -0.781 & 2 & 0.052 & 0.030 \\
3 & $p_x$(C) $s$(H) $p_x$(N) $p_y$(C) & -0.777 & 2 & 0.059 & 0.021 \\
4 & $p_x$(C) $s$(H) $p_y$(C) $p_x$(N) & -0.773 & 2 & 0.057 & 0.025 \\
5 & $p_z$(C) $p_y$(C) $s$(H) & -0.772 & 2 & 0.054 & 0.021 \\
6 & $s$(H) $p_y$(C) $p_z$(C) $p_x$(C) & -0.764 & 2 & 0.044 & 0.043 \\
7 & $p_x$(C) $s$(H) $d_{yz}$(Ni) $p_y$(C) & -0.743 & 2 & 0.034 & 0.053 \\
8 & $d_{xz}$(Ni) $d_{x^2-y^2}$(Ni) $p_x$(C) $s$(H) & -0.720 & 2 & 0.061 & \textbf{0.108} \\
\end{longtable}

\clearpage

\begin{longtable}{@{}rp{6.2cm}rcrr@{}}
\caption{Single-orbital entropies for complex \textbf{4}, \ce{[Ni($\eta^3$-C4H7)(NCCH3)2]+}, triplet ($2S+1 = 3$): DMRG reference and SpinAware prediction. $E_\mathrm{UHF} = -1925.329155$~$E_\mathrm{h}$; $\langle \hat S^2\rangle = 2.032$ (exact 2.000). Values of $s_1 \ge 0.10$ are in bold. Atomic-orbital character is given for the $\alpha$ component of each orbital (L\"owdin weights $\ge 0.08$).}\label{tab:si_s1_4_triplet}\\
\toprule
No. & Atomic-orbital character ($\alpha$) & $\varepsilon$ ($E_\mathrm{h}$) & Occ. & $s_1$ DMRG & $s_1$ SpinAware \\
\midrule\endfirsthead
\multicolumn{6}{l}{\small\itshape continued from previous page}\\\toprule
No. & Atomic-orbital character ($\alpha$) & $\varepsilon$ ($E_\mathrm{h}$) & Occ. & $s_1$ DMRG & $s_1$ SpinAware \\
\midrule\endhead
\bottomrule\endlastfoot
1 & $p_y$(C) $p_z$(C) $s$(H) $s$(C) & -0.835 & 2 & 0.037 & 0.039 \\
2 & $p_y$(C) $d_{yz}$(Ni) $s$(H) $d_{x^2-y^2}$(Ni) & -0.817 & 2 & 0.046 & 0.034 \\
3 & $p_z$(C) $p_y$(C) $s$(H) $p_z$(N) & -0.811 & 2 & 0.038 & 0.030 \\
4 & $p_z$(C) $s$(H) & -0.780 & 2 & 0.057 & 0.024 \\
5 & $p_x$(C) $s$(H) $p_z$(C) $p_y$(C) & -0.778 & 2 & 0.043 & 0.042 \\
6 & $p_x$(C) $s$(H) $p_y$(C) $p_z$(C) & -0.776 & 2 & 0.042 & 0.047 \\
7 & $p_y$(C) $s$(H) $p_x$(C) & -0.772 & 2 & 0.044 & 0.033 \\
8 & $p_x$(C) $s$(H) $p_z$(C) & -0.771 & 2 & 0.040 & 0.027 \\
\end{longtable}

\clearpage

\begin{longtable}{@{}rp{6.2cm}rcrr@{}}
\caption{Single-orbital entropies for complex \textbf{5}, \ce{[Cr(CO)5(HNC4H8O)]}, singlet ($2S+1 = 1$): DMRG reference and SpinAware prediction. $E_\mathrm{UHF} = -1892.070711$~$E_\mathrm{h}$; $\langle \hat S^2\rangle = 0.000$ (exact 6.000). Values of $s_1 \ge 0.10$ are in bold. Atomic-orbital character is given for the $\alpha$ component of each orbital (L\"owdin weights $\ge 0.08$).}\label{tab:si_s1_5_singlet}\\
\toprule
No. & Atomic-orbital character ($\alpha$) & $\varepsilon$ ($E_\mathrm{h}$) & Occ. & $s_1$ DMRG & $s_1$ SpinAware \\
\midrule\endfirsthead
\multicolumn{6}{l}{\small\itshape continued from previous page}\\\toprule
No. & Atomic-orbital character ($\alpha$) & $\varepsilon$ ($E_\mathrm{h}$) & Occ. & $s_1$ DMRG & $s_1$ SpinAware \\
\midrule\endhead
\bottomrule\endlastfoot
1 & $p_z$(O) $p_x$(O) $p_z$(C) $p_x$(C) & -0.637 & 2 & 0.081 & 0.072 \\
2 & $p_z$(O) $p_x$(O) $p_z$(C) $p_x$(C) & -0.636 & 2 & 0.074 & 0.069 \\
3 & $p_y$(O) $p_z$(O) $p_x$(O) $p_y$(C) & -0.627 & 2 & 0.085 & 0.080 \\
4 & $p_y$(O) $p_x$(O) $p_z$(O) $p_y$(C) & -0.622 & 2 & 0.082 & 0.074 \\
5 & $p_z$(O) $p_y$(O) $p_x$(O) $p_z$(C) & -0.609 & 2 & 0.093 & 0.074 \\
6 & $p_x$(O) $p_y$(O) $p_z$(O) $p_x$(C) & -0.608 & 2 & 0.096 & 0.083 \\
7 & $s$(C) $d_{x^2-y^2}$(Cr) $d_{xz}$(Cr) $p_z$(C) & -0.596 & 2 & 0.063 & 0.045 \\
8 & $p_x$(C) $p_y$(C) $s$(H) $p_z$(C) & -0.584 & 2 & 0.034 & 0.028 \\
\end{longtable}

\clearpage

\begin{longtable}{@{}rp{6.2cm}rcrr@{}}
\caption{Single-orbital entropies for complex \textbf{5}, \ce{[Cr(CO)5(HNC4H8O)]}, triplet ($2S+1 = 3$): DMRG reference and SpinAware prediction. $E_\mathrm{UHF} = -1891.990451$~$E_\mathrm{h}$; $\langle \hat S^2\rangle = 2.276$ (exact 2.000). Values of $s_1 \ge 0.10$ are in bold. Atomic-orbital character is given for the $\alpha$ component of each orbital (L\"owdin weights $\ge 0.08$). The reference is spin-contaminated; this state is excluded from all averages.}\label{tab:si_s1_5_triplet}\\
\toprule
No. & Atomic-orbital character ($\alpha$) & $\varepsilon$ ($E_\mathrm{h}$) & Occ. & $s_1$ DMRG & $s_1$ SpinAware \\
\midrule\endfirsthead
\multicolumn{6}{l}{\small\itshape continued from previous page}\\\toprule
No. & Atomic-orbital character ($\alpha$) & $\varepsilon$ ($E_\mathrm{h}$) & Occ. & $s_1$ DMRG & $s_1$ SpinAware \\
\midrule\endhead
\bottomrule\endlastfoot
1 & $p_y$(C) $p_x$(C) $p_y$(O) $p_x$(O) & -0.647 & 2 & 0.056 & 0.047 \\
2 & $p_z$(O) $p_x$(C) $p_x$(O) $p_y$(O) & -0.643 & 2 & 0.095 & 0.050 \\
3 & $p_x$(O) $p_z$(O) $p_y$(O) $p_y$(C) & -0.634 & 2 & 0.092 & 0.082 \\
4 & $p_z$(O) $s$(C) $p_y$(C) $p_x$(C) & -0.624 & 2 & 0.093 & 0.060 \\
5 & $p_z$(O) $s$(C) $p_x$(C) $d_{xz}$(Cr) & -0.621 & 2 & 0.097 & 0.078 \\
6 & $p_y$(O) $p_x$(O) $p_x$(C) $p_y$(C) & -0.616 & 2 & 0.076 & 0.058 \\
7 & $p_x$(O) $p_z$(C) $p_y$(O) $p_z$(O) & -0.597 & 2 & 0.086 & 0.086 \\
8 & $p_y$(O) $p_z$(C) $p_y$(C) $s$(C) & -0.584 & 2 & 0.085 & 0.077 \\
\end{longtable}

\clearpage

\begin{longtable}{@{}rp{6.2cm}rcrr@{}}
\caption{Single-orbital entropies for complex \textbf{5}, \ce{[Cr(CO)5(HNC4H8O)]}, quintet ($2S+1 = 5$): DMRG reference and SpinAware prediction. $E_\mathrm{UHF} = -1891.945629$~$E_\mathrm{h}$; $\langle \hat S^2\rangle = 6.462$ (exact 6.000). Values of $s_1 \ge 0.10$ are in bold. Atomic-orbital character is given for the $\alpha$ component of each orbital (L\"owdin weights $\ge 0.08$).}\label{tab:si_s1_5_quintet}\\
\toprule
No. & Atomic-orbital character ($\alpha$) & $\varepsilon$ ($E_\mathrm{h}$) & Occ. & $s_1$ DMRG & $s_1$ SpinAware \\
\midrule\endfirsthead
\multicolumn{6}{l}{\small\itshape continued from previous page}\\\toprule
No. & Atomic-orbital character ($\alpha$) & $\varepsilon$ ($E_\mathrm{h}$) & Occ. & $s_1$ DMRG & $s_1$ SpinAware \\
\midrule\endhead
\bottomrule\endlastfoot
1 & $p_y$(O) $p_y$(C) $p_x$(O) $p_x$(C) & -0.644 & 2 & 0.066 & 0.049 \\
2 & $s$(C) $d_{yz}$(Cr) $p_z$(O) $p_y$(C) & -0.633 & 2 & 0.092 & 0.059 \\
3 & $p_x$(O) $p_y$(O) $p_z$(O) $p_x$(C) & -0.631 & 2 & 0.080 & 0.062 \\
4 & $d_{xz}$(Cr) $p_x$(O) $p_y$(O) $p_z$(C) & -0.626 & 2 & 0.071 & 0.069 \\
5 & $p_x$(O) $p_z$(O) $p_y$(O) $p_x$(C) & -0.624 & 2 & 0.067 & 0.052 \\
6 & $p_y$(O) $p_y$(C) $s$(C) $p_z$(O) & -0.605 & 2 & 0.054 & 0.041 \\
7 & $p_x$(O) $s$(C) $p_x$(C) $p_z$(O) & -0.595 & 2 & 0.038 & 0.051 \\
8 & $p_x$(C) $s$(H) $p_y$(C) $p_z$(C) & -0.585 & 2 & 0.021 & 0.023 \\
\end{longtable}

\clearpage

\begin{longtable}{@{}rp{6.2cm}rcrr@{}}
\caption{Single-orbital entropies for complex \textbf{5}, \ce{[Cr(CO)5(HNC4H8O)]}, septet ($2S+1 = 7$): DMRG reference and SpinAware prediction. $E_\mathrm{UHF} = -1891.871014$~$E_\mathrm{h}$; $\langle \hat S^2\rangle = 12.071$ (exact 12.000). Values of $s_1 \ge 0.10$ are in bold. Atomic-orbital character is given for the $\alpha$ component of each orbital (L\"owdin weights $\ge 0.08$).}\label{tab:si_s1_5_septet}\\
\toprule
No. & Atomic-orbital character ($\alpha$) & $\varepsilon$ ($E_\mathrm{h}$) & Occ. & $s_1$ DMRG & $s_1$ SpinAware \\
\midrule\endfirsthead
\multicolumn{6}{l}{\small\itshape continued from previous page}\\\toprule
No. & Atomic-orbital character ($\alpha$) & $\varepsilon$ ($E_\mathrm{h}$) & Occ. & $s_1$ DMRG & $s_1$ SpinAware \\
\midrule\endhead
\bottomrule\endlastfoot
1 & $p_x$(O) $p_z$(O) $p_y$(O) $p_x$(C) & -0.651 & 2 & 0.042 & 0.036 \\
2 & $p_x$(C) $p_y$(C) $p_y$(N) $p_x$(N) & -0.647 & 2 & 0.019 & 0.035 \\
3 & $p_x$(O) $p_z$(O) $p_x$(C) $s$(C) & -0.626 & 2 & 0.035 & 0.053 \\
4 & $p_y$(C) $s$(C) $p_y$(O) $p_z$(O) & -0.611 & 2 & 0.087 & 0.061 \\
5 & $p_x$(O) $p_y$(O) $p_x$(N) & -0.593 & 2 & 0.066 & 0.075 \\
6 & $p_x$(C) $p_y$(C) $p_z$(C) $s$(H) & -0.588 & 2 & 0.095 & 0.086 \\
7 & $d_{xy}$(Cr) $d_{yz}$(Cr) $d_{z^2}$(Cr) $p_x$(O) & -0.584 & 2 & 0.044 & 0.039 \\
8 & $p_y$(C) $p_y$(O) $p_x$(C) $p_z$(O) & -0.579 & 2 & 0.098 & 0.097 \\
\end{longtable}

\clearpage

\begin{longtable}{@{}rp{6.2cm}rcrr@{}}
\caption{Single-orbital entropies for complex \textbf{6}, \ce{[Cr(CO)4(H2NCH2CO2)]-}, singlet ($2S+1 = 1$): DMRG reference and SpinAware prediction. $E_\mathrm{UHF} = -1775.684255$~$E_\mathrm{h}$; $\langle \hat S^2\rangle = 0.000$ (exact 0.000). Values of $s_1 \ge 0.10$ are in bold. Atomic-orbital character is given for the $\alpha$ component of each orbital (L\"owdin weights $\ge 0.08$).}\label{tab:si_s1_6_singlet}\\
\toprule
No. & Atomic-orbital character ($\alpha$) & $\varepsilon$ ($E_\mathrm{h}$) & Occ. & $s_1$ DMRG & $s_1$ SpinAware \\
\midrule\endfirsthead
\multicolumn{6}{l}{\small\itshape continued from previous page}\\\toprule
No. & Atomic-orbital character ($\alpha$) & $\varepsilon$ ($E_\mathrm{h}$) & Occ. & $s_1$ DMRG & $s_1$ SpinAware \\
\midrule\endhead
\bottomrule\endlastfoot
1 & $p_z$(O) $p_x$(O) $p_y$(C) $p_y$(O) & -0.504 & 2 & 0.062 & 0.064 \\
2 & $p_x$(O) $p_z$(O) $p_y$(C) $p_y$(O) & -0.490 & 2 & 0.062 & 0.054 \\
3 & $p_z$(O) $p_x$(O) $p_y$(C) $p_y$(O) & -0.485 & 2 & \textbf{0.103} & 0.092 \\
4 & $p_z$(O) $p_y$(O) $p_x$(O) $p_z$(C) & -0.477 & 2 & 0.094 & 0.093 \\
5 & $p_x$(C) $s$(H) $p_x$(O) $p_y$(C) & -0.470 & 2 & 0.042 & 0.042 \\
6 & $p_y$(O) $p_z$(O) $p_x$(C) $p_z$(C) & -0.463 & 2 & \textbf{0.104} & 0.080 \\
7 & $p_x$(O) $p_z$(O) $p_x$(C) $p_y$(O) & -0.459 & 2 & 0.098 & 0.094 \\
8 & $p_z$(O) $p_y$(O) $p_x$(O) $p_y$(C) & -0.453 & 2 & \textbf{0.113} & \textbf{0.104} \\
\end{longtable}

\clearpage

\begin{longtable}{@{}rp{6.2cm}rcrr@{}}
\caption{Single-orbital entropies for complex \textbf{7}, \ce{[Ti($\eta^6$-C6H6)(AlCl4)2]}, singlet ($2S+1 = 1$): DMRG reference and SpinAware prediction. $E_\mathrm{UHF} = -5238.034090$~$E_\mathrm{h}$; $\langle \hat S^2\rangle = 0.000$ (exact 0.000). Values of $s_1 \ge 0.10$ are in bold. Atomic-orbital character is given for the $\alpha$ component of each orbital (L\"owdin weights $\ge 0.08$).}\label{tab:si_s1_7_singlet}\\
\toprule
No. & Atomic-orbital character ($\alpha$) & $\varepsilon$ ($E_\mathrm{h}$) & Occ. & $s_1$ DMRG & $s_1$ SpinAware \\
\midrule\endfirsthead
\multicolumn{6}{l}{\small\itshape continued from previous page}\\\toprule
No. & Atomic-orbital character ($\alpha$) & $\varepsilon$ ($E_\mathrm{h}$) & Occ. & $s_1$ DMRG & $s_1$ SpinAware \\
\midrule\endhead
\bottomrule\endlastfoot
1 & $p_x$(Cl) $p_z$(Cl) $p_y$(Cl) & -0.529 & 2 & 0.029 & 0.052 \\
2 & $p_x$(Cl) $p_z$(Cl) $p_y$(Cl) & -0.515 & 2 & 0.021 & 0.046 \\
3 & $p_z$(Cl) $p_x$(Cl) $p_y$(Cl) & -0.511 & 2 & 0.046 & 0.055 \\
4 & $p_x$(Cl) $p_z$(Cl) $p_y$(Cl) & -0.501 & 2 & 0.016 & 0.039 \\
5 & $p_y$(Cl) $p_z$(Cl) $p_x$(Cl) & -0.493 & 2 & 0.015 & 0.030 \\
6 & $p_y$(Cl) $p_z$(Cl) & -0.487 & 2 & 0.027 & 0.017 \\
7 & $p_x$(Cl) $p_z$(Cl) $p_y$(Cl) & -0.483 & 2 & 0.022 & 0.053 \\
8 & $p_y$(Cl) $p_x$(Cl) $p_z$(Cl) $p_z$(C) & -0.481 & 2 & 0.082 & \textbf{0.133} \\
\end{longtable}

\clearpage

\begin{longtable}{@{}rp{6.2cm}rcrr@{}}
\caption{Single-orbital entropies for complex \textbf{7}, \ce{[Ti($\eta^6$-C6H6)(AlCl4)2]}, triplet ($2S+1 = 3$): DMRG reference and SpinAware prediction. $E_\mathrm{UHF} = -5238.101363$~$E_\mathrm{h}$; $\langle \hat S^2\rangle = 2.139$ (exact 2.000). Values of $s_1 \ge 0.10$ are in bold. Atomic-orbital character is given for the $\alpha$ component of each orbital (L\"owdin weights $\ge 0.08$).}\label{tab:si_s1_7_triplet}\\
\toprule
No. & Atomic-orbital character ($\alpha$) & $\varepsilon$ ($E_\mathrm{h}$) & Occ. & $s_1$ DMRG & $s_1$ SpinAware \\
\midrule\endfirsthead
\multicolumn{6}{l}{\small\itshape continued from previous page}\\\toprule
No. & Atomic-orbital character ($\alpha$) & $\varepsilon$ ($E_\mathrm{h}$) & Occ. & $s_1$ DMRG & $s_1$ SpinAware \\
\midrule\endhead
\bottomrule\endlastfoot
1 & $p_x$(Cl) $p_z$(Cl) $p_y$(Cl) & -0.520 & 2 & 0.018 & 0.039 \\
2 & $p_z$(Cl) $p_x$(Cl) $p_y$(Cl) & -0.511 & 2 & 0.039 & 0.034 \\
3 & $p_x$(Cl) $p_z$(Cl) $p_y$(Cl) & -0.500 & 2 & 0.021 & 0.034 \\
4 & $p_y$(Cl) $p_z$(Cl) $p_x$(Cl) & -0.492 & 2 & 0.018 & 0.022 \\
5 & $p_y$(Cl) $p_z$(Cl) & -0.487 & 2 & 0.024 & 0.022 \\
6 & $p_z$(Cl) $p_x$(Cl) $p_y$(Cl) & -0.486 & 2 & 0.054 & \textbf{0.107} \\
7 & $p_x$(Cl) $p_y$(Cl) $p_z$(Cl) & -0.481 & 2 & 0.020 & 0.029 \\
8 & $p_x$(Cl) $p_z$(C) $p_z$(Cl) $p_y$(Cl) & -0.478 & 2 & 0.057 & \textbf{0.108} \\
\end{longtable}

\clearpage

\begin{longtable}{@{}rp{6.2cm}rcrr@{}}
\caption{Single-orbital entropies for complex \textbf{9}, \ce{[CrCl4]^2-}, quintet ($2S+1 = 5$): DMRG reference and SpinAware prediction. $E_\mathrm{UHF} = -2880.846959$~$E_\mathrm{h}$; $\langle \hat S^2\rangle = 6.012$ (exact 6.000). Values of $s_1 \ge 0.10$ are in bold. Atomic-orbital character is given for the $\alpha$ component of each orbital (L\"owdin weights $\ge 0.08$).}\label{tab:si_s1_9_quintet}\\
\toprule
No. & Atomic-orbital character ($\alpha$) & $\varepsilon$ ($E_\mathrm{h}$) & Occ. & $s_1$ DMRG & $s_1$ SpinAware \\
\midrule\endfirsthead
\multicolumn{6}{l}{\small\itshape continued from previous page}\\\toprule
No. & Atomic-orbital character ($\alpha$) & $\varepsilon$ ($E_\mathrm{h}$) & Occ. & $s_1$ DMRG & $s_1$ SpinAware \\
\midrule\endhead
\bottomrule\endlastfoot
1 & $s$(Cl) $s$(Cr) & -0.681 & 2 & 0.023 & 0.024 \\
2 & $s$(Cl) & -0.668 & 2 & 0.025 & 0.024 \\
3 & $s$(Cl) & -0.664 & 2 & 0.023 & 0.024 \\
4 & $s$(Cl) & -0.660 & 2 & 0.026 & 0.024 \\
5 & $d_{xz}$(Cr) $d_{xy}$(Cr) $d_{x^2-y^2}$(Cr) & -0.252 & 2 & 0.074 & 0.072 \\
6 & $d_{xy}$(Cr) $d_{xz}$(Cr) & -0.214 & 2 & 0.067 & 0.068 \\
7 & $d_{z^2}$(Cr) $d_{x^2-y^2}$(Cr) & -0.211 & 2 & 0.067 & 0.068 \\
8 & $d_{yz}$(Cr) $p_x$(Cl) $d_{x^2-y^2}$(Cr) & -0.200 & 2 & 0.074 & 0.078 \\
\end{longtable}

\clearpage

\begin{longtable}{@{}rp{6.2cm}rcrr@{}}
\caption{Single-orbital entropies for complex \textbf{10}, \ce{[VF6]^3-}, singlet ($2S+1 = 1$): DMRG reference and SpinAware prediction. $E_\mathrm{UHF} = -1538.704175$~$E_\mathrm{h}$; $\langle \hat S^2\rangle = -0.000$ (exact 0.000). Values of $s_1 \ge 0.10$ are in bold. Atomic-orbital character is given for the $\alpha$ component of each orbital (L\"owdin weights $\ge 0.08$).}\label{tab:si_s1_10_singlet}\\
\toprule
No. & Atomic-orbital character ($\alpha$) & $\varepsilon$ ($E_\mathrm{h}$) & Occ. & $s_1$ DMRG & $s_1$ SpinAware \\
\midrule\endfirsthead
\multicolumn{6}{l}{\small\itshape continued from previous page}\\\toprule
No. & Atomic-orbital character ($\alpha$) & $\varepsilon$ ($E_\mathrm{h}$) & Occ. & $s_1$ DMRG & $s_1$ SpinAware \\
\midrule\endhead
\bottomrule\endlastfoot
1 & $s$(F) $p_z$(V) & -0.819 & 2 & 0.021 & 0.040 \\
2 & $p_x$(F) & -0.034 & 2 & 0.052 & 0.064 \\
3 & $p_z$(F) $p_y$(F) $d_{yz}$(V) & -0.013 & 2 & \textbf{0.156} & \textbf{0.158} \\
4 & $p_y$(F) $p_x$(F) & -0.006 & 2 & 0.059 & 0.044 \\
5 & $p_z$(F) $p_x$(F) & -0.006 & 2 & 0.047 & 0.049 \\
6 & $p_y$(F) $p_z$(F) & -0.006 & 2 & 0.056 & 0.083 \\
7 & $p_x$(F) & -0.003 & 2 & 0.041 & 0.046 \\
8 & $p_y$(F) & 0.005 & 2 & 0.037 & 0.054 \\
\end{longtable}

\clearpage

\begin{longtable}{@{}rp{6.2cm}rcrr@{}}
\caption{Single-orbital entropies for complex \textbf{10}, \ce{[VF6]^3-}, triplet ($2S+1 = 3$): DMRG reference and SpinAware prediction. $E_\mathrm{UHF} = -1538.802165$~$E_\mathrm{h}$; $\langle \hat S^2\rangle = 2.004$ (exact 2.000). Values of $s_1 \ge 0.10$ are in bold. Atomic-orbital character is given for the $\alpha$ component of each orbital (L\"owdin weights $\ge 0.08$).}\label{tab:si_s1_10_triplet}\\
\toprule
No. & Atomic-orbital character ($\alpha$) & $\varepsilon$ ($E_\mathrm{h}$) & Occ. & $s_1$ DMRG & $s_1$ SpinAware \\
\midrule\endfirsthead
\multicolumn{6}{l}{\small\itshape continued from previous page}\\\toprule
No. & Atomic-orbital character ($\alpha$) & $\varepsilon$ ($E_\mathrm{h}$) & Occ. & $s_1$ DMRG & $s_1$ SpinAware \\
\midrule\endhead
\bottomrule\endlastfoot
1 & $p_y$(F) $d_{yz}$(V) $d_{xy}$(V) $p_z$(F) & -0.048 & 2 & 0.040 & 0.082 \\
2 & $d_{xz}$(V) $p_x$(F) $p_z$(F) $p_y$(F) & -0.048 & 2 & 0.053 & 0.081 \\
3 & $p_x$(F) $p_z$(F) $p_y$(F) & -0.023 & 2 & 0.063 & 0.065 \\
4 & $p_x$(F) $p_z$(F) $d_{z^2}$(V) & -0.003 & 2 & 0.062 & 0.065 \\
5 & $p_y$(F) $p_z$(F) $p_x$(F) $d_{x^2-y^2}$(V) & -0.003 & 2 & 0.060 & 0.083 \\
6 & $p_x$(F) $p_z$(F) $p_y$(F) & -0.000 & 2 & 0.047 & 0.060 \\
7 & $p_x$(F) $p_z$(F) $p_y$(F) & 0.006 & 2 & 0.043 & 0.068 \\
8 & $p_x$(F) $p_z$(F) & 0.006 & 2 & 0.042 & 0.061 \\
\end{longtable}

\clearpage

\begin{longtable}{@{}rp{6.2cm}rcrr@{}}
\caption{Single-orbital entropies for complex \textbf{11}, \ce{[CoBr2Cl(NCCH3)]-}, quartet ($2S+1 = 4$): DMRG reference and SpinAware prediction. $E_\mathrm{UHF} = -7116.882956$~$E_\mathrm{h}$; $\langle \hat S^2\rangle = 3.757$ (exact 3.750). Values of $s_1 \ge 0.10$ are in bold. Atomic-orbital character is given for the $\alpha$ component of each orbital (L\"owdin weights $\ge 0.08$).}\label{tab:si_s1_11_quartet}\\
\toprule
No. & Atomic-orbital character ($\alpha$) & $\varepsilon$ ($E_\mathrm{h}$) & Occ. & $s_1$ DMRG & $s_1$ SpinAware \\
\midrule\endfirsthead
\multicolumn{6}{l}{\small\itshape continued from previous page}\\\toprule
No. & Atomic-orbital character ($\alpha$) & $\varepsilon$ ($E_\mathrm{h}$) & Occ. & $s_1$ DMRG & $s_1$ SpinAware \\
\midrule\endhead
\bottomrule\endlastfoot
1 & $p_z$(C) $p_z$(N) $d_{z^2}$(Co) $s$(C) & -0.655 & 2 & 0.084 & 0.069 \\
2 & $p_z$(C) $d_{z^2}$(Co) $p_z$(N) & -0.609 & 2 & 0.080 & 0.083 \\
3 & $p_x$(C) $s$(H) $p_y$(C) & -0.574 & 2 & 0.092 & 0.063 \\
4 & $p_y$(C) $s$(H) $p_x$(C) & -0.574 & 2 & 0.085 & 0.097 \\
5 & $d_{xy}$(Co) $d_{xz}$(Co) & -0.556 & 2 & 0.094 & \textbf{0.100} \\
6 & $d_{x^2-y^2}$(Co) $d_{yz}$(Co) & -0.555 & 2 & 0.094 & \textbf{0.101} \\
7 & $d_{yz}$(Co) $d_{xz}$(Co) $d_{x^2-y^2}$(Co) & -0.507 & 2 & 0.049 & 0.076 \\
8 & $d_{xz}$(Co) $d_{xy}$(Co) $d_{yz}$(Co) & -0.506 & 2 & 0.050 & 0.074 \\
\end{longtable}

\clearpage

\begin{longtable}{@{}rp{6.2cm}rcrr@{}}
\caption{Single-orbital entropies for complex \textbf{12}, \ce{[MnBr(C13H10N2O4)]}, singlet ($2S+1 = 1$): DMRG reference and SpinAware prediction. $E_\mathrm{UHF} = -4627.372240$~$E_\mathrm{h}$; $\langle \hat S^2\rangle = 0.000$ (exact 0.000). Values of $s_1 \ge 0.10$ are in bold. Atomic-orbital character is given for the $\alpha$ component of each orbital (L\"owdin weights $\ge 0.08$).}\label{tab:si_s1_12_singlet}\\
\toprule
No. & Atomic-orbital character ($\alpha$) & $\varepsilon$ ($E_\mathrm{h}$) & Occ. & $s_1$ DMRG & $s_1$ SpinAware \\
\midrule\endfirsthead
\multicolumn{6}{l}{\small\itshape continued from previous page}\\\toprule
No. & Atomic-orbital character ($\alpha$) & $\varepsilon$ ($E_\mathrm{h}$) & Occ. & $s_1$ DMRG & $s_1$ SpinAware \\
\midrule\endhead
\bottomrule\endlastfoot
1 & $p_y$(C) $s$(H) $p_x$(C) & -0.599 & 2 & 0.014 & 0.013 \\
2 & $p_z$(C) $s$(H) & -0.591 & 2 & 0.024 & 0.020 \\
3 & $p_y$(C) $s$(H) $p_x$(C) & -0.582 & 2 & 0.008 & 0.011 \\
4 & $p_x$(C) $p_z$(O) $s$(C) $s$(H) & -0.578 & 2 & 0.034 & 0.026 \\
5 & $p_x$(C) $s$(H) $p_z$(C) $p_y$(C) & -0.575 & 2 & 0.020 & 0.022 \\
6 & $p_x$(C) $s$(C) $p_x$(O) $p_y$(C) & -0.567 & 2 & 0.035 & 0.031 \\
7 & $p_z$(C) $p_z$(O) $s$(H) $p_z$(N) & -0.559 & 2 & 0.052 & 0.065 \\
8 & $p_x$(C) $s$(H) $p_y$(C) $s$(C) & -0.557 & 2 & 0.025 & 0.025 \\
\end{longtable}

\clearpage

\begin{longtable}{@{}rp{6.2cm}rcrr@{}}
\caption{Single-orbital entropies for complex \textbf{13}, \ce{[Ti($\eta^6${-}C6H6)(AlCl4)2]}, singlet ($2S+1 = 1$): DMRG reference and SpinAware prediction. $E_\mathrm{UHF} = -5238.044961$~$E_\mathrm{h}$; $\langle \hat S^2\rangle = 0.000$ (exact 0.000). Values of $s_1 \ge 0.10$ are in bold. Atomic-orbital character is given for the $\alpha$ component of each orbital (L\"owdin weights $\ge 0.08$).}\label{tab:si_s1_13_singlet}\\
\toprule
No. & Atomic-orbital character ($\alpha$) & $\varepsilon$ ($E_\mathrm{h}$) & Occ. & $s_1$ DMRG & $s_1$ SpinAware \\
\midrule\endfirsthead
\multicolumn{6}{l}{\small\itshape continued from previous page}\\\toprule
No. & Atomic-orbital character ($\alpha$) & $\varepsilon$ ($E_\mathrm{h}$) & Occ. & $s_1$ DMRG & $s_1$ SpinAware \\
\midrule\endhead
\bottomrule\endlastfoot
1 & $p_x$(Cl) $p_z$(Cl) $p_y$(Cl) & -0.523 & 2 & 0.029 & 0.057 \\
2 & $p_x$(Cl) $p_z$(Cl) $p_y$(Cl) & -0.513 & 2 & 0.018 & 0.046 \\
3 & $p_x$(Cl) $p_z$(Cl) $p_y$(Cl) & -0.511 & 2 & 0.047 & 0.053 \\
4 & $p_x$(Cl) $p_z$(Cl) $p_y$(Cl) & -0.500 & 2 & 0.016 & 0.051 \\
5 & $p_y$(Cl) $p_z$(Cl) $p_x$(Cl) & -0.487 & 2 & 0.027 & 0.022 \\
6 & $p_y$(Cl) $p_z$(Cl) $p_x$(Cl) & -0.483 & 2 & 0.030 & 0.017 \\
7 & $p_x$(Cl) $p_y$(Cl) $p_z$(Cl) & -0.479 & 2 & 0.040 & 0.080 \\
8 & $p_x$(Cl) $p_z$(C) $p_z$(Cl) $p_y$(Cl) & -0.476 & 2 & \textbf{0.113} & \textbf{0.152} \\
\end{longtable}

\clearpage

\begin{longtable}{@{}rp{6.2cm}rcrr@{}}
\caption{Single-orbital entropies for complex \textbf{14}, \ce{[NiCl4]^2-}, singlet ($2S+1 = 1$): DMRG reference and SpinAware prediction. $E_\mathrm{UHF} = -3344.166809$~$E_\mathrm{h}$; $\langle \hat S^2\rangle = 0.000$ (exact 0.000). Values of $s_1 \ge 0.10$ are in bold. Atomic-orbital character is given for the $\alpha$ component of each orbital (L\"owdin weights $\ge 0.08$).}\label{tab:si_s1_14_singlet}\\
\toprule
No. & Atomic-orbital character ($\alpha$) & $\varepsilon$ ($E_\mathrm{h}$) & Occ. & $s_1$ DMRG & $s_1$ SpinAware \\
\midrule\endfirsthead
\multicolumn{6}{l}{\small\itshape continued from previous page}\\\toprule
No. & Atomic-orbital character ($\alpha$) & $\varepsilon$ ($E_\mathrm{h}$) & Occ. & $s_1$ DMRG & $s_1$ SpinAware \\
\midrule\endhead
\bottomrule\endlastfoot
1 & $s$(Cl) $s$(Ni) & -0.687 & 2 & 0.025 & 0.027 \\
2 & $s$(Cl) & -0.677 & 2 & 0.026 & 0.029 \\
3 & $s$(Cl) & -0.675 & 2 & 0.027 & 0.029 \\
4 & $s$(Cl) & -0.673 & 2 & 0.027 & 0.028 \\
5 & $d_{z^2}$(Ni) $d_{x^2-y^2}$(Ni) & -0.396 & 2 & 0.029 & 0.063 \\
6 & $d_{yz}$(Ni) $d_{xy}$(Ni) & -0.317 & 2 & \textbf{0.177} & 0.090 \\
7 & $d_{xy}$(Ni) $d_{yz}$(Ni) & -0.312 & 2 & \textbf{0.225} & \textbf{0.109} \\
8 & $d_{x^2-y^2}$(Ni) $d_{z^2}$(Ni) & -0.279 & 2 & \textbf{0.207} & \textbf{0.115} \\
\end{longtable}

\clearpage

\begin{longtable}{@{}rp{6.2cm}rcrr@{}}
\caption{Single-orbital entropies for complex \textbf{14}, \ce{[NiCl4]^2-}, triplet ($2S+1 = 3$): DMRG reference and SpinAware prediction. $E_\mathrm{UHF} = -3344.290676$~$E_\mathrm{h}$; $\langle \hat S^2\rangle = 2.004$ (exact 2.000). Values of $s_1 \ge 0.10$ are in bold. Atomic-orbital character is given for the $\alpha$ component of each orbital (L\"owdin weights $\ge 0.08$).}\label{tab:si_s1_14_triplet}\\
\toprule
No. & Atomic-orbital character ($\alpha$) & $\varepsilon$ ($E_\mathrm{h}$) & Occ. & $s_1$ DMRG & $s_1$ SpinAware \\
\midrule\endfirsthead
\multicolumn{6}{l}{\small\itshape continued from previous page}\\\toprule
No. & Atomic-orbital character ($\alpha$) & $\varepsilon$ ($E_\mathrm{h}$) & Occ. & $s_1$ DMRG & $s_1$ SpinAware \\
\midrule\endhead
\bottomrule\endlastfoot
1 & $s$(Cl) & -0.679 & 2 & 0.024 & 0.026 \\
2 & $s$(Cl) & -0.677 & 2 & 0.024 & 0.026 \\
3 & $s$(Cl) & -0.677 & 2 & 0.024 & 0.026 \\
4 & $d_{xz}$(Ni) $d_{x^2-y^2}$(Ni) & -0.428 & 2 & 0.022 & 0.028 \\
5 & $d_{yz}$(Ni) $d_{xy}$(Ni) & -0.427 & 2 & 0.026 & 0.029 \\
6 & $d_{z^2}$(Ni) & -0.376 & 2 & 0.020 & 0.029 \\
7 & $d_{xy}$(Ni) $d_{yz}$(Ni) & -0.375 & 2 & 0.061 & 0.063 \\
8 & $d_{x^2-y^2}$(Ni) $d_{xz}$(Ni) & -0.374 & 2 & 0.051 & 0.060 \\
\end{longtable}

\clearpage